\documentclass[onecolumn,notitlepage,floats,floatfix,amssymb,aps,prd,showpacs,superscriptaddress,groupedaddress,nofootinbib]{revtex4-2}  % for review and submission
\usepackage[T1]{fontenc}
\usepackage{lmodern} 
\usepackage{graphicx}  % needed for figures
\usepackage{dcolumn}   % needed for some tables
\usepackage{bm}        % for math
\usepackage{amssymb}   % for math
\usepackage{amsmath}
\usepackage{bbold}
\usepackage{array}
\usepackage{makecell}
\usepackage{braket}
\usepackage{longtable}
\usepackage{supertabular,booktabs}

\usepackage{titlesec} %for hyperref to detect names of section etc
\usepackage[colorlinks,citecolor=blue,urlcolor=blue,hypertexnames=true]{hyperref}
\usepackage{subfigure}
\usepackage[usenames,dvipsnames,svgnames,table]{xcolor} % use color 

\renewcommand{\arraystretch}{2}

\newcommand{\be}{\begin{equation}}
\newcommand{\ee}{\end{equation}}
\newcommand{\bea}{\begin{eqnarray}}
\newcommand{\eea}{\end{eqnarray}}
\newcommand{\bml}{\begin{subequations}}
\newcommand{\eml}{\end{subequations}}
\newcommand{\bfig}{\begin{figure}}
\newcommand{\efig}{\end{figure}}

\newcommand{\slashed}{\hspace{-1.1ex}/}

\newcommand{\bmat}{\begin{pmatrix}}
\newcommand{\emat}{\end{pmatrix}}
\usepackage{graphicx,slashed,booktabs,xcolor,multirow,float,
amsfonts,bbold,mathtools,sidecap,tikz,bm,enumitem}
\usepackage{multirow}
\usepackage{bbding}
\usepackage{titlesec}
\usepackage{hyperref}
\usepackage{cancel}
\usepackage{wasysym}
\usepackage{amssymb}% http://ctan.org/pkg/amssymb
\usepackage{pifont}% http://ctan.org/pkg/pifont

\renewcommand{\leq}{\leqslant}
\renewcommand{\geq}{\geqslant}

\usepackage[dvipsnames, usenames]{xcolor}

\definecolor{linkcolor}{rgb}{0.55, 0.13, .32}

\definecolor{oucrimsonred}{rgb}{0.6, 0.0, 0.0}
\definecolor{persianblue}{rgb}{0.11, 0.22, 0.73}
\definecolor{forestgreen}{rgb}{0.13,0.35,0.13}
\definecolor{lightgray}{rgb}{0.83, 0.83, 0.83}
 \hypersetup{colorlinks, citecolor=oucrimsonred, linkcolor=persianblue, urlcolor=oucrimsonred}
\definecolor{cornellred}{rgb}{0.7, 0.11, 0.11}
\definecolor{navyblue}{rgb}{0.0, 0.0, 0.5}
\definecolor{amethyst}{rgb}{0.6, 0.4, 0.8}
\definecolor{yellow}{rgb}{1.0, 1.0, 0.0}
\definecolor{firebrick}{rgb}{0.7, 0.13, 0.13}
\definecolor{tangerineyellow}{rgb}{1.0, 0.8, 0.0}
\definecolor{deepfuchsia}{rgb}{0.76, 0.33, 0.76}
\definecolor{amber}{rgb}{1.0, 0.75, 0.0}
\definecolor{VioletRed4}{rgb}{0.55, 0.13, .32}
\definecolor{indiagreen}{rgb}{0.07, 0.53, 0.03}
\definecolor{VioletRed4}{rgb}{0.55, 0.13, .32}

\usepackage{hyperref}

\usepackage{graphics,appendix,afterpage,makecell}

\usepackage{bbold}
\usepackage{tikz}
\usepackage{adjustbox}
\usepackage{tikz-feynman}
\usepackage{tcolorbox}

\definecolor{oucrimsonred}{rgb}{0.6, 0.0, 0.0}
\definecolor{persianblue}{rgb}{0.11, 0.22, 0.73}
\definecolor{forestgreen}{rgb}{0.13,0.35,0.13}
\definecolor{lightgray}{rgb}{0.83, 0.83, 0.83}
 \hypersetup{colorlinks, citecolor=oucrimsonred, linkcolor=persianblue, urlcolor=oucrimsonred}
\definecolor{cornellred}{rgb}{0.7, 0.11, 0.11}
\definecolor{navyblue}{rgb}{0.0, 0.0, 0.5}
\definecolor{amethyst}{rgb}{0.6, 0.4, 0.8}
\definecolor{yellow}{rgb}{1.0, 1.0, 0.0}
\definecolor{firebrick}{rgb}{0.7, 0.13, 0.13}
\definecolor{tangerineyellow}{rgb}{1.0, 0.8, 0.0}
\definecolor{deepfuchsia}{rgb}{0.76, 0.33, 0.76}
\definecolor{amber}{rgb}{1.0, 0.75, 0.0}
\definecolor{VioletRed4}{rgb}{0.55, 0.13, .32}
\definecolor{indiagreen}{rgb}{0.07, 0.53, 0.03}
\definecolor{VioletRed4}{rgb}{0.55, 0.13, .32}

\definecolor{oucrimsonred}{rgb}{0.6, 0.0, 0.0}
\newcommand\vertarrowbox[3][6ex]{%
  \begin{array}[t]{@{}c@{}} #2 \\
  \left\uparrow\vcenter{\hrule height #1}\right.\kern-\nulldelimiterspace\\
  \makebox[0pt]{\scriptsize#3}
  \end{array}%
}

\definecolor{mtcolor}{rgb}{.8,.3,.1}

\definecolor{violachiaro}{rgb}{1,0.6,1}

\definecolor{gbcolor}{rgb}{.43,.22,.12}
 
\definecolor{gbcolor2}{rgb}{.9,.2,.6}
\definecolor{gbcolor3}{rgb}{.3,.2,.6}

\definecolor{verdechiaro}{rgb}{0.6,1,0.6}
\definecolor{giallochiaro}{rgb}{1,1,0.6}
\definecolor{bluscuro}{rgb}{0.15, 0.2, 0.9}
\definecolor{verdes}{rgb}{0.1, 0.5, 0.1}%
\definecolor{tangerineyellow}{rgb}{1.0, 0.8, 0.0}
\definecolor{smokyblack}{rgb}{0.06, 0.05, 0.03}

\definecolor{americanrose}{rgb}{1.0, 0.01, 0.24}
\definecolor{cobalt}{rgb}{0.0, 0.28, 0.67}
\definecolor{brandeisblue}{rgb}{0.0, 0.44, 1.0}
\definecolor{mycolor}{rgb}{0.0, 0.0, 0.5}%navyblue
\definecolor{oxfordblue}{rgb}{0.0, 0.13, 0.28}
\definecolor{azure}{rgb}{0.0, 0.5, 1.0}
\definecolor{turquoiseblue}{rgb}{0.0, 1.0, 0.94}
\newtcolorbox{mynewbox}[1]{colback=white!5!white,colframe=azure!75!black,fonttitle=\bfseries,title=#1}
\newtcolorbox{mybox}{colback=mycolor!5!white,colframe=azure!75!black}
\newtcolorbox{mynamedbox}[1]{colback=mycolor!5!white,colframe=azure!75!black,title=#1}
\definecolor{venetianred}{rgb}{0.78, 0.03, 0.08}
\newtcolorbox{mynamedbox1}[1]{colback=venetianred!5!white,colframe=venetianred!80!black,title=#1}
\newtcolorbox{mynamedbox2}[1]{colback=azure!5!white,colframe=azure!80!black,title=#1}

\definecolor{rossocorsa}{rgb}{0.83, 0.0, 0.0}

\tikzset{->-/.style={decoration={
  markings,
  mark=at position #1 with {\arrow{>}}},postaction={decorate}}}
\tikzset{-<-/.style={decoration={
  markings,
  mark=at position #1 with {\arrow{<}}},postaction={decorate}}} 
\def\be{\begin{equation}}
\def\ee{\end{equation}}
\def\ba{\begin{eqnarray}}
\def\ea{\end{eqnarray}}

\def\L*{{\cal L}_*}
\def\L{\mathcal{L}}
\def\({\left(}
\def\){\right)}

\def\<{\langle}
\def\>{\rangle}

 \def\neq {\not\equiv}

\def\cs2{c_{s}^{2}}

 \def\be   {\begin{equation}}   \def\ee   {\end{equation}}
 \def\ba   {\begin{array}}      \def\ea   {\end{array}}
 \def\bea  {\begin{eqnarray}}   \def\eea  {\end{eqnarray}}
 \def\bean {\begin{eqnarray*}}  \def\eean {\end{eqnarray*}}
\titleclass{\subsubsubsection}{straight}[\subsection]

\newcounter{subsubsubsection}[subsubsection]
\renewcommand\thesubsubsubsection{\thesubsubsection.\arabic{subsubsubsection}}
\titleformat{\subsubsubsection}
  {\normalfont\normalsize\bfseries}{\thesubsubsubsection}{1em}{}
\titlespacing*{\subsubsubsection}
{0pt}{3.25ex plus 1ex minus .2ex}{1.5ex plus .2ex}

\makeatletter
\renewcommand\paragraph{\@startsection{paragraph}{5}{\z@}%
  {3.25ex \@plus1ex \@minus.2ex}%
  {-1em}%
  {\normalfont\normalsize\bfseries}}
\renewcommand\subparagraph{\@startsection{subparagraph}{6}{\parindent}%
  {3.25ex \@plus1ex \@minus .2ex}%
  {-1em}%
  {\normalfont\normalsize\bfseries}}
\def\toclevel@subsubsubsection{4}
\def\toclevel@paragraph{5}
\def\toclevel@paragraph{6}
\def\l@subsubsubsection{\@dottedtocline{4}{7em}{4em}}
\def\l@paragraph{\@dottedtocline{5}{10em}{5em}}
\def\l@subparagraph{\@dottedtocline{6}{14em}{6em}}
\makeatother

\begin{document}

% The following information is for internal review, please remove them for submission
%\widetext

\definecolor{lime}{HTML}{A6CE39}
\DeclareRobustCommand{\orcidicon}{\hspace{-2.1mm}
\begin{tikzpicture}
\draw[lime,fill=lime] (0,0.0) circle [radius=0.13] node[white] {{\fontfamily{qag}\selectfont \tiny \,ID}}; \draw[white, fill=white] (-0.0525,0.095) circle [radius=0.007]; 
\end{tikzpicture} \hspace{-3.7mm} }
\foreach \x in {A, ..., Z}{\expandafter\xdef\csname orcid\x\endcsname{\noexpand\href{https://orcid.org/\csname orcidauthor\x\endcsname} {\noexpand\orcidicon}}}
\newcommand{\orcidauthorA}{0000-0002-0459-3873}
\newcommand{\orcidauthorD}{0009-0003-9227-8615}
\newcommand{\orcidauthorB}{0000-0001-9434-0505}

% the following line is for submission, including submission to the arXiv!!
%\hspace{5.2in} \mbox{Fermilab-Pub-04/xxx-E}

\title{\textcolor{Sepia}{\textbf \Large\LARGE
%Loop corrections on the power spectrum in Effective Field Theory of non-singular bounce with Primordial Black Hole formation
Regularized-Renormalized-Resummed \\loop corrected power spectrum of non-singular bounce with Primordial Black Hole formation
%and NANOGrav 15-year data
}}
%Tree-level Bispectrum under squeezed limit in Galileon inflation and Large non-Gaussianity.

%\input author_list.tex       % D0 authors (remove the first 3 lines
                             % of this file prior to submission, they
                             % contain a time stamp for the authorlist)
                             % (includes institutions and visitors)
%\date{\today}

%\date{\today}
\author{{\large  Sayantan Choudhury\orcidA{}${}^{1}$}}
\email{sayantan\_ccsp@sgtuniversity.org, sayantan.choudhury@nanograv.org, sayanphysicsisi@gmail.com (Corresponding author)}
\author{{\large  Ahaskar Karde\orcidD{}${}^{1}$}}
\email{kardeahaskar@gmail.com}
\author{\large Sudhakar~Panda\orcidB{}${}^{1,2}$}
\email{panda@niser.ac.in }
\author{\large Soumitra SenGupta${}^{3}$}
\email{tpssg@iacs.res.in, soumitraiacs@gmail.com}

\affiliation{ ${}^{1}$Centre For Cosmology and Science Popularization (CCSP),\\
        SGT University, Gurugram, Delhi- NCR, Haryana- 122505, India.}
\affiliation{${}^{2}$School of Physical Sciences,  National Institute of Science Education and Research, Bhubaneswar, Odisha - 752050, India,}
\affiliation{${}^{3}$School of Physical Sciences, Indian Association for the Cultivation of Science,
2A \& 2B Raja S.C Mullick Road, Kolkata-700032, India.}

\begin{abstract}
%%%%%%%%%%%%%%%%%%%%%%%%%%%%%%%%%%%%%%%%%%% 

We present a complete and consistent exposition of the regularization, renormalization, and resummation procedures in the setup of having a contraction and then non-singular bounce followed by inflation with a sharp transition from slow-roll (SR) to ultra-slow roll (USR) phase for generating primordial black holes (PBHs). We consider following an effective field theory (EFT) approach and study the quantum loop corrections to the power spectrum from each phase. We demonstrate the complete removal of quadratic UV divergences after renormalization and softened logarithmic IR divergences after resummation and illustrate the scheme-independent nature of our renormalization approach. We further show that the addition of a contracting and bouncing phase allows us to successfully generate PBHs of solar-mass order, $M_{\rm PBH}\sim {\cal O}(M_{\odot})$, by achieving the minimum e-folds during inflation to be $\Delta N_{\rm Total}\sim {\cal O}(60)$ and in this process successfully evading the strict no-go theorem. We notice that varying the effective sound speed between $0.88\leq c_{s}\leq 1$, allows the peak spectrum amplitude to lie within $10^{-3}\leq A \leq 10^{-2}$, indicating that causality and unitarity remain protected in the theory. We analyse PBHs in the extremely small, $M_{\rm PBH}\sim {\cal O}(10^{-33}-10^{-27})M_{\odot}$, and the large, $M_{\rm PBH}\sim {\cal O}(10^{-6}-10^{-1})M_{\odot}$, mass limits and confront the PBH abundance results with the latest microlensing constraints. We also study the cosmological beta functions across all phases and find their interpretation consistent in the context of bouncing and inflationary scenarios while satisfying the pivot scale normalization requirement. Further, we estimate the spectral distortion effects and shed light on controlling PBH overproduction.

\end{abstract}

\maketitle
\tableofcontents
\newpage

\section{Introduction}
\label{s1}

In the recent times, the interest in studying primordial black holes (PBHs) \cite{Zeldovich:1967lct,Hawking:1974rv,Carr:1974nx,Carr:1975qj,Chapline:1975ojl,Carr:1993aq,Choudhury:2011jt,Yokoyama:1998pt,Kawasaki:1998vx,Rubin:2001yw,Khlopov:2002yi,Khlopov:2004sc,Saito:2008em,Khlopov:2008qy,Carr:2009jm,Choudhury:2011jt,Lyth:2011kj,Drees:2011yz,Drees:2011hb,Ezquiaga:2017fvi,Bezrukov:2017dyv,Kannike:2017bxn,Hertzberg:2017dkh,Pi:2017gih,Gao:2018pvq,Dalianis:2018frf,Cicoli:2018asa,Ozsoy:2018flq,Byrnes:2018txb,Ballesteros:2018wlw,Belotsky:2018wph,Martin:2019nuw,Ezquiaga:2019ftu,Motohashi:2019rhu,Fu:2019ttf,Ashoorioon:2019xqc,Auclair:2020csm,Vennin:2020kng,Nanopoulos:2020nnh,Inomata:2021uqj,Stamou:2021qdk,Ng:2021hll,Wang:2021kbh,Kawai:2021edk,Solbi:2021rse,Ballesteros:2021fsp,Rigopoulos:2021nhv,Animali:2022otk,Frolovsky:2022ewg,Escriva:2022duf,Ozsoy:2023ryl,Ivanov:1994pa,Afshordi:2003zb,Frampton:2010sw,Carr:2016drx,Kawasaki:2016pql,Inomata:2017okj,Espinosa:2017sgp,Ballesteros:2017fsr,Sasaki:2018dmp,Ballesteros:2019hus,Dalianis:2019asr,Cheong:2019vzl,Green:2020jor,Carr:2020xqk,Ballesteros:2020qam,Carr:2020gox,Ozsoy:2020kat,Baumann:2007zm,Saito:2008jc,Saito:2009jt,Choudhury:2013woa,Sasaki:2016jop,Raidal:2017mfl,Papanikolaou:2020qtd,Ali-Haimoud:2017rtz,Di:2017ndc,Raidal:2018bbj,Cheng:2018yyr,Vaskonen:2019jpv,Drees:2019xpp,Hall:2020daa,Ballesteros:2020qam,Carr:2020gox,Ozsoy:2020kat,Ashoorioon:2020hln,Papanikolaou:2020qtd,Wu:2021zta,Kimura:2021sqz,Solbi:2021wbo,Teimoori:2021pte,Cicoli:2022sih,Ashoorioon:2022raz,Papanikolaou:2022chm,Papanikolaou:2023crz,Wang:2022nml,ZhengRuiFeng:2021zoz,Cohen:2022clv,Cicoli:2022sih,Brown:2017osf,Palma:2020ejf,Geller:2022nkr,Braglia:2022phb,Frolovsky:2023xid,Aldabergenov:2023yrk,Aoki:2022bvj,Frolovsky:2022qpg,Aldabergenov:2022rfc,Ishikawa:2021xya,Gundhi:2020kzm,Aldabergenov:2020bpt,Cai:2018dig,Cheng:2021lif,Balaji:2022rsy,Qin:2023lgo,Riotto:2023hoz,Riotto:2023gpm,Papanikolaou:2022did,Choudhury:2011jt,Choudhury:2023vuj, Choudhury:2023jlt, Choudhury:2023rks,Choudhury:2023hvf,Choudhury:2023kdb,Choudhury:2023hfm,Bhattacharya:2023ysp,Choudhury:2023fwk,Choudhury:2023fjs,Choudhury:2024one,Choudhury:2024ybk,Choudhury:2024jlz,Harada:2013epa,Harada:2017fjm,Kokubu:2018fxy,Gu:2023mmd,Saburov:2023buy,Stamou:2023vxu,Libanore:2023ovr,Friedlander:2023qmc,Chen:2023lou,Cai:2023uhc,Karam:2023haj,Iacconi:2023slv,Gehrman:2023esa,Padilla:2023lbv,Xie:2023cwi,Meng:2022low,Qiu:2022klm,Mu:2022dku,Fu:2022ypp,Davies:2023hhn,Firouzjahi:2023ahg,Firouzjahi:2023aum, Iacconi:2023ggt,Davies:2023hhn,Jackson:2023obv,Riotto:2024ayo,Maity:2023qzw,Ragavendra:2024yfp,Papanikolaou:2024rlq,Papanikolaou:2024kjb,Banerjee:2021lqu,Choudhury:2023kam,Heydari:2021gea,Heydari:2021qsr,Heydari:2023xts,Heydari:2023rmq,Caravano:2024tlp,Rezazadeh:2021clf,Solbi:2024zhl,Ashrafzadeh:2023ndt,Heydari:2024bxj,Solbi:2021rse,Choudhury:2024dzw,Choudhury:2024aji} has seen a tremendous growth in the ideas to produce them in different contexts of early Universe physics. In this huge expanse of approaches, the prospect of generating PBHs from single field inflation has witnessed the emergence of a crucial and ongoing debate \cite{Kristiano:2022maq,Kristiano:2023scm,Riotto:2023hoz,Riotto:2023gpm,Firouzjahi:2023aum,Firouzjahi:2023ahg,Firouzjahi:2023bkt,Choudhury:2023vuj,Choudhury:2023jlt,Choudhury:2023rks,Motohashi:2023syh,Franciolini:2023agm,Cheng:2023ikq,Tasinato:2023ukp,Tasinato:2023ioq, Iacconi:2023ggt,Davies:2023hhn}. This debate is fundamentally focused on the possibility to produce PBHs, especially in the solar-mass regime, in the presence of large quantum corrections coming from the short scales during inflation. The general idea requires the presence of an instance during inflation where the primordial fluctuations suffer a rapid enhancement in their strength from where generating PBHs can become possible. One of the easiest ways to achieve the above conditions is by having an ultra-slow roll (USR) regime in addition to the usual slow-roll (SR) features during inflation, that is preferably short-lived so as to control the enhancements. However, the search for a proper theoretical analysis to study the impact of quantum loop corrections to the scalar power spectrum and handling the SR to USR transition has fueled the debate till now.  

The idea of producing black holes from the gravitational collapse of large primordial fluctuations in the early Universe, after the corresponding mode re-enters the Hubble horizon during radiation domination, is already quite old. With the recent debate questioning their formation prospects, it becomes all the more important to take a careful look into understanding the underlying physics and develop concrete ways to tackle this issue. The authors of \cite{Choudhury:2023vuj,Choudhury:2023jlt,Choudhury:2023rks} took one such step towards this method, where they introduced a new line of procedure to correct the power spectrum calculations from any troubling divergences brought up by quantum loop corrections and, in the end, produce an expression that represents the physically relevant one-loop corrected two-point correlation function. The result also led to a no-go theorem that strictly restricts formation of any PBHs above $M_{\rm PBH}\sim {\cal O}(10^{2}){\rm gm}$ in single-field models of inflation. The no-go theorem also implies that if one wishes to generate solar-mass, ${\cal O}(M_{\odot})$, PBHs then one cannot achieve successful inflation as the total e-foldings of expansion halts with $\Delta N_{\rm Total}\sim {\cal O}(25)$. Since the above procedure of regularization-renormalization-resummation that leads to the no-go theorem is developed in an entirely model-independent manner, and the renormalization procedure employed for the quantum loop corrections is also scheme-independent, we find it essential to construct alternatives that can evade this strong no-go bound on PBH mass. Some of the interesting alternatives can already be found in refs.\cite{Choudhury:2023hvf,Choudhury:2023kdb,Choudhury:2023hfm,Choudhury:2023fwk,Choudhury:2024one,Bhattacharya:2023ysp,Choudhury:2023fjs}.  

In this work, we consider the theory of contraction and non-singular bounce \cite{Khoury:2001wf,Khoury:2001zk,Khoury:2001bz,Buchbinder:2007ad,Lehners:2007ac,Lehners:2008vx,Raveendran:2018yyh,Brandenberger:2012zb,Raveendran:2017vfx,Chowdhury:2015cma,Cai:2011tc,Brandenberger:2016vhg,Boyle:2004gv,Wands:1998yp,Peter:2002cn,Allen:2004vz,Martin:2003sf,Papanikolaou:2024fzf,Raveendran:2023auh,Raveendran:2019idj,Brustein:1998kq,Starobinsky:1980te,Mukhanov:1991zn,Brandenberger:1993ef,Novello:2008ra,Lilley:2015ksa,Battefeld:2014uga,Peter:2008qz,Biswas:2005qr,Bamba:2013fha,Nojiri:2014zqa,Bajardi:2020fxh,Bhargava:2020fhl,Cai:2009in,Cai:2012ag,Shtanov:2002mb,Ilyas:2020qja,Ilyas:2020zcb,Zhu:2021whu,Banerjee:2016hom,Saridakis:2018fth,Barca:2021qdn,Wilson-Ewing:2012lmx,K:2023gsi,Agullo:2020cvg,Agullo:2020fbw,Agullo:2020wur,Chowdhury:2018blx,Chowdhury:2016aet,Nandi:2019xag,Raveendran:2018why,Raveendran:2018yyh,Stargen:2016cft,Sriramkumar:2015yza,Banerjee:2022gpy,Paul:2022mup,Odintsov:2021yva,Banerjee:2020uil,Das:2017jrl,Pan:2024ydt,Colas:2024xjy,Piao:2003zm,Cai:2017dyi,Cai:2017pga,Cai:2015nya,Cai:2019hge,Battista:2020lqv,Zhu:2023lbf,Banerjee:2022xft}, followed by usual inflation in the presence of a USR phase and ending with another SR phase, as a new framework to study the formation of large mass PBHs. For this, we aim to provide a detailed exposition of the three necessary stages of regularization, renormalization, and resummation of the quantum loop contributions in the current context of having contraction and bounce before inflation. Our present analysis, done also in a model-independent manner, extends the previously developed theoretical techniques to a much broader sense of study, going beyond inflation. During the renormalization of the scalar power spectrum, we utilise the methods from two different schemes, namely the Late time renormalization and the Adiabatic/Wave-function renormalization scheme to arrive at similar results along with showing that any harmful quadratic or other power law divergences from the sub-horizon physics are now fully removed from the calculations of the theory. We follow this by working out the smoothening of the remaining logarithmic IR divergences. After the power law divergent contributions completely die out with only the smoothened logarithmic IR divergences still surviving, we perform a rigorous Dynamical Renormalization Group (DRG) resummation analysis to systematically package the quantum effects coming at all loop orders to yield the final regularized-renormalized-resummed version of one-loop corrected power spectrum. The above analysis also includes a detailed justification of the independence from renormalization schemes when studying quantum loop contributions from each phase of our setup. The other requirement to satisfy, keeping in mind the no-go theorem, is the constraint on e-folds that it implies for generating large mass PBHs, and thus, with the current theory, we further aim to illustrate the possibility of achieving the minimum requirement of $\Delta N_{\rm Total}\sim {\cal O}(60)$.

For the later part of this work, we study the characteristics of the power spectrum obtained from the above analysis, including both the tree-level and the final resummed version, and move towards producing PBHs. In our analysis for both the power spectra, we draw some interesting inferences when the effective sound speed parameter, $c_{s}$, is treated as another variable. The impact of having different values of $c_{s}$ can be felt directly on the peak amplitude of the power spectrum, which ultimately affects how PBH abundance gets estimated as we consider the maximum for the amplitude to produce PBHs be of the order, $A\sim {\cal O}(10^{-2})$. If the conditions of $c_{s} \leq 1 $ remain satisfied, it signals that the causality and unitarity properties within the underlying EFT framework are maintained, and we demonstrate this to be the case here.
Next, we examine the PBH formation following the Press-Schechter (or threshold statistics) approach and obtain their present-day fraction of energy density contained as dark matter, also known as abundance, $f_{\rm PBH}$. To support our results in light of the latest numerical analysis of the observational data, we confront our PBH abundance estimates with the constraints coming from microlensing experiments and highlight the predicted mass windows with significant abundance, $10^{-3}\leq f_{\rm PBH}< 1$. During our estimations of PBH abundance we also consider the impact of changing $c_{s}$ and observe how much the corresponding change in power spectrum amplitude affect the allowed mass windows. Some other critical observational effects in cosmology that are now also related to the properties of the scalar power spectrum consist of the spectral distortion effects. These tiny deviations in the CMB energy spectrum have been found crucial to understanding our Universe's thermal history and can also provide significant constraints to the power spectrum at the larger scales. In this work, we consider these strong limits and estimate the magnitudes of the two types of spectral distortions, namely the $\mu-$ and $y$ type, using the power spectrum developed here. Lastly, we touch upon another issue of serious measure regarding the overproduction of PBHs. This issue has come to light recently \cite{Inomata:2023zup,Balaji:2023ehk,Franciolini:2023pbf,Gorji:2023sil,DeLuca:2023tun} from the analysis of the latest stochastic gravitational wave background (SGWB) signal captured by the pulsar timing array (PTA) collaborations, which includes the NANOGrav \cite{NANOGrav:2023gor, NANOGrav:2023hde, NANOGrav:2023ctt, NANOGrav:2023hvm, NANOGrav:2023hfp, NANOGrav:2023tcn, NANOGrav:2023pdq, NANOGrav:2023icp}, EPTA \cite{EPTA:2023fyk, EPTA:2023sfo, EPTA:2023akd, EPTA:2023gyr, EPTA:2023xxk, EPTA:2023xiy}, PPTA \cite{Reardon:2023gzh, Reardon:2023zen, Zic:2023gta}, and CPTA \cite{Xu:2023wog}. Due to the abundance of PBHs, that are associated with the enhancements at the wavenumbers also sensitive to the SGWB signal, crossing $f_{\rm PBH}\sim 1$ with the power spectrum amplitudes measured by the signal, this becomes a question of immediate concern at the present moment.

%changes in the power spectrum before and after accounting for the one-loop corrections.  We also consider to estimate the spectral distortion effects, namely the $\mu-$ and $y$ distortion effects, and  

The outline of this work is as follows: In section \ref{s2}, we concentrate on a few primary motivation points that we consider throughout this work. In section \ref{s3}, we start with presenting the effective field theory (EFT) of bounce and using this build the second-order perturbed action for the Goldstone modes. In section \ref{s4}, we examine the construction of the five consecutive phases, which includes the contraction and bouncing phases followed by the sequence of SRI-USR-SRII phases and describe how the slow-roll and Hubble parameter vary with the e-folds for each phase. In section \ref{s5}, we present the numerical outcomes for the variation of these parameters with e-folds. In section \ref{s6}, we conduct a semi-analytic study of the curvature perturbation modes during each phase of our setup and follow this with a numerical study of their evolution in section \ref{s7}. In section \ref{s8}, we construct the tree-level scalar power spectrum after performing the mode quantization. In section \ref{s9}, we study the regularization procedure of the scalar power spectrum after considering quantum one-loop corrections in each phase of our setup. In section \ref{s10}, we outline in detail the exact renormalization procedure followed in this work to show complete removal of the quadratic and other power law divergences coming from the sub-horizon modes and demonstrate the softening of the logarithmic IR divergences. In section \ref{s11}, we follow the previous steps with the resummation procedure that allows us to get a finite result after convergence of the logarithmic IR divergent contributions at all loop orders in the horizon-crossing and super-horizon scales. In section \ref{s12}, we make detailed comments on the scheme dependence of various renormalization schemes. In section \ref{s13}, we begin with studying the behaviour of the tree-level and the regularized-renormalized-resummed scalar power spectrum and what they imply in the context of PBH formation. In section \ref{s14}, we study the behaviour of the cosmological beta functions in their renormalized and resummed versions separately. In section \ref{s15}, we study the Press-Schechter mechanism for PBH formation and introduce the relevant statistical variables for calculating PBH abundance. In section \ref{s16}, we present the numerical outcomes on the PBH abundance study done for both the high mass (near solar-mass) and extremely small mass PBHs and use the tree-level as well as the resummed version of the scalar power spectrum. In section \ref{s17}, we discuss the spectral distortion effects and the PBH overproduction problem. In section \ref{s18}, we present the numerical estimates of the two different spectral distortions, namely the $\mu-$ and $y-$ type distortions and also analyse the resolution of the PBH overproduction issue. In section \ref{s19}, we highlight the major new points of this work in comparison with the works done previously into generating large mass PBHs. In section \ref{s20}, we shed light on the crucial issue of the smooth and sharp nature of the transition into the USR. We conclude this work summarizing our findings in section \ref{s21}.

\section{Motivation and Approach}
\label{s2}

%In this section we concentrate on the primary motivation points which cover the main ideas and goals of this work.

\subsection{Formation of large mass PBHs by evading no-go theorem on PBH mass}

The prospects of forming large mass PBHs in single-field inflation models have been debated heavily for a while now. The challenges posed due to large quantum corrections when studying PBH formation and finding methods to mitigate them have invited many attempts to solve this issue. One such attempt to remedy this was introduced in \cite{Choudhury:2023vuj,Choudhury:2023jlt,Choudhury:2023rks} by exposing tools such as renormalization and resummation to a degree where the results led to a no-go theorem on the PBH mass. The no-go theorem was derived using an EFT approach for single-field models of inflation, and it provides strict limits to obtain PBH with mass not more than within a few ${\cal O}(10^{2}){\rm gm}$ if one desires successful inflation with the total number of e-foldings of expansion reaching at least, $\Delta {N}_{\rm Total}\sim {\cal O}(60)$. Our primary motivation to pursue this setup of having a non-singular bouncing feature before getting into the slow-roll (SR) and ultra-slow-roll (USR) regimes stems from finding an alternative approach to evade the no-go theorem on PBH mass. Since the predicted mass window is small and not attractive from the cosmology perspective, it also becomes an important task to find suitable methods by which one can evade such a strict bound. In refs.\cite{Choudhury:2023hvf,Choudhury:2023kdb,Choudhury:2023hfm,Choudhury:2023fwk,Choudhury:2024one,Bhattacharya:2023ysp,Choudhury:2023fjs}, a few such approaches towards evading this no-go bound are studied in detail, which include multiple sharp transitions and Galileon inflation. The current work aims to study yet another exciting approach of adding a non-singular bouncing scenario that will also allow us to evade the PBH mass bound, for which we utilize the theoretical tools of renormalization and resummation developed in the previous studies establishing the no-go theorem and extend them to study the underlying framework presently necessary. The current construction does not introduce any different underlying features to the standard single-field inflationary setup of SR-USR-SR for PBH formation, unlike in the attempts stated previously to evade the no-go bound.

\subsection{Preserving Perturbativity}

Since the PBH formation issue began in single-field models of inflation, all the analysis hinged on utilizing the proper methods to control the impact of one-loop quantum corrections successfully. If not done correctly, the USR phase contributions can quickly break perturbativity features within the calculations, leading to questionable conclusions. Following the renormalization and resummation procedures, we understand that such enhanced quantum contributions to the scalar power spectrum are controllable and allow PBH generation, albeit in the extremely small sub-solar regime. An important constraint on keeping perturbativity intact found via proper application of the above procedures is for the duration of the USR to be within ${\cal O}(2)$ e-folds. With this condition, the power spectrum amplitude in the USR does not exceed the magnitude of ${\cal O}(10^{-2})$ which is another necessary condition for forming PBHs and respect perturbativity. However, to manage near solar mass of PBHs, ${\cal O}(M_{\odot})$, a transition into the USR is required at the smaller wavenumbers, $k\sim {\cal O}(10^6){\rm Mpc^{-1}}$ which facilitates this condition but the remaining e-folds to complete inflation cannot be achieved via another prolonged SR regime. This constraint comes from following the renormalization and resummation procedures, which limits the total e-folds to ${\cal O}(25)$ in a scenario of large mass PBHs. With our present setup, we manage to show how keeping the mentioned perturbative arguments intact, one can generate PBHs both in the large and small mass regimes and for this, we need to elaborate further on the concept of e-foldings in presence of the bounce.

\subsection{Can bounce provide sufficient e-foldings?}

The total e-folding requirement during inflation is essential when focusing on large mass PBHs. As remarked before, we need to incorporate ways to extend the ${\cal O}(25)$ bound on e-folds during inflation, and one such idea comes from the current theory of non-singular bounce before commencing the slow-roll conditions. We focus on the fact that the minimum requirement on the e-folds of expansion during inflation is within roughly, $\Delta N_{\rm Total}\sim {\cal O}(60)$. By including the contraction and bouncing scenarios, we can add the magnitude of their total duration to the succeeding SR-USR-SR phases that, in effect, compensate for the total minimum duration stated above. After we fix the counting of e-folds in SR and USR along the positive or increasing magnitude, then for the contraction and bounce phases, the e-folds increase in magnitude as we go backwards, starting from the end of the bounce and going to the beginning of the contraction phase. We choose the total duration in e-folds elapsed before bouncing ends in the order of ${\cal O}(35)$ that ultimately allows us to satisfy the minimum requirement after totaling the magnitude of e-folds from each phase. In this sense, this setup allows us to place the transition scales at the lower wavenumbers, $k\sim {\cal O}(10^{6}){\rm Mpc^{-1}}$ for producing large near solar mass PBHs and achieve the minimum requirement on e-folds during inflation.

\subsection{Correct interpretation of regularization-renormalization-resummation with bounce}

During our study of quantum fluctuations from examining higher-order interaction terms in the action, large quantum corrections from a USR-like phase invite the need to perform a careful renormalization procedure. This procedure completely eradicates harmful quadratic UV divergences after appropriately adding suitable counter-terms into the action.
Such divergences can also impact the contributions in effect in the late-time limit after horizon crossing. Thus, properly removing these quadratic UV divergences and softening the remaining logarithmic IR divergences is required, which must undergo a further resummation process to give a physically relevant answer for the scalar power spectrum that includes the quantum effects. The previous studies in \cite{Choudhury:2023vuj,Choudhury:2023jlt,Choudhury:2023rks} have intensely focused on developing this whole procedure from the ground up and also talked about how the renormalization procedure there is completely independent of the chosen renormalization scheme. However, with new features related to the bounce and our focus being generating large mass PBHs, we must expand our understanding of handling the quantum divergences with the regularization-renormalization-resummation procedures when phases like contraction and bounce are also present in a study and thus to a broader framework beyond inflation. The conclusions obtained previously under inflation with an SR-USR-SR setup remain the same, with the one-loop corrected and DRG resummed power spectrum falling rapidly soon after USR ceases while maintaining perturbativity. Since the first construction of the regularized-renormalized-resummed scalar power spectrum was done under the framework of the EFT of inflation, our aim for the following sections would be to expand the same to give us an EFT of bounce.

\section{The Effective Field Theory (EFT) of bounce}
\label{s3}

The primary idea behind the framework under consideration is to begin with a model-independent, effective action that is valid below the UV cut-off scale. Symmetry constrains the structure of the EFT action. Using EFT settings, we may restrict the speed of sound ($c_s$).
We use the unitary gauge for the St$\ddot{u}$ckelberg method, which involves scalar perturbation known as Goldstone modes. The Goldstone mode, also known as the UV-completed version of linearized gauge symmetry, may be integrated into the non-linear sigma model framework, similar to how the Standard Model Higgs sector works. See refs.\cite{Weinberg:2008hq,Cheung:2007st,Choudhury:2017glj,Choudhury:2024ybk,Choudhury:2024jlz,Choudhury:2021brg,Adhikari:2022oxr,Banerjee:2021lqu,Naskar:2017ekm,Choudhury:2015pqa,Choudhury:2015eua,Choudhury:2015zlc,Choudhury:2015hvr,Choudhury:2014sua,Choudhury:2013iaa,Choudhury:2013jya,Choudhury:2013zna,Choudhury:2011sq,Choudhury:2011jt,Choudhury:2012yh,Choudhury:2012whm,Choudhury:2014sxa,Choudhury:2014uxa,Choudhury:2014kma,Choudhury:2016cso,Choudhury:2016pfr,Choudhury:2017cos,Bohra:2019wxu,Akhtar:2019qdn,Choudhury:2020yaa,Choudhury:2021tuu,Choudhury:2016wlj,Cabass:2022avo,Cai:2016thi,Cai:2017tku,Agarwal:2012mq,Piazza:2013coa,Delacretaz:2016nhw,Salcedo:2024smn,Colas:2023wxa,Senatore:2010wk,Noumi:2012vr,Tong:2017iat,Noumi:2012vr,Arkani-Hamed:2015bza,Kim:2021pbr,Baumann:2018muz,Choudhury:2018glz,Hongo:2018ant,Baumann:2017jvh,An:2017hlx,Gong:2017yih,Liu:2016aaf} for more details on various implications of EFT.

\subsection{The underlying bouncing EFT setup}

It is worth noting that, if we describe the formulation in terms of scalar field degrees of freedom, such contribution becomes a scalar under the complete diffeomorphism symmetry:
\be x^{\nu}\longrightarrow x^{\nu}+\xi^{\nu}(t,{\bf x})~~~\forall~ \nu=0,1,2,3~.\ee 
In this context, $\xi^{\nu}(t,{\bf x})$ refers to the diffeomorphism parameter. Instead of applying complete symmetry, the perturbation on the field $\delta \phi$ changes in the following two parallel scenarios:
\begin{enumerate}
    \item Under the implementation of spatial component of the space-time diffeomorphism symmetry, solely as a scalar degrees of freedom, and

    \item Similarly under the implementation of the temporal component of the space-time diffeomorphism symmetry in a non-linear fashion. 
\end{enumerate}
The following is one approach to convey these specific modifications:
\bea
	t&&\longrightarrow t,~x^{i}\longrightarrow   x^{i}+\xi^{i}(t,{\bf x})~~~\forall~ i=1,2,3
~~~\delta\phi\longrightarrow \delta\phi,\\	
	t&&\longrightarrow t+ \xi^{0}(t,{\bf x}),~x^{i}\longrightarrow x^{i}~~~\forall~ i=1,2,3
	~~~\delta\phi\longrightarrow \delta\phi +\dot{\phi}_{0}(t)\xi^{0}(t,{\bf x}).\quad\quad
\eea
The spatial and temporal diffeomorphism parameters are defined as $\xi^{0}(t,{\bf x})$ and $\xi^{i}(t,{\bf x})\forall i=1,2,3$, respectively. We employ the gravitational gauge $\phi(t,{\bf x})=\phi_{0}(t)$ in this situation. The background time-dependent scalar field is embedded in homogeneous, isotropic, and spatially flat FLRW space-time, which is reflected in $\phi_0(t)$. Moreover, in this gauge selection, $\delta \phi(t,{\bf x})=0$.

The key components required to create this EFT setup are the following: 
\begin{itemize}
    \item EFT operators require a function of the gravitational space-time metric $g_{\mu\nu}$.  The space-time metric derivatives may now be used to calculate the Ricci tensor $R_{\mu\nu}$, the Riemann tensor $R_{\mu\nu\alpha\beta}$, and the Ricci scalar $R$,  all of which become the EFT action components.

    \item To create the EFT action, first acquire the polynomial powers of the temporally perturbed component of the metric ($\delta g^{00}$), which are written as $\delta g^{00}= \left(g^{00}+1\right)$. In this scenario, $g^{00}$ represents the time component of the background metric.  This operator must be invariant under the symmetry transformation of spatial diffeomorphisms.

    \item The spatially flat FLRW space-time metric represents the background geometry, which is described by the infinitesimal line element, \bea ds^2=a^2(\tau)\left(-d\tau^2+d{\bf x}^2\right),\eea
    where the scale factor is represented by the symbol $a(\tau)$, whose adequate solutions are given by the following expression:
\bea \label{scalef} a(\tau) =
\left\{
	\begin{array}{ll}
		a_0\left(\displaystyle\frac{\tau}{\tau_0}\right)^{\frac{1}{\epsilon-1}}  & \mbox{for } {\bf Case-I} \\
		a_0\left[1+\left(\displaystyle\frac{\tau}{\tau_0}\right)^2\right]^{\frac{1}{2(\epsilon-1)}} & \mbox{for } {\bf Case-II}
	\end{array}
\right.\eea
The mathematical structure of the scale factors characterizes the following physical situations:
\begin{itemize}
    \item Here {\bf Case-I} represents the power law solution of the scale factor \cite{Khoury:2001wf,Khoury:2001zk,Khoury:2001bz,Buchbinder:2007ad,Lehners:2007ac,Lehners:2008vx,Raveendran:2018yyh,Brandenberger:2012zb,Raveendran:2017vfx,Chowdhury:2015cma,Cai:2011tc,Brandenberger:2016vhg,Boyle:2004gv,Wands:1998yp,Peter:2002cn,Allen:2004vz,Martin:2003sf,Papanikolaou:2024fzf,Raveendran:2023auh,Raveendran:2019idj,Brustein:1998kq,Starobinsky:1980te,Mukhanov:1991zn,Brandenberger:1993ef,Novello:2008ra,Lilley:2015ksa,Battefeld:2014uga,Peter:2008qz,Biswas:2005qr,Bamba:2013fha,Nojiri:2014zqa,Bajardi:2020fxh,Bhargava:2020fhl,Cai:2009in,Cai:2012ag,Shtanov:2002mb,Ilyas:2020qja,Ilyas:2020zcb,Zhu:2021whu,Banerjee:2016hom,Saridakis:2018fth,Barca:2021qdn,Wilson-Ewing:2012lmx,K:2023gsi,Agullo:2020cvg,Agullo:2020fbw,Agullo:2020wur,Chowdhury:2018blx,Chowdhury:2016aet,Nandi:2019xag,Raveendran:2018why,Raveendran:2018yyh,Stargen:2016cft,Sriramkumar:2015yza,Banerjee:2022gpy,Paul:2022mup,Odintsov:2021yva,Banerjee:2020uil,Das:2017jrl,Pan:2024ydt,Colas:2024xjy,Piao:2003zm,Cai:2017dyi,Cai:2017pga,Cai:2015nya,Cai:2019hge,Battista:2020lqv} 
    During inflation the expansion is described by a quasi de Sitter phase where the parameter $\epsilon<1$. One can further consider scenarios, where $1<\epsilon<3$ out of which $\epsilon=3/2$ represents the matter contracting phase solution from {\bf Case I}. One can further consider another interesting situation out of this power law solution when $\epsilon>3$, which actually represents the ekpyrotic contracting phase in this context. Additionally, it is essential to note that at the conformal time scale $\tau=\tau_0$, the scale factor is denoted by the symbol $a(\tau_0)=a_0$ which sets the reference for the {\bf Case-I}. When we try to explain inflation, ekpyrotic contraction and matter contraction scenarios with these mentioned form of power law parametrization of the scale factor, the corresponding reference scale factors are given by $a_0=a_i$, $a_0=a_{ec}$, $a_0=a_{mc}$ for the corresponding conformal time scales, $\tau_0=\tau_i$, $\tau_0=\tau_{ec}$ and $\tau_0=\tau_{mc}$ respectively.

    \item Here {\bf Case-II} represents the scale factor that describes the bouncing solution \cite{Khoury:2001wf,Khoury:2001zk,Khoury:2001bz,Buchbinder:2007ad,Lehners:2007ac,Lehners:2008vx,Raveendran:2018yyh,Brandenberger:2012zb,Raveendran:2017vfx,Chowdhury:2015cma,Cai:2011tc,Brandenberger:2016vhg,Boyle:2004gv,Wands:1998yp,Peter:2002cn,Allen:2004vz,Martin:2003sf,Papanikolaou:2024fzf,Raveendran:2023auh,Raveendran:2019idj,Brustein:1998kq,Starobinsky:1980te,Mukhanov:1991zn,Brandenberger:1993ef,Novello:2008ra,Lilley:2015ksa,Battefeld:2014uga,Peter:2008qz,Biswas:2005qr,Bamba:2013fha,Nojiri:2014zqa,Bajardi:2020fxh,Bhargava:2020fhl,Cai:2009in,Cai:2012ag,Shtanov:2002mb,Ilyas:2020qja,Ilyas:2020zcb,Zhu:2021whu,Banerjee:2016hom,Saridakis:2018fth,Barca:2021qdn,Wilson-Ewing:2012lmx,K:2023gsi,Agullo:2020cvg,Agullo:2020fbw,Agullo:2020wur,Chowdhury:2018blx,Chowdhury:2016aet,Nandi:2019xag,Raveendran:2018why,Raveendran:2018yyh,Stargen:2016cft,Sriramkumar:2015yza,Banerjee:2022gpy,Paul:2022mup,Odintsov:2021yva,Banerjee:2020uil,Das:2017jrl,Pan:2024ydt,Colas:2024xjy,Piao:2003zm,Cai:2017dyi,Cai:2017pga,Cai:2015nya,Cai:2019hge,Battista:2020lqv} Specifically, $\epsilon=3/2$ corresponds to the matter bounce solution in this context. On the other hand, the situation that describes $\epsilon>3$ physically represents the ekpyrotic bounce solution in this related discussion. Following this discussion, it is further necessary to note that at the conformal time scale, $\tau=\tau_0$ the corresponding scale factor is represented by $a(\tau_0)=a_0$, which also sets the reference scale for this specific {\bf Case-II}. Utilizing this solution of the scale factor, when we try to explain ekpyrotic and matter bounce scenarios, the corresponding characteristic scale factors are given by $a_0=a_{eb}$ and $a_0=a_{mb}$ respectively.
\end{itemize}
 \item Furthermore, the polynomial powers of the fluctuation in the extrinsic curvature computed at constant time slice ($\delta K_{\mu\nu}$) are necessarily required to generate the ultimate structure of the representative EFT action, which is described in terms of, $\delta K_{\mu\nu}=\left(K_{\mu\nu}-a^2Hh_{\mu\nu}\right),$ where the extrinsic curvature ($K_{\mu\nu}$), unit normal vector ($n_{\mu}$), and the induced metric ($h_{\mu\nu}$) is defined by the following expressions:
 \bea K_{\mu \nu}&=&h^{\sigma}_{\mu}\nabla_{\sigma} n_{\nu}
=\left[\frac{\delta^{0}_{\mu}\partial_{\nu}g^{00}+\delta^{0}_{\nu}\partial_{\mu}g^{00}}{2(-g^{00})^{3/2}}
+\frac{\delta^{0}_{\mu}\delta^{0}_{\nu}g^{0\sigma}\partial_{\sigma}g^{00}}{2(-g^{00})^{5/2}}-\frac{g^{0\rho}\left(\partial_{\mu}g_{\rho\nu}+\partial_{\nu}g_{\rho\mu}-\partial_{\rho}g_{\mu\nu}\right)}{2(-g^{00})^{1/2}}\right],\nonumber\\
h_{\mu \nu}&=&g_{\mu \nu}+n_{\mu} n_{\nu},\quad
n_{\mu}=\frac{\partial_{\mu}t}{\sqrt{-g^{\mu \nu}\partial_{\mu}t \partial_{\nu}t}}
=\frac{\delta_{\mu}^0}{\sqrt{-g^{00}}},\eea

\end{itemize}

Therefore, the generic form of the EFT action can be described by the following simplified expression:
\bea
	S&=&\displaystyle\int d^{4}x \sqrt{-g}\left[\frac{M^2_{pl}}{2}R-c(t)g^{00}-\Lambda(t)+{\cal F}(\delta g^{00},\delta K^{\mu\nu},\cdots)\right],\eea
 where $c(t)$ and $\Lambda(t)$ are two time-dependent factors which we need to fix from the corresponding field equations, which are Friedmann equations in the present context of discussion. Additionally, it is important to note that, the last term in the above-mentioned expression ${\cal F}(\delta g^{00},\delta K^{\mu\nu},\cdots)$ captures all contributions from the small perturbation as described in both Case-I and Case-II scenarios and quantified by the following simplified mathematical form:
 \bea {\cal F}(\delta g^{00},\delta K^{\mu\nu},\cdots):&=&\left[\frac{M^{4}_{2}(t)}{2!}\left(\delta g^{00}\right)^2+\frac{M^{4}_{3}(t)}{3!}\left(\delta g^{00}\right)^3-\frac{\bar{M}^{3}_{1}(t)}{2}\left(\delta g^{00}\right)\delta K^{\mu}_{\mu}-\frac{\bar{M}^{2}_{2}(t)}{2}(\delta K^{\mu}_{\mu})^2-\frac{\bar{M}^{2}_{3}(t)}{2}\delta K^{\mu}_{\nu}\delta K^{\nu}_{\mu}\right].\quad\eea
In this expression
the time-dependent coefficients $M_1(t)$, $M_3(t)$,  $\bar{M}_1(t)$,  $\bar{M}_2(t)$ and $\bar{M}_3(t)$ are replicating the role of Wilson coefficients which one needs to fix from the analysis presented in the work.

If we focus only on the background contributions in the generic EFT action, the accompanying Friedmann equations may be expressed as follows:
\bea
\left(\frac{\dot{a}}{{a}}\right)^2=H^2&=&\frac{1}{3M^2_{ pl}}\Bigg(c(t)+\Lambda(t)\Bigg)=\frac{{\cal H}^2}{a^2},\\
\frac{\ddot{a}}{{a}}=\dot{H}+H^2&=&-\frac{1}{3M^2_{ pl}}\Bigg(2c(t)-\Lambda(t)\Bigg)
=\frac{{\cal H}^{'}}{a^2},
\eea
where $'$ describes the conformal time derivative. Also, $\displaystyle {\cal H}=\frac{a^{'}}{a}=aH$ describes the relationship between conformal time-dependent Hubble parameter ${\cal H}$ and physical time-dependent Hubble parameter $H$. Finally, we obtain the following formulas for the background-level time-dependent parameters, $\Lambda(t)$ and $c(t)$:
\bea c(t)&=&-M^2_{pl} \dot{H}=-\frac{M^2_{pl}}{a^2}\Bigg({\cal H}^{'}-{\cal H}^2\Bigg),\\
\Lambda(t)&=&M^2_{pl} \left(3H^2+\dot{H}\right)=\frac{M^2_{pl}}{a^2}\Bigg(2{\cal H}^2+{\cal H}^{'}\Bigg).\eea
Finally, substituting the explicit form of these
background-level time-dependent parameters, $\Lambda(t)$ and $c(t)$ in the representative EFT action we get the following reduced form:
\bea
	S&=&\displaystyle\int d^{4}x \sqrt{-g}\left[\frac{M^2_{pl}}{2}R+M^2_{pl} \dot{H}g^{00}-M^2_{pl} \left(3H^2+\dot{H}\right)+{\cal F}(\delta g^{00},\delta K^{\mu\nu},\cdots)\right],\eea
 It is also required to incorporate the following important parameters, which are also referred to as the slow-roll parameters:
\bea
\epsilon &=&\bigg(1-\frac{{\cal H}^{'}}{{\cal H}^2}\bigg),\quad\quad
\eta =\frac{\epsilon^{'}}{\epsilon {\cal H}}.
\eea
\subsection{Goldstone EFT after implementing decoupling limiting approximation}
The Goldstone mode ($\pi(t, {\bf x})$) changes, $\pi(t, {\bf x})\rightarrow\tilde{\pi}(t, {\bf x})=\pi(t, {\bf x})-\xi^{0}(t,{\bf x})$ under the temporal diffeomorphism symmetry. Here $\xi^{0}(t,{\bf x})$ is the local parameter. The function of these Goldstone modes is compared to that of the scalar modes in cosmic perturbation in this work. This gives $\pi(t,{\bf x})=0$ as the fixing criterion for the applicable unitary gauge.This suggests that $\tilde{\pi}(t,{\bf x})=-\xi^{0}(t,{\bf x})~.$

Now it is crucial to discuss how the broken time diffeomorphism symmetry affects the space-time metric, the Ricci tensor, the Ricci scalar, the perturbation on the extrinsic curvature, the time-dependent coefficients, and the slowly growing Hubble parameter:
\bea
		&&{g}^{00}\longrightarrow
	(1+\dot{\pi}(t,{\bf x}))^2 {g}^{00}+2(1+\dot{\pi}(t,{\bf x})){g}^{0 i}\partial_{i}\pi(t,{\bf x})+{g}^{ij}\partial_{i}\pi(t,{\bf x})\partial_{j}\pi(t,{\bf x}),\quad\quad\quad \nonumber\\ 
		&&{g}^{0i}\longrightarrow
		(1+\dot{\pi}(t,{\bf x})){g}^{0i}+{g}^{ij}\partial_j \pi(t,{\bf x}),\quad
	{g}^{ij}\longrightarrow{g}^{ij},\nonumber\\
			&&g_{00}\longrightarrow  (1+\dot{\pi}(t,{\bf x}))^2 g_{00},\quad
		g_{0i}\longrightarrow (1+\dot{\pi}(t,{\bf x})){g}_{0i}+{g}_{00}\dot{\pi}(t,{\bf x})\partial_{i}\pi(t,{\bf x}),\nonumber \\ && g_{ij}\longrightarrow{g}_{ij}+{g}_{0j}\partial_{i}\pi(t,{\bf x})+{g}_{i0}\partial_{j}\pi(t,{\bf x}),\nonumber\\
			 &&{}^{(3)}R\longrightarrow\displaystyle {}^{(3)}R+\frac{4}{a^2}H(\partial^2\pi(t,{\bf x})),\quad
			 {}^{(3)}R_{ij}\longrightarrow{}^{(3)}R_{ij}+H(\partial_{i}\partial_{j}\pi(t,{\bf x})+\delta_{ij}\partial^2\pi(t,{\bf x})),\quad
			\quad \nonumber\\
   &&\delta K\longrightarrow \displaystyle \delta K-3\pi\dot{H}-\frac{1}{a^2}(\partial^2\pi(t,{\bf x})),\quad
		\delta K_{ij}\longrightarrow\delta K_{ij}-\pi(t,{\bf x})\dot{H}h_{ij}-\partial_{i}\partial_{j}\pi(t,{\bf x}),\nonumber\\
		&&\delta K^{0}_{0}\longrightarrow\delta K^{0}_{0},\quad
		\delta K^{0}_{i}\longrightarrow\delta K^{0}_{i},\quad
		\delta K^{i}_{0}\longrightarrow\delta K^{i}_{0}+2Hg^{ij}\partial_j\pi(t,{\bf x}),\nonumber\\
  &&Q(t)\longrightarrow\displaystyle \sum^{\infty}_{n=0}\frac{\pi^{n}_c(t,{\bf x})}{n!Q^{2n}(t)}\frac{d^{n}Q(t)}{dt^n}\approx Q(t),\quad
   H(t)\longrightarrow \displaystyle \sum^{\infty}_{n=0}\frac{\pi^{n}}{n!}\frac{d^{n}H(t)}{dt^n}=\displaystyle\left[1-\pi(t,{\bf x}) H(t) \epsilon\right]H(t)\approx H(t).
	\eea
To generate the most generic EFT action, we must first have a better grasp of the decoupling limit. In this limit, the mixing contributions of the Goldstone modes and gravity may be easily ignored. To establish this conclusion, let's begin with the EFT operator $-\dot{H}M_{pl}^2g^{00}$, which is needed for upcoming calculations. This operator undergoes the following transformation when the broken time diffeomorphism symmetry is applied:
\bea &&-\dot{H}M_{pl}^2g^{00}\longrightarrow -\dot{H}M_{pl}^2\bigg[ (1+\dot{\pi}(t,{\bf x}))^2g^{00}%\nonumber\\  &&\quad\quad\quad\quad\quad\quad\quad\quad\quad\quad
+\left(2(1+\dot{\pi}(t,{\bf x}))\partial_i \pi(t,{\bf x}) g^{0i}+g^{ij}\partial_i\pi(t,{\bf x}) \partial_j \pi(t,{\bf x})\right)\bigg].\eea
The temporal component of the metric after perturbation may be represented as $g^{00}=\bar{g}^{00}+\delta g^{00},$ where the perturbation is denoted by $\delta g^{00}$ and we have $\bar{g}^{00}=-1$. The remaining contributions include a mixing contribution $M_{pl}^2\dot{H}\dot{\pi}\delta g^{00}$ and a kinetic contribution $M_{pl}^2\dot{H}\dot{\pi}^2\bar{g^{00}}$. Furthermore, $\delta g^{00}_c=M_{pl}\delta g^{00},$ is used as a standard normalized metric perturbation mixing contribution following canonical normalization, which is provided by, $M_{pl}^2\dot{H}\dot{\pi}\delta g^{00}= \sqrt{\dot{H}}\dot{\pi}_c\delta g^{00}_c$.The mixing term above the characteristic energy scale, $E_{mix}=\sqrt{\dot{H}},$ may be easily neglected in the decoupling limit. Following this treatment further two important factors, which after implementing the canonical normalization can be recast as, $M_{pl}^2\dot{H}\dot{\pi}^2\delta{g^{00}}=\dot{\pi}_c^2\delta{g^{00}_c}/M_{pl},$ and 
$\pi M_{pl}^2\ddot{H}\dot{\pi}\bar{g}^{00}=\ddot{H}\pi_c\dot{\pi}_c\bar{g}^{00}/\dot{H}$. This clearly shows that out of these two possibilities, one can safely neglect the first one after implementing the decoupling limiting approximation and one can finally write the following simplified expression:
\bea -\dot{H}M_{pl}^2g^{00}\rightarrow -\dot{H}M_{pl}^2g^{00}\left[\dot{\pi}^2-\frac{1}{a^2}(\partial_i\pi)^2\right].\eea

\subsection{Implication of Goldstone EFT on perturbations}
In the decoupling limit, the second-order perturbed action for the Goldstone modes can now be expressed in the following fashion:
\bea 
  	S^{(2)}_{\pi}&\approx&\displaystyle \int d^{4}x ~a^3\left[-M^2_{pl}\dot{H}\left(\dot{\pi}^2-\frac{1}{a^2}(\partial_{i}\pi)^2\right)
   	+2M^4_2 \dot{\pi}^2\right]=\displaystyle \int d^{4}x ~a^3\left(\frac{-M^2_{pl}\dot{H}}{c^2_s}\right)\Bigg(\dot{\pi}^2-c^2_s\frac{\left(\partial_{i}\pi\right)^2}{a^2}\Bigg).~~~~~\quad\quad\eea 
Now the definition of the effective sound speed in terms of the EFT Wilson coefficient:
 \bea c_{s}\equiv \frac{1}{\displaystyle \sqrt{1-\frac{2M^4_2}{\dot{H}M^2_{pl}}}},\eea
 It is crucial to keep in mind that the following defines the spatial component of the metric fluctuation:
\bea g_{ij}=a^{2}(t)\exp(2\zeta(t,{\bf x}))\delta_{ij}\sim a^{2}(t)\left[\left(1+2\zeta(t,{\bf x})\right)\delta_{ij}\right]~~\forall~~~i=1,2,3,\eea
When we explain the background geometry in terms of the conformal time coordinates instead of using the physical time coordinate the corresponding scale factor is denoted by the symbol $a(\tau)$, where in this paper we are interested in two specific types of the solutions described in the parametrizations mentioned in the Case-I and Case-II as stated in equation(\ref{scalef}). In addition, the notation $\zeta(t,{\bf x})$ is used to represent the scalar comoving curvature perturbation and we have truncated the perturbation in the first order in $\zeta(t,{\bf x})$. In this case, the broken temporal diffeomorphism causes the scale factor to transform as follows:
\bea a(t)\rightarrow  a(t)\left(1-H\pi(t,{\bf x})\right)~\Longrightarrow ~a^2(t)\left(1-H\pi(t,{\bf x})\right)^2&\approx & a^{2}(t)\left(1+2\zeta(t,{\bf x})\right)~\Longrightarrow~\zeta(t,{\bf x})=-H\pi(t,{\bf x}),\eea
using which we can now write the rest of the discussions and their outcomes in terms of the gauge invariant comoving curvature perturbation variable rather than expressing quantities in terms of the Goldstone modes. As an immediate outcome, the second order perturbed action now can be written in terms of the variable $\zeta(t,{\bf x})$ as:
 \bea \label{seconda}
  	S^{(2)}_{\zeta}&=&\displaystyle \int d^{4}x ~a^3\left(\frac{M^2_{pl}\epsilon}{c^2_s}\right)\Bigg(\dot{\zeta}^2-c^2_s\frac{\left(\partial_{i}\zeta\right)^2}{a^2}\Bigg)=M^2_{pl}\displaystyle \int d\tau\;  d^3x\;  a^2\;\left(\frac{\epsilon}{c^2_s}\right)\Bigg(\zeta^{'2}-c^2_s\left(\partial_i\zeta\right)^2\Bigg).~~~~\quad\eea 
   Here in the last step, we have utilized the transformation, $d\tau=dt/a(t)$. In the later part of this utilizing this perturbed action we are going to evaluate the scalar mode functions in the Fourier space for the five distinctive phases, 1) ekpyrotic/matter contraction, 2) ekpyrotic/matter bounce, 3) first slow roll, 4) ultra slow roll, and 5) second slow roll utilizing which in the rest of the paper we will carry forward the analysis within the framework of Goldstone EFT.
   
\section{Realization of ultra-slow-roll phase within the framework of EFT of bounce}
\label{s4}

 \subsection{Phase I: Ekpyrotic/Matter contraction}

In this section we focus on the contraction phase of our setup which is associated with the scale factor for Case-I in equation(\ref{scalef}). Given this we can immediately determine the conformal Hubble parameter following its standard definition as follows: 
\bea \label{hubbleC}
{\cal H}&=&\frac{a'}{a} = \frac{1}{\tau(\epsilon-1)}.
\eea
where the prime denotes derivative with respect to conformal-time. Using the above we can determine the expression relating the conformal time and the number of e-foldings $N$ from the definition:
\bea
dN= Hdt = aH\frac{dt}{a} = {\cal H}d\tau,
\eea
which after integrating gives us the required relation
\bea \label{timeC}
\int_{N_{0}}^{N}dN = \int_{\tau_{0}}^{\tau}\frac{d\tau}{\tau(\epsilon-1)}\implies \tau = \tau_{0}\exp{[(N-N_{0})(\epsilon-1)]}.
\eea
and this remain valid for $N< N_c$ where at $N=N_c$ the contraction phase stops. Substitution of the above into the equation(\ref{hubbleC}) provides us with the conformal Hubble as a function of the number of e-foldings
\bea
{\cal H}(N) = \frac{1}{\tau_{0}(\epsilon-1)}\exp{[-(N-N_{0})(\epsilon-1)]}.
\eea
The conformal Hubble here will allow us to evaluate the behaviour of the slow-roll parameter also as a function of the e-foldings and for the same we now proceed with the first slow-roll parameter. From its definition we have 
\bea
\epsilon &=& 1-\frac{{\cal H}'}{{\cal H}^2},\nonumber\\
&=& 1+\frac{\tau^2(\epsilon-1)^2}{\tau^2(\epsilon-1)},\nonumber\\
&=& \epsilon\sim\frac{3}{2},
\eea
which leads us to verify that the first slow-roll parameter $\epsilon$ remains same throughout this phase irrespective of whether its a matter or ekpyrotic contraction. Similarly, for the second slow-roll parameter we can determine its relation as follows
\bea
\eta = \frac{\epsilon'}{\epsilon{\cal H}} = 0,
\eea
which remains zero throughout since the parameter $\epsilon$ remains a constant.

 \subsection{Phase II: Ekpyrotic/Matter bounce}

In this section we focus on the bounce phase of our setup which is associated with the scale factor for Case-II in equation(\ref{scalef}). Following the procedure similar to the contraction phase we determine first the conformal Hubble parameter using its standard definition as: 
\bea \label{hubbleB}
{\cal H}&=&\frac{a'}{a} = \frac{\tau}{\tau_{0}^2 (\epsilon-1)}\bigg(1+\bigg(\frac{\tau}{\tau_0}\bigg)^2 \bigg)^{-1},
\eea
Using this we find the conformal time and e-folding relation in the bounce phase from the definition as follows
\bea \label{timeB}
\int_{N_{0}}^{N}dN &=& \int_{\tau_{0}}^{\tau}d\tau\frac{\tau}{\tau_{0}^2 (\epsilon-1)}\bigg(1+\bigg(\frac{\tau}{\tau_0}\bigg)^2 \bigg)^{-1},\nonumber\\
\Longrightarrow N-N_{0} &=& \frac{\ln{[\tau^2 + \tau_{0}^2]}}{2(\epsilon-1)}\Biggr|_{\tau_{0}}^{\tau},\nonumber\\
&=& \frac{1}{2(\epsilon-1)}\ln{\bigg[\frac{1}{2}\bigg(1+\bigg(\frac{\tau}{\tau_0}\bigg)^2\bigg)\bigg]}.
\eea
and it remains valid during the interval, $N_c< N< N_b$, after which at $N=N_b$ the bounce phase stops. Substituting the above in equation(\ref{hubbleB}) provides us with the conformal Hubble in the bounce phase in terms of e-folds as
\bea
{\cal H}(N) &=& \frac{1}{\tau_{0}(\epsilon-1)}\frac{\tau}{\tau_{0}}\bigg(1+\bigg(\frac{\tau}{\tau_0}\bigg)^2 \bigg)^{-1},\nonumber\\
&=& \frac{1}{2\tau_{0}(\epsilon-1)}\exp{[-2(N-N_{0})(\epsilon-1)]}\sqrt{2\exp{[2(N-N_{0})(\epsilon-1)]}-1}.
\eea
We proceed as before with the above definition starting with the first slow-roll parameter and its defining relation here as
\bea
\epsilon(N) &=& 1-\frac{{\cal H}'}{{\cal H}^2},\nonumber\\
&=& 1- \frac{(\epsilon-1)(\tau_{0}^2-\tau^2)}{\tau^2},\nonumber\\
&=& \epsilon + (1-\epsilon)\frac{\tau_{0}^2}{\tau^2},\nonumber\\
&=& \epsilon + \frac{(1-\epsilon)}{2\exp{[2(N-N_{0})(\epsilon-1)]}-1}.
\eea
from which we can again verify that due to having an increasing exponential in the denominator of the expression the the first slow-roll parameter remains almost as a constant with value $\epsilon$ in both the matter and ekpyrotic bounce phase, except for a short while near the transition from contraction to bounce phase.  For the second slow-roll parameter we use its definition, the third relation from above, and equation(\ref{timeB}) to get the following relation with e-foldings,
\bea
\eta(N) &=& \frac{\epsilon'}{\epsilon{\cal H}},\nonumber\\
&=& 2(\epsilon-1)^{2}\times \frac{\tau_{0}^2}{\tau^2}\frac{1+\big(\frac{\tau}{\tau_0}\big)^2}{1+ \epsilon\big(\frac{\tau^2}{\tau_0^2}-1\big)},\nonumber\\
&=& 2(\epsilon-1)^{2}\times \frac{1}{(2\exp{[2(N-N_{0})(\epsilon-1)]}-1)}\frac{2\exp{[2(N-N_{0})(\epsilon-1)]}}{1+2\epsilon(\exp{[2(N-N_{0})(\epsilon-1)]}-1)}.
\eea

 \subsection{Phase III: First slow-roll (SRI)}

We now focus on the phases in our setup after the contraction and bounce. Here we discuss the first slow-roll phase and behaviour of the slow-roll parameters with changing e-folding number. We start with assuming a very small, and negative, value of the second slow-roll parameter, $\eta_{\rm SRI} \rightarrow 0$, that is almost a constant. Upon using the definition of $\eta$ we can determine how the first slow-roll parameter varies with e-folds $N$,
\bea
\eta &=& \frac{\epsilon'}{\epsilon{\cal H}} = \frac{1}{\epsilon}\frac{d\epsilon}{dN},
\eea
which tells us that $\epsilon(N)$ behaves almost as a constant in SRI for $\eta$ to satisfy the above condition of being close to zero. Thus we have for $N_{*}\leq N\leq N_s$,
\bea
\epsilon(N)=\epsilon,
\eea
where $N_{*}$ refers to the reference value of e-folds that corresponds to the beginning of SRI phase, with choosing $\epsilon(N=N_{*})={\cal O}(10^{-3})$. The behaviour of the Hubble parameter can also be found from the definition of the first slow-roll parameter as follows
\bea
\epsilon=-\frac{\dot{H}}{H^{2}} = -\frac{d\ln H}{dN},\nonumber\\
\eea
which after integration leads to $H(N)$ for the interval, $N_{*}\leq N\leq N_s$, as
\bea \label{hubbleSRI}
-\int_{H_{i}}^{H}\frac{dH}{H} &=& \int_{N_{*}}^{N}\epsilon(N) dN,\nonumber\\
\Longrightarrow H(N) &=& H_{i}\exp{\big(-\int_{N_{*}}^{N}\epsilon(N) dN\big)},
\eea
where $H_i = H(N_{*})$ is its initial value.

 \subsection{Phase IV: Ultra slow-roll (USR)}

After exiting from the SRI phase at $N=N_{s}$, we enter the phase of ultra-slow roll following a sharp transition at $N=N_{s}$, and the USR continues for $N_{s}\leq N\leq N_{e}$. This phase is characterized by a sharp increase in the magnitude of the second SR parameter, with $\eta\sim {\cal O}(-6)$ having a negative signature. This remains a constant for the duration of the e-foldings in the USR and using this the first slow-roll parameter can be found to behave as follows
\bea
\eta &=& \frac{1}{\epsilon}\frac{d\epsilon}{dN},\nonumber\\
\Longrightarrow \int_{N_s}^{N}\eta dN &=& \int_{\epsilon(N_s)}^{\epsilon}d\ln \epsilon,\nonumber\\
\Longrightarrow \epsilon(N) &=& \epsilon(N_s)\exp{(\eta (N-N_{s}))},
\eea
where the negative signature of $\eta$ leads to an exponential decrease in the value of $\epsilon$ in USR. For the initial condition we perform its matching with the SRI value, $\epsilon(N_{s}) = \epsilon$. Similarly, we again determine the Hubble parameter for the USR phase whose analytic expression remains the same as in equation(\ref{hubbleSRI}) except with the following change
\bea 
H(N) &=& H_{i}\exp{\big(-\int_{N_{s}}^{N}\epsilon(N) dN\big)}
\eea
where now the initial condition becomes $H_{i} = H(N_{s})$ from the continuity between SRI and USR.

 \subsection{Phase V: Second slow-roll (SRII)}

The last phase in our setup comes after the end of USR, with a sharp transition into the second SR phase at $N=N_e$. During this phase, the parameter $\eta$ climbs from the large negative value in the USR to becoming almost $\eta\sim {\cal O}(-1)$, marking the end of inflation. Again throughout this phase, we choose $\eta$ to remain almost a constant right after the sharp transition. The behaviour of the first SR parameter can be found in a similar manner as before to give:
\bea
\eta &=& \frac{1}{\epsilon}\frac{d\epsilon}{dN},\nonumber\\
\Longrightarrow \int_{N}^{N_{\rm end}}\eta dN &=& \int_{\epsilon}^{\epsilon(N_{\rm end})}d\ln \epsilon,\nonumber\\
\Longrightarrow \epsilon(N) &=& \epsilon(N_{\rm end})\exp{(-\eta (N_{\rm end}-N))},
\eea
where $N_{e}\leq N\leq N_{\rm end}$, with $\epsilon(N=N_{\rm end})\sim {\cal O}(1)$ as inflation ends. With $\eta$ also negative for this phase this formula shows how $\epsilon$ now rises in magnitude throughout the SRII. For the Hubble parameter, following the same procedure as in the previous two phases gives us with the following expression:
\bea
H(N) &=& H_{i}\exp{\big(-\int_{N_{e}}^{N}\epsilon(N) dN\big)}
\eea
where the initial condition from the continuity between SRII and USR gives $H_{i}=H(N_e)$.

 %%%%%%%%%%%%%%%%%%%%%%%%%%%%%%%%%%%%%%%%%%%%%
\begin{figure*}[htb!]
    	\centering
    \subfigure[]{
      	\includegraphics[width=8.5cm,height=7.5cm]{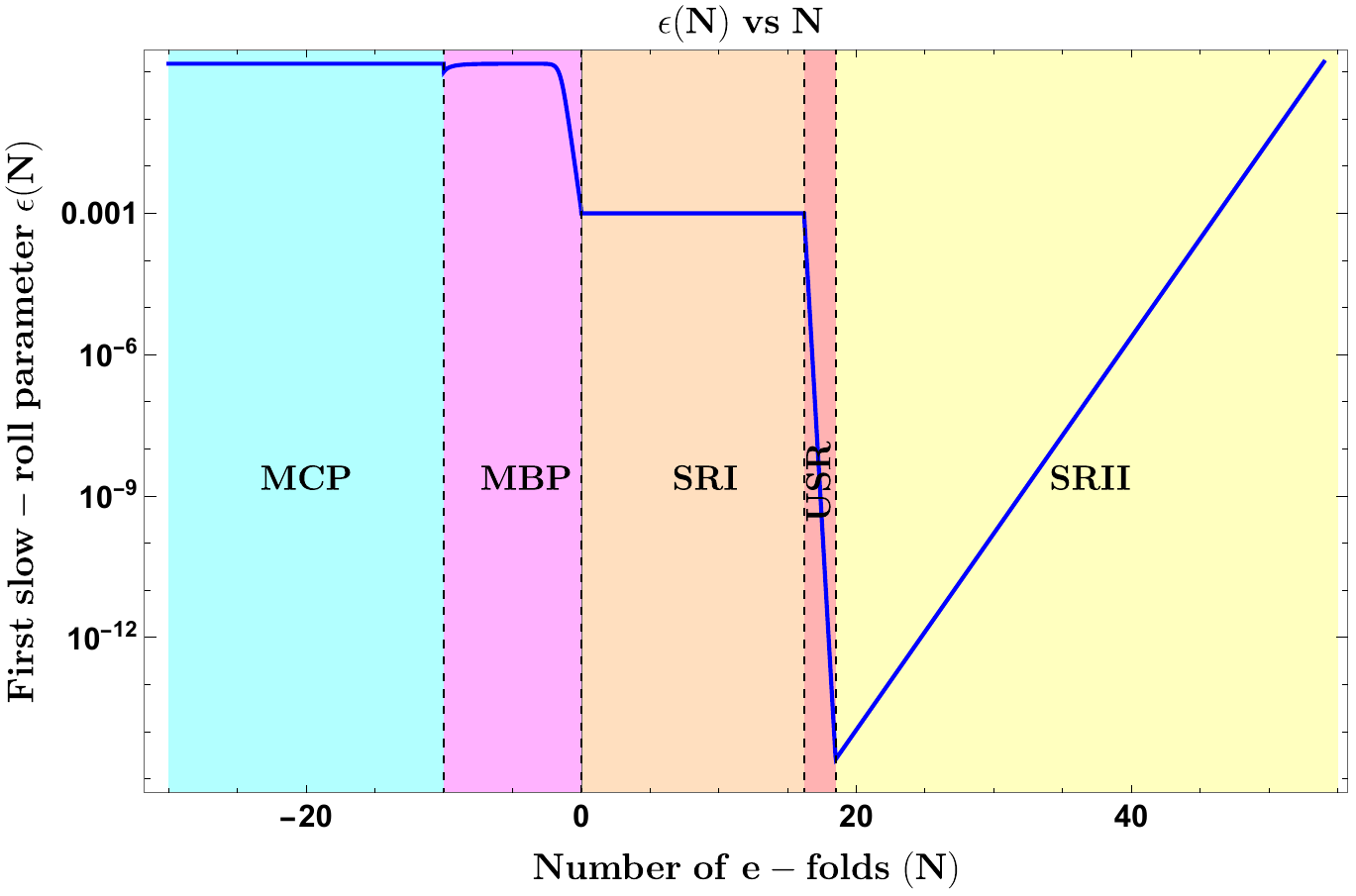}
        \label{epsilonMatter}
    }
    \subfigure[]{
       \includegraphics[width=8.5cm,height=7.5cm]{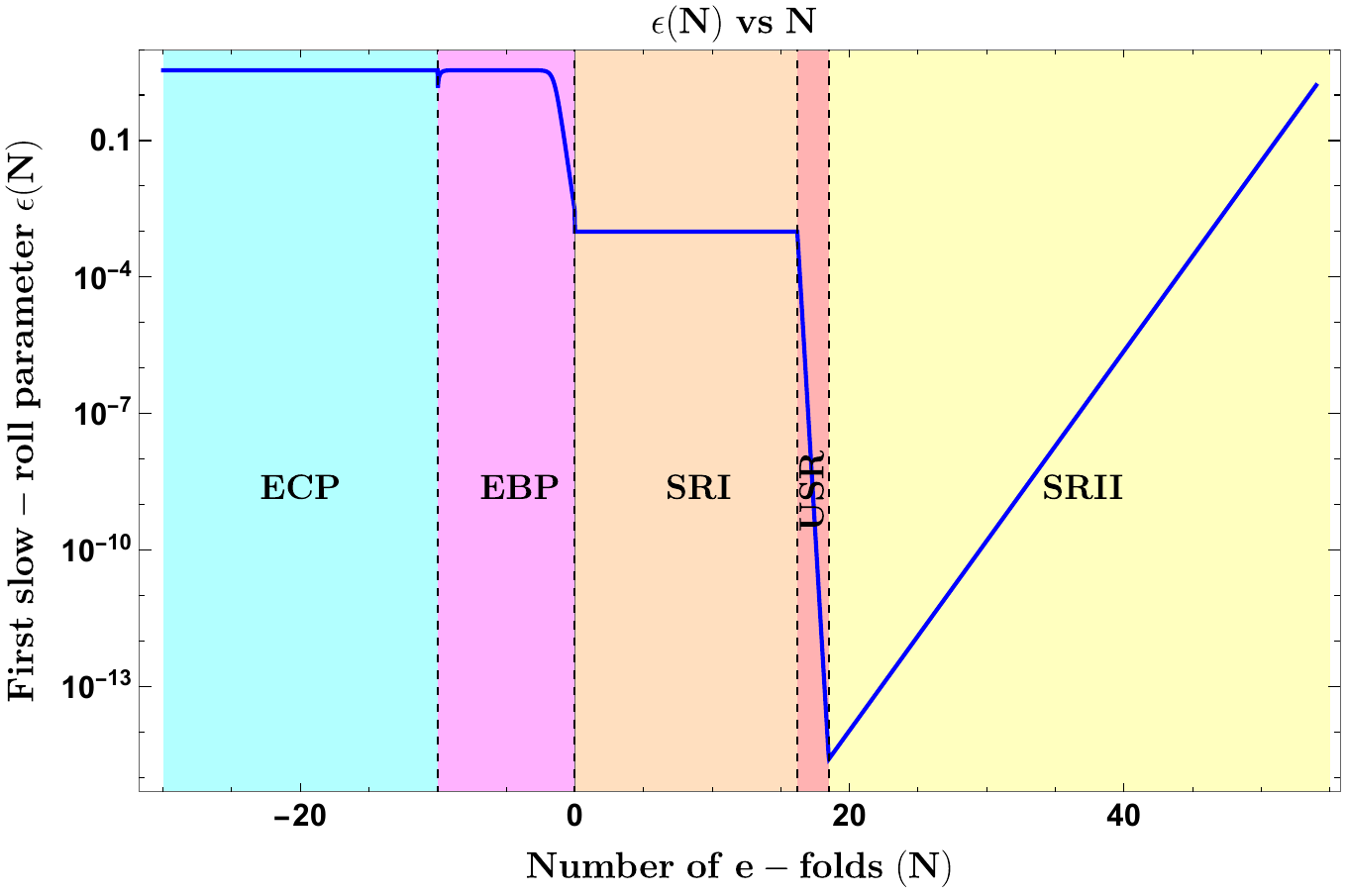}
        \label{epsilonEkpy}
    }
    	\caption[Optional caption for list of figures]{Plot shows behaviour of the first slow-roll parameter $\epsilon(N)$ as a function of the e-folds $N$ in \ref{epsilonMatter} with the matter contraction (MCP) and matter bounce (MBP) phases and in \ref{epsilonEkpy} with the ekpyrotic contraction (ECP) and ekpkyrotic bounce (EBP) phases followed by the SRI, USR, and SRII phases. } 
    	\label{epsilonplots }
    \end{figure*}
%%%%%%%%%%%%%%%%%%%%%%%%%%%%%%%%%%%%%%%%%%%%%

 \section{Numerical outcomes I: Behaviour of dynamical parameters in the five consecutive phases}
 \label{s5}

 %%%%%%%%%%%%%%%%%%%%%%%%%%%%%%%%%%%%%%%%%%%%%
\begin{figure*}[htb!]
    	\centering
    \subfigure[]{
      	\includegraphics[width=8.5cm,height=7.5cm]{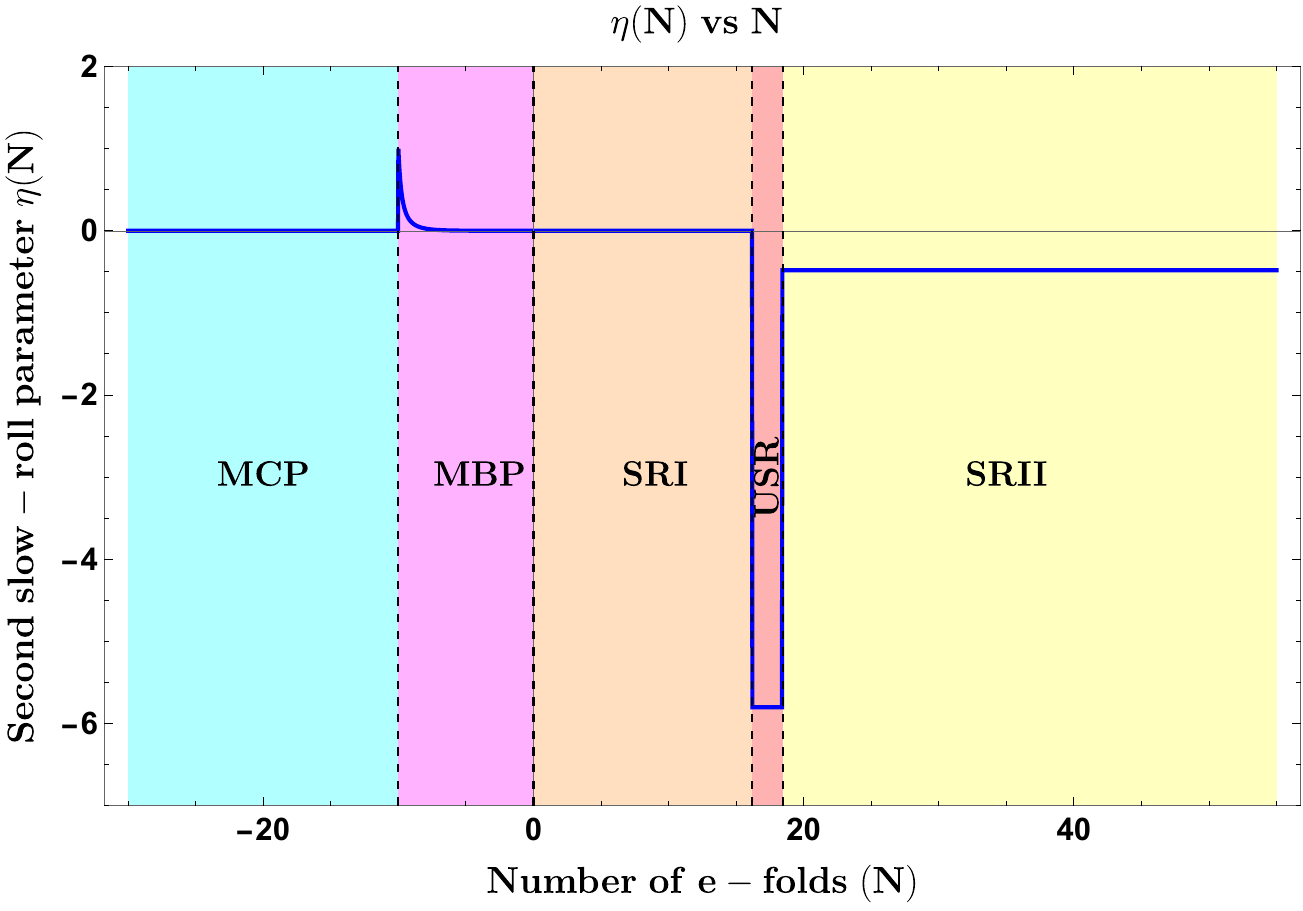}
        \label{etaMatter}
    }
    \subfigure[]{
       \includegraphics[width=8.5cm,height=7.5cm]{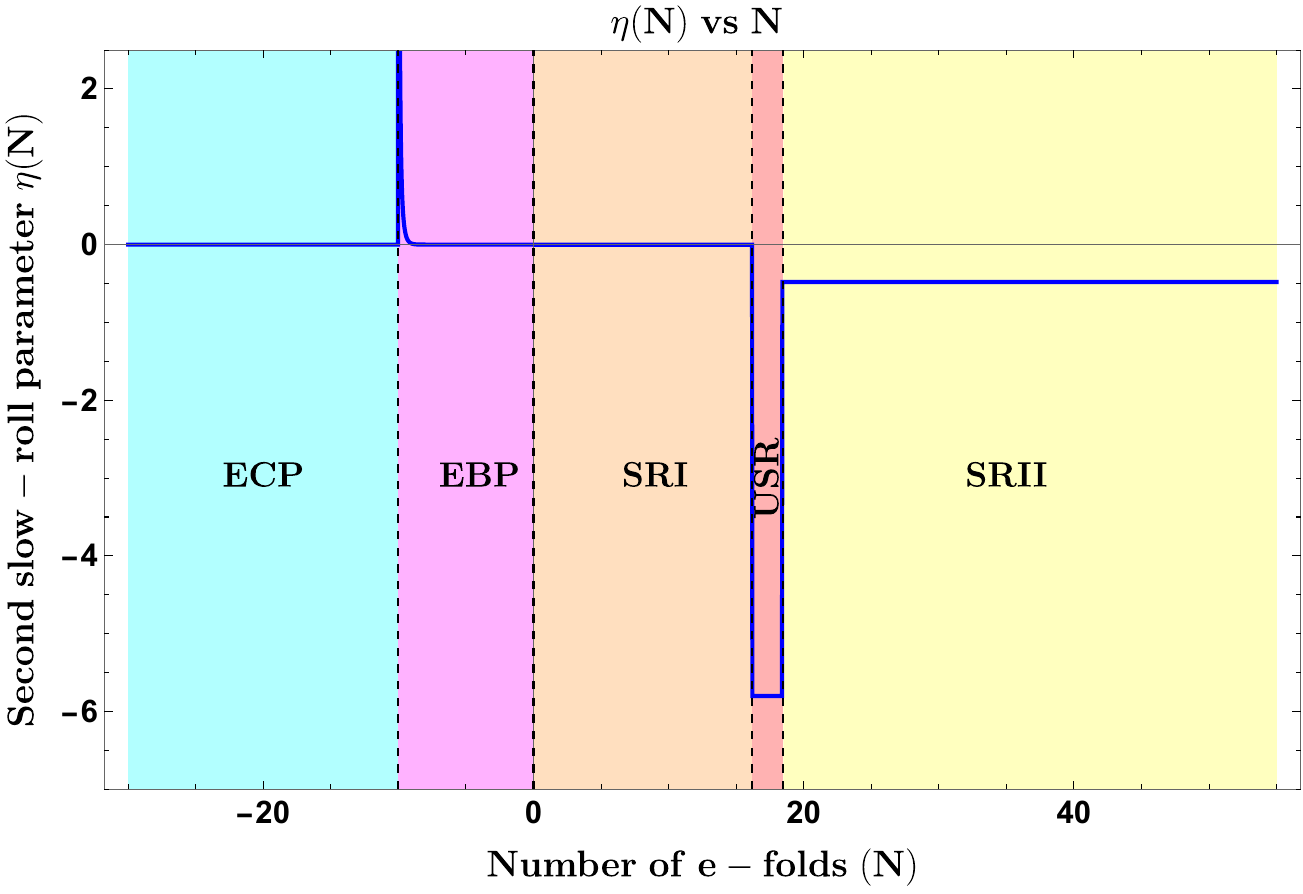}
        \label{etaEkpy}
    }
    	\caption[Optional caption for list of figures]{Plot shows behaviour of the second slow-roll parameter $\eta(N)$ as a function of the e-folds $N$ in \ref{etaMatter} with the matter contraction (MCP) and matter bounce (MBP) phases and in \ref{etaEkpy} with the ekpyrotic contraction (ECP) and ekpkyrotic bounce (EBP) phases followed by the SRI, USR, and SRII phases. } 
    	\label{etaplots }
    \end{figure*}
%%%%%%%%%%%%%%%%%%%%%%%%%%%%%%%%%%%%%%%%%%%%%

In the figures \ref{epsilonMatter} and \ref{epsilonEkpy} we can see the how the first slow-roll parameter behaves throughout the consecutive five phases of our setup. The left figure describes the evolution in with the matter type contraction and bounce phases, where the parameter satisfies $\epsilon=3/2$ in both as is also clear from the plot. In the right plot we observe the behaviour when the ekpyrotic contraction and bounce phases are involved where $\epsilon>3$ is the condition, here we choose $\epsilon=7/2$. The SR parameter value at the transition from the bounce into SRI phase shows a smooth joining as it connects and continues with the value of ${\cal O}(10^{-3})$ in the SRI. At the instant of SRI to USR phase $\epsilon$ starts to drop exponentially for the few e-folds, $\Delta N_{\rm USR}\sim {\cal O}(2)$, of the USR and climbs back up while in the SRII phase to reach the value of ${\cal O}(1)$ at the end of inflation.

In figures \ref{etaMatter} and \ref{etaEkpy}, we can see the how the second SR parameter behaves throughout the five consecutive phases in our setup with the matter contraction and bounce phases present in the left and the ekpyrotic contraction and bounce phases present on the right. Its value remains zero in the contraction phases and at the instance of transitioning into the next bounce phase observes a sudden rise in its value with $\eta=1$ in the matter contraction to bounce scenario, and $\eta>2$ in the ekpyrotic contraction to bounce scenario. In the bouncing phase the $\eta$ parameter keeps on decreasing close to zero till it joins with the SRI phase and there it continues with value of ${\cal O}(10^{-3})$ in the negative. The sharp nature of the transition is clearly displayed by the sudden changes in $\eta$ occurring at the instances between SRI-USR  and USR-SRII transitions. 

\begin{figure*}[htb!]
    	\centering
    \subfigure[]{
      	\includegraphics[width=8.5cm,height=7.5cm]{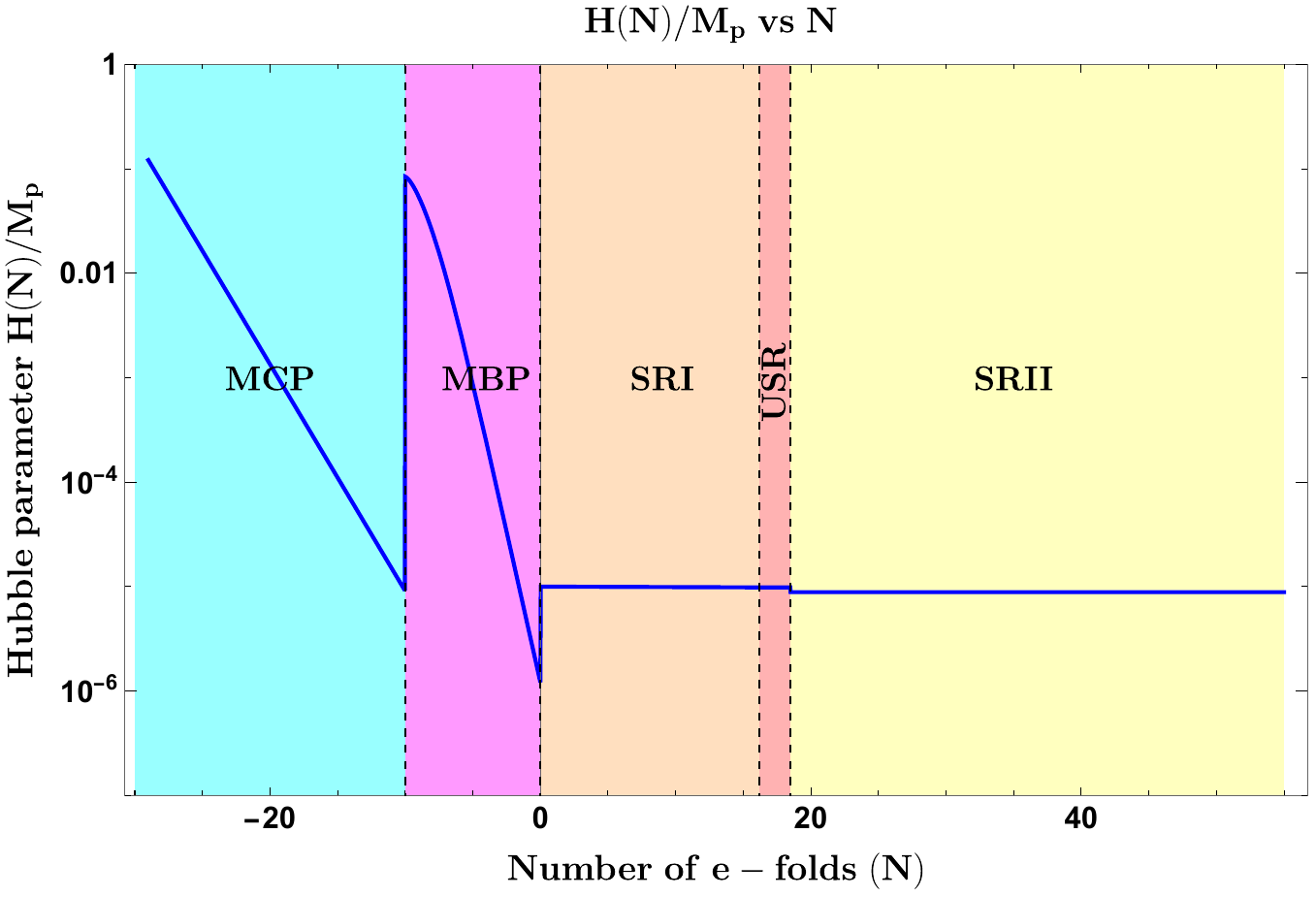}
        \label{hubbleMatter}
    }
    \subfigure[]{
       \includegraphics[width=8.5cm,height=7.5cm]{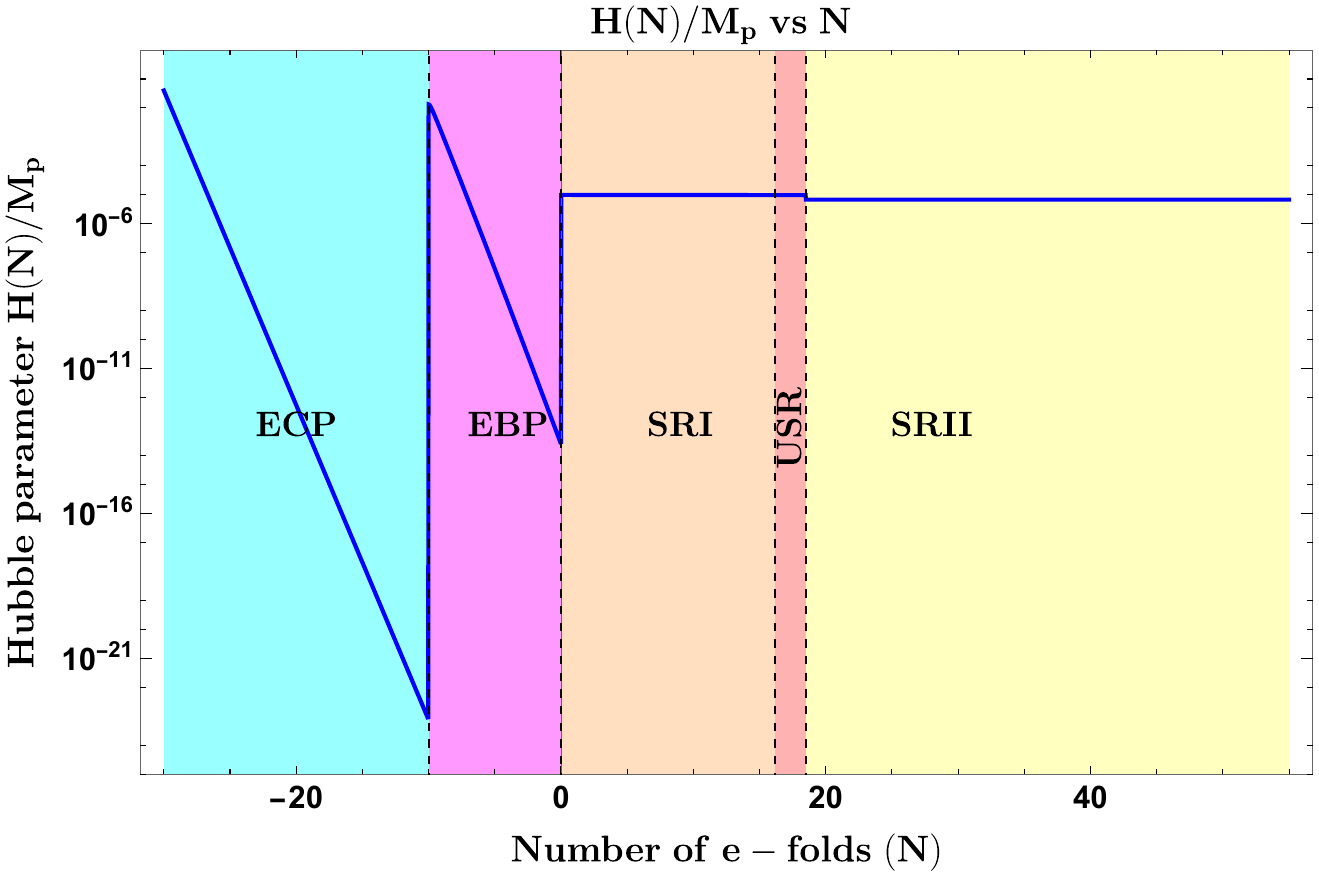}
        \label{hubbleEkpy}
    }
    	\caption[Optional caption for list of figures]{Plot shows behaviour of the Hubble parameter $H(N)$ in reduced Planck units as a function of the e-folds $N$ in \ref{hubbleMatter} with the matter contraction (MCP) and matter bounce (MBP) phases and in \ref{hubbleEkpy} with the ekpyrotic contraction (ECP) and ekpkyrotic bounce (EBP) phases followed by the SRI, USR, and SRII phases. } 
    	\label{hubbleplots }
    \end{figure*}

In figures \ref{hubbleMatter} and \ref{hubbleEkpy} we observe the behaviour of the Hubble parameter as it changes with e-foldings, and its value scaled according to the reduced Planck units, throughout the five consecutive phases in our setup. In the left panel, with the matter contraction and bounce phases, the drop in value for both is comparable in orders of magnitude with a discontinuity that occurs at the instant of transitioning in between the two phases. The discontinuity remains when moving from the matter bounce to the SRI phase where it starts with magnitude of ${\cal O}(1)$ and decreases by a very small amount going into the USR and the SRII phases. In the right panel, with the ekpyrotic contraction and bounce phases, the amount of decrease in $H$ is greatly enhanced in the contraction phase as compared to the bounce phase which is opposite to the behaviour in the matter scenario in left. The rest of the evolution remains the same in both scenarios. In both cases, the initial values of the Hubble remains much larger, with the largest in the matter scenario, when compared to its value through the phases after the commencing of slow-roll from $N=0$.  

 %%%%%%%%%%%%%%%%%%%%%%%%%%%%%%%%%%%%%%%%%%%%%
%%%%%%%%%%%%%%%%%%%%%%%%%%%%%%%%%%%%%%%%%%%%%

 \section{Comoving scalar curvature perturbation modes: The semi-analytical study}
 \label{s6}

Using the underlying Goldstone EFT framework, this part concentrates on establishing the solutions for the comoving curvature perturbation in a spatially flat FLRW backdrop. Our setup consists of five phases: ekpyrotic/matter contraction, ekpyrotic/matter bounce, first slow roll (SRI), ultra slow roll (USR), and last but not least, the second slow roll phase (SRII). We utilize the decoupling limit to safely analyze how the mode solutions behave in the five phases. Discovering connections between the comoving curvature perturbation and its conjugate momentum variable will subsequently aid in constructing the different components of the power spectrum of the scalar modes. This section examines the Mukhanov-Sasaki equation solutions for each of the five phases that were previously described.
In the current context of our investigation, the answers found will be crucial for our examination of the power spectrum at the tree and loop levels. After obtaining the Mukhanov-Sasaki (MS) equation by variation of the action in equation (\ref{seconda}), we may solve it in Fourier space to obtain the curvature perturbation modes for various phases. The Fourier space MS equation is expressed as follows: 
\bea \label{MSfourier}
\bigg[\frac{d^2}{d\tau^2}+2 \frac{z'(\tau)}{z(\tau)}\partial_{\tau}+c_s ^2 k^2\bigg]\zeta_{\bf k}(\tau) = 0.
\eea 
 The {\it Mukhanov-Sasaki} variable, denoted by $z(\tau)$ in this case, may be found in the following expression:
\bea z(\tau):=\frac{a(\tau)\sqrt{2\epsilon}}{c_s},\eea
where the conformal time-dependent scale factor $a(\tau)$ is explicitly defined before using which one can easily explain the existence of the previously mentioned five consecutive phases. To avoid further confusion, it is also essential to note that the first slow roll parameter $\epsilon$ takes different values during different phases, and its conformal time-dependent behaviour also turns out to be different in all of these five phases. This is extremely crucial information that will be frequently used in the remaining analysis presented in this work.

We now use the following finding, which will be very helpful in solving the second-order differential equation stated above:
\bea  \frac{z'(\tau)}{z(\tau)}={\cal H}\left(1-\eta+\epsilon-s\right)\quad\quad\quad{\rm where}\quad\quad\quad s:=\frac{c^{'}_s}{{\cal H}c_s}.\eea
Additionally, we are going to introduce another parameter, $\nu$, which sometimes in the corresponding contexts is identified as an effective mass parameter and defined by the following expression:
\bea \nu:\equiv \frac{1}{2}-\frac{1}{\epsilon-1}+\frac{\eta}{\epsilon-1}-\frac{3s}{\epsilon-1}.\eea
Let us now discuss the implications of these newly defined mass parameters in the consecutive five phases which are appended below point-wise:
\begin{itemize}
    \item In the case of the inflationary phase i.e. for SRI, USR, and SRII phases the first SR parameter is very small where $\epsilon\ll 1$. In those cases, the effective mass parameter is described by the following simplified expression:
    \bea \nu\approx\frac{3}{2}+\epsilon-\eta-3s.\eea

    \item In the case of matter contraction and bouncing scenarios the first slow-roll parameter takes the value $\epsilon=3/2$. In these cases, the effective mass parameter is described by the following simplified expression:
     \bea \nu= \frac{1}{2}+2+2\eta+6s=\frac{5}{2}+2\eta+6s.\eea

     \item In the case of the ekpyrotic contraction and bounce scenarios, the first SR parameter has to be $\epsilon>3$ i.e. say $\epsilon=7/2(=3.5)$. Consequently, the effective mass parameter is described by the following simplified expression:
     \bea \nu= \frac{1}{2}+\frac{2}{5}+\frac{2}{5}\eta+\frac{6}{5}s=\frac{9}{10}+\frac{2}{5}\eta+\frac{6}{5}s.\eea

\end{itemize}

The physical framework under consideration in this paper consists of the five phases listed below, which are discussed in chronological order point-wise:
 \begin{enumerate}
        \item \underline{\bf Phase I:} We first examine the region where we are realizing the ekpyrotic/matter contraction scenarios that appear over the conformal time scale window $\tau_{c}<\tau<\tau_b$. The construction period starts at $\tau=\tau_c$ and has the transition to the next bouncing phase at the scale $\tau=\tau_b$. This suggests that in this arrangement, Phase I terminates at $\tau=\tau_b$. The corresponding momentum scale associated with this window is given by, $k_c<k<k_b$. Within this interval, 
        the first and second SR parameters satisfy the following constraints for matter contraction:
            \bea \epsilon=\frac{3}{2} \quad \quad {\rm and}\quad \quad \eta=0.\eea
            Similarly, for the ekpyrotic contraction, we have the following constraints:
 \bea \epsilon>3 \quad \quad {\rm and}\quad \quad \eta=0.\eea
 
          \item \underline{\bf Phase II:} Further, we examine the region where we are realizing the ekpyrotic/matter bouncing scenarios that appear over the conformal time scale window $\tau_{b}<\tau<\tau_i$. The construction period starts at $\tau=\tau_b$ where the bounce occurs and has the transition to the next first slow-roll phase at the scale $\tau=\tau_i$. This suggests that in this arrangement, Phase II terminates at $\tau=\tau_i$. The corresponding momentum scale associated with this window is given by, $k_b<k<k_i$. Within this interval, 
        the first and second slow-roll parameters satisfy the following constraints for matter bounce:
            \bea \epsilon=\frac{3}{2} \quad \quad {\rm and}\quad \quad 0<\eta<1.\eea
 Similarly, for the ekpyrotic bounce, we have the following constraints:
 \bea \epsilon>3 \quad \quad {\rm and}\quad \quad 0<\eta<2.\eea
 
       \item \underline{\bf Phase III:} Next, we examine an area where we are realizing the first Slow Roll (SRI) that endures over the conformal time scale $\tau_i<\tau<\tau_s$. The SRI starts at $\tau=\tau_i$ and transits to an Ultra Slow Roll (USR) phase at the scale $\tau=\tau_s$. This suggests that in this arrangement, Phase III terminates at $\tau=\tau_s$. The corresponding momentum scale associated with this window is given by, $k_i<k<k_s$. Within this interval, 
        the first and second SR parameters satisfy the following constraints for SRI phase:
            \bea \epsilon\sim 10^{-3} \quad \quad {\rm and}\quad \quad \eta\sim -10^{-3}.\eea

       \item \underline{\bf Phase IV:} Next, we analyse a region known as the Ultra Slow Roll (USR), which ends at the scale $\tau=\tau_e$ and begins at the conformal time scale $\tau=\tau_s$. On the other hand, $\tau=\tau_e$ represents the second steep transition scale from USR to SRII. The corresponding momentum scale associated with this window is given by, $k_s<k<k_e$. Within this interval, 
        the first and second SR parameters satisfy the following constraints for the USR phase:
            \bea 10^{-15}<\epsilon< 10^{-3} \quad \quad {\rm and}\quad \quad \eta\sim -6.\eea
In the USR phase, $\epsilon $ becomes extremely suppressed and the conformal time-dependence can be described in terms of SRI counterpart as, 
\be \epsilon(\tau)=\epsilon \;\left(\frac{a(\tau_s)}{a(\tau)}\right)^{6}=\epsilon  \;\left(\frac{\tau}{\tau_s}\right)^{6}.\ee

       \item \underline{\bf Phase V:} Finally, we investigate the second Slow Roll (SRII) region, where inflation finishes at $\tau=\tau_{\rm end}$ after a short amount of time, commencing at $\tau=\tau_e$. The corresponding momentum scale associated with this window is given by, $k_e<k<k_{\rm end}$. Within this interval, 
        the first and second slow-roll parameters satisfy the following constraints for the USR phase:
            \bea 10^{-15}<\epsilon< 1 \quad \quad {\rm and}\quad \quad \eta\sim -1.\eea
In this phase the explicit conformal time dependence of the first slow-roll parameter can be further expressed in terms of its SRI counterpart as:
\bea 
\epsilon(\tau)= \epsilon \left(\frac{a(\tau_s)}{a(\tau_e)}\right)^6 = \epsilon \left(\frac{\tau_e}{\tau_s}\right)^6.
\eea

\end{enumerate}
In the ensuing subsections of this review, our task is to explicitly investigate the classical solution and its quantum consequences from these three locations independently.

Before going into further details of the computations, let us first outline some crucial facts we have maintained in the rest of the analysis. We have considered that the vacuum structure remains the same in the contraction, bounce, and SRI phases. Further vacuum structure changes in the USR and SRII stages due to having two consecutive phase transitions which are implemented in terms of sharp transitions in the boundaries of the USR phase. At the starting point in principle, one can choose any arbitrary vacuum state that preserves the norm. However, for practical purposes in the contraction, bounce, and SRI phases we have considered the well-known Bunch Davies type of vacuum which changes its structure in the USR and SRII phases due to the implementation of the sharp transition. Another important point to be noted here is that the conclusions derived in this paper will be completely unchanged if we insert smooth transitions instead of the sharp ones. Recently such possibility is explicitly pointed in ref \cite{Choudhury:2024ybk}.

 \subsection{Phase I: Ekpyrotic/Matter contraction}
For Phase I which describes ekpyrotic and matter contraction scenarios depending on the previously mentioned values of the first slow roll parameter $\epsilon$, we get the following general solution:
 \bea \label{modezetacon}
{\bf \zeta}_{\bf C}&=&\frac{2^{\nu-\frac{3}{2}} c_s (-k c_s \tau )^{\frac{3}{2}-\nu}}{ia_0\tau \sqrt{2 \epsilon_*}(k c_s)^{\frac{3}{2}}\sqrt{2}  M_{pl}}\left(\frac{\tau}{\tau_0}\right)^{-\frac{1}{(\epsilon-1)}}\sqrt{\left(\frac{\epsilon_*}{\epsilon_c}\right)}\Bigg|\frac{\Gamma(\nu)}{\Gamma(\frac{3}{2})}\Bigg|\nonumber\\
&&\quad\quad\quad\times\Bigg\{\alpha_1 (1+i k c_s\tau) e^{-i\left(k c_s\tau+\frac{\pi}{2}\left(\nu+\frac{1}{2}\right)\right)}-\beta_1(1-i k c_s \tau)e^{i\left(k c_s\tau+\frac{\pi}{2}\left(\nu+\frac{1}{2}\right)\right)}\Bigg\}.\quad\quad
\eea
Here $\alpha_1$ and $\beta_1$ represent the Bogoliubov coefficients that characterize the vacuum structure of the corresponding contraction scenarios. The above-mentioned general solution of the MS equation for the ekpyrotic and matter contraction is written for a general vacuum state, which preserves the normalization condition, $|\alpha_1|^2-|\beta_1|^2=1$. Henceforth, for the sake of simplicity, we consider that the Bunch Davies vacuum state describes the initial vacuum state during the contraction phase. In terms of the Bogoliubov coefficients, it is described by, $\alpha_1=1$ and $\beta_1=0$, which satisfy the previously mentioned normalization criteria. Substituting these values in equation (\ref{modezetacon}) in the presence of Bunch Davies quantum initial state we get the following simplified result for the gauge invariant comoving curvature perturbation:
\bea \label{modezetacon1}
{\bf \zeta}_{\bf C}=\frac{2^{\nu-\frac{3}{2}} c_s (-k c_s \tau )^{\frac{3}{2}-\nu}}{ia_0\tau \sqrt{2 \epsilon_*}(k c_s)^{\frac{3}{2}}\sqrt{2} M_{pl}}\left(\frac{\tau}{\tau_0}\right)^{-\frac{1}{(\epsilon-1)}}\sqrt{\left(\frac{\epsilon_*}{\epsilon_c}\right)}\Bigg|\frac{\Gamma(\nu)}{\Gamma(\frac{3}{2})}\Bigg | (1+i k c_s\tau) e^{-i\left(k c_s\tau+\frac{\pi}{2}\left(\nu+\frac{1}{2}\right)\right)}.\quad\quad
\eea

 \subsection{Phase II: Ekpyrotic/Matter bounce}
For Phase II which describes ekpyrotic and matter bounce scenarios depending on the previously mentioned values of the first slow roll parameter $\epsilon$, we get the following general solution:
 \bea \label{modezetabou}
&&{\bf \zeta}_{\bf B}=\frac{2^{\nu-\frac{3}{2}} c_s (-k c_s \tau )^{\frac{3}{2}-\nu}}{ia_0\tau \sqrt{2 \epsilon_*}(k c_s)^{\frac{3}{2}}\sqrt{2}  M_{pl}}\left[1+\left(\frac{\tau}{\tau_0}\right)^2\right]^{-\frac{1}{2(\epsilon-1)}}\sqrt{\left(\frac{\epsilon_*}{\epsilon_b}\right)}\Bigg|\frac{\Gamma(\nu)}{\Gamma(\frac{3}{2})}\Bigg |\nonumber\\
&&\quad\quad\quad\quad\quad\quad\quad\quad\quad\quad\quad\quad\quad\quad\times\Bigg\{\alpha_1 (1+i k c_s\tau) e^{-i\left(k c_s\tau+\frac{\pi}{2}\left(\nu+\frac{1}{2}\right)\right)}-\beta_1(1-i k c_s \tau)e^{i\left(k c_s\tau+\frac{\pi}{2}\left(\nu+\frac{1}{2}\right)\right)}\Bigg\}.\quad\quad
\eea
Here $\alpha_1$ and $\beta_1$ represent the Bogoliubov coefficients which characterize the vacuum structure of the corresponding bouncing scenarios. Also, it is important to note that we have assumed that the vacuum structure remains the same in the bouncing scenario with the initial contraction phase. The above-mentioned general solution of the MS equation for the ekpyrotic and matter bounce is written for a general vacuum state, which preserves the normalization condition, $|\alpha_1|^2-|\beta_1|^2=1$. Henceforth, for the sake of simplicity, we consider that the Bunch Davies vacuum state describes the initial vacuum state during the bouncing phase. In terms of the Bogoliubov coefficients, it is described by, $\alpha_1=1$ and $\beta_1=0$, which satisfy the previously mentioned normalization criteria. Substituting these values in equation (\ref{modezetabou}) in the presence of Bunch Davies quantum initial state we get the following simplified result for the gauge invariant comoving curvature perturbation:
\bea \label{modezetabou1}
{\bf \zeta}_{\bf B}=\frac{2^{\nu-\frac{3}{2}} c_s (-k c_s \tau )^{\frac{3}{2}-\nu}}{ia_0\tau \sqrt{2 \epsilon_*}(k c_s)^{\frac{3}{2}}\sqrt{2}  M_{pl}}\left[1+\left(\frac{\tau}{\tau_0}\right)^2\right]^{-\frac{1}{2(\epsilon-1)}}\Bigg|\frac{\Gamma(\nu)}{\Gamma(\frac{3}{2})}\Bigg |\sqrt{\left(\frac{\epsilon_*}{\epsilon_b}\right)}(1+i k c_s\tau) e^{-i\left(k c_s\tau+\frac{\pi}{2}\left(\nu+\frac{1}{2}\right)\right)}.\quad\quad
\eea

 \subsection{Phase III: First slow-roll (SRI)}

 For Phase III which describes the first slow-roll scenario, we get the following general solution:
 \bea \label{modezetasr1}
{\bf \zeta}_{\bf SRI}=\frac{2^{\nu-\frac{3}{2}} c_s H (-k c_s \tau )^{\frac{3}{2}-\nu}}{i \sqrt{2 \epsilon_*}(k c_s)^{\frac{3}{2}}\sqrt{2} M_{pl}}\Bigg|\frac{\Gamma(\nu)}{\Gamma(\frac{3}{2})}\Bigg |\Bigg\{\alpha_1 (1+i k c_s\tau) e^{-i\left(k c_s\tau+\frac{\pi}{2}\left(\nu+\frac{1}{2}\right)\right)}-\beta_1(1-i k c_s \tau)e^{i\left(k c_s\tau+\frac{\pi}{2}\left(\nu+\frac{1}{2}\right)\right)}\Bigg\}.
\eea
Here $\alpha_1$ and $\beta_1$ represent the Bogoliubov coefficients that characterize the vacuum structure of the corresponding SRI region. Here we have assumed that the vacuum structure remains the same in the SRI region with the previously discussed contraction and bouncing phase. The above-mentioned general solution of the MS equation for the SRI phase is written for a general vacuum state, which preserves the normalization condition, $|\alpha_1|^2-|\beta_1|^2=1$. Henceforth, for the sake of simplicity, we consider that the Bunch Davies vacuum state describes the initial vacuum state during the SRI phase. In terms of the Bogoliubov coefficients, it is described by, $\alpha_1=1$ and $\beta_1=0$, which satisfy the previously mentioned normalization criteria. Substituting these values in equation (\ref{modezetasr1}) in the presence of Bunch Davies quantum initial state we get the following simplified result for the gauge invariant comoving curvature perturbation:
\bea \label{modezetasr1a}
{\bf \zeta}_{\bf SRI}=\frac{2^{\nu-\frac{3}{2}} c_s H (-k c_s \tau )^{\frac{3}{2}-\nu}}{i \sqrt{2 \epsilon_*}(k c_s)^{\frac{3}{2}}\sqrt{2} M_{pl}}\Bigg|\frac{\Gamma(\nu)}{\Gamma(\frac{3}{2})}\Bigg | (1+i k c_s\tau) e^{-i\left(k c_s\tau+\frac{\pi}{2}\left(\nu+\frac{1}{2}\right)\right)}.
\eea

 \subsection{Phase IV: Ultra slow-roll (USR)}
  For Phase III which describes the first slow-roll scenario, we get the following general solution:
 \bea \label{modezetausr}
{\bf \zeta}_{\bf USR}=\frac{2^{\nu-\frac{3}{2}} c_s   H (-k c_s \tau )^{\frac{3}{2}-\nu}}{i \sqrt{2 \epsilon_*}(k c_s)^{\frac{3}{2}}\sqrt{2} M_{pl}}\bigg(\frac{\tau_s}{\tau}\bigg)^3\Bigg|\frac{\Gamma(\nu)}{\Gamma(\frac{3}{2})}\Bigg |\Bigg\{\alpha_2 (1+i k c_s\tau) e^{-i\left(k c_s\tau+\frac{\pi}{2}\left(\nu+\frac{1}{2}\right)\right)}-\beta_2(1-i k c_s \tau)e^{i\left(k c_s\tau+\frac{\pi}{2}\left(\nu+\frac{1}{2}\right)\right)}\Bigg\}.\quad\quad
\eea
Here, the Bogoliubov coefficients in the USR region are $\alpha_2$ and $\beta_2$, which we can obtain in terms of the Bogoliubov coefficient of the previous phases, $\alpha_1$ and $\beta_1$ by applying the two boundary conditions, which reads as Israel junction conditions implemented at the sharp transition instance $\tau = \tau_s$. These constraint conditions are given by:
\begin{enumerate}
    \item First condition states that the comoving curvature modes as obtained for the scalar perturbation become continuous at the scale $\tau=\tau_s$, where the sharp transition in implemented i.e. $\left[\zeta_{\bf SRI}\right]_{\tau=\tau_s}=\left[\zeta_{\bf USR}\right]_{\tau=\tau_s}$.

    \item Second condition states that the canonically conjugate momenta obtained from the comoving curvature modes become continuous at the scale $\tau=\tau_s$, where the sharp transition in implemented i.e. $\left[\Pi_{\zeta_{\bf SRI}}\right]_{\tau=\tau_s}=\left[\Pi_{\zeta_{\bf USR}}\right]_{\tau=\tau_s}$. Here the canonically conjugate momenta corresponding to the computed modes are defined as, $\Pi_{\zeta}=\zeta^{'}(\tau)$.
\end{enumerate}
Before going to the further details of the computation here, it is essential to note that, due to the implementation of the sharp transitions at the instant $\tau=\tau_s$, where the USR phase starts, and SRI phase transits to the USR phase, the corresponding vacuum structure changes in the USR phase. Such a changing vacuum is the direct consequence of the phase transition that happened at the SRI and USR boundary. 

We now derive two determining equations for Bogoliubov coefficients in the USR phase after applying the two junction conditions mentioned above:
\bea \label{alpha2a}
&& \alpha_2 = \frac{1}{2 k^3 \tau_s^3 c_s^3} \Bigg\{ \Bigg(3 i + 3 i  k^2 c_s^2 \tau_s ^2  + 2  k^3 c_s^3  \tau_s ^3  \Bigg  )\alpha_1 - \Bigg(3 i +6  k c_s \tau_s  -3 i k^2  c_s^2 \tau_s ^2    \Bigg) \beta_1 e^{i \left(2  k \tau_s  c_s +  \pi  \left(\nu +\frac{1}{2}\right)\right)}\Bigg\},
 \\
&&  \label{beta2a} \beta_2 = \frac{1}{2 k ^3 c_s ^3 \tau_s ^3} \Bigg\{ \Bigg( 3i -6 k c_s \tau_s -3i k^2 c_s ^2 \tau_s^2 \Bigg) \alpha_1 e^{-i\left(\pi\left(\nu+\frac{1}{2}\right)+ 2k c_s \tau_s\right)}- \Bigg(3i +3 i k^2 c_s ^2 \tau_s ^ 2 - 2 k^3c_s^3 \tau_s ^ 3  \Bigg )\beta_1
 \Bigg\}.
\eea 
Here from the above-mentioned expressions, it is clearly visible that the Bogolibov coefficients that characterize the USR phase i.e. $\alpha_2$ and $\beta_2$ are now expressed in terms of the Bogoliubov coefficients,  $\alpha_1$ and $\beta_1$ which describe the contraction, bounce, and SRI phases in terms of general vacuum state. In the case, where we start with the Bunch Davies initial condition, we fix $\alpha_1=1$ and $\beta_1=0$, which further gives the following simplified results:
\bea \label{alpha2b}
&& \alpha_{{2}} =  \frac{1}{2 k^3 \tau_s^3 c_s^3}  \Bigg(3 i + 3 i  k^2 c_s^2 \tau_s ^2  + 2  k^3 c_s^3  \tau_s ^3  \Bigg  ), \\
&& \label{beta2b}  \beta_{{2}} = \frac{1}{2 k ^3 c_s ^3 \tau_s ^3}  \Bigg( 3i -6 k c_s \tau_s -3i k^2 c_s ^2 \tau_s^2 \Bigg) e^{-i\left(\pi\left(\nu+\frac{1}{2}\right)+ 2k c_s \tau_s\right)}.
\eea 

 \subsection{Phase V: Second slow-roll (SRII)}

  For Phase III which describes the first slow-roll scenario, we get the following general solution:
 \bea \label{modezetasr2}
{\bf \zeta}_{\bf SRII}=\frac{2^{\nu-\frac{3}{2}} c_s   H (-k c_s \tau )^{\frac{3}{2}-\nu}}{i \sqrt{2 \epsilon_*}(k c_s)^{\frac{3}{2}}\sqrt{2} M_p}\left(\frac{\tau_s}{\tau_e}\right)^3\Bigg|\frac{\Gamma(\nu)}{\Gamma(\frac{3}{2})}\Bigg |\Bigg\{\alpha_3 (1+i k c_s\tau) e^{-i\left (kc_s\tau+\frac{\pi}{2}(\nu+\frac{1}{2})\right )}-\beta_3(1-i k c_s \tau)e^{i\left(k c_s\tau+\frac{\pi}{2}(\nu+\frac{1}{2})\right)}\Bigg\}.\quad\quad
\eea
Here, the Bogoliubov coefficients in the SRII region are $\alpha_3$ and $\beta_3$, which we can obtain in terms of the Bogoliubov coefficient of the previous phases, $\alpha_2$ and $\beta_2$ by applying the two boundary conditions, which reads as Israel junction conditions implemented at sharp transition instance $\tau = \tau_e$. These constraint conditions are given by:
\begin{enumerate}
    \item First condition states that the comoving curvature modes as obtained for the scalar perturbation become continuous at the scale $\tau=\tau_s$, where the sharp transition in implemented i.e. $\left[\zeta_{\bf USR}\right]_{\tau=\tau_e}=\left[\zeta_{\bf SRII}\right]_{\tau=\tau_e}$.

    \item Second condition states that the canonically conjugate momenta obtained from the comoving curvature modes become continuous at the scale $\tau=\tau_e$, where the sharp transition in implemented i.e. $\left[\Pi_{\zeta_{\bf USR}}\right]_{\tau=\tau_e}=\left[\Pi_{\zeta_{\bf SRII}}\right]_{\tau=\tau_e}$. Here the canonically conjugate momenta corresponding to the computed modes are defined as, $\Pi_{\zeta}=\zeta^{'}(\tau)$.
\end{enumerate}
Before going to the further details of the computation here, it is essential to note that, due to the implementation of the sharp transitions at the instant $\tau=\tau_e$, where the USR phase ends and USR phase transits to the SRII phase, the corresponding vacuum structure changes in the SRII phase. Such a changing vacuum is the direct consequence of the phase transition that happened at the USR and SRII boundary. Here, it is essential to note that the inflation ends as the SRII phase ends at the instant $\tau=\tau_{\rm end}$ where the magnitude of the second SR parameter $\eta$ reaches unity, i.e., $|\eta(\tau_{\rm end})|=1$.

We now derive two determining equations for Bogoliubov coefficients in the SRII phase after applying the two junction conditions mentioned above:
\bea \label{alpha3a}
 && \alpha _3 =  \frac{1}{2 k^3 \tau_e^3 c_s^3}\Bigg\{\left(-3 i -3 i  k^2 \tau_e^2 c_s^2 +2  k^3 \tau_e^3 c_s^3 \right)\alpha _2 - \left(-3 i -6  k \tau_e c_s   +3 i  k^2 \tau_e^2 c_s^2 \right)\beta _2 e^{\left(2 i k \tau_e c_s+i \pi  \left(\nu +\frac{1}{2}\right)\right)}\Bigg\},\\
&&  \label{beta3a}  \beta _3 =  \frac{1}{2 k^3 \tau_e^3 c_s^3}\Bigg\{ \left(-3 i  +6  k \tau_e c_s +3 i k^2 \tau_e^2 c_s^2 \right)\alpha _2  e^{-\left(2 i k \tau_e c_s + i \pi  \left(\nu +\frac{1}{2}\right)\right)}+ \left(2  k^3 \tau_e^3 c_s^3 +3 i  k^2 \tau_e^2 c_s^2 +3 i  \right)\beta _2\Bigg\}, 
\eea 
Here from the above-mentioned expressions, it is clearly visible that the Bogolibov coefficients that characterize the SRII phase i.e. $\alpha_3$ and $\beta_3$ are now expressed in terms of the Bogoliubov coefficients $\alpha_2$ and $\beta_2$ which describe the USR phase in terms of shifted new vacuum state. Substituting the expressions of $\alpha_2$ and $\beta_2$ in eq.(\ref{alpha3a}) and eq.(\ref{beta3a}) we can express the Bogolibov coefficients $\alpha_3$ and $\beta_3$ in terms of the Bogoliubov coefficients $\alpha_1$ and $\beta_1$ that characterize the initial vacuum structure:
\bea
  \alpha _3 &=&  \frac{1}{(2 k^3 \tau_e^3 c_s^3)(2 k^3 \tau_s^3 c_s^3)}\Bigg[\left(-3 i -3 i  k^2 \tau_e^2 c_s^2 +2  k^3 \tau_e^3 c_s^3 \right) \Bigg\{ \left(3 i + 3 i  k^2 c_s^2 \tau_s ^2  + 2  k^3 c_s^3  \tau_s ^3   \right )\alpha_1   - \left(3 i +6  k c_s \tau_s  -3 i k^2  c_s^2 \tau_s ^2   \right)
 \nonumber\\ 
&& \quad\quad\quad\quad\quad\quad\quad\quad\quad\quad\quad \times \beta_1  e^{i \left(2  k \tau_s  c_s +  \pi  \left(\nu +\frac{1}{2}\right)\right)}\Bigg\}
 \quad  - \left(-3 i -6  k \tau_e c_s   +3 i  k^2 \tau_e^2 c_s^2 \right) \quad \Bigg\{ \left( 3i -6 k c_s \tau_s -3i k^2 c_s ^2 \tau_s^2 \right)  
 \nonumber \\
  &&\quad\quad\quad \quad\quad\quad \quad\quad\quad\quad\quad \times \alpha_1 e^{-i(\pi(\nu+\frac{1}{2})+ 2k c_s \tau_s)} - \left (3i +3 i k^2 c_s ^2 \tau_s ^ 2 - 2 k^3c_s^3 \tau_s ^ 3  \right )\beta_1 \quad 
 \Bigg\} e^{\left(2 i k \tau_e c_s+i \pi  \left(\nu +\frac{1}{2}\right)\right)}\Bigg],
 \\
 \beta_3 &=& \frac{1}{(2 k^3 \tau_e^3 c_s^3)(2 k^3 \tau_s^3 c_s^3)}\Bigg[] \left(-3 i  +6  k \tau_e c_s +3 i k^2 \tau_e^2 c_s^2\right) \Bigg\{ \left(3 i + 3 i  k^2 c_s^2 \tau_s ^2  + 2  k^3 c_s^3  \tau_s ^3  \right)\alpha_1 - \left(3 i +6  k c_s \tau_s  -3 i k^2  c_s^2 \tau_s ^2    \right) \nonumber \\
 && \quad \quad\quad\quad\quad\quad\quad \quad\quad\times \beta_1 e^{i \left(2  k \tau_s  c_s +  \pi  \left(\nu +\frac{1}{2}\right)\right)}\Bigg\}  e^{-\left(2 i k \tau_e c_s + i \pi  \left(\nu +\frac{1}{2}\right)\right)}+ \left(2  k^3 \tau_e^3 c_s^3 +3 i  k^2 \tau_e^2 c_s^2 +3 i  \right )  \nonumber \\
 &&  \quad \quad\quad\quad\quad\quad\quad\quad\quad\times\Bigg\{ \left( 3i -6 k c_s \tau_s -3i k^2 c_s ^2 \tau_s^2 \right) \alpha_1 e^{-i\left(\pi(\nu+\frac{1}{2})+ 2k c_s \tau_s\right)}
 - \left(3i +3 i k^2 c_s ^2 \tau_s ^ 2 - 2 k^3c_s^3 \tau_s ^ 3  \right )\beta_1
 \Bigg\}\Bigg].
\quad\quad\quad \eea 
Further choosing the initial vacuum state is described by Bunch Davies states i.e. fixing $\alpha_1=1$ and $\beta_1=0$ we get the following simplified expressions for the Bogoliubov coefficients which characterize the SRII phase:
\bea
 \alpha _{3} &=& \frac{1}{(2 k^3 \tau_e^3 c_s^3)(2 k^3 \tau_s^3 c_s^3)}\Bigg\{\left(-3 i -3 i  k^2 \tau_e^2 c_s^2 +2  k^3 \tau_e^3 c_s^3 \right) \left(3 i + 3 i  k^2 c_s^2 \tau_s ^2  + 2  k^3 c_s^3  \tau_s ^3  \right  )  \nonumber\\ &&
\quad\quad\quad\quad\quad\quad\quad\quad\quad\quad - \left(-3 i -6  k \tau_e c_s   +3 i  k^2 \tau_e^2 c_s^2 \right)
 \left( 3i -6 k c_s \tau_s -3i k^2 c_s ^2 \tau_s^2 \right) e^{2 i k c_s( \tau_e -\tau_s) }\Bigg\},
 \\
 \beta _{3} &=&  \frac{1}{(2 k^3 \tau_e^3 c_s^3)(2 k ^3 c_s ^3 \tau_s ^3)}\Bigg\{ \left(-3 i  +6  k \tau_e c_s +3 i k^2 \tau_e^2 c_s^2\right) \left(3 i + 3 i  k^2 c_s^2 \tau_s ^2  + 2  k^3 c_s^3  \tau_s ^3  \right) e^{-\left(2 i k \tau_e c_s + i \pi  \left(\nu +\frac{1}{2}\right)\right)} \nonumber \\ 
&&\quad\quad\quad\quad\quad\quad\quad\quad\quad\quad +\left(2  k^3 \tau_e^3 c_s^3 + 3 i  k^2 \tau_e^2 c_s^2 +3 i \right )  \left( 3i -6 k c_s \tau_s -3i k^2 c_s ^2 \tau_s^2 \right) e^{-i\left(\pi(\nu+\frac{1}{2})+ 2k c_s \tau_s \right)}\Bigg\}.
\eea 

\section{Numerical outcomes II: Behaviour of modes and associated momenta in the five consecutive phases}
\label{s7}

  %%%%%%%%%%%%%%%%%%%%%%%%%%%%%%%%%%%%%%%%%%%%%
\begin{figure*}[htb!]
    	\centering
    \subfigure[]{
      	\includegraphics[width=8.5cm,height=7.5cm]{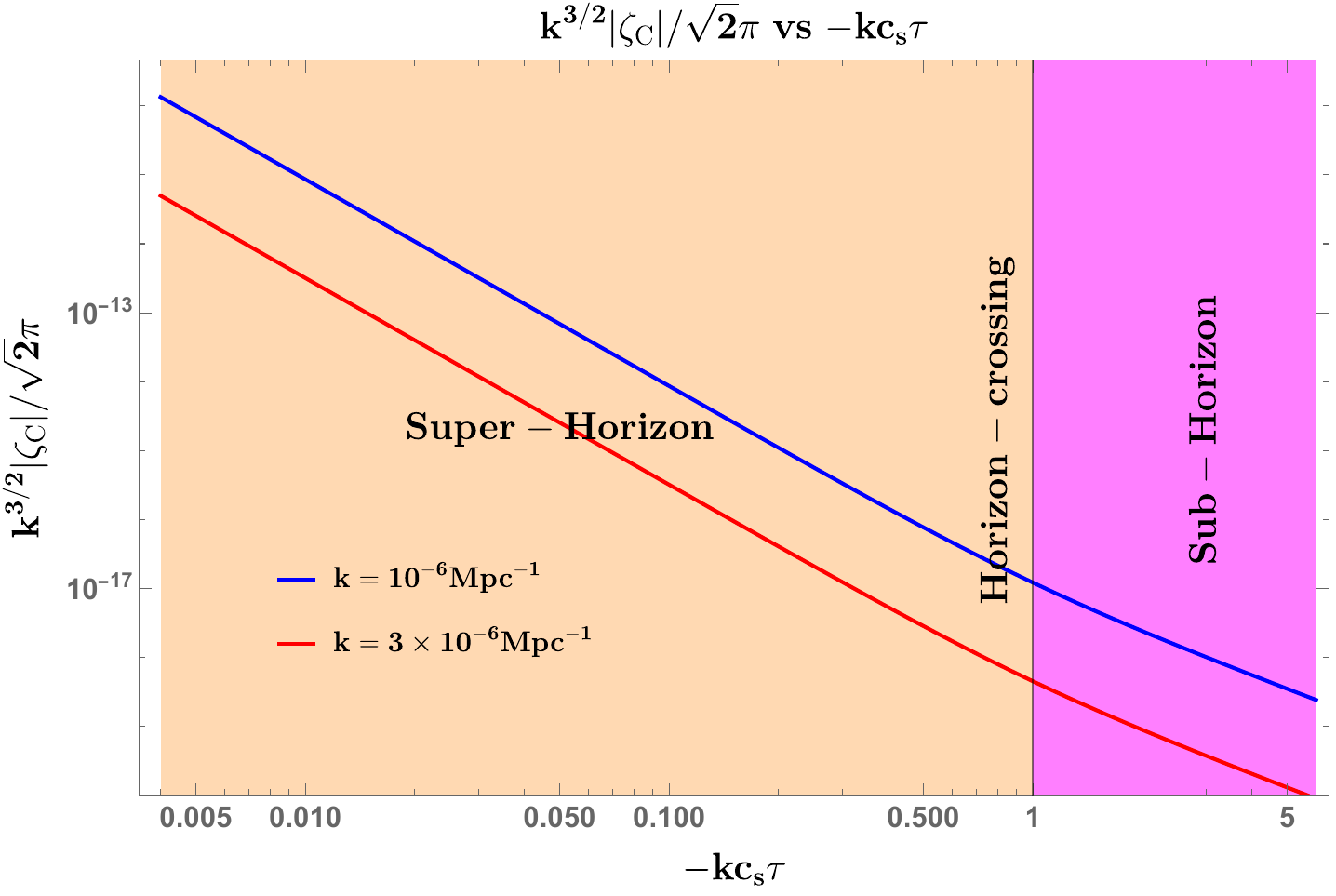}
        \label{zetacontract}
    }
    \subfigure[]{
       \includegraphics[width=8.5cm,height=7.5cm]{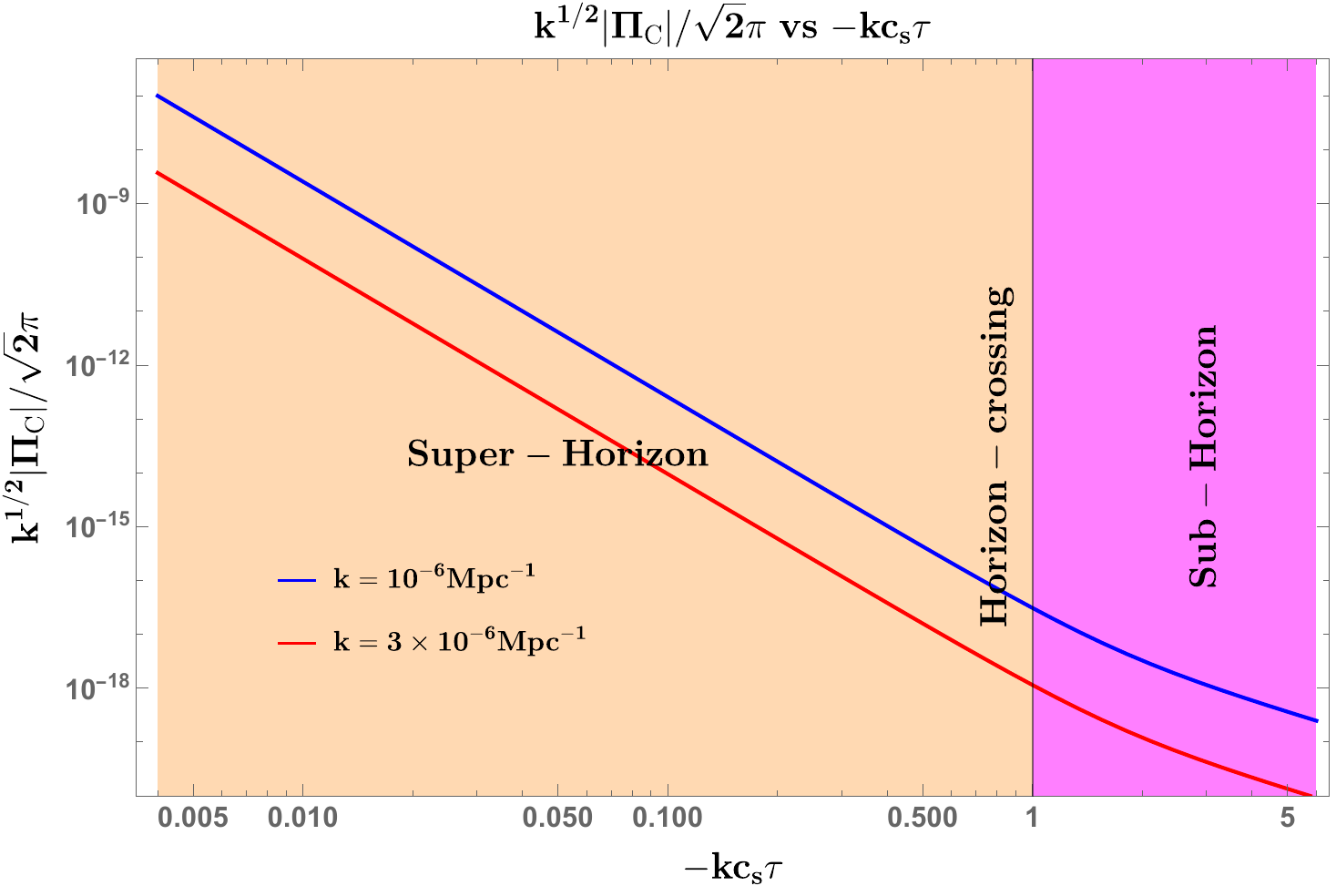}
        \label{picontract}
    }
    	\caption[Optional caption for list of figures]{ Plots show evolution of scalar modes and their associated momenta as a function of $-kc_{s}\tau$ for the contraction phase. In the left, the function $k^{3/2}|\zeta_{C}|/\sqrt{2}\pi$ is plotted and in the right function $k^{1/2}|\Pi_{C}|/\sqrt{2}\pi$. The orange and magenta regions highlight the Super-Horizon ($-kc_{s}\tau\ll 1$) and Sub-Horizon ($-kc_{s}\tau\gg 1$) regions, respectively. } 
    	\label{contractionmodes }
    \end{figure*}
%%%%%%%%%%%%%%%%%%%%%%%%%%%%%%%%%%%%%%%%%%%%%

In fig.(\ref{zetacontract},\ref{picontract}), we introduce the evolution of the scalar mode and its conjugate momenta as they go from the sub-horizon and cross into the super-horizon during the contraction phase of our setup. Both the modes show extremely suppressed strengths when deep inside the Horizon. To study this we encounter the dimensionless number $-k_{0c}c_{s}\tau_{c}$, that we choose to have magnitude ${\cal O}(10^{-2})$, where $k_{0c}$ is the wavenumber corresponding to the reference conformal time $\tau_{0}$ for either matter $(\tau_{mc})$ or ekpyrotic $(\tau_{ec})$ contraction. This combination arises due to our choice of variable $-kc_{s}\tau$, which then brings in a dependence on the wavenumber $k$. We show behaviour for two wavenumbers $k=10^{-6}{\rm Mpc^{-1}}$ (blue) and $k=3\times 10^{-6}{\rm Mpc^{-1}}$ (red) and choice on the range of wavenumbers will become more apparent once we study the scalar power spectrum in the upcoming sections. When coming from inside the Horizon, both the mode and its conjugate momenta evolve in a similar fashion, and a small change in the wavenumber magnitude leads to a difference of almost two orders of magnitude their strengths. After crossing and throughout their remaining evolution in the super-horizon, the modes follow the same behaviour, with the growth of the conjugate momenta always remaining comparable to that of the scalar modes.

  %%%%%%%%%%%%%%%%%%%%%%%%%%%%%%%%%%%%%%%%%%%%%
\begin{figure*}[htb!]
    	\centering
    \subfigure[]{
      	\includegraphics[width=8.5cm,height=7.5cm]{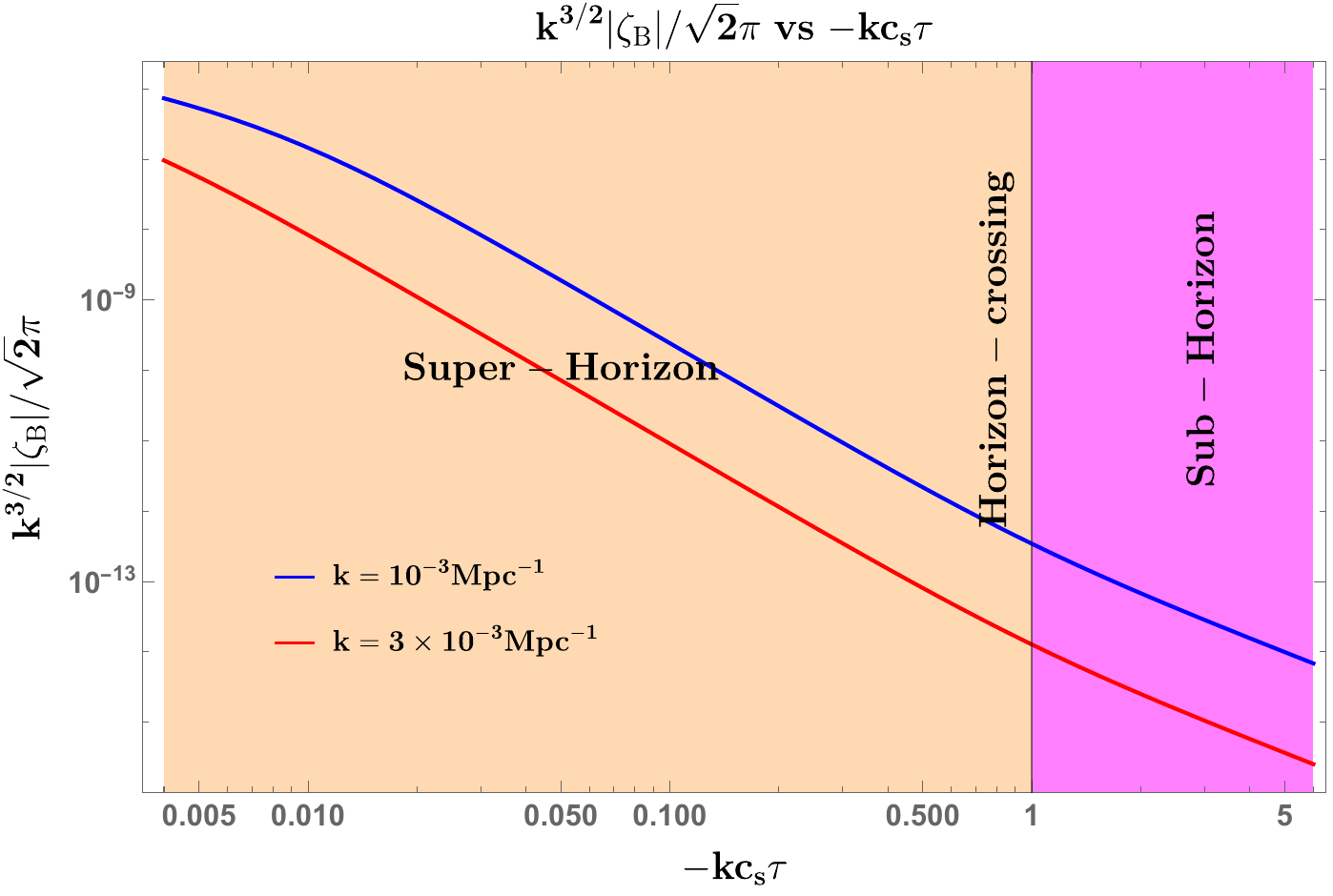}
        \label{zetabounce}
    }
    \subfigure[]{
       \includegraphics[width=8.5cm,height=7.5cm]{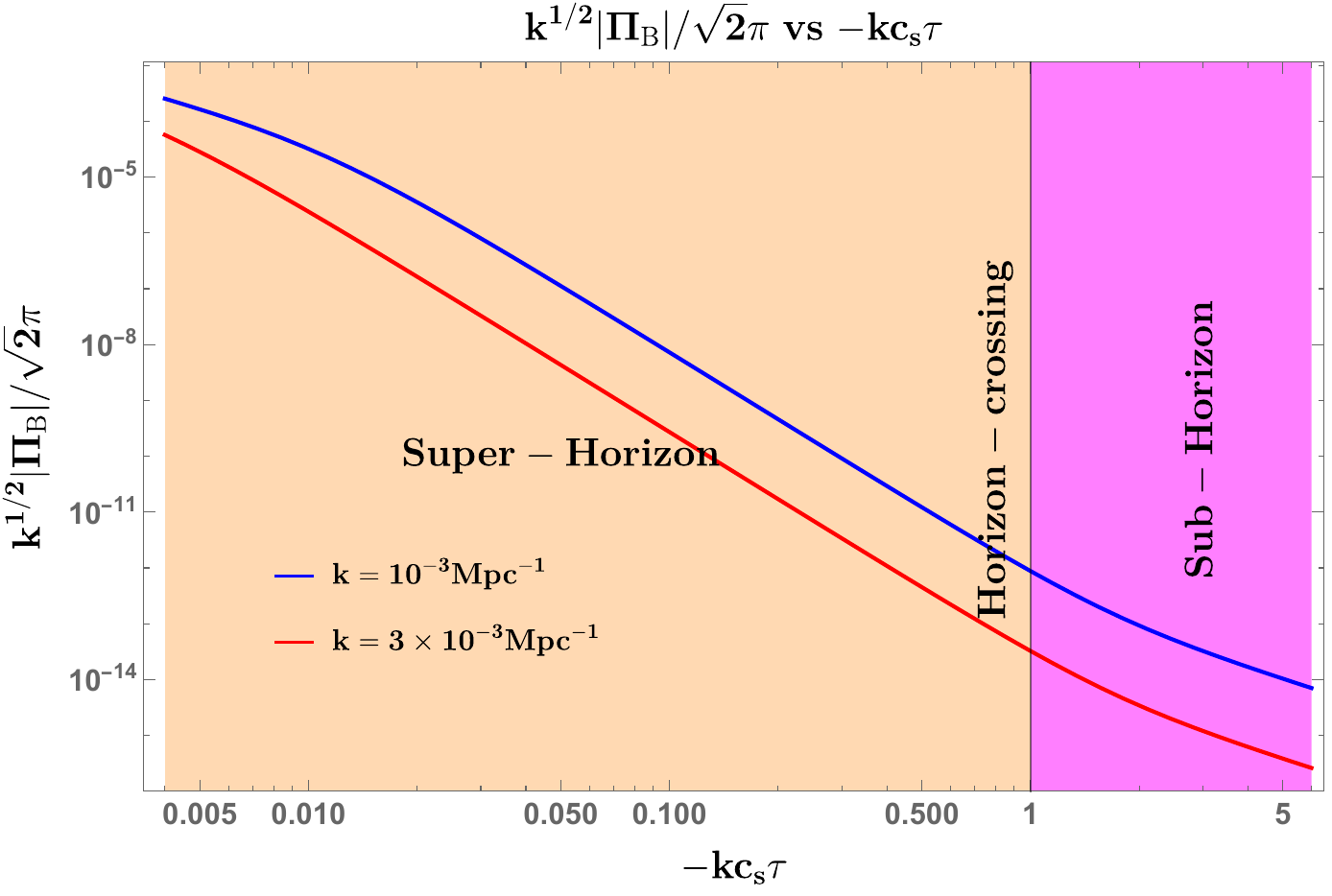}
        \label{pibounce}
    }
    	\caption[Optional caption for list of figures]{ Plots show evolution of scalar modes and their associated momenta as a function of $-kc_{s}\tau$ for the bouncing phase. In the left, the function $k^{3/2}|\zeta_{B}|/\sqrt{2}\pi$ is plotted and in the right function $k^{1/2}|\Pi_{B}|/\sqrt{2}\pi$. The orange and magenta regions highlight the Super-Horizon ($-kc_{s}\tau\ll 1$) and Sub-Horizon ($-kc_{s}\tau\gg 1$) regions, respectively. } 
    	\label{bouncemodes }
    \end{figure*}
%%%%%%%%%%%%%%%%%%%%%%%%%%%%%%%%%%%%%%%%%%%%%

  %%%%%%%%%%%%%%%%%%%%%%%%%%%%%%%%%%%%%%%%%%%%%
\begin{figure*}[htb!]
    	\centering
    \subfigure[]{
      	\includegraphics[width=8.5cm,height=7.5cm]{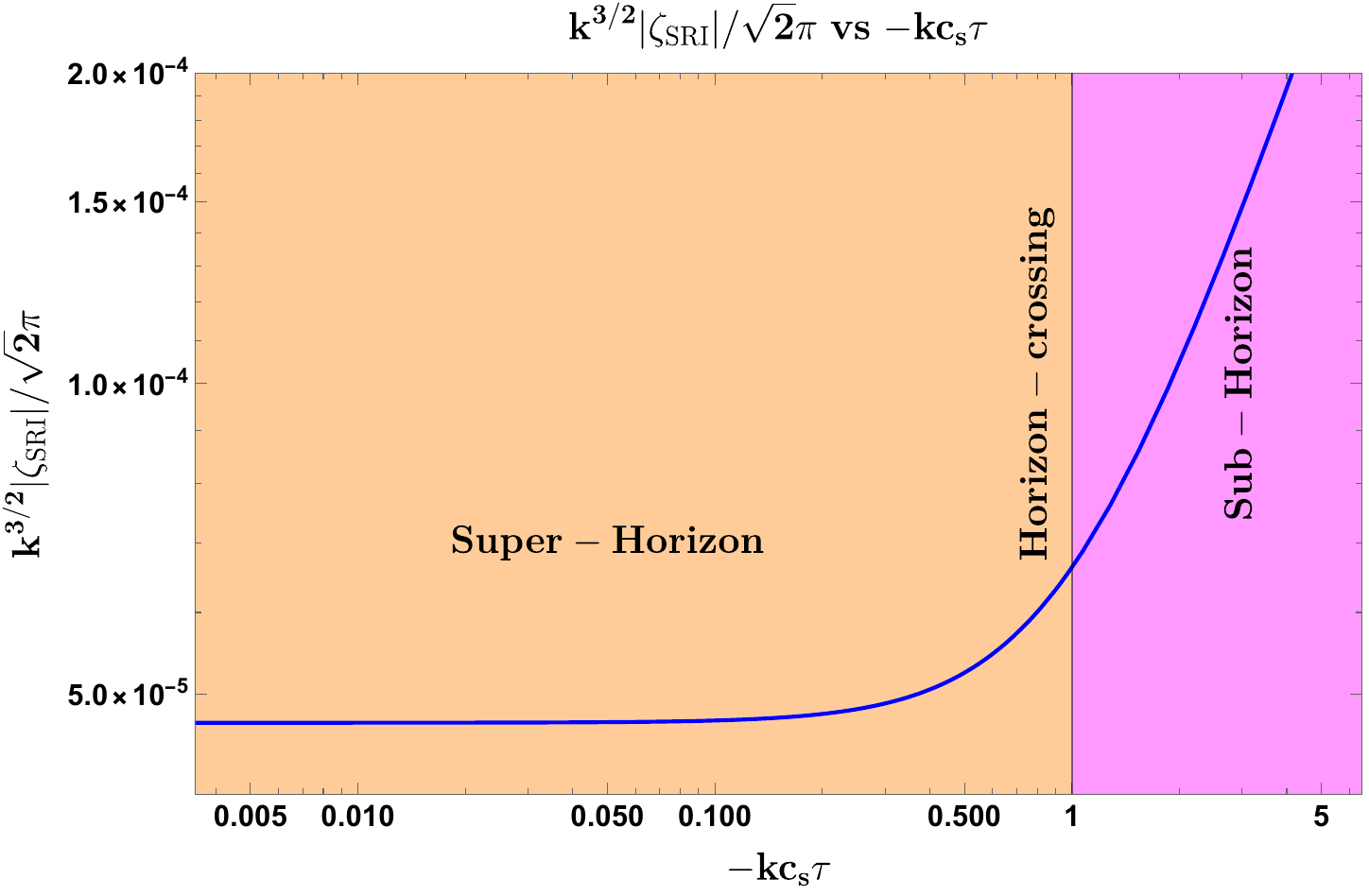}
        \label{zetasr1}
    }
    \subfigure[]{
       \includegraphics[width=8.5cm,height=7.5cm]{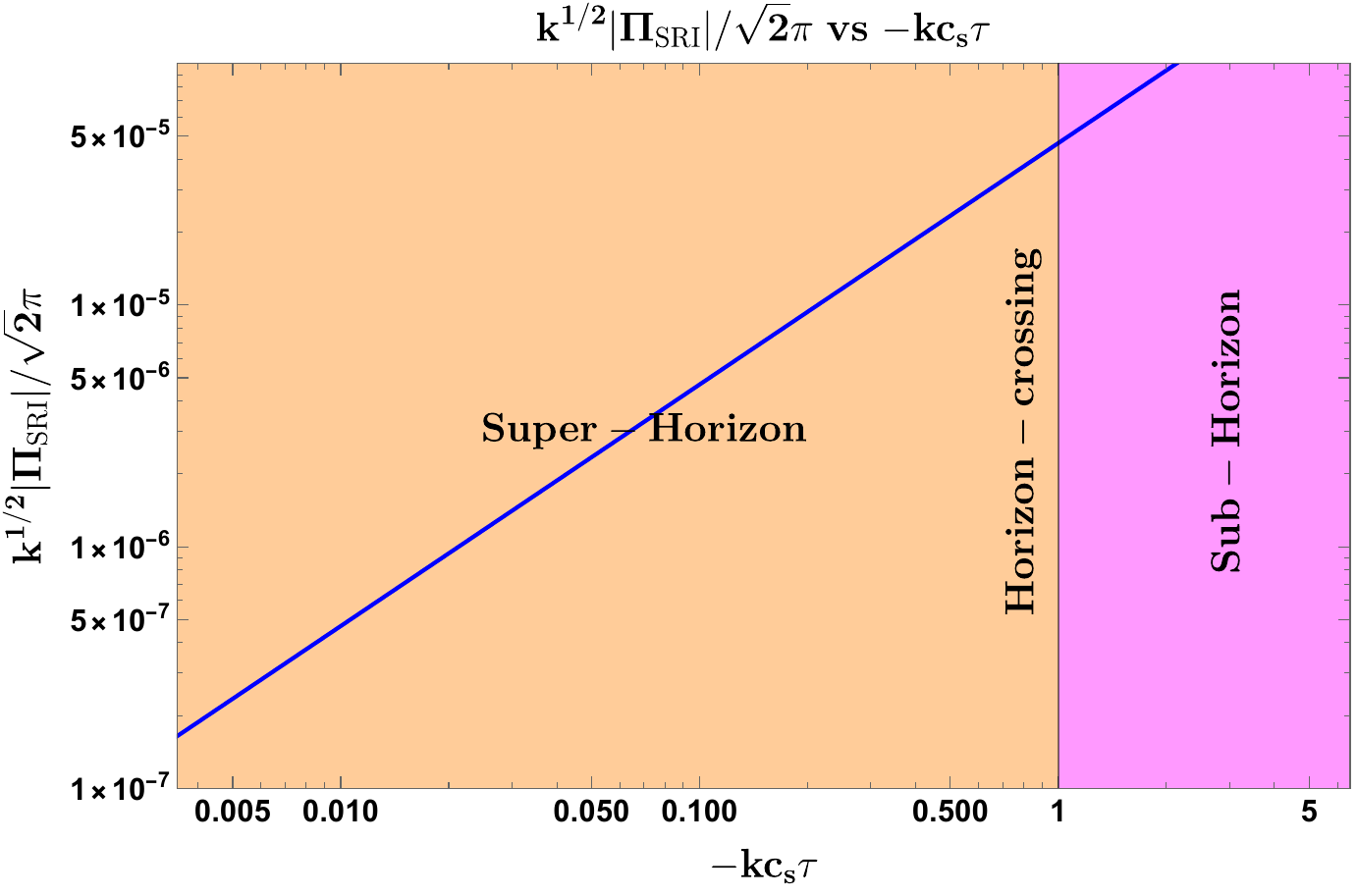}
        \label{pisr1}
    }
    	\caption[Optional caption for list of figures]{ Plots show evolution of scalar modes and their associated momenta as a function of $-kc_{s}\tau$ for the first slow-roll (SRI) phase. In the left, the function $k^{3/2}|\zeta_{\rm SRI}|/\sqrt{2}\pi$ is plotted and in the right function $k^{1/2}|\Pi_{\rm SRI}|/\sqrt{2}\pi$. The orange and magenta regions highlight the Super-Horizon ($-kc_{s}\tau\ll 1$) and Sub-Horizon ($-kc_{s}\tau\gg 1$) regions, respectively. } 
    	\label{sr1modes }
    \end{figure*}
%%%%%%%%%%%%%%%%%%%%%%%%%%%%%%%%%%%%%%%%%%%%%

We move to fig.(\ref{zetabounce},\ref{pibounce}) showing evolution of the scalar modes and their conjugate momenta in the bouncing phase. The behaviour for both type of modes remain almost similar during their journey from the sub-horizon and into the super-horizon regimes. Similar to the analysis in the contraction phase here we require to set another dimensionless variable $-k_{0b}c_{s}\tau_{0}$, which we set to ${\cal O}(10^{-2})$, where $k_{0b}$ is the wavenumber corresponding to the reference conformal time $\tau_{0}$ for either matter $(\tau_{mb})$ or ekpyrotic $(\tau_{eb})$ bounce. This time the wavenumbers used are $k=10^{-3}{\rm Mpc^{-1}}$ (blue) and $k=3\times 10^{-3}{\rm Mpc^{-1}}$ (red) and such choice of wavenumbers will become clear in the upcoming power spectrum sections. The evolution remains similar without any significant changes for both categories of modes, except in the change in amplitude for the higher wavenumbers, and, in comparison with the contraction scenario, the amplitude has risen quite significantly and continues to do so with the modes going outside the Horizon. Also, the scalar modes and their momenta remain comparable in their growth, similar to what one observes in the contraction phase.

%%%%%%%%%%%%%%%%%%%%%%%%%%%%%%%%%%%%%%%%%%%%%
\begin{figure*}[htb!]
    	\centering
    \subfigure[]{
      	\includegraphics[width=8.5cm,height=7.5cm]{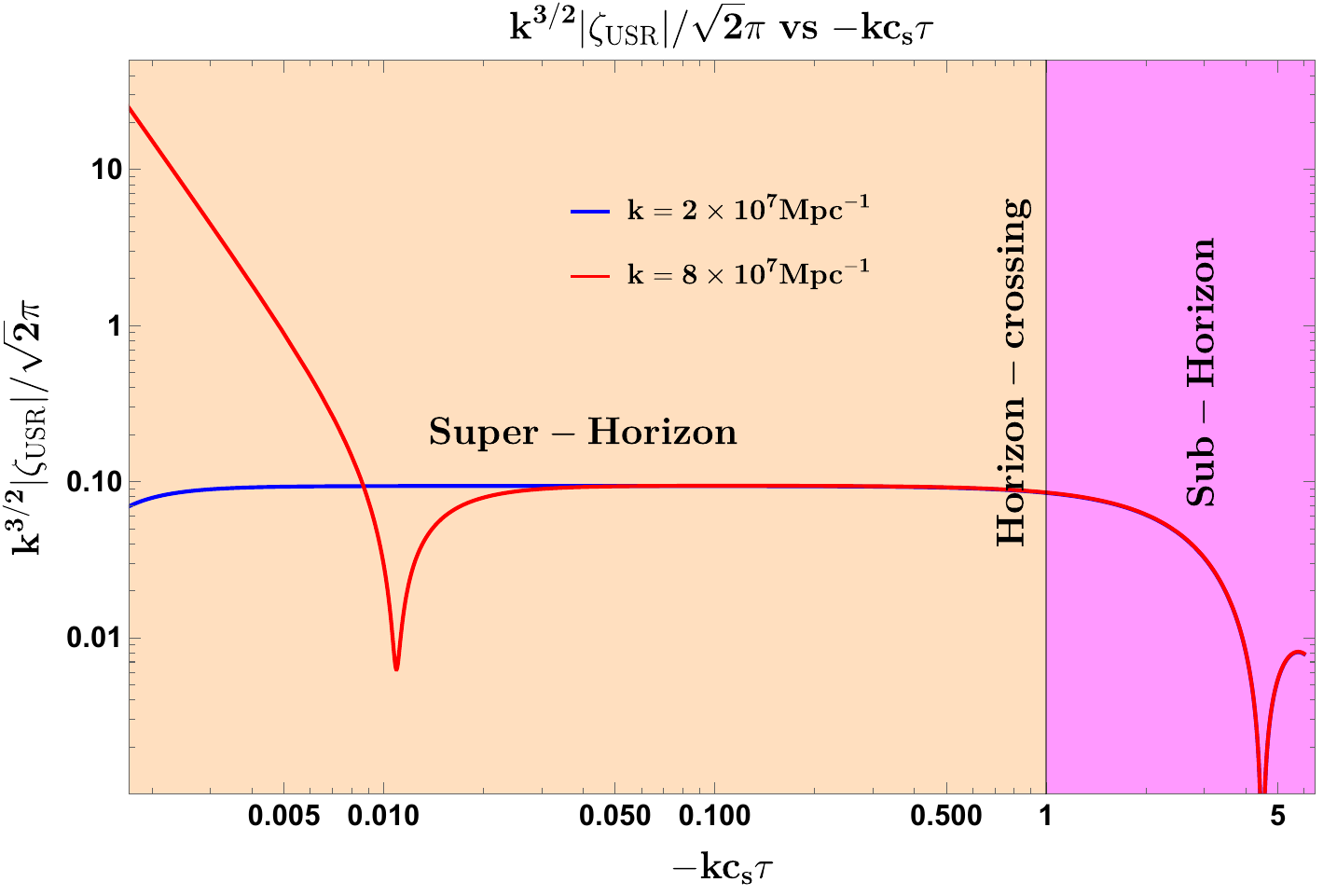}
        \label{zetausr}
    }
    \subfigure[]{
       \includegraphics[width=8.5cm,height=7.5cm]{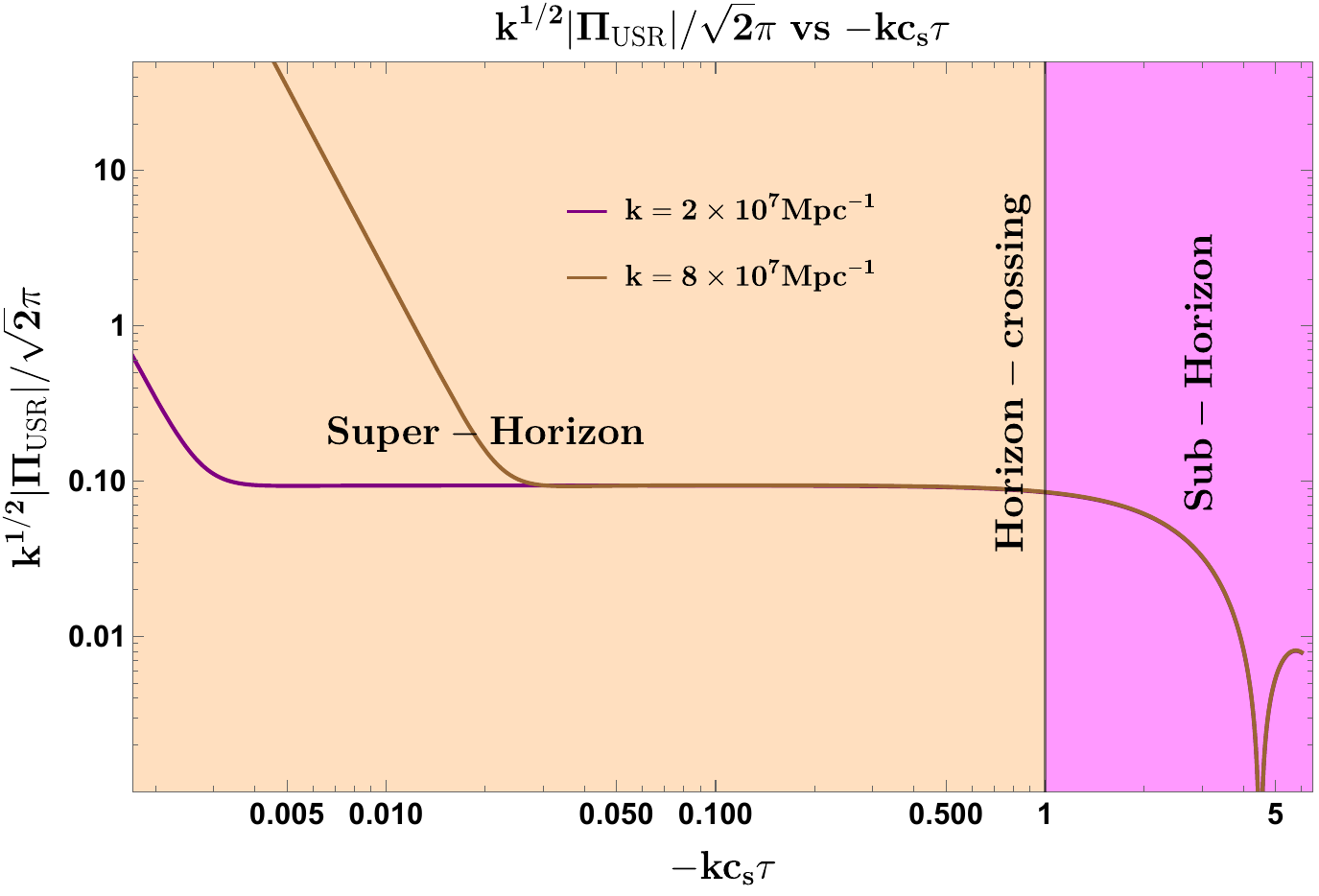}
        \label{piusr}
    }
    	\caption[Optional caption for list of figures]{ Plots show evolution of scalar modes and their associated momenta as a function of $-kc_{s}\tau$ for the ultra slow-roll (USR) phase. In the left, the function $k^{3/2}|\zeta_{\rm USR}|/\sqrt{2}\pi$ is plotted and in the right function $k^{1/2}|\Pi_{\rm USR}|/\sqrt{2}\pi$. The orange and magenta regions highlight the Super-Horizon ($-kc_{s}\tau\ll 1$) and Sub-Horizon ($-kc_{s}\tau\gg 1$) regions, respectively. } 
    	\label{usrmodes }
    \end{figure*}
%%%%%%%%%%%%%%%%%%%%%%%%%%%%%%%%%%%%%%%%%%%%%

After the bouncing phases ceases we continue with the inflationary stage of our setup starting with the mode evolution in the first slow-roll phase. The fig.(\ref{zetasr1},\ref{pisr1}) describes this behaviour. During this phase the scalar mode evolution remains constant soon we go far outside the Horizon while inside the Horizon it shows gradual increase in amplitude. There does not come any wavenumber dependence for this phase and thus the dynamics visible remain common to all the modes. The associated conjugate momentum modes steadily decreases in amplitude upon venturing beyond the Horizon. This time a clear distinction between evolution of the mode and its momenta can be seen and in the super-horizon the strength of the conjugate momentum component remains highly suppressed compared with their scalar modes counterpart.

Moving forward, we now consider the evolution of the small-scale modes involved in the USR phase, as shown in fig.(\ref{zetausr},\ref{piusr}). When inside the Horizon, the modes show similar behaviour in their evolution for different wavenumbers and the type of mode considered. Here, the wavenumber dependence comes again, with different behaviour for each mode and the dimensionless variable is chosen to be of order $-k_{s}c_{s}\tau_{s}\lesssim {\cal O}(10^{-2})$. We consider two different wavenumbers for comparison with $k=2\times 10^{7}{\rm Mpc^{-1}}$ (blue) and $k=8\times 10^{7}{\rm Mpc^{-1}}$ (red). As these modes move far outside the Horizon, the smaller wavenumber increases in strength later compared to the larger one. After a certain time, the strength keeps increasing for the momenta modes, but the scalar modes experience a dip in the amplitude, after which it also becomes too large. The  amplitude for most part of interest after horizon crossing remains sufficient enough for our purposes. In this phase, both the scalar mode and its conjugate momenta remain comparable in amplitude, unlike in the slow-roll phase.

  %%%%%%%%%%%%%%%%%%%%%%%%%%%%%%%%%%%%%%%%%%%%%
\begin{figure*}[htb!]
    	\centering
    \subfigure[]{
      	\includegraphics[width=8.5cm,height=7.5cm]{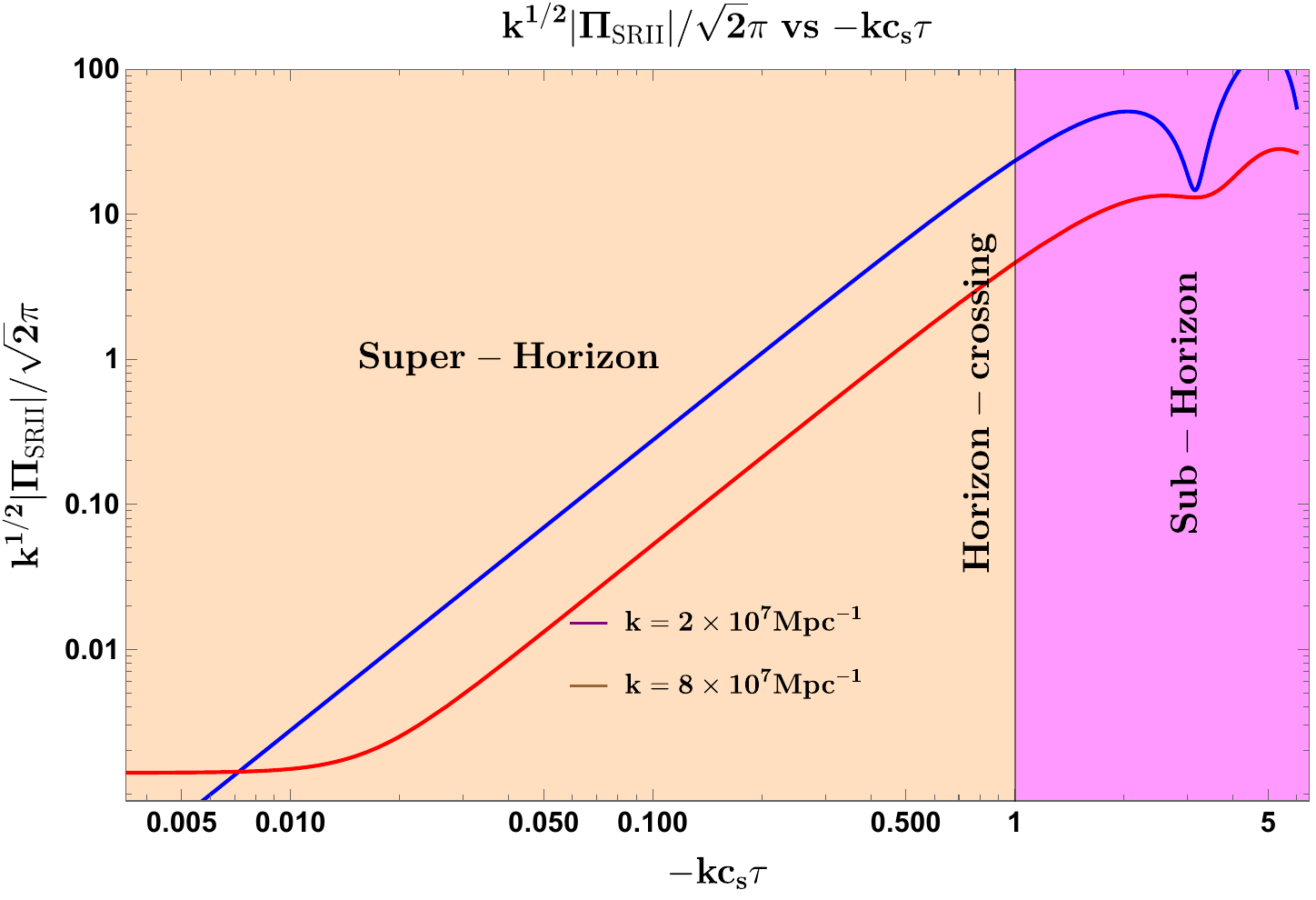}
        \label{zetasr2}
    }
    \subfigure[]{
       \includegraphics[width=8.5cm,height=7.5cm]{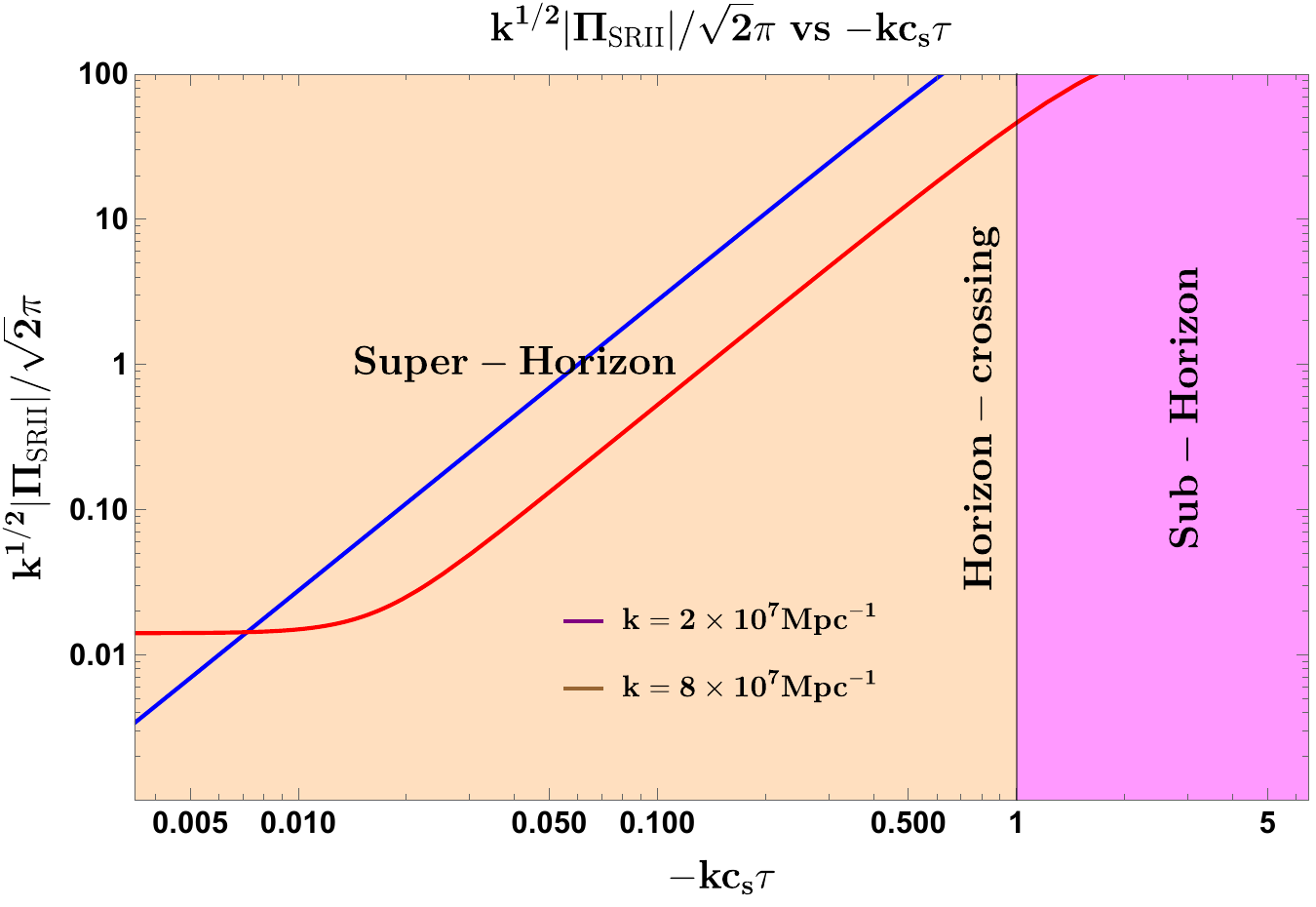}
        \label{pisr2}
    }
    	\caption[Optional caption for list of figures]{ Plots show evolution of scalar modes and their associated momenta as a function of $-kc_{s}\tau$ for the second slow-roll (SRII) phase. In the left, the function $k^{3/2}|\zeta_{\rm SRII}|/\sqrt{2}\pi$ is plotted and in the right function $k^{1/2}|\Pi_{\rm SRII}|/\sqrt{2}\pi$. The orange and magenta regions highlight the Super-Horizon ($-kc_{s}\tau\ll 1$) and Sub-Horizon ($-kc_{s}\tau\gg 1$) regions, respectively. } 
    	\label{sr2modes }
    \end{figure*}
%%%%%%%%%%%%%%%%%%%%%%%%%%%%%%%%%%%%%%%%%%%%%

Lastly, we enter in to the final phase of our setup which is the SRII phase and the relevant mode evolution is shown in fig.(\ref{zetasr2},\ref{pisr2}). A wavenumber dependence also appear in this phase as well and we examine the behaviour for the same two modes considered previously in the USR. The dimensionless variable is chosen to be of order $-k_{e}c_{s}\tau_{e}\lesssim {\cal O}(10^{-1})$. Inside the Horizon the scalar modes appear to have increased strength which quickly decreases after horizon crossing and in far outside the horizon becomes almost constant with periods of dip in the amplitude for different wavenumbers. The conjugate momenta modes continue to lose strength in the super-horizon and after a period become constant in amplitude with the lower wavenumbers becoming weaker as they move farther outside and have their overall magnitudes relatively smaller than the corresponding scalar modes.  

Our next step is to analyze the structure of the total scalar power spectrum and the contributions from each of the five phases in our setup, building from the scalar mode solution we discussed in the previous and current sections.

 \section{Quantifying tree-level scalar power spectrum}
\label{s8}

\subsection{Quantization of scalar modes}

Properly quantizing the scalar modes is necessary to generate the two-point correlation function expression and the related power spectrum in Fourier space, which are required to compute the cosmic correlations. Firstly, we construct the generation operator, $\hat{a}^{\dagger}_{\bf k}$, and the annihilation operator, $\hat{a}_{\bf k}$, which will generate and eliminate an excited state from the initial Bunch Davies state, respectively. Hereafter, $|0\rangle$ is designated as Bunch Davies initial state for the remaining reasons. It must adhere to the subsequent constraint condition: 
\be \hat{a}_{\bf k}|0\rangle=0\quad\forall {\bf k}.\ee
It is now necessary for the scalar perturbed mode and its canonically conjugate momenta to meet the equal time commutation relations:
\bea &&\left[\hat{\zeta}_{\bf k}(\tau),\hat{\Pi_{\zeta}}_{{\bf k}^{'}}(\tau)\right]_{\rm ETCR}=i\;\delta^{3}\left({\bf k}+{\bf k}^{'}\right),~~~~~
\left[\hat{\zeta}_{\bf k}(\tau),\hat{\zeta}_{{\bf k}^{'}}(\tau)\right]_{\rm ETCR}=0,~~~~~
\left[\hat{\Pi_{\zeta}}_{\bf k}(\tau),\hat{\Pi_{\zeta}}_{{\bf k}^{'}}(\tau)\right]_{\rm ETCR}=0.\eea
The corresponding quantum mechanical operators for the scalar mode and its conjugate momenta are indicated by the following formulas:
\bea \hat{\zeta}_{\bf k}(\tau)&=&\bigg[{\zeta}_{\bf k}(\tau)\hat{a}_{\bf k}+{\zeta}^{*}_{\bf k}(\tau)\hat{a}^{\dagger}_{-{\bf k}}\bigg],~~~~~~~
\hat{\Pi_{\zeta}}_{\bf k}(\tau)=\bigg[{\Pi_{\zeta}}_{\bf k}(\tau)\hat{a}_{\bf k}+{\Pi_{\zeta}}^{*}_{\bf k}(\tau)\hat{a}^{\dagger}_{-{\bf k}}\bigg].\quad\quad \eea
It will be very helpful in the following section when we utilize the in-in formalism to complete the one-loop calculation. In addition to the representation described above, this may also be expressed using all possible commutation relations between the creation and annihilation operators:
\bea \left[\hat{a}_{\bf k},\hat{a}^{\dagger}_{{\bf k}^{'}}\right]&=&\delta^{3}\left({\bf k}+{\bf k}^{'}\right),~~~~
 \left[\hat{a}_{\bf k},\hat{a}_{{\bf k}^{'}}\right]=0=\left[\hat{a}^{\dagger}_{\bf k},\hat{a}^{\dagger}_{{\bf k}^{'}}\right].\eea

\subsection{Calculation of tree-level spectrum}

The late time scale, $\tau\rightarrow 0$, is recognized as the location of the comoving curvature perturbation. Based on this information, the tree-level contribution to the two-point cosmic correlation function for the comoving curvature perturbation is represented as:
\bea \langle \hat{\zeta}_{\bf k}\hat{\zeta}_{{\bf k}^{'}}\rangle_{{\bf Tree}} &=&\lim_{\tau\rightarrow 0}\langle \hat{\zeta}_{\bf k}(\tau)\hat{\zeta}_{{\bf k}^{'}}(\tau)\rangle_{{\bf Tree}}=(2\pi)^{3}\;\delta^{3}\left({\bf k}+{\bf k}^{'}\right)P^{\bf Tree}_{\zeta}(k),\quad\eea
The dimensionful power spectrum in Fourier space is denoted using the notation $P^{\bf Tree}_{\zeta}(k)$ and can be expressed as:
\bea P^{\bf Tree}_{\zeta}(k)=\langle \hat{\zeta}_{\bf k}\hat{\zeta}_{-{\bf k}}\rangle_{(0,0)}=\left[{\zeta}_{\bf k}(\tau){\zeta}_{-{\bf k}}(\tau)\right]_{\tau\rightarrow 0}=|{\zeta}_{\bf k}(\tau)|^{2}_{\tau\rightarrow 0}.\quad\eea
However, for practical purposes and to relate to cosmic observations, it becomes relevant to use the dimensionless form of the power spectrum in Fourier space, and is represented as follows:
\bea \label{tree} \Delta^{2}_{\zeta,{\bf Tree}}(k)=\frac{k^{3}}{2\pi^{2}}P^{\bf Tree}_{\zeta}(k)=\frac{k^{3}}{2\pi^{2}}|{\zeta}_{\bf k}(\tau)|^{2}_{\tau\rightarrow 0}.\eea
From the detailed computations performed in the previously mentioned five distinctive phases, it is evident that the solutions obtained for the gauge invariant comoving curvature perturbation are different in nature. Utilizing these obtained results and further substituting in equation(\ref{tree}), we get the following simplified form of the dimensionless power spectrum in each of the phases:
\begin{enumerate}
    \item \underline{\bf Phase I: Ekpyrotic/matter contraction in $k_c\leq k<k_b$:}\\ \\
    In Phase I, which physically describes the ekpyrotic/matter contraction scenarios depending on the values of the first slow roll parameter $\epsilon$, we have the following result for the tree-level power spectrum for the scalar modes for the initial general vacuum states:
    \bea \Delta^{2}_{\zeta,{\bf Tree}}(k)
&=& \left(\frac{\epsilon_*}{\epsilon_c}\right)\times\left(\frac{2^{2\nu-3}H^{2}}{8\pi^{2}M^{2}_{ pl}\epsilon c_s}\left|\frac{\Gamma(\nu)}{\Gamma\left(\frac{3}{2}\right)}\right|^2\right)_*\times\left(\frac{k}{k_*}\right)^{\frac{2\epsilon_c}{\epsilon_c-1}}\nonumber\\
&&\quad\quad\quad\quad\quad\quad\quad\quad\times \left|\alpha_1 (1+i k c_s\tau) e^{-i\left(k c_s\tau+\frac{\pi}{2}\left(\nu+\frac{1}{2}\right)\right)}-\beta_1(1-i k c_s \tau)e^{i\left(k c_s\tau+\frac{\pi}{2}\left(\nu+\frac{1}{2}\right)\right)}\right|^2.\quad\quad\eea
In the super-horizon scale considering $-kc_s\tau\ll 1$ we get further the following simplified result for the scalar power spectrum in Phase I for the initial general vacuum states:
\bea \Delta^{2}_{\zeta,{\bf Tree}}(k)
&=& \left(\frac{\epsilon_*}{\epsilon_c}\right)\times\left(\frac{2^{2\nu-3}H^{2}}{8\pi^{2}M^{2}_{ pl}\epsilon c_s}\left|\frac{\Gamma(\nu)}{\Gamma\left(\frac{3}{2}\right)}\right|^2\right)_*\times\left(\frac{k}{k_*}\right)^{\frac{2\epsilon_c}{\epsilon_c-1}}\times \left|\alpha_1-\beta_1\right|^2.\quad\quad\eea
If we fix the initial vacuum state as Bunch Davies state which is described by $\alpha_1=1$ and $\beta_1=0$ we get the following simplified result:
\bea \Delta^{2}_{\zeta,{\bf Tree}}(k)
&=& \left(\frac{\epsilon_*}{\epsilon_c}\right)\times\left(\frac{2^{2\nu-3}H^{2}}{8\pi^{2}M^{2}_{ pl}\epsilon c_s}\left|\frac{\Gamma(\nu)}{\Gamma\left(\frac{3}{2}\right)}\right|^2\right)_*\times\left(\frac{k}{k_*}\right)^{\frac{2\epsilon_c}{\epsilon_c-1}}.\quad\quad\eea
Further, for better understanding purposes, the contribution computed to describe the scalar power spectrum of the contraction phase can be written in terms of the SRI (Phase III) counterpart by the following simplified expression:
\bea \Delta^{2}_{\zeta,{\bf Tree}}(k)
&=& \Delta^{2}_{\zeta,{\bf SRI}}(k_*)\times\left(\frac{\epsilon_*}{\epsilon_c}\right)\times\left(\frac{k}{k_*}\right)^{\frac{2\epsilon_c}{\epsilon_c-1}},\quad\quad\eea
where the SRI counterpart of the scalar power spectrum in the presence of initial state can be written as:
\bea \label{sr1}\Delta^{2}_{\zeta,{\bf SRI}}(k_*)=\left(\frac{2^{2\nu-3}H^{2}}{8\pi^{2}M^{2}_{ pl}\epsilon c_s}\left|\frac{\Gamma(\nu)}{\Gamma\left(\frac{3}{2}\right)}\right|^2\right)_*\times \left|\alpha_1-\beta_1\right|^2\xrightarrow[\alpha_1=1,\beta_1=0]{{\bf Bunch\;Davies}}\left(\frac{2^{2\nu-3}H^{2}}{8\pi^{2}M^{2}_{ pl}\epsilon c_s}\left|\frac{\Gamma(\nu)}{\Gamma\left(\frac{3}{2}\right)}\right|^2\right)_*.\eea
In terms of the number of e-foldings this expression can be recast in the following simplified form:
\bea \bigg[\Delta^{2}_{\zeta,{\bf Tree}}(N)\bigg]_{\bf CONTRACTION}
&=& \Delta^{2}_{\zeta,{\bf SRI}}(N_*)\times\left(\frac{\epsilon_*}{\epsilon_c}\right)\times\exp\left(\frac{2\epsilon_c}{\epsilon_c-1}(N-N_*)\right)\quad\quad{\rm where}\quad\quad N_c\leq N<N_b.\quad\quad\eea

    \item \underline{\bf Phase II: Ekpyrotic/matter bounce in $k_b\leq k<k_*$:}\\ \\
    In Phase II, which physically describes the ekpyrotic/matter bounce scenarios depending on the values of the first slow roll parameter $\epsilon$, we have the following result for the tree-level power spectrum for the scalar modes for the initial general vacuum states:
    \bea \Delta^{2}_{\zeta,{\bf Tree}}(k)
&=& \left(\frac{\epsilon_*}{\epsilon_b}\right)\times\left(\frac{2^{2\nu-3}H^{2}}{8\pi^{2}M^{2}_{ pl}\epsilon c_s}\left|\frac{\Gamma(\nu)}{\Gamma\left(\frac{3}{2}\right)}\right|^2\right)_*\times\left(\frac{k}{k_*}\right)^{2}\times\left[1+\left(\frac{k_*}{k}\right)^2\right]^{-\frac{1}{(\epsilon_b-1)}}\nonumber\\
&&\quad\quad\quad\quad\quad\quad\quad\quad\times \left|\alpha_1 (1+i k c_s\tau) e^{-i\left(k c_s\tau+\frac{\pi}{2}\left(\nu+\frac{1}{2}\right)\right)}-\beta_1(1-i k c_s \tau)e^{i\left(k c_s\tau+\frac{\pi}{2}\left(\nu+\frac{1}{2}\right)\right)}\right|^2.\quad\quad\eea
In the super-horizon scale considering $-kc_s\tau\ll 1$ we get further the following simplified result for the scalar power spectrum in Phase II for the initial general vacuum states:
\bea \Delta^{2}_{\zeta,{\bf Tree}}(k)
&=& \left(\frac{\epsilon_*}{\epsilon_b}\right)\times\left(\frac{2^{2\nu-3}H^{2}}{8\pi^{2}M^{2}_{ pl}\epsilon c_s}\left|\frac{\Gamma(\nu)}{\Gamma\left(\frac{3}{2}\right)}\right|^2\right)_*\times\left(\frac{k}{k_*}\right)^{2}\times\left[1+\left(\frac{k_*}{k}\right)^2\right]^{-\frac{1}{(\epsilon_b-1)}}\times \left|\alpha_1-\beta_1\right|^2.\quad\quad\eea
If we fix the initial vacuum state as Bunch Davies state which is described by $\alpha_1=1$ and $\beta_1=0$ we get the following simplified result:
\bea \Delta^{2}_{\zeta,{\bf Tree}}(k)
&=& \left(\frac{\epsilon_*}{\epsilon_b}\right)\times\left(\frac{2^{2\nu-3}H^{2}}{8\pi^{2}M^{2}_{ pl}\epsilon c_s}\left|\frac{\Gamma(\nu)}{\Gamma\left(\frac{3}{2}\right)}\right|^2\right)_*\times\left(\frac{k}{k_*}\right)^{2}\times\left[1+\left(\frac{k_*}{k}\right)^2\right]^{-\frac{1}{(\epsilon_b-1)}}.\quad\quad\eea
Further, for better understanding purposes, the contribution computed to describe the scalar power spectrum of the contraction phase can be written in terms of the SRI (Phase III) counterpart by the following simplified expression:
\bea \Delta^{2}_{\zeta,{\bf Tree}}(k)
&=& \Delta^{2}_{\zeta,{\bf SRI}}(k_*)\times\left(\frac{\epsilon_*}{\epsilon_b}\right)\times\left(\frac{k}{k_*}\right)^{2}\times\left[1+\left(\frac{k_*}{k}\right)^2\right]^{-\frac{1}{(\epsilon_b-1)}},\quad\quad\eea
where the SRI phase counterpart of the spectrum is written explicitly in equation(\ref{sr1}).

In terms of the number of e-foldings this expression can be recast in the following simplified form:
\bea  \bigg[\Delta^{2}_{\zeta,{\bf Tree}}(N)\bigg]_{\bf BOUNCE}
&=& \Delta^{2}_{\zeta,{\bf SRI}}(N_*)\times\left(\frac{\epsilon_*}{\epsilon_b}\right)\nonumber\\
&&\quad\quad\times\exp(2(N-N_*))\times\left[1+\exp(-2(N-N_*))\right]^{-\frac{1}{(\epsilon_b-1)}}\quad{\rm where}\quad N_b\leq N<N_*.\quad\quad\eea

    \item \underline{\bf Phase III: SRI in $k_*\leq k<k_s$:}\\ \\
    In Phase III, which physically describes the SRI, we have the following result for the tree-level power spectrum for the scalar modes for the initial general vacuum states:
    \bea \Delta^{2}_{\zeta,{\bf Tree}}(k)
&=& \left(\frac{2^{2\nu-3}H^{2}}{8\pi^{2}M^{2}_{ pl}\epsilon c_s}\left|\frac{\Gamma(\nu)}{\Gamma\left(\frac{3}{2}\right)}\right|^2\right)_*\times \left|\alpha_1 (1+i k c_s\tau) e^{-i\left(k c_s\tau+\frac{\pi}{2}\left(\nu+\frac{1}{2}\right)\right)}-\beta_1(1-i k c_s \tau)e^{i\left(k c_s\tau+\frac{\pi}{2}\left(\nu+\frac{1}{2}\right)\right)}\right|^2.\quad\quad\eea
In the super-horizon scale considering $-kc_s\tau\ll 1$ we get further the following simplified result for the scalar power spectrum in Phase III for the initial general vacuum states:
\bea \Delta^{2}_{\zeta,{\bf Tree}}(k)
&=& \left(\frac{2^{2\nu-3}H^{2}}{8\pi^{2}M^{2}_{ pl}\epsilon c_s}\left|\frac{\Gamma(\nu)}{\Gamma\left(\frac{3}{2}\right)}\right|^2\right)_*\times \left|\alpha_1-\beta_1\right|^2=\Delta^{2}_{\zeta,{\bf SRI}}(k_*).\quad\quad\eea
If we fix the initial vacuum state as Bunch Davies state which is described by $\alpha_1=1$ and $\beta_1=0$ we get the following simplified result:
\bea \Delta^{2}_{\zeta,{\bf Tree}}(k)
&=& \left(\frac{2^{2\nu-3}H^{2}}{8\pi^{2}M^{2}_{ pl}\epsilon c_s}\left|\frac{\Gamma(\nu)}{\Gamma\left(\frac{3}{2}\right)}\right|^2\right)_*=\Delta^{2}_{\zeta,{\bf SRI}}(k_*)\approx \bigg[\Delta^{2}_{\zeta,{\bf Tree}}(k)\bigg]_{\bf SRI}.\quad\quad\eea
where we have:
\bea \bigg[\Delta^{2}_{\zeta,{\bf Tree}}(k)\bigg]_{\bf SRI}&=&\Delta^{2}_{\zeta,{\bf SRI}}(k_*)\times\bigg[1+\left(\frac{k}{k_s}\right)^2\Bigg].\eea
In terms of the number of e-foldings, we can further write this expression as:
\bea \bigg[\Delta^{2}_{\zeta,{\bf Tree}}(N)\bigg]_{\bf SRI}&=&\Delta^{2}_{\zeta,{\bf SRI}}(N_*)\times\bigg[1+\exp(2(N-N_s))\Bigg]\quad\quad {\rm where}\quad\quad N_*\leq N<N_s.\quad\quad\eea

    \item \underline{\bf Phase IV: USR in $k_s\leq k<k_e$:}\\ \\
    In Phase IV, which physically describes the USR scenario, we have the following result for the tree-level power spectrum for the scalar modes for the initial general vacuum states:
    \bea \Delta^{2}_{\zeta,{\bf Tree}}(k)
&=& \left(\frac{2^{2\nu-3}H^{2}}{8\pi^{2}M^{2}_{ pl}\epsilon c_s}\left|\frac{\Gamma(\nu)}{\Gamma\left(\frac{3}{2}\right)}\right|^2\right)_*\times\left(\frac{k}{k_s}\right)^{6}\nonumber\\
&&\quad\quad\quad\quad\quad\quad\quad\quad\times \left|\alpha_2 (1+i k c_s\tau) e^{-i\left(k c_s\tau+\frac{\pi}{2}\left(\nu+\frac{1}{2}\right)\right)}-\beta_2(1-i k c_s \tau)e^{i\left(k c_s\tau+\frac{\pi}{2}\left(\nu+\frac{1}{2}\right)\right)}\right|^2.\quad\quad\eea
In the super-horizon scale considering $-kc_s\tau\ll 1$ we get further the following simplified result for the scalar power spectrum in Phase IV for the initial general vacuum states:
\bea \Delta^{2}_{\zeta,{\bf Tree}}(k)
&=& \left(\frac{2^{2\nu-3}H^{2}}{8\pi^{2}M^{2}_{ pl}\epsilon c_s}\left|\frac{\Gamma(\nu)}{\Gamma\left(\frac{3}{2}\right)}\right|^2\right)_*\times\left(\frac{k}{k_s}\right)^{6}\times \left|\alpha_2-\beta_2\right|^2.\quad\quad\eea
Further, for better understanding purposes, the contribution computed to describe the scalar power spectrum of the contraction phase can be written in terms of the SRI (Phase III) counterpart by the following simplified expression:
\bea \Delta^{2}_{\zeta,{\bf Tree}}(k)
&=& \Delta^{2}_{\zeta,{\bf SRI}}(k_*)\times\left(\frac{k}{k_s}\right)^{6}\times\frac{\left|\alpha_2-\beta_2\right|^2}{\left|\alpha_1-\beta_1\right|^2},\quad\quad\eea
where the SRI phase counterpart of the spectrum is written explicitly in equation(\ref{sr1}). If we further choose Bunch Davies initial condition described by, $\alpha_1=1$ and $\beta_1=0$, we get:
\bea \Delta^{2}_{\zeta,{\bf Tree}}(k)
&=& \Delta^{2}_{\zeta,{\bf SRI}}(k_*)\times\left(\frac{k}{k_s}\right)^{6}\times\left|\alpha_2-\beta_2\right|^2,\quad\quad\eea
In terms of the number of e-foldings, we can further write this expression as:
\bea \bigg[\Delta^{2}_{\zeta,{\bf Tree}}(N)\bigg]_{\bf USR}&=&\Delta^{2}_{\zeta,{\bf SRI}}(N_*)\times\exp(6(N-N_s))\times\left|\alpha_2-\beta_2\right|^2\quad\quad {\rm where}\quad\quad N_s\leq N<N_e.\quad\quad\eea

    \item \underline{\bf Phase V: SRII in $k_e\leq k<k_{\rm end}$:}\\ \\
    In Phase V, which physically describes the SRII scenario, we have the following result for the tree-level power spectrum for the scalar modes for the initial general vacuum states:
    \bea \Delta^{2}_{\zeta,{\bf Tree}}(k)
&=& \left(\frac{2^{2\nu-3}H^{2}}{8\pi^{2}M^{2}_{ pl}\epsilon c_s}\left|\frac{\Gamma(\nu)}{\Gamma\left(\frac{3}{2}\right)}\right|^2\right)_*\times\left(\frac{k_e}{k_s}\right)^{6}\nonumber\\
&&\quad\quad\quad\quad\quad\quad\quad\quad\times \left|\alpha_3 (1+i k c_s\tau) e^{-i\left(k c_s\tau+\frac{\pi}{2}\left(\nu+\frac{1}{2}\right)\right)}-\beta_3(1-i k c_s \tau)e^{i\left(k c_s\tau+\frac{\pi}{2}\left(\nu+\frac{1}{2}\right)\right)}\right|^2.\quad\quad\eea
In the super-horizon scale considering $-kc_s\tau\ll 1$ we get further the following simplified result for the scalar power spectrum in Phase V for the initial general vacuum states:
\bea \Delta^{2}_{\zeta,{\bf Tree}}(k)
&=& \left(\frac{2^{2\nu-3}H^{2}}{8\pi^{2}M^{2}_{ pl}\epsilon c_s}\left|\frac{\Gamma(\nu)}{\Gamma\left(\frac{3}{2}\right)}\right|^2\right)_*\times\left(\frac{k_e}{k_s}\right)^{6}\times \left|\alpha_3-\beta_3\right|^2.\quad\quad\eea
Further, for better understanding purposes, the contribution computed to describe the scalar power spectrum of the contraction phase can be written in terms of the SRI (Phase III) counterpart by the following simplified expression:
\bea \Delta^{2}_{\zeta,{\bf Tree}}(k)
&=& \Delta^{2}_{\zeta,{\bf SRI}}(k_*)\times\left(\frac{k_e}{k_s}\right)^{6}\times\frac{\left|\alpha_3-\beta_3\right|^2}{\left|\alpha_1-\beta_1\right|^2},\quad\quad\eea
where the SRI phase counterpart of the spectrum is written explicitly in equation(\ref{sr1}). If we further choose Bunch Davies initial condition described by, $\alpha_1=1$ and $\beta_1=0$, we get:
\bea \Delta^{2}_{\zeta,{\bf Tree}}(k)
&=& \Delta^{2}_{\zeta,{\bf SRI}}(k_*)\times\left(\frac{k}{k_s}\right)^{6}\times\left|\alpha_3-\beta_3\right|^2,\quad\quad\eea
In terms of the number of e-foldings, we can further write this expression as:
\bea \bigg[\Delta^{2}_{\zeta,{\bf Tree}}(N)\bigg]_{\bf SRII}&=&\Delta^{2}_{\zeta,{\bf SRI}}(N_*)\times\exp(6(N_e-N_s))\times\left|\alpha_3-\beta_3\right|^2\quad\quad {\rm where}\quad\quad N_e\leq N<N_{\rm end}.\quad\quad\eea
\end{enumerate}
After combining all the results of the level power spectrum for the scalar modes as obtained for the five consecutive phases we get the following result for the total power spectrum when we choose the general initial vacuum state:
\bea \Delta^{2}_{\zeta,{\bf Tree-Total}}(k)&=&\Delta^{2}_{\zeta,{\bf SRI}}(k_*)\times\bigg[1+\left(\frac{\epsilon_*}{\epsilon_c}\right)\times\left(\frac{k}{k_*}\right)^{\frac{2\epsilon_c}{\epsilon_c-1}}+\left(\frac{\epsilon_*}{\epsilon_b}\right)\times\left(\frac{k}{k_*}\right)^{2}\times\left[1+\left(\frac{k_*}{k}\right)^2\right]^{-\frac{1}{(\epsilon_b-1)}}\nonumber\\
&&\quad\quad\quad\quad\quad\quad\quad\quad+\Theta(k-k_s)\left(\frac{k}{k_s}\right)^{6}\times\frac{\left|\alpha_2-\beta_2\right|^2}{\left|\alpha_1-\beta_1\right|^2}+\Theta(k-k_e)\left(\frac{k_e}{k_s}\right)^{6}\times\frac{\left|\alpha_3-\beta_3\right|^2}{\left|\alpha_1-\beta_1\right|^2}\bigg].\quad\eea
Here the two Heaviside Theta functions characterize the SRI to USR and USR to SRII sharp transitions. Further, if we choose the initial Bunch Davies condition then we get the following simplified result for the total power spectrum for the scalar modes: 
\bea \label{treetotalpspec} \Delta^{2}_{\zeta,{\bf Tree-Total}}(k)&=&\Delta^{2}_{\zeta,{\bf SRI}}(k_*)\times\bigg[1+\left(\frac{\epsilon_*}{\epsilon_c}\right)\times\left(\frac{k}{k_*}\right)^{\frac{2\epsilon_c}{\epsilon_c-1}}+\left(\frac{\epsilon_*}{\epsilon_b}\right)\times\left(\frac{k}{k_*}\right)^{2}\times\left[1+\left(\frac{k_*}{k}\right)^2\right]^{-\frac{1}{(\epsilon_b-1)}}\nonumber\\
&&\quad\quad\quad\quad\quad\quad\quad\quad+\Theta(k-k_s)\left(\frac{k}{k_s}\right)^{6}\times\left|\alpha_2-\beta_2\right|^2+\Theta(k-k_e)\left(\frac{k_e}{k_s}\right)^{6}\times\left|\alpha_3-\beta_3\right|^2\bigg].\quad\eea
In terms of the number of e-foldings, we get the following simplified result in the context of having sharp transitions:
\bea \Delta^{2}_{\zeta,{\bf Tree-Total}}(N)&=&\Delta^{2}_{\zeta,{\bf SRI}}(N_*)\times\bigg[1+\left(\frac{\epsilon_*}{\epsilon_c}\right)\times\exp\left(\frac{2\epsilon_c}{\epsilon_c-1}(N-N_*)\right)\nonumber\\
&&\quad\quad\quad\quad\quad\quad\quad\quad+\left(\frac{\epsilon_*}{\epsilon_b}\right)\times\exp(2(N-N_*))\times\left[1+\exp(-2(N-N_*))\right]^{-\frac{1}{(\epsilon_b-1)}}\nonumber\\
&&\quad\quad\quad\quad\quad\quad\quad\quad+\Theta(N-N_s)\exp(6(N-N_s))\times\left|\alpha_2-\beta_2\right|^2\nonumber\\
&&\quad\quad\quad\quad\quad\quad\quad\quad+\Theta(N-N_e)\exp(6(N_e-N_s))\times\left|\alpha_3-\beta_3\right|^2\bigg].\quad\eea
If we use a smooth transition in the current context of discussion rather than a sharp one, the expression is changed to the following expression in the presence of a general initial condition:
\bea \Delta^{2}_{\zeta,{\bf Tree-Total}}(k)&=&\Delta^{2}_{\zeta,{\bf SRI}}(k_*)\times\bigg[1+\left(\frac{\epsilon_*}{\epsilon_c}\right)\times\left(\frac{k}{k_*}\right)^{\frac{2\epsilon_c}{\epsilon_c-1}}+\left(\frac{\epsilon_*}{\epsilon_b}\right)\times\left(\frac{k}{k_*}\right)^{2}\times\left[1+\left(\frac{k_*}{k}\right)^2\right]^{-\frac{1}{(\epsilon_b-1)}}\nonumber\\
&&\quad\quad\quad\quad+{\rm tanh}\left(\frac{k-k_s}{\Delta k}\right)\left(\frac{k}{k_s}\right)^{6}\times\frac{\left|\alpha_2-\beta_2\right|^2}{\left|\alpha_1-\beta_1\right|^2}+{\rm tanh}\left(\frac{k-k_e}{\Delta k}\right)\left(\frac{k_e}{k_s}\right)^{6}\times\frac{\left|\alpha_3-\beta_3\right|^2}{\left|\alpha_1-\beta_1\right|^2}\bigg].\quad\eea
Thus, in the event of a smooth transition, $\Delta k=k_e-k_s$ denotes the breadth of the USR phase. The result for the sharp transition may be obtained by taking the limit $\Delta k$ to be very tiny because:
\bea &&\lim_{\Delta k\rightarrow 0}{\rm tanh}\left(\frac{k-k_s}{\Delta k}\right)=\Theta(k-k_s),\\
&&\lim_{\Delta k\rightarrow 0}{\rm tanh}\left(\frac{k-k_e}{\Delta k}\right)=\Theta(k-k_e).\eea
It is noteworthy, although, that for the tree-level power spectrum, we obtain nearly identical physical results, save from smoothing the behavior of the joining functions at the transition points in both cases.
 Further, if we choose the initial Bunch Davies condition then we get the following simplified result for the total power spectrum for the scalar modes: 
 \bea \Delta^{2}_{\zeta,{\bf Tree-Total}}(k)&=&\Delta^{2}_{\zeta,{\bf SRI}}(k_*)\times\bigg[1+\left(\frac{\epsilon_*}{\epsilon_c}\right)\times\left(\frac{k}{k_*}\right)^{\frac{2\epsilon_c}{\epsilon_c-1}}+\left(\frac{\epsilon_*}{\epsilon_b}\right)\times\left(\frac{k}{k_*}\right)^{2}\times\left[1+\left(\frac{k_*}{k}\right)^2\right]^{-\frac{1}{(\epsilon_b-1)}}\nonumber\\
&&\quad\quad\quad\quad+{\rm tanh}\left(\frac{k-k_s}{\Delta k}\right)\left(\frac{k}{k_s}\right)^{6}\times\left|\alpha_2-\beta_2\right|^2+{\rm tanh}\left(\frac{k-k_e}{\Delta k}\right)\left(\frac{k_e}{k_s}\right)^{6}\times\left|\alpha_3-\beta_3\right|^2\bigg].\quad\eea
In terms of the number of e-foldings, we get the following simplified result in the context of having sharp transitions:
\bea \Delta^{2}_{\zeta,{\bf Tree-Total}}(N)&=&\Delta^{2}_{\zeta,{\bf SRI}}(N_*)\times\bigg[1+\left(\frac{\epsilon_*}{\epsilon_c}\right)\times\exp\left(\frac{2\epsilon_c}{\epsilon_c-1}(N-N_*)\right)\nonumber\\
&&\quad\quad\quad\quad\quad\quad\quad\quad+\left(\frac{\epsilon_*}{\epsilon_b}\right)\times\exp(2(N-N_*))\times\left[1+\exp(-2(N-N_*))\right]^{-\frac{1}{(\epsilon_b-1)}}\nonumber\\
&&\quad\quad\quad\quad\quad\quad\quad\quad+{\rm tanh}\left(\frac{N-k_s}{\Delta N}\right)\exp(6(N-N_s))\times\left|\alpha_2-\beta_2\right|^2\nonumber\\
&&\quad\quad\quad\quad\quad\quad\quad\quad+{\rm tanh}\left(\frac{N-N_e}{\Delta N}\right)\exp(6(N_e-N_s))\times\left|\alpha_3-\beta_3\right|^2\bigg].\quad\eea
Here we have $\Delta N=N_e-N_s$ which is equivalent to $\Delta k$ and also we have here:
\bea &&\lim_{\Delta N\rightarrow 0}{\rm tanh}\left(\frac{N-N_s}{\Delta N}\right)=\Theta(N-N_s),\\
&&\lim_{\Delta N\rightarrow 0}{\rm tanh}\left(\frac{N-N_e}{\Delta N}\right)=\Theta(N-N_e).\eea
\section{Regularization in loop corrected scalar power spectrum}
\label{s9}

\subsection{The third order perturbed action}
Next, we will make a straightforward calculation of the effect of one-loop corrections on the power spectrum from scalar perturbation modes. The computation that follows will be done under the assumption that the curvature perturbation widens the typical EFT action in third order:
\bea &&S^{(3)}_{\zeta}=\int d\tau\;  d^3x\;  M^2_{ pl}a^2\; \bigg[\left(3\left(c^2_s-1\right)\epsilon+\epsilon^2-\frac{1}{2}\epsilon^3\right)\zeta^{'2}\zeta+\frac{\epsilon}{c^2_s}\bigg(\epsilon-2s+1-c^2_s\bigg)\left(\partial_i\zeta\right)^2\zeta\nonumber\\ 
&&\quad\quad\quad\quad\quad\quad\quad\quad\quad\quad\quad-\frac{2\epsilon}{c^2_s}\zeta^{'}\left(\partial_i\zeta\right)\left(\partial_i\partial^{-2}\left(\frac{\epsilon\zeta^{'}}{c^2_s}\right)\right)-\frac{1}{aH}\left(1-\frac{1}{c^2_{s}}\right)\epsilon \bigg(\zeta^{'3}+\zeta^{'}(\partial_{i}\zeta)^2\bigg)
     \nonumber\\
&& \quad\quad\quad\quad\quad\quad\quad\quad\quad\quad\quad+\frac{1}{2}\epsilon\zeta\left(\partial_i\partial_j\partial^{-2}\left(\frac{\epsilon\zeta^{'}}{c^2_s}\right)\right)^2
+\underbrace{\frac{1}{2c^2_s}\epsilon\partial_{\tau}\left(\frac{\eta}{c^2_s}\right)\zeta^{'}\zeta^{2}}_{\bf Most~dominant ~term~in~USR}\nonumber\\ 
&&\quad\quad\quad\quad\quad\quad\quad\quad\quad\quad\quad+\frac{3}{2}\frac{1}{aH}\frac{\bar{M}^3_1}{ HM^2_{ pl}}
	  \zeta^{'}(\partial_{i}\zeta)^2+\frac{9}{2}\frac{\bar{M}^3_1}{ HM^2_{ pl}}\zeta \zeta^{'2}\nonumber\\ 
&&\quad\quad\quad\quad\quad\quad\quad\quad\quad\quad\quad+\bigg(\frac{3}{2}\frac{\bar{M}^3_1}{ HM^2_{ pl}}-\frac{4}{3}\frac{M^4_3}{H^2M^2_{ pl}}\bigg)\zeta^{'3}-\frac{3}{2}\frac{1}{aH}\frac{\bar{M}^3_1}{ HM^2_{ pl}} \zeta\partial_{\tau}\left(\partial_{i}\zeta\right)^2
  \bigg],\quad\quad\eea
  It is important to keep in mind that in this instance, putting $M_2=0$ yields $c_s=1$. Additionally, if we put $M_3=0$, $\bar{M}_1=0$, the third-order action for the conventional single-field slow roll model is returned. The one-loop corrected equation for the scalar power spectrum will not be considerably changed by the contributions with EFT coefficients other than $M_2$ since they are severely suppressed in the contracting, bouncing, and SR phases of the one-loop contribution. The provided EFT coefficients for the different types of non-minimal or non-canonical single field $P(X,\phi)$ models might not be exactly zero. Having said that, we must limit our computations to modest values of these parameters in order to accurately preserve the perturbation theory for the scalar modes during the one-loop computation in the current scenario. To extract the one-loop quantum effect corrections or to detect the substantial amplification in the primordial non-Gaussian amplitudes, we will explicitly apply all of the contributions mentioned earlier in the presence of abrupt transitions. It is also crucial to note that the conclusions we get in the remaining calculations will not alter for smooth transitions, only the tedious mathematical statements we must deal with throughout the computations in those instances. It is also worth noting that the conclusions we reach in the remaining calculations will not alter for a single or numerous smooth transitions, simply the burdensome mathematical expressions we must deal with throughout the computations in those cases. The highlighted term contributes the most to the first SR (SRI) region, second SR (SRII) region, and USR region as ${\cal O}(\epsilon^{3})$, ${\cal O}(\epsilon^{3})$, and ${\cal O}(\epsilon)$, respectively, for sharp or smooth transitions. Additionally, it is essential to consider that, the above highlighted term has a vanishing contribution in the contraction phase and is highly suppressed in the bouncing phase. The last five contributions are highly Planck-suppressed operators which give rise to a very small correction to the one-loop power spectrum for all the previously mentioned five phases. For more completeness, we are going to explicitly evaluate the cumulative effects of all of these mentioned contributions to the one-loop corrected result computed in the five phases. The five contributions provided in the final line of the above calculation will benefit the one-loop adjusted power spectrum only in the contracting, bouncing, and SRI, SRII regions. These terms show more suppression in the USR phase of the one-loop correction than in the contracting, bouncing, and SRI, SRII regions. As a result, the enhanced one-loop corrected power spectrum of scalar modes will remain unchanged by these factors in the final expression. Once again, in the analogous one-loop result, the remaining contributions exhibit more suppression in the USR phase compared to the contracting, bouncing, and SRI, SRII regions.

\subsection{ In-In formalism and one-loop computation}

In this subsection, we will explicitly analyze the contributions of each term as appearing in the third-order action, with a focus on the most significant highlighted term that arises as a result of the EFT framework that we have chosen for our present study. To this goal, we employ the well-known in-in formalism.  This implies that the two-point correlation function of the next quantum operator at $\tau\rightarrow 0$, the late time scale, may be represented as follows:
 \bea \label{Hamx}\langle\hat{\zeta}_{\bf p}\hat{\zeta}_{-{\bf p}}\rangle:&=&\lim_{\tau\rightarrow 0}\langle\hat{\zeta}_{\bf p}(\tau)\hat{\zeta}_{-{\bf p}}(\tau)\rangle\nonumber\\
    &=&\lim_{\tau\rightarrow 0}\left\langle\bigg[\overline{T}\exp\bigg(i\int^{\tau}_{-\infty(1-i\epsilon)}d\tau^{'}\;H_{\rm int}(\tau^{'})\bigg)\bigg]\;\;\hat{\zeta}_{\bf p}(\tau)\hat{\zeta}_{-{\bf p}}(\tau)%\right.\nonumber\\
%&& \left.
\;\;\bigg[{T}\exp\bigg(-i\int^{\tau}_{-\infty(1+i\epsilon)}d\tau^{''}\;H_{\rm int}(\tau^{''})\bigg)\bigg]\right\rangle,\quad\quad \eea
where $\overline{T}$ and $T$ characterize the time ordering and anti-time operations, respectively. Here $H_{\rm int}(\tau)$ represents the interaction Hamiltonian, which may be obtained from the third-order EFT action as follows:
\bea && H_{\rm int}(\tau)=-\int d^3x\;  M^2_{ pl}a^2\; \bigg[\left(3\left(c^2_s-1\right)\epsilon+\epsilon^2-\frac{1}{2}\epsilon^3\right)\zeta^{'2}\zeta+\frac{\epsilon}{c^2_s}\bigg(\epsilon-2s+1-c^2_s\bigg)\left(\partial_i\zeta\right)^2\zeta\nonumber\\ 
&&\quad\quad\quad\quad\quad\quad\quad\quad\quad\quad\quad-\frac{2\epsilon}{c^2_s}\zeta^{'}\left(\partial_i\zeta\right)\left(\partial_i\partial^{-2}\left(\frac{\epsilon\zeta^{'}}{c^2_s}\right)\right)-\frac{1}{aH}\left(1-\frac{1}{c^2_{s}}\right)\epsilon \bigg(\zeta^{'3}+\zeta^{'}(\partial_{i}\zeta)^2\bigg)
     \nonumber\\
&& \quad\quad\quad\quad\quad\quad\quad\quad\quad\quad\quad+\frac{1}{2}\epsilon\zeta\left(\partial_i\partial_j\partial^{-2}\left(\frac{\epsilon\zeta^{'}}{c^2_s}\right)\right)^2
+\underbrace{\frac{1}{2c^2_s}\epsilon\partial_{\tau}\left(\frac{\eta}{c^2_s}\right)\zeta^{'}\zeta^{2}}_{\bf Most~dominant ~term~in~USR}\nonumber\\ 
&&\quad\quad\quad\quad\quad\quad\quad\quad\quad\quad\quad+\frac{3}{2}\frac{1}{aH}\frac{\bar{M}^3_1}{ HM^2_{ pl}}
	  \zeta^{'}(\partial_{i}\zeta)^2+\frac{9}{2}\frac{\bar{M}^3_1}{ HM^2_{ pl}}\zeta \zeta^{'2}\nonumber\\ 
&&\quad\quad\quad\quad\quad\quad\quad\quad\quad\quad\quad+\bigg(\frac{3}{2}\frac{\bar{M}^3_1}{ HM^2_{ pl}}-\frac{4}{3}\frac{M^4_3}{H^2M^2_{ pl}}\bigg)\zeta^{'3}-\frac{3}{2}\frac{1}{aH}\frac{\bar{M}^3_1}{ HM^2_{ pl}} \zeta\partial_{\tau}\left(\partial_{i}\zeta\right)^2
  \bigg],\quad\quad\eea
where we explicitly utilized the fact that the interaction Hamiltonian is related to the Lagrangian density that describes the third-order perturbation via the Legendre transformation, $H_{\rm int}=-{\cal L}_3$. Further using the Dyson Swinger series one can explicitly write down the expression for the two-point correlation function of the scalar modes by the following expression:
\bea\label{g}\langle\hat{\zeta}_{\bf p}\hat{\zeta}_{-{\bf p}}\rangle=\langle\hat{\zeta}_{\bf p}\hat{\zeta}_{-{\bf p}}\rangle_{\bf Tree}+\langle\hat{\zeta}_{\bf p}\hat{\zeta}_{-{\bf p}}\rangle_{\bf One-loop},\eea
where the tree and one-loop level contributions are given by the following results:
 \bea  &&\label{g1}\langle\hat{\zeta}_{\bf p}\hat{\zeta}_{-{\bf p}}\rangle_{\bf Tree}= \langle\hat{\zeta}_{\bf p}\hat{\zeta}_{-{\bf p}}\rangle_{(0,0)},\\
 &&\label{g2}\langle\hat{\zeta}_{\bf p}\hat{\zeta}_{-{\bf p}}\rangle_{\bf One-loop}=\langle\hat{\zeta}_{\bf p}\hat{\zeta}_{-{\bf p}}\rangle_{(0,1)}+\langle\hat{\zeta}_{\bf p}\hat{\zeta}_{-{\bf p}}\rangle^{\dagger}_{(0,1)}+\langle\hat{\zeta}_{\bf p}\hat{\zeta}_{-{\bf p}}\rangle_{(0,2)}+\langle\hat{\zeta}_{\bf p}\hat{\zeta}_{-{\bf p}}\rangle^{\dagger}_{(0,2)}+\langle\hat{\zeta}_{\bf p}\hat{\zeta}_{-{\bf p}}\rangle_{(1,1)}.
\eea
Explicit contribution that are appearing in the tree level and in the one-loop level result of the two point correlation function of the scalar modes are appended below, which we need to evaluate for the previously mentioned five consecutive phases in this paper:
\bea
     &&\label{c0}\langle\hat{\zeta}_{\bf p}\hat{\zeta}_{-{\bf p}}\rangle_{(0,0)}=\lim_{\tau\rightarrow 0}\langle \hat{\zeta}_{\bf p}(\tau)\hat{\zeta}_{-{\bf p}}(\tau)\rangle,\\
    &&\label{c1}\langle\hat{\zeta}_{\bf p}\hat{\zeta}_{-{\bf p}}\rangle_{(0,1)}=-i\lim_{\tau\rightarrow 0}\int^{\tau}_{-\infty}d\tau_1\;\langle \hat{\zeta}_{\bf p}(\tau)\hat{\zeta}_{-{\bf p}}(\tau)H_{\rm int}(\tau_1)\rangle,\\
 &&\label{c2}\langle\hat{\zeta}_{\bf p}\hat{\zeta}_{-{\bf p}}\rangle^{\dagger}_{(0,1)}=-i\lim_{\tau\rightarrow 0}\int^{\tau}_{-\infty}d\tau_1\;\langle \hat{\zeta}_{\bf p}(\tau)\hat{\zeta}_{-{\bf p}}(\tau)H_{\rm int}(\tau_1)\rangle^{\dagger},\\
 &&\label{c3}\langle\hat{\zeta}_{\bf p}\hat{\zeta}_{-{\bf p}}\rangle_{(0,2)}=\lim_{\tau\rightarrow 0}\int^{\tau}_{-\infty}d\tau_1\;\int^{\tau}_{-\infty}d\tau_2\;\langle \hat{\zeta}_{\bf p}(\tau)\hat{\zeta}_{-{\bf p}}(\tau)H_{\rm int}(\tau_1)H_{\rm int}(\tau_2)\rangle,\\
 &&\label{c4}\langle\hat{\zeta}_{\bf p}\hat{\zeta}_{-{\bf p}}\rangle^{\dagger}_{(0,2)}=\lim_{\tau\rightarrow 0}\int^{\tau}_{-\infty}d\tau_1\;\int^{\tau}_{-\infty}d\tau_2\;\langle \hat{\zeta}_{\bf p}(\tau)\hat{\zeta}_{-{\bf p}}(\tau)H_{\rm int}(\tau_1)H_{\rm int}(\tau_2)\rangle^{\dagger},\\
  &&\label{c5}\langle\hat{\zeta}_{\bf p}\hat{\zeta}_{-{\bf p}}\rangle^{\dagger}_{(1,1)}=\lim_{\tau\rightarrow 0}\int^{\tau}_{-\infty}d\tau_1\;\int^{\tau}_{-\infty}d\tau_2\;\langle H_{\rm int}(\tau_1)\hat{\zeta}_{\bf p}(\tau)\hat{\zeta}_{-{\bf p}}(\tau)H_{\rm int}(\tau_2)\rangle^{\dagger}.\eea

\subsection{Calculation of one-loop corrected power spectrum: Individual phases}
\label{s9c}

 \subsubsection{Phase I: Ekpyrotic/Matter contraction}
 
 After implementing the cut-off regularization in the ekpyrotic/matter contraction scenario, which is applied within the momentum interval $k_c<k<k_b$ we get the following result for the one-loop contribution that appears from Phase I:
 \bea \bigg[\Delta^{2}_{\zeta, {\bf One-loop}}(p)\bigg]_{\bf CONTRACTION}&=&\bigg[\Delta^{2}_{\zeta,{\bf Tree}}(p)\bigg]^2_{\bf SRI}\times\Bigg(1+\frac{2}{15\pi^2}\frac{1}{c^2_{s}k^2_c}\bigg(-\left(1-\frac{1}{c^2_{s}}\right)\epsilon_c+6\frac{\bar{M}^3_1}{ HM^2_{ pl}}-\frac{4}{3}\frac{M^4_3}{H^2M^2_{ pl}}\bigg)\Bigg)\nonumber\\&&\quad\quad\quad\quad\quad\quad\quad\quad\quad\quad\quad\quad\quad\quad\quad\quad\quad\quad\quad\quad\quad\quad\quad\times\Bigg({\cal K}_{\bf C}-\frac{4}{3}{\bf I}_{\bf C}\Bigg),\eea
 where ${\cal K}_{\bf C}$ is the counter-term contribution in Phase I which we need to explicitly determine while performing the renormalization in the present context of the computation. Additionally, it is essential to consider that the prime contribution from the one-loop cut-off regulated momentum integral ${\bf I}_{\bf C}$ during the ekpyrotic/matter contraction is described by the following expression:
 \bea \label{eqx1} {\bf I}_{\bf C}:&=& \left(\frac{\epsilon_*}{\epsilon_c}\right)\times\int^{k_b}_{k_c}\frac{dk}{k}\;\left(\frac{k}{k_*}\right)^{\delta_{\bf C}}\;\left(1+k^2c^2_s\tau^2\right),\nonumber\\
 &=& \left(\frac{\epsilon_*}{\epsilon_c}\right)\times\int^{k_b/k_*}_{k_c/k_*}d\left(\frac{k}{k_*}\right)\;\left(\frac{k}{k_*}\right)^{\delta_{\bf C}-1}\;\left(1+\left(\frac{k}{k_*}\right)^{2}\right),\nonumber\\
 &=&\left(\frac{\epsilon_*}{\epsilon_c}\right)\times\Bigg[\frac{1}{\delta_{\bf C}}\left\{\left(\frac{k_b}{k_*}\right)^{\delta_{\bf C}}-\left(\frac{k_c}{k_*}\right)^{\delta_{\bf C}}\right\}+\frac{1}{\delta_{\bf C}+2}\left\{\left(\frac{k_b}{k_*}\right)^{\delta_{\bf C}+2}-\left(\frac{k_c}{k_*}\right)^{\delta_{\bf C}+2}\right\}\Bigg],\nonumber\\
  &=&\left(\frac{\epsilon_*}{\epsilon_c}\right)\times\Bigg[\frac{1}{\left(3-2\nu+\frac{2\epsilon_c}{\epsilon_c-1}\right)}\left\{\left(\frac{k_b}{k_*}\right)^{\left(3-2\nu+\frac{2\epsilon_c}{\epsilon_c-1}\right)}-\left(\frac{k_c}{k_*}\right)^{\left(3-2\nu+\frac{2\epsilon_c}{\epsilon_c-1}\right)}\right\},\nonumber\\
  &&\quad\quad\quad+\frac{1}{\left(5-2\nu+\frac{2\epsilon_c}{\epsilon_c-1}\right)}\left\{\left(\frac{k_b}{k_*}\right)^{\left(5-2\nu+\frac{2\epsilon_c}{\epsilon_c-1}\right)}-\left(\frac{k_c}{k_*}\right)^{\left(5-2\nu+\frac{2\epsilon_c}{\epsilon_c-1}\right)}\right\}\Bigg].\eea
  In terms of the number of e-foldings, the final result of the loop integral in the contracting phase can be further simplified as:
  \bea {\bf I}_{\bf C}:&=&\left(\frac{\epsilon_*}{\epsilon_c}\right)\times\Bigg[\frac{1}{\delta_{\bf C}}\bigg\{\exp(-\delta_{\bf C}\Delta N_{\bf B})-\exp(-\delta_{\bf C}(\Delta N_{\bf B}+\Delta N_{\bf B}))\bigg\}\nonumber\\
  &&\quad\quad\quad\quad\quad\quad\quad\quad\quad+\frac{1}{\delta_{\bf C}+2}\bigg\{\exp(-(\delta_{\bf C}+2)\Delta N_{\bf B})-\exp(-(\delta_{\bf C}+2)(\Delta N_{\bf B}+\Delta N_{\bf B}))\bigg\}\Bigg].\eea
   Here we define:
 \bea \Delta N_{\bf B}:=(N_*-N_b),\quad\quad\quad {\rm and}\quad\quad\quad \Delta N_{\bf C}:=(N_b-N_c).\eea
 Here we have introduced a new symbol $\delta_{\bf C}$ for Phase I which is defined by the following expression:
 \bea \delta_{\bf C}:=\left(3-2\nu+\frac{2\epsilon_c}{\epsilon_c-1}\right).\eea
 To avoid any further confusion it is important to note that the effective mass parameter $\nu$ here is evaluated at the ekpyrotic/matter contraction phase. Depending on the values of the first and second slow-roll parameters one can separately quantify the one-loop correction in both the ekpyrotic and matter contracting phases which are part of Phase I. Here to arrive at the final one-loop corrected result we have utilized the fact that the initial vacuum state is described by the Bunch Davies initial condition and the corresponding Bogolibov coefficient is given by $\alpha_1=1$ and $\beta_1=0$.

 \subsubsection{Phase II: Ekpyrotic/Matter bounce}

 Further implementing the cut-off regularization in the ekpyrotic/matter bounce scenario, which is applied within the momentum interval $k_b<k<k_*$ we get the following result for the one-loop contribution that appears from Phase II:
 \bea \bigg[\Delta^{2}_{\zeta, {\bf One-loop}}(p)\bigg]_{\bf BOUNCE}&=&\bigg[\Delta^{2}_{\zeta,{\bf Tree}}(p)\bigg]^2_{\bf SRI}\times\Bigg(1+\frac{2}{15\pi^2}\frac{1}{c^2_{s}k^2_b}\bigg(-\left(1-\frac{1}{c^2_{s}}\right)\epsilon_b+6\frac{\bar{M}^3_1}{ HM^2_{ pl}}-\frac{4}{3}\frac{M^4_3}{H^2M^2_{ pl}}\bigg)\Bigg)\nonumber\\&&\quad\quad\quad\quad\quad\quad\quad\quad\quad\quad\quad\quad\quad\quad\quad\quad\quad\quad\quad\quad\quad\quad\quad\quad\quad\quad\times\Bigg({\cal K}_{\bf B}-\frac{4}{3}{\bf I}_{\bf B}\Bigg),\eea
 where ${\cal K}_{\bf B}$ is the counter-term contribution in Phase II which we need to explicitly determine while performing the renormalization in the present context of the computation. Additionally, it is essential to consider that the prime contribution from the one-loop cut-off regulated momentum integral ${\bf I}_{\bf B}$ during the ekpyrotic/matter bounce is described by the following expression:
 \bea \label{eqx2}{\bf I}_{\bf B}:&=& \left(\frac{\epsilon_*}{\epsilon_b}\right)\times\int^{k_*}_{k_b}\frac{dk}{k}\;\left(\frac{k}{k_*}\right)^{2}\left(1+\left(\frac{k_*}{k}\right)^2\right)^{-\frac{1}{(\epsilon_b-1)}}\;\left(\frac{k}{k_*}\right)^{3-2\nu}\left(1+k^2c^2_s\tau^2\right),\nonumber\\
 &=& \left(\frac{\epsilon_*}{\epsilon_b}\right)\times\int^{1}_{k_b/k_*}d\left(\frac{k}{k_*}\right)\;\left(\frac{k}{k_*}\right)^{\frac{\epsilon_b+1}{\epsilon_b-1}}\;\left(\frac{k}{k_*}\right)^{3-2\nu}\left(1+\left(\frac{k}{k_*}\right)^{2}\right)^{\frac{\epsilon_b-2}{\epsilon_b-1}},\nonumber\\
 &=& \left(\frac{\epsilon_*}{\epsilon_b}\right)\times\int^{1}_{k_b/k_*}d\left(\frac{k}{k_*}\right)\;\left(\frac{k}{k_*}\right)^{1+\delta_{\bf B}}\left(1+\left(\frac{k}{k_*}\right)^{2}\right)^{\frac{\epsilon_b-2}{\epsilon_b-1}},\nonumber\\
 &=&\left(\frac{\epsilon_*}{\epsilon_b}\right)\times\frac{1}{\delta_{\bf B} +2}\bigg[\, _2F_1\left(\frac{\delta_{\bf B}+2}{2},\frac{1}{\epsilon_b-1}-1;\frac{\delta_{\bf B}+4}{2};-1\right)\nonumber\\
 &&\quad\quad\quad\quad\quad\quad\quad\quad\quad\quad-\left(\frac{k_{b}}{k_{*}}\right)^{\delta_{\bf B}+2}\, _2F_1\left(\frac{\delta_{\bf B}+2}{2},\frac{1}{\epsilon_b-1}-1;\frac{\delta_{\bf B}+4}{2};-\left(\frac{k_{b}}{k_{*}}\right)^{2}\right)\bigg],\nonumber\\
 &=&\left(\frac{\epsilon_*}{\epsilon_b}\right)\times\frac{1}{\left(5-2\nu+\frac{2}{\epsilon_b-1}\right)}\bigg[\, _2F_1\left(\left(\frac{5}{2}-\nu+\frac{1}{\epsilon_b-1}\right),\frac{1}{\epsilon_b-1}-1;\left(\frac{7}{2}-\nu+\frac{1}{\epsilon_b-1}\right);-1\right)\nonumber\\
 &&\quad\quad\quad\quad\quad\quad-\left(\frac{k_b}{k_*}\right)^{5-2\nu+\frac{2}{\epsilon_b-1}}\, _2F_1\left(\left(\frac{5}{2}-\nu+\frac{1}{\epsilon_b-1}\right),\frac{1}{\epsilon_b-1}-1;\left(\frac{7}{2}-\nu+\frac{1}{\epsilon_b-1}\right);-\left(\frac{k_b}{k_*}\right)^{2}\right)\bigg].\quad\quad\eea
  In terms of the number of e-foldings, the final result of the loop integral in the bouncing phase can be further simplified as:
  \bea {\bf I}_{\bf B}:&=&\left(\frac{\epsilon_*}{\epsilon_b}\right)\times\frac{1}{\delta_{\bf B} +2}\bigg[\, _2F_1\left(\frac{\delta_{\bf B}+2}{2},\frac{1}{\epsilon_b-1}-1;\frac{\delta_{\bf B}+4}{2};-1\right)\nonumber\\
 &&-\exp(-(\delta_{\bf B}+2)\Delta N_{\bf B})\, _2F_1\left(\frac{\delta_{\bf B}+2}{2},\frac{1}{\epsilon_b-1}-1;\frac{\delta_{\bf B}+4}{2};-\exp(-2\Delta N_{\bf B})\right)\bigg].\eea
 Additionally, we have introduced a new symbol $\delta_{\bf B}$ for Phase I which is defined by the following expression:
 \bea \delta_{\bf B}:=\left(3-2\nu+\frac{2}{\epsilon_b-1}\right).\quad\quad\eea
 To avoid any further confusion it is important to note that the effective mass parameter $\nu$ here is evaluated at the ekpyrotic/matter bounce phase. Depending on the values of the first and second slow-roll parameters one can separately quantify the one-loop correction in both the ekpyrotic and matter-bouncing phases which are part of Phase II. Here to arrive at the final one-loop corrected result we have utilized the fact that the initial vacuum state is described by the Bunch Davies initial condition and the corresponding Bogoliubov coefficient is given by $\alpha_1=1$ and $\beta_1=0$. We have also used the fact that the vacuum structure remains unchanged in Phase II.

 \subsubsection{Phase III: First slow-roll (SRI)}

 Next implementing the cut-off regularization in the first slow-roll i.e. SRI scenario, which is applied within the momentum interval $k_*<k<k_s$ we get the following result for the one-loop contribution that appears from Phase III:
 \bea \bigg[\Delta^{2}_{\zeta, {\bf One-loop}}(p)\bigg]_{\bf SRI}&=&\bigg[\Delta^{2}_{\zeta,{\bf Tree}}(p)\bigg]^2_{\bf SRI}\times\Bigg(1+\frac{2}{15\pi^2}\frac{1}{c^2_{s}k^2_*}\bigg(-\left(1-\frac{1}{c^2_{s}}\right)\epsilon_*+6\frac{\bar{M}^3_1}{ HM^2_{ pl}}-\frac{4}{3}\frac{M^4_3}{H^2M^2_{ pl}}\bigg)\Bigg)\nonumber\\&&\quad\quad\quad\quad\quad\quad\quad\quad\quad\quad\quad\quad\quad\quad\quad\quad\quad\quad\quad\quad\quad\quad\quad\quad\quad\quad\times\Bigg({\cal K}_{\bf SRI}-\frac{4}{3}{\bf I}_{\bf SRI}\Bigg),\eea
 where ${\cal K}_{\bf SRI}$ is the counter-term contribution in Phase III which we need to explicitly determine while performing the renormalization in the present context of the computation. Additionally, it is essential to consider that the prime contribution from the one-loop cut-off regulated momentum integral ${\bf I}_{\bf SRI}$ during the SRI is described by the following relation:
 \bea\label{eqx3} {\bf I}_{\bf SRI}:&=& \int^{k_s}_{k_*}\frac{dk}{k}\;\left(\frac{k}{k_*}\right)^{3-2\nu}\left(1+k^2c^2_s\tau^2\right),\nonumber\\
 &=& \int^{k_s/k_*}_{1}d\left(\frac{k}{k_*}\right)\;\left(\frac{k}{k_*}\right)^{2\delta_{\bf SRI}-1}\left(1+\left(\frac{k}{k_*}\right)^{2}\right),\nonumber\\
 &=& \bigg[\frac{1}{2\delta_{\bf SRI}}\left\{\left(\frac{k_s}{k_*}\right)^{2\delta_{\bf SRI}}-1\right\}+\frac{1}{2\left(\delta_{\bf SRI}+1\right)}\left\{\left(\frac{k_s}{k_*}\right)^{2\left(\delta_{\bf SRI}+1\right)}-1\right\}\bigg],\nonumber\\
 &=&\bigg[\frac{1}{2\delta_{\bf SRI}}+\ln\left(\frac{k_s}{k_*}\right)-\frac{1}{2\delta_{\bf SRI}}+\cdots+\frac{1}{2\left(\delta_{\bf SRI}+1\right)}\left\{\left(\frac{k_s}{k_*}\right)^{2\left(\delta_{\bf SRI}+1\right)}-1\right\}\bigg],\nonumber\\
 &=&\bigg[\ln\left(\frac{k_s}{k_*}\right)+\frac{1}{5-2\nu}\left\{\left(\frac{k_s}{k_*}\right)^{5-2\nu}-1\right\}+\cdots\bigg],\nonumber\\
 &=&\bigg[\ln\left(\frac{k_s}{k_*}\right)+\frac{1}{2+2\eta+6s-2\epsilon}\left\{\left(\frac{k_s}{k_*}\right)^{2+2\eta+6s-2\epsilon}-1\right\}+\cdots\bigg].\eea
 In terms of the number of e-foldings, the final result of the loop integral in the SRI phase can be further simplified as:
  \bea {\bf I}_{\bf SRI}:&=&\bigg[\Delta N_{\bf SRI}+\frac{1}{2\left(\delta_{\bf SRI}+1\right)}\bigg\{\exp(2\left(\delta_{\bf SRI}+1\right)\Delta N_{\bf SRI})-1\bigg\}\bigg]\quad\quad{\rm with}\quad\Delta N_{\bf SRI}:=(N_s-N_*).\eea
 Here we have introduced a new symbol $\delta_{\bf SRI}$ for Phase III which is defined by the following expression:
 \bea \delta_{\bf SRI}:=\left(\frac{3}{2}-\nu\right).\eea
 To avoid any further confusion it is essential to recall that the effective mass parameter $\nu$ here is evaluated at the SRI phase. Depending on the values of the first and second SR parameters one can quantify the one-loop correction in the SRI region which is part of Phase III. Here to arrive at the final one-loop corrected result we have utilized the fact that the initial vacuum state is described by the Bunch Davies initial condition and the corresponding Bogolibov coefficient is given by $\alpha_1=1$ and $\beta_1=0$. We have also used the fact that the vacuum structure remains unchanged in Phase III. Additionally, it is important to note that to avoid the singularity in the loop integral ${\bf I}_{\bf SRI}$ at $\nu=3/2$ i.e. $\delta_{\bf SRI}=0$, we have used dimensional regularization to remove the dependence on the parameter $\delta_{\bf SRI}$ and after implementing this successfully we have obtained the logarithmic contribution which is in principle IR divergent. Also, it is evident from the last two terms that exactly at $\nu=3/2$ there is no such previously mentioned singularity appears in the final result. However, from the derived result for the one-loop contribution from the SRI region, particularly from the last two terms it is clearly visible that if we consider exactly, $\nu=3/2$ with $\delta_{\bf SRI}=0$, then quadratic contribution $(k_s/k_*)^{2}$ appears which is technically a UV divergent term in this computation. But if we slightly deviate from the de Sitter limit with $\nu=3/2$ i.e. if we consider the quasi de Sitter approximation holds good perfectly in the SRI region then instead of getting a quadratic divergence just like above, we get here, in this case, the contribution, $(k_s/k_*)^{2+2\eta+6s-2\epsilon}$, which basically deviates from its quadratic power law behaviour. this further implies that the quasi de Sitter approximation in SRI automatically provides a regulator in the quadratic UV divergence term which appears in the exponent and is expressed in terms of the slow-roll parameters $\epsilon$, $\eta$ and $s$, evaluated in Phase III.

 \subsubsection{Phase IV: Ultra slow-roll (USR)}
 The highlighted cubic self-interaction will be examined next since it will also have an influence on the two-point correlation function of the scalar modes at the one-loop level during the insertion of a single USR period with the sharp transition. This function can be expressed in the following ways:
\bea   \langle\hat{\zeta}_{\bf p}\hat{\zeta}_{-{\bf p}}\rangle_{(0,1)}&=& -\frac{iM^2_{ pl}}{2}\int^{0}_{-\infty}d\tau\frac{a^2(\tau)}{c^2_s(\tau)}\epsilon(\tau)\partial_{\tau}\left(\frac{\eta(\tau)}{c^2_s(\tau)}\right)\nonumber\\
  &&\times\int \frac{d^{3}{\bf k}_1}{(2\pi)^3} \int \frac{d^{3}{\bf k}_2}{(2\pi)^3} \int \frac{d^{3}{\bf k}_3}{(2\pi)^3} \nonumber\\
  && \times\delta^3\bigg({\bf k}_1+{\bf k}_2+{\bf k}_3\bigg) \times \langle \hat{\zeta}_{\bf p}\hat{\zeta}_{-{\bf p}}\hat{\zeta}^{'}_{{\bf k}_1}(\tau)\hat{\zeta}_{{\bf k}_2}(\tau)\hat{\zeta}_{{\bf k}_3}(\tau)\rangle,\\
   \langle\hat{\zeta}_{\bf p}\hat{\zeta}_{-{\bf p}}\rangle_{(0,1)}&=& -\frac{iM^2_{ pl}}{2}\int^{0}_{-\infty}d\tau\frac{a^2(\tau)}{c^2_s(\tau)}\epsilon(\tau)\partial_{\tau}\left(\frac{\eta(\tau)}{c^2_s(\tau)}\right)\nonumber\\
  &&\times\int \frac{d^{3}{\bf k}_1}{(2\pi)^3} \int \frac{d^{3}{\bf k}_2}{(2\pi)^3} \int \frac{d^{3}{\bf k}_3}{(2\pi)^3} \nonumber\\
  && \times\delta^3\bigg({\bf k}_1+{\bf k}_2+{\bf k}_3\bigg) \times \langle \hat{\zeta}_{\bf p}\hat{\zeta}_{-{\bf p}}\hat{\zeta}^{'}_{{\bf k}_1}(\tau)\hat{\zeta}_{{\bf k}_2}(\tau)\hat{\zeta}_{{\bf k}_3}(\tau)\rangle^{\dagger},\\
     \langle\hat{\zeta}_{\bf p}\hat{\zeta}_{-{\bf p}}\rangle_{(0,2)}&=& -\frac{M^4_{ pl}}{4}\int^{0}_{-\infty}d\tau_1\frac{a^2(\tau_1)}{c^2_s(\tau_1)}\epsilon(\tau_1)\partial_{\tau_1}\left(\frac{\eta(\tau_1)}{c^2_s(\tau_1)}\right)\;\int^{0}_{-\infty}d\tau_2\;\frac{a^2(\tau_2)}{c^2_s(\tau_2)}\epsilon(\tau_2)\partial_{\tau_2}\left(\frac{\eta(\tau_2)}{c^2_s(\tau_2)}\right)\nonumber\\
  &&\times\int \frac{d^{3}{\bf k}_1}{(2\pi)^3} \int \frac{d^{3}{\bf k}_2}{(2\pi)^3} \int \frac{d^{3}{\bf k}_3}{(2\pi)^3} \int \frac{d^{3}{\bf k}_4}{(2\pi)^3} \int \frac{d^{3}{\bf k}_5}{(2\pi)^3} \int \frac{d^{3}{\bf k}_6}{(2\pi)^3}\nonumber\\
  &&\times \delta^3\bigg({\bf k}_1+{\bf k}_2+{\bf k}_3\bigg) \delta^3\bigg({\bf k}_4+{\bf k}_5+{\bf k}_6\bigg)\nonumber\\
  &&\times \langle \hat{\zeta}_{\bf p}\hat{\zeta}_{-{\bf p}}\hat{\zeta}^{'}_{{\bf k}_1}(\tau_1)\hat{\zeta}_{{\bf k}_2}(\tau_1)\hat{\zeta}_{{\bf k}_3}(\tau_1)\hat{\zeta}^{'}_{{\bf k}_4}(\tau_2)\hat{\zeta}_{{\bf k}_5}(\tau_2)\hat{\zeta}_{{\bf k}_6}(\tau_2)\rangle,\\
  \langle\hat{\zeta}_{\bf p}\hat{\zeta}_{-{\bf p}}\rangle^{\dagger}_{(0,2)}&=& -\frac{M^4_{ pl}}{4}\int^{0}_{-\infty}d\tau_1\frac{a^2(\tau_1)}{c^2_s(\tau_1)}\epsilon(\tau_1)\partial_{\tau_1}\left(\frac{\eta(\tau_1)}{c^2_s(\tau_1)}\right)\;\int^{0}_{-\infty}d\tau_2\;\frac{a^2(\tau_2)}{c^2_s(\tau_2)}\epsilon(\tau_2)\partial_{\tau_2}\left(\frac{\eta(\tau_2)}{c^2_s(\tau_2)}\right)\nonumber\\
  &&\times\int \frac{d^{3}{\bf k}_1}{(2\pi)^3} \int \frac{d^{3}{\bf k}_2}{(2\pi)^3} \int \frac{d^{3}{\bf k}_3}{(2\pi)^3} \int \frac{d^{3}{\bf k}_4}{(2\pi)^3} \int \frac{d^{3}{\bf k}_5}{(2\pi)^3} \int \frac{d^{3}{\bf k}_6}{(2\pi)^3}\nonumber\\
  &&\times \delta^3\bigg({\bf k}_1+{\bf k}_2+{\bf k}_3\bigg) \delta^3\bigg({\bf k}_4+{\bf k}_5+{\bf k}_6\bigg)\nonumber\\
  &&\times \langle \hat{\zeta}_{\bf p}\hat{\zeta}_{-{\bf p}}\hat{\zeta}^{'}_{{\bf k}_1}(\tau_1)\hat{\zeta}_{{\bf k}_2}(\tau_1)\hat{\zeta}_{{\bf k}_3}(\tau_1)\hat{\zeta}^{'}_{{\bf k}_4}(\tau_2)\hat{\zeta}_{{\bf k}_5}(\tau_2)\hat{\zeta}_{{\bf k}_6}(\tau_2)\rangle^{\dagger},\\
 \langle\hat{\zeta}_{\bf p}\hat{\zeta}_{-{\bf p}}\rangle_{(1,1)}&=& \frac{M^4_{ pl}}{4}\int^{0}_{-\infty}d\tau_1\frac{a^2(\tau_1)}{c^2_s(\tau_1)}\epsilon(\tau_1)\partial_{\tau_1}\left(\frac{\eta(\tau_1)}{c^2_s(\tau_1)}\right)\;\int^{0}_{-\infty}d\tau_2\;\frac{a^2(\tau_2)}{c^2_s(\tau_2)}\epsilon(\tau_2)\partial_{\tau_2}\left(\frac{\eta(\tau_2)}{c^2_s(\tau_2)}\right)\nonumber\\
  &&\times\int \frac{d^{3}{\bf k}_1}{(2\pi)^3} \int \frac{d^{3}{\bf k}_2}{(2\pi)^3} \int \frac{d^{3}{\bf k}_3}{(2\pi)^3} \int \frac{d^{3}{\bf k}_4}{(2\pi)^3} \int \frac{d^{3}{\bf k}_5}{(2\pi)^3} \int \frac{d^{3}{\bf k}_6}{(2\pi)^3}\nonumber\\
  &&\times \delta^3\bigg({\bf k}_1+{\bf k}_2+{\bf k}_3\bigg) \delta^3\bigg({\bf k}_4+{\bf k}_5+{\bf k}_6\bigg)\nonumber\\
  &&\times \langle \hat{\zeta}^{'}_{{\bf k}_1}(\tau_1)\hat{\zeta}_{{\bf k}_2}(\tau_1)\hat{\zeta}_{{\bf k}_3}(\tau_1)\hat{\zeta}_{\bf p}\hat{\zeta}_{-{\bf p}}\hat{\zeta}^{'}_{{\bf k}_4}(\tau_2)\hat{\zeta}_{{\bf k}_5}(\tau_2)\hat{\zeta}_{{\bf k}_6}(\tau_2)\rangle. \eea
In the presence of sharp transitions at the conformal time scales $\tau=\tau_s$ (from SRII to USR) and $\tau=\tau_e$ (from USR to SRII), we have the following formula:
\bea \left(\frac{\eta(\tau)}{c^2_s(\tau)}\right)^{'}\approx\frac{\Delta \eta(\tau)}{c^2_s(\tau)}\bigg(\delta(\tau-\tau_e)-\delta(\tau-\tau_s)\bigg).\eea
As a result for $\tau_s<\tau<\tau_e$ region we get:
\be \displaystyle \left(\frac{\eta(\tau)}{c^2_s(\tau)}\right)^{'}\approx 0,\ee 
In the current context, let us investigate the time-dependent parametrization of the effective sound speed parameter with the purpose of improving comprehension. Throughout the time scale evolution, the effective sound speed parameter is fixed at the CMB pivot scale value, $c_s(\tau_*)=c_{s}$. The parameterization is as follows, but for abrupt transition points: $c_s(\tau_e)=c_s(\tau_s)=\tilde{c}_s=1\pm \delta$. $\delta$ represents a fine-tuning parameter that is kept constant by retaining $\delta\ll 1$. This additional information will be extremely useful for rest of the computation performed in this paper.

 Further implementing the cut-off regularization in USR region, which is applied within the momentum interval $k_s<k<k_e$ we get the following result for the one-loop contribution that appears from Phase IV:
 \bea \bigg[\Delta^{2}_{\zeta, {\bf One-loop}}(p)\bigg]_{\bf USR}&=&\bigg\{\bigg[\frac{1}{4}\bigg[\Delta^{2}_{\zeta,{\bf Tree}}(p)\bigg]^2_{\bf SRI}\times\bigg(\frac{\left(\Delta\eta(\tau_e)\right)^2}{c^8_s} \left(\frac{k_e}{k_s}\right)^{6}{\bf I}^{(1)}_{\bf USR}(\tau_e)- \frac{\left(\Delta\eta(\tau_s)\right)^2}{c^8_s}{\bf I}^{(1)}_{\bf USR}(\tau_s)\bigg)\nonumber\\
&&\quad\quad+\frac{1}{2}\bigg[\Delta^{2}_{\zeta,{\bf Tree}}(p)\bigg]^2_{\bf SRI}\times\bigg(\frac{\left(\Delta\eta(\tau_e)\right)}{c^4_s}\left(\frac{k_e}{k_s}\right)^{6}{\bf I}^{(2)}_{\bf USR}(\tau_e)-\frac{\left(\Delta\eta(\tau_s)\right)}{c^4_s}{\bf I}^{(2)}_{\bf USR}(\tau_s)\bigg)\bigg]\nonumber\\
&&\quad\quad\quad\quad\quad\quad\quad\quad\quad\quad\quad\quad-\bigg[\Delta^{2}_{\zeta,{\bf Tree}}(p)\bigg]^2_{\bf SRI}\times{\cal K}_{\bf USR}\bigg\},\quad\quad\quad\eea
 where ${\cal K}_{\bf USR}$ is the counter-term contribution in Phase IV which we need to explicitly determine while performing the renormalization in the present context of the computation. Additionally, it is important to note that the prime contribution from the one-loop cut-off regulated momentum integrals ${\bf I}^{(1)}_{\bf USR}$ and ${\bf I}^{(2)}_{\bf USR}$ during the USR phase is described by the following expression:
\bea {\bf I}^{(1)}_{\bf USR}(\tau)&=&\int^{k_e}_{k_s}\frac{dk}{k}\;\left|{\cal J}_{\bf k}(\tau)\right|^{2},\eea
In this instance, a new function ${\cal J}_{\bf k}(\tau)$ is defined as follows:
\bea \label{hhgxee} {\cal J}_{\bf k}(\tau)&=&\left(-kc_s\tau\right)^{2\delta_{\bf USR}}\Bigg\{\alpha_2 (1+i k c_s\tau) e^{-i\left(k c_s\tau+\frac{\pi}{2}\left(\nu+\frac{1}{2}\right)\right)}-\beta_2(1-i k c_s \tau)e^{i\left(k c_s\tau+\frac{\pi}{2}\left(\nu+\frac{1}{2}\right)\right)}\Bigg\}.\eea
In the USR phase, the Bogoliubov coefficients $\alpha_{2}$ and $\beta_{2}$ are denoted by the following expressions:
\bea 
 &&\alpha_{{2}} =  \frac{1}{2 k^3 \tau_s^3 c_s^3}  \Bigg(3 i + 3 i  k^2 c_s^2 \tau_s ^2  + 2  k^3 c_s^3  \tau_s ^3  \Bigg  ), \\  &&\beta_{{2}} = \frac{1}{2 k ^3 c_s ^3 \tau_s ^3}  \Bigg( 3i -6 k c_s \tau_s -3i k^2 c_s ^2 \tau_s^2 \Bigg) e^{-i\left(\pi\left(\nu+\frac{1}{2}\right)+ 2k c_s \tau_s\right)}.
\eea 
using the above-mentioned expressions in the loop integral we get the following simplified result:
\bea\label{eqx4}{\bf I}^{(1)}_{\bf USR}(\tau)&=&\bigg[\frac{1}{2\delta_{\bf USR}}\left\{\left(\frac{k_e}{k_s}\right)^{2\delta_{\bf USR}}-1\right\}+\frac{1}{2\left(\delta_{\bf USR}+1\right)}\left\{\left(\frac{k_e}{k_s}\right)^{2\left(\delta_{\bf USR}+1\right)}-1\right\}\bigg],\nonumber\\
 &=&\bigg[\frac{1}{2\delta_{\bf USR}}+\ln\left(\frac{k_e}{k_s}\right)-\frac{1}{2\delta_{\bf USR}}+\cdots+\frac{1}{2\left(\delta_{\bf USR}+1\right)}\left\{\left(\frac{k_e}{k_s}\right)^{2\left(\delta_{\bf USR}+1\right)}-1\right\}\bigg],\nonumber\\
 &=&\bigg[\ln\left(\frac{k_e}{k_s}\right)+\frac{1}{5-2\nu}\left\{\left(\frac{k_e}{k_s}\right)^{5-2\nu}-1\right\}+\cdots\bigg],\nonumber\\
 &=&\bigg[\ln\left(\frac{k_e}{k_s}\right)+\frac{1}{2+2\eta+6s-2\epsilon}\left\{\left(\frac{k_e}{k_s}\right)^{2+2\eta+6s-2\epsilon}-1\right\}+\cdots\bigg].\eea
  In terms of the number of e-foldings, the final result of the first loop integral in the USR phase can be further simplified as:
  \bea {\bf I}_{\bf USR}^{(1)}(N):&=&\bigg[\Delta N_{\bf USR}+\frac{1}{2\left(\delta_{\bf USR}+1\right)}\bigg\{\exp(2\left(\delta_{\bf USR}+1\right)\Delta N_{\bf USR})-1\bigg\}\bigg]\quad{\rm with}\quad\Delta N_{\bf USR}:=(N_e-N_s).\eea
 Here we have introduced a new symbol $\delta_{\bf USR}$ for Phase IV which is defined by the following expression:
 \bea \delta_{\bf USR}:=\left(\frac{3}{2}-\nu\right).\eea
Here the $\cdots$ represents the suppressed contribution in which we are not interested in the present discussion. 
Also from the derived result, it is clearly evident that the one-loop integral at the conformal time scales $\tau=\tau_s$ and $\tau=\tau_e$ turning out to be the same because of having sharp transitions at these mentioned points, which are the precise boundaries of the USR region. In a more technical language we then have:
\bea {\bf I}^{(1)}_{\bf USR}(\tau)={\bf I}^{(1)}_{\bf USR}(\tau_s)={\bf I}^{(1)}_{\bf USR}(\tau_e)={\bf I}^{(1)}_{\bf USR}.\eea
To avoid any further confusion it is essential to recall that the effective mass parameter $\nu$ here is evaluated at the USR phase. Depending on the values of the first and second slow-roll parameters one can quantify the one-loop correction in the USR region which is part of Phase IV. Here the expressions for $\delta_{\bf SRI}$ and $\delta_{\bf USR}$ look similar, but due to having different values of the mass parameter $\nu$ in SRI and USR phase we have actually $\delta_{\bf SRI}\neq \delta_{\bf USR}$. Here to arrive at the final one-loop corrected result we have utilized the fact that the initial vacuum state is described by the Bunch Davies initial condition and the corresponding Bogolibov coefficient is given by $\alpha_1=1$ and $\beta_1=0$. We have also used the fact that the vacuum structure remains unchanged up to Phase III. After that, a sharp transition takes place at the conformal time scale $\tau=\tau_s$ for which a phase transition occurs at this point, and as an immediate outcome Bunch Davies vacuum changes to a new state. It persists up to the end of the USR phase which is in our discussion and is identified with Phase IV. Further, this new vacuum state transforms to another vacuum state due to having another sharp transition at the scale $\tau=\tau_e$ from USR to SRII phase transition. Additionally, it is essential to consider that to avoid the singularity in the loop integral ${\bf I}^{(1)}_{\bf USR}$ at $\nu=3/2$ i.e. $\delta_{\bf USR}=0$, we have used dimensional regularization to remove the dependence on the parameter $\delta_{\bf USR}$ and after implementing this successfully we have obtained the logarithmic contribution which is in principle IR divergent. Also, it is evident from the last two terms that exactly at $\nu=3/2$ no such previously mentioned singularity appears in the final result. However, from the derived result for the one-loop contribution from the USR region, particularly from the last two terms it is clearly visible that if we consider exactly, $\nu=3/2$ with $\delta_{\bf USR}=0$, then quadratic contribution $(k_e/k_s)^{2}$ appears which is technically a UV divergent term in this computation. But if we slightly deviate from the de Sitter limit with $\nu=3/2$ i.e. If we consider the quasi de Sitter approximation holds good perfectly in the USR region then instead of getting a quadratic divergence just like above, we get here, in this case, the contribution, $(k_e/k_s)^{2+2\eta+6s-2\epsilon}$, which basically deviates from its quadratic power law behaviour. This further implies that the quasi de Sitter approximation in USR automatically provides a regulator in the quadratic UV divergence term which appears in the exponent and is expressed in terms of the slow-roll parameters $\epsilon$, $\eta$ and $s$, evaluated in Phase IV.

Another contribution to the one-loop contribution can be expressed as follows:
\bea  \label{gslk1cz} &&{\bf I}^{(2)}_{\bf USR}(\tau)=\int^{k_e}_{k_s}\frac{dk}{k}\;|{\cal J}_{\bf k}(\tau)|^2\bigg(\frac{d\ln|{\cal J}_{\bf k}(\tau)|^2 }{d\ln k}\bigg)=\int^{k_e}_{k_s}d\ln k\;\bigg(\frac{d|{\cal J}_{\bf k}(\tau)|^2 }{d\ln k}\bigg)=\Bigg[|{\cal J}_{\bf k}(\tau)|^2\Bigg]^{k_e}_{k_s},\eea
which further implies that:
\bea {\bf I}^{(2)}_{\bf USR}(\tau)={\bf I}^{(2)}_{\bf USR}(\tau_s)={\bf I}^{(2)}_{\bf USR}(\tau_e)={\bf I}^{(2)}_{\bf USR}=-{\cal O}(1).\eea
Utilizing the derived results for the loop integrals we can further write down the following expression:
\bea \bigg[\Delta^{2}_{\zeta, {\bf One-loop}}(p)\bigg]_{\bf USR}&=&\bigg\{\bigg[\frac{1}{4}\bigg[\Delta^{2}_{\zeta,{\bf Tree}}(p)\bigg]^2_{\bf SRI}\times\bigg(\frac{\left(\Delta\eta(\tau_e)\right)^2}{c^8_s} \left(\frac{k_e}{k_s}\right)^{6}- \frac{\left(\Delta\eta(\tau_s)\right)^2}{c^8_s}\bigg){\bf I}^{(1)}_{\bf USR}\nonumber\\
&&\quad\quad+\frac{1}{2}\bigg[\Delta^{2}_{\zeta,{\bf Tree}}(p)\bigg]^2_{\bf SRI}\times\bigg(\frac{\left(\Delta\eta(\tau_e)\right)}{c^4_s}\left(\frac{k_e}{k_s}\right)^{6}-\frac{\left(\Delta\eta(\tau_s)\right)}{c^4_s}\bigg){\bf I}^{(2)}_{\bf USR}\bigg]-\bigg[\Delta^{2}_{\zeta,{\bf Tree}}(p)\bigg]^2_{\bf SRI}\times{\cal K}_{\bf USR}\bigg\}\nonumber\\
&=&\left({\bf Y}^{(1)}_{\bf USR}+{\bf Y}^{(2)}_{\bf USR}-\bigg[\Delta^{2}_{\zeta,{\bf Tree}}(p)\bigg]^2_{\bf SRI}\times{\cal K}_{\bf USR}\right),\quad\quad\quad\eea
where we define ${\bf Y}^{(1)}_{\bf USR}$ and ${\bf Y}^{(2)}_{\bf USR}$ by the following expressions:
\bea && {\bf Y}^{(1)}_{\bf USR}=\frac{1}{4}\bigg[\Delta^{2}_{\zeta,{\bf Tree}}(p)\bigg]^2_{\bf SRI}\times\bigg(\frac{\left(\Delta\eta(\tau_e)\right)^2}{c^8_s} \left(\frac{k_e}{k_s}\right)^{6}- \frac{\left(\Delta\eta(\tau_s)\right)^2}{c^8_s}\bigg){\bf I}^{(1)}_{\bf USR},\\
&& {\bf Y}^{(2)}_{\bf USR}=\frac{1}{2}\bigg[\Delta^{2}_{\zeta,{\bf Tree}}(p)\bigg]^2_{\bf SRI}\times\bigg(\frac{\left(\Delta\eta(\tau_e)\right)}{c^4_s}\left(\frac{k_e}{k_s}\right)^{6}-\frac{\left(\Delta\eta(\tau_s)\right)}{c^4_s}\bigg){\bf I}^{(2)}_{\bf USR}.\eea
Now we need to study the relative contributions from both the factors, ${\bf Y}^{(1)}_{\bf USR}$ and ${\bf Y}^{(2)}_{\bf USR}$ and to serve the purpose we consider the following ratio:
\bea \frac{{\bf Y}^{(2)}_{\bf USR}}{{\bf Y}^{(1)}_{\bf USR}}=2\times\frac{\displaystyle \bigg(\frac{\left(\Delta\eta(\tau_e)\right)}{c^4_s}\left(\frac{k_e}{k_s}\right)^{6}-\frac{\left(\Delta\eta(\tau_s)\right)}{c^4_s}\bigg)}{\displaystyle \bigg(\frac{\left(\Delta\eta(\tau_e)\right)^2}{c^8_s} \left(\frac{k_e}{k_s}\right)^{6}- \frac{\left(\Delta\eta(\tau_s)\right)^2}{c^8_s}\bigg)}\times \frac{{\bf I}^{(2)}_{\bf USR}}{{\bf I}^{(1)}_{\bf USR}},\eea
where we have:
\bea \frac{{\bf I}^{(2)}_{\bf USR}}{{\bf I}^{(1)}_{\bf USR}}=-\frac{{\cal O}(1)}{\displaystyle \bigg(\ln\left(\frac{k_e}{k_s}\right)+\frac{1}{2+2\eta+6s-2\epsilon}\left\{\left(\frac{k_e}{k_s}\right)^{2+2\eta+6s-2\epsilon}-1\right\}\bigg)}\sim -0.009<1.\eea
To arrive at the numerical estimate we fix here $k_s=10^6 {\rm Mpc}^{-1}$, $k_e=10^7 {\rm Mpc}^{-1}$, $\epsilon=10^{-4}$, $\eta=-6$ and $s=10^{-4}$. Further fixing $\Delta\eta(\tau_e)=1$ and $\Delta\eta(\tau_s)=-6$ and confining the effective sound speed inside the interval $0.028<c_s<1$ we get: 
\bea -1.07\times 10^{-8}~({\rm for}~c_s=0.028)<\frac{{\bf Y}^{(2)}_{\bf USR}}{{\bf Y}^{(1)}_{\bf USR}}<-0.017~({\rm for}~c_s=1)\quad\quad\Longrightarrow\quad\quad\left|\frac{{\bf Y}^{(2)}_{\bf USR}}{{\bf Y}^{(1)}_{\bf USR}}\right|\ll 1.\eea
This further implies:
\bea \left({\bf Y}^{(1)}_{\bf USR}+{\bf Y}^{(2)}_{\bf USR}\right)={\bf Y}^{(1)}_{\bf USR}\left(1+\frac{{\bf Y}^{(2)}_{\bf USR}}{{\bf Y}^{(1)}_{\bf USR}}\right)\approx{\bf Y}^{(1)}_{\bf USR}.\eea
As a consequence, the final one-loop contribution from the USR period is given by:
\bea \bigg[\Delta^{2}_{\zeta, {\bf One-loop}}(p)\bigg]_{\bf USR}&=&\frac{1}{4}\bigg[\Delta^{2}_{\zeta,{\bf Tree}}(p)\bigg]^2_{\bf SRI}\times\bigg\{\bigg(\frac{\left(\Delta\eta(\tau_e)\right)^2}{c^8_s} \left(\frac{k_e}{k_s}\right)^{6}- \frac{\left(\Delta\eta(\tau_s)\right)^2}{c^8_s}\bigg){\bf I}_{\bf USR}-{\cal K}_{\bf USR}\bigg\}.\eea
Here it is important to note that in the last step, we have redefined the counter-term factor $4{\cal K}_{\bf USR}$ with the factor ${\cal K}_{\bf USR}$, which is implemented to do the rest of computation in a simpler fashion. Additionally we have introduced a new redefinition, ${\bf I}^{(1)}_{\bf USR}={\bf I}_{\bf USR}$.

 In terms of the number of e-foldings, the final result of the one-loop correction in the USR phase can be further simplified as:
 \bea \bigg[\Delta^{2}_{\zeta, {\bf One-loop}}(N)\bigg]_{\bf USR}&=&\frac{1}{4}\bigg[\Delta^{2}_{\zeta,{\bf Tree}}(N)\bigg]^2_{\bf SRI}\times\bigg\{\bigg(\frac{\left(\Delta\eta(\tau_e)\right)^2}{c^8_s} \exp(6\Delta N_{\bf USR})- \frac{\left(\Delta\eta(\tau_s)\right)^2}{c^8_s}\bigg)\nonumber\\
 &&\quad\quad\times\bigg[\Delta N_{\bf USR}+\frac{1}{2\left(\delta_{\bf USR}+1\right)}\bigg\{\exp(2\left(\delta_{\bf USR}+1\right)\Delta N_{\bf USR})-1\bigg\}\bigg]-{\cal K}_{\bf USR}\bigg\}.\eea

 \subsubsection{Phase V: Second slow-roll (SRII)}
Next implementing the cut-off regularization in the second slow-roll i.e. SRII scenario, which is applied within the momentum interval $k_e<k<k_{\rm end}$ we get the following result for the one-loop contribution that appears from Phase V:
 \bea \bigg[\Delta^{2}_{\zeta, {\bf One-loop}}(p)\bigg]_{\bf SRII}&=&\bigg[\Delta^{2}_{\zeta,{\bf Tree}}(p)\bigg]^2_{\bf SRI}\times\Bigg(1+\frac{2}{15\pi^2}\frac{1}{c^2_{s}k^2_*}\bigg(-\left(1-\frac{1}{c^2_{s}}\right)\epsilon_*+6\frac{\bar{M}^3_1}{ HM^2_{ pl}}-\frac{4}{3}\frac{M^4_3}{H^2M^2_{ pl}}\bigg)\Bigg)\nonumber\\&&\quad\quad\quad\quad\quad\quad\quad\quad\quad\quad\quad\quad\quad\quad\quad\quad\quad\quad\quad\quad\quad\quad\quad\quad\quad\quad\times\Bigg({\cal K}_{\bf SRII}+{\bf I}_{\bf SRII}\Bigg),\eea
 where ${\cal K}_{\bf SRII}$ is the counter-term contribution in Phase V which we need to explicitly determine while performing the renormalization in the present context of the computation. Additionally, it is essential to consider that the prime contribution from the one-loop cut-off regulated momentum integral ${\bf I}_{\bf SRII}$ during the SRII is described by the following expression:
 \bea {\bf I}_{\bf SRII}(\tau)&=&\left(\frac{k_e}{k_s}\right)^6\int^{k_{\rm end}}_{k_e}\frac{dk}{k}\;\left|{\cal P}_{\bf k}(\tau)\right|^{2},\eea
In this instance, a new function ${\cal J}_{\bf k}(\tau)$ is defined as follows:
\bea \label{hhgxee} {\cal P}_{\bf k}(\tau)&=&\left(-kc_s\tau\right)^{2\delta_{\bf SRII}}\Bigg\{\alpha_3 (1+i k c_s\tau) e^{-i\left(k c_s\tau+\frac{\pi}{2}\left(\nu+\frac{1}{2}\right)\right)}-\beta_3(1-i k c_s \tau)e^{i\left(k c_s\tau+\frac{\pi}{2}\left(\nu+\frac{1}{2}\right)\right)}\Bigg\}.\eea
In the SRII phase, the Bogoliubov coefficients $\alpha_{3}$ and $\beta_{3}$ are denoted by the following expressions:
\bea
 \alpha _{3} &=& \frac{1}{(2 k^3 \tau_e^3 c_s^3)(2 k^3 \tau_s^3 c_s^3)}\Bigg\{\left(-3 i -3 i  k^2 \tau_e^2 c_s^2 +2  k^3 \tau_e^3 c_s^3 \right) \left(3 i + 3 i  k^2 c_s^2 \tau_s ^2  + 2  k^3 c_s^3  \tau_s ^3  \right  )  \nonumber\\ &&
\quad\quad\quad\quad\quad\quad\quad\quad\quad\quad - \left(-3 i -6  k \tau_e c_s   +3 i  k^2 \tau_e^2 c_s^2 \right)
 \left( 3i -6 k c_s \tau_s -3i k^2 c_s ^2 \tau_s^2 \right) e^{2 i k c_s( \tau_e -\tau_s) }\Bigg\},
 \\
 \beta _{3} &=&  \frac{1}{(2 k^3 \tau_e^3 c_s^3)(2 k ^3 c_s ^3 \tau_s ^3)}\Bigg\{ \left(-3 i  +6  k \tau_e c_s +3 i k^2 \tau_e^2 c_s^2\right) \left(3 i + 3 i  k^2 c_s^2 \tau_s ^2  + 2  k^3 c_s^3  \tau_s ^3  \right) e^{-\left(2 i k \tau_e c_s + i \pi  \left(\nu +\frac{1}{2}\right)\right)} \nonumber \\ 
&&\quad\quad\quad\quad\quad\quad\quad\quad\quad\quad +\left(2  k^3 \tau_e^3 c_s^3 + 3 i  k^2 \tau_e^2 c_s^2 +3 i \right )  \left( 3i -6 k c_s \tau_s -3i k^2 c_s ^2 \tau_s^2 \right) e^{-i\left(\pi(\nu+\frac{1}{2})+ 2k c_s \tau_s \right)}\Bigg\}.
\eea 
using the above-mentioned expressions in the loop integral we get the following simplified result:
\bea \label{eqx5}{\bf I}_{\bf SRII}(\tau)&=&\frac{1}{2\delta_{\bf SRII}}\left\{\left(\frac{k_{\rm end}}{k_e}\right)^{2\delta_{\bf SRII}}-1\right\}=\frac{1}{2\delta_{\bf SRII}}+\ln\left(\frac{k_{\rm end}}{k_e}\right)-\frac{1}{2\delta_{\bf SRII}}+\cdots\approx\ln\left(\frac{k_{\rm end}}{k_e}\right).\eea
In terms of the number of e-foldings, the final result of the loop integral in the contracting phase can be further simplified as:
\bea \label{eqx5}{\bf I}_{\bf SRII}(N)&=&\Delta N_{\bf SRII}=(N_{\rm end}-N_e).\eea
 Here we have introduced a new symbol $\delta_{\bf SRII}$ for Phase V which is defined by the following expression:
 \bea \delta_{\bf SRII}:=\left(\frac{3}{2}-\nu\right).\eea
Here the $\cdots$ represents the suppressed contribution in which we are not interested in the present discussion. Also from the derived result, it is clearly evident that the one-loop integral at the conformal time scales $\tau=\tau_e$ and $\tau=\tau_{\rm end}$ turns out to be the same because the final result is completely independent of arbitrary conformal time scale, which are the precise boundaries of the SRII region. In a more technical language we then have:
\bea {\bf I}_{\bf SRII}(\tau)={\bf I}_{\bf SRII}(\tau_e)={\bf I}_{\bf SRII}(\tau_{\rm end})={\bf I}_{\bf SRII}.\eea
To avoid any further confusion it is essential to recall that the effective mass parameter $\nu$ here is evaluated at the SRII phase. Depending on the values of the first and second slow-roll parameters one can quantify the one-loop correction in the SRII region which is part of Phase V. Here the expressions for $\delta_{\bf SRI}$, $\delta_{\bf USR}$ and $\delta_{\bf SRII}$ look similar, but due to having different values of the mass parameter $\nu$ in SRI, USR and SRII phases we have $\delta_{\bf SRI}\neq \delta_{\bf USR}\neq \delta_{\bf SRII}$. Here to arrive at the final one-loop corrected result we have utilized the fact that the initial vacuum state is described by the Bunch Davies initial condition and the corresponding Bogolibov coefficient is given by $\alpha_1=1$ and $\beta_1=0$. We have also used the fact that the vacuum structure remains unchanged up to Phase III. After that, a sharp transition takes place at the conformal time scale $\tau=\tau_s$ for which a phase transition occurs at this point, and as an immediate outcome Bunch Davies vacuum changes to a new state. It persists up to the end of the USR phase which is in our discussion and is identified with Phase V. Further, this new vacuum state transforms to another vacuum state due to having another sharp transition at the scale $\tau=\tau_e$ from USR to SRII phase transition. Additionally, it is important to note that to avoid the singularity in the loop integral ${\bf I}_{\bf SRII}$ at $\nu=3/2$ i.e. $\delta_{\bf SRII}=0$, we have used dimensional regularization to remove the dependence on the parameter $\delta_{\bf SRII}$ and after implementing this successfully we have obtained the logarithmic contribution which is in principle IR divergent. Also, it is evident from this result that one-loop contribution in the SRII region is completely UV divergence-free, and other terms in this integral are highly suppressed.  

 \subsection{Total one-loop regularized but unrenormalized result}
 Considering all the derived results for one-loop correction to the power spectrum for scalar modes in the consecutive five phases and summing them over we get the following expression for the total one-loop correction:
 \bea  \bigg[\Delta^{2}_{\zeta, {\bf One-loop}}(p)\bigg]_{\bf Total}&=&\bigg[\Delta^{2}_{\zeta,{\bf Tree}}(p)\bigg]_{\bf SRI}\times\bigg({\bf W}_{\bf C}+{\bf W}_{\bf B}+{\bf W}_{\bf SRI}+{\bf W}_{\bf USR}+{\bf W}_{\bf SRII}\bigg),\eea
    \textcolor{black}{ where the symbols ${\bf W}_{\bf C}$ (for contraction), ${\bf W}_{\bf B}$ (for bounce), ${\bf W}_{\bf SRI}$ (for SRI), ${\bf W}_{\bf USR}$ (for USR) and ${\bf W}_{\bf SRII}$ (for SRII) are defined explicitly in the appendix \ref{appA}.}
     Hence, the total unrenormalized but regularized one-loop corrected power spectrum for scalar modes can be expressed by the following expression:
   \bea \label{one-loopR} \Delta^{2}_{\zeta, {\bf EFT}}(p)&=&\Delta^{2}_{\zeta, {\bf R}}(p)\nonumber\\
   &=&\bigg[\Delta^{2}_{\zeta,{\bf Tree}}(p)\bigg]_{\bf SRI}+\bigg[\Delta^{2}_{\zeta, {\bf One-loop}}(p)\bigg]_{\bf CONTRACTION}+\bigg[\Delta^{2}_{\zeta, {\bf One-loop}}(p)\bigg]_{\bf BOUNCE}\nonumber\\
   &&\quad\quad\quad\quad+\bigg[\Delta^{2}_{\zeta, {\bf One-loop}}(p)\bigg]_{\bf SRI}+\bigg[\Delta^{2}_{\zeta, {\bf One-loop}}(p)\bigg]_{\bf USR}+\bigg[\Delta^{2}_{\zeta, {\bf One-loop}}(p)\bigg]_{\bf SRII}\nonumber\\
   &=&\bigg[\Delta^{2}_{\zeta,{\bf Tree}}(p)\bigg]_{\bf SRI}\times\bigg(1+\underbrace{{\bf W}_{\bf C}+{\bf W}_{\bf B}+{\bf W}_{\bf SRI}+{\bf W}_{\bf USR}+{\bf W}_{\bf SRII}}_{\textbf{Regularized one-loop correction}}\bigg),\nonumber\\
   &=&\begin{tikzpicture}[baseline={([yshift=-3.5ex]current bounding box.center)},very thick]
  
    % Loop
  \def\radius{1}
  \scalebox{1}{\draw[violet,very thick] (0,\radius) circle (\radius);
  \draw[violet,very thick] (4.5*\radius,0) circle (\radius);}

  % External lines
  %\filldraw;
  \draw[black, very thick] (-4*\radius,0) -- 
  (-2.5*\radius,0);
  \node at (-2*\radius,0) {+};
  \draw[black, very thick] (-1.5*\radius,0) -- (0,0);
  \draw[blue,fill=blue] (0,0) circle (.5ex);
  \draw[black, very thick] (0,0)  -- (1.5*\radius,0);
  \node at (2*\radius,0) {+};
  \draw[black, very thick] (2.5*\radius,0) -- (3.5*\radius,0); 
  \draw[blue,fill=blue] (3.5*\radius,0) circle (.5ex);
  \draw[blue,fill=blue] (5.5*\radius,0) circle (.5ex);
  \draw[black, very thick] (5.5*\radius,0) -- (6.5*\radius,0);
\end{tikzpicture},\eea
   which we are going to renormalize in the next section of this paper. Here the symbol ${\bf R}$ is used to specify that the result that we have computed is regularized. 

 \section{Renormalization in loop corrected scalar power spectrum}
\label{s10}
In this section, our prime objective is to renormalize the underlying theory and the associated power spectrum in the Fourier space. The analysis performed in this section is going to help us to completely remove the power law divergences including the quadratic UV divergences and smoothen the logarithmic IR divergences that appeared in the one-loop corrected spectrum as mentioned in equation (\ref{one-loopR}). To serve the present purpose in this section we are going to follow the following steps which will be helpful in extracting the renormalized result from the present computation:
\begin{enumerate}
    \item In the first step we try to design and further implement at the level of action which describes the underlying theoretical framework using which we describe the associated primordial cosmological set-up. While doing this we try to implement renormalization in the Quantum Field Theory in a non-trivial curved background having the solution of the space-time described by contracting, bouncing, first slow-roll, ultra slow-roll, and second slow-roll phases in the backdrop of FLRW metric. \textcolor{black}{The technical details are described in the appendix \ref{appB}.}

    \item In the second step we will be going to use the late-time renormalization technique using which we are interested in removing the contributions from all types of power-law divergences including the quadratic UV divergences as appearing in the $\nu=3/2$ limiting situations. The technical details are described in the remaining section.

    \item In the third step we will be going to use the adiabatic/wave-function renormalization technique using which we are interested in removing the contributions from all types of power-law divergences including the quadratic UV divergences as appearing in the $\nu=3/2$ limiting situations. We will try to justify with proper arguments that adiabatic and late-time renormalization schemes give the same result in the present context of the computation. The technical details are described in the remaining section.

    \item In the fourth and last step we will use the renormalization scheme at the level of the cosmological two-point function and its associated power spectrum of the scalar modes. We identify this as the power spectrum renormalization scheme using which we coarse grain or soften the hardness of the logarithmic IR divergent contributions which are appearing in the computation of the one-loop corrected power spectrum. The technical details are described in the remaining section.
\end{enumerate}

\subsection{Step II: Late time renormalization and removal of UV/power law divergence}

In this calculation, two distinct types of divergences appear: UV and IR. The IR divergence appears as logarithmic contributions and the UV divergence as quadratic contributions resulting from cut-off regularization. The main issue arises in UV regularization as compared to the IR. In particular, we have used a constant UV cutoff scale while doing the loop integrals for this time-varying calculation. Therefore, it can be immediately assumed that the momentum integration at any given instant may contain superfluous momentum modes, and that an erroneous conclusion may come from this type of mode overcounting. We will now discuss this theory in greater detail and show why it is erroneous. To elucidate our stance, let us start with the representative momentum integrals, which are explicitly present in the contracting, bouncing, SRI, USR, and SRII phases—these five successive phases—when we are dealing with situations involving abrupt transitions at the USR boundary:
\begin{itemize}
    \item[\ding{43}] \underline{\textbf{Type-I: Contraction phase integral}}
    \begin{itemize}
         \item[$\blacksquare$] \underline{\bf Method-I:}\\ \\  The representative one-loop momentum integral which describes the contraction phase is written down as follows:
\bea {\bf E}_1(\tau):&=&\int^{k_{\bf UV}}_{k_{\bf IR}}\frac{dk}{k}\;\left(\frac{k}{k_*}\right)^{\delta_{\bf C}}\;\left({\bf A}+{\bf B}k^2c^2_s\tau^2\right)+{\bf C}\nonumber\\
&=&\Bigg[\frac{{\bf A}}{\delta_{\bf C}}\bigg\{\left(\frac{k_{\bf UV}}{k_{*}}\right)^{\delta_{\bf C}}-\left(\frac{k_{\bf IR}}{k_{*}}\right)^{\delta_{\bf C}}\bigg\}+\frac{{\bf B}}{\left(\delta_{\bf C}+2\right)}k^2_*c^2_s\tau^2\bigg\{\left(\frac{k_{\bf UV}}{k_{*}}\right)^{\delta_{\bf C}+2}-\left(\frac{k_{\bf IR}}{k_{*}}\right)^{\delta_{\bf C}+2}\bigg\}\Bigg]+{\bf C},\quad\quad\quad\eea
where we identify the UV and IR cut-offs of the contracting phase as, $k_{\bf UV}=k_b$ and $k_{\bf IR}=k_c$ and also the symbols ${\bf A}$, ${\bf B}$ are two constants and $\delta_{\bf C}$ is defined by, $\delta_{\bf C}:=\left(3-2\nu+\frac{2\epsilon_c}{\epsilon_c-1}\right)$.
Here ${\bf C}$ is the counter term which we need to determine explicitly. In the limit $-k_*c_s\tau\rightarrow 1$ we have:
\bea {\bf E}_1(-k_*c_s\tau\rightarrow 1):
&=&\Bigg[\frac{{\bf A}}{\delta_{\bf C}}\bigg\{\left(\frac{k_{\bf UV}}{k_{*}}\right)^{\delta_{\bf C}}-\left(\frac{k_{\bf IR}}{k_{*}}\right)^{\delta_{\bf C}}\bigg\}+\frac{{\bf B}}{\left(\delta_{\bf C}+2\right)}\bigg\{\left(\frac{k_{\bf UV}}{k_{*}}\right)^{\delta_{\bf C}+2}-\left(\frac{k_{\bf IR}}{k_{*}}\right)^{\delta_{\bf C}+2}\bigg\}\Bigg],\quad\quad\eea 
which further implies that here the counter term is fixed at:
\bea {\bf C}={\cal K}_{\bf C}=0.\eea
We have explicitly used the fact that $k_{\bf UV}\neq k_{\bf IR}$ and $k_{\bf UV}\gg k_{\bf IR}$ to build the counter term ${\bf C}$ which in our previous calculation is directly associated with ${\cal K}_{\bf C}$.
IR divergence is not harmful in the present circumstances, and even with the suggested counter-term constraint relation in the super-horizon limit, one cannot eliminate such a contribution coming from the IR. This constraint relation is the only one that allows us to modify the IR logarithmic divergence behaviour in the current computing environment. Upon completing momentum integration and accounting for the late time limit (which is critical from an observational perspective), we can clearly observe that all of these unnecessary, superfluous momentum modes completely washed out of the computation at the super-horizon scale and during the re-entry through the horizon, resulting in a smoother version of the softened IR result.

 \item[$\blacksquare$] \underline{\bf Method-II:}\\ \\
 In order to provide a full-proof technical justification and to explicitly show that no overcounting of momentum modes will occur during such computations, we have reevaluated the aforementioned integral while keeping the UV and IR cut-off values unfixed. We did this by taking into account the following trick:
\bea \int^{k_{\bf UV}}_{k_{\bf IR}}:=\bigg(\int^{k_{\bf INT}}_{k_{\bf IR}}+\int^{k_{\bf UV}=\frac{\Lambda_{\bf UV} a(\tau)}{c_s}}_{k_{\bf INT}}\bigg),\eea
Since the related wave number $k_{\bf UV}$ is represented in the comoving scale and all the results calculated in the super-horizon scale at the moment of horizon re-entry, the UV limit of the integration can be explained using the conformal time dependency, which is reasonably physically consistent. In order to improve understanding, we have specifically included the time-dependent data to the UV limit of the integration $k_{\bf UV}=\frac{\Lambda_{\bf UV} a(\tau)}{c_s}$. When the conformal time dependence in the current context has been removed, the UV cutoff contribution is applied, and in this example, it is represented as $\Lambda_{\bf UV}$. The fact that the momentum integral boundaries have been divided into two halves is also interesting. The first half essentially deals with the finite contribution of the integral, while the second half deals with the quadratic UV divergence in the event that time dependency exists at the upper bound of the momentum-dependent loop integration. Using the previously mentioned integration decomposition, we set down the following simplified formulation:
\bea \label{yhxx} {\bf E}_{1}:&=&\bigg(\int^{k_{\bf INT}}_{k_{\bf IR}}+\int^{k_{\bf UV}=\frac{\Lambda_{\bf UV} a(\tau)}{c_s}}_{k_{\bf INT}}\bigg)\frac{dk}{k}\;\left(\frac{k}{k_*}\right)^{\delta_{\bf C}}\;\left({\bf A}+{\bf B}k^2c^2_s\tau^2\right)+{\bf C}\nonumber\\
&=&\bigg[\frac{{\bf A}}{\delta_{\bf C}}\bigg(\bigg\{\left(\frac{k_{\bf INT}}{k_*}\right)^{\delta_{\bf C}}-\left(\frac{k_{\bf IR}}{k_*}\right)^{\delta_{\bf C}}\bigg\}+\left(\frac{k_{\bf INT}}{k_*}\right)^{\delta_{\bf C}}\bigg\{\left(\frac{\Lambda_{\bf UV}a(\tau)}{k_{\bf INT}c_s}\right)^{\delta_{\bf C}}-1\bigg\}\bigg)\nonumber\\
&&\quad\quad\quad\quad\quad\quad+\frac{{\bf B}}{\left(\delta_{\bf C}+2\right)}k^2_*c^2_s\tau^2\bigg(\bigg\{\left(\frac{k_{\bf INT}}{k_*}\right)^{\delta_{\bf C}+2}-\left(\frac{k_{\bf IR}}{k_*}\right)^{\delta_{\bf C}+2}\bigg\}\nonumber\\&&\quad\quad\quad\quad\quad\quad\quad\quad\quad\quad\quad\quad
+\left(\frac{k_{\bf INT}}{k_*}\right)^{\delta_{\bf C}+2}\bigg\{\left(\frac{\Lambda_{\bf UV}a(\tau)}{k_{\bf INT}c_s}\right)^{\delta_{\bf C}+2}-1\bigg\}\bigg)\bigg]+{\bf C},\nonumber\\
&=&\bigg[\frac{{\bf A}}{\delta_{\bf C}}\bigg(\bigg\{\left(\frac{k_{\bf INT}}{k_*}\right)^{\delta_{\bf C}}-\left(\frac{k_{\bf IR}}{k_*}\right)^{\delta_{\bf C}}\bigg\}+\left(\frac{k_{\bf INT}}{k_*}\right)^{\delta_{\bf C}}\bigg\{\left(\frac{\Lambda_{\bf UV}}{H}\right)^{\delta_{\bf C}}-1\bigg\}\bigg)\nonumber\\
&&\quad\quad\quad\quad\quad\quad+\frac{{\bf B}}{\left(\delta_{\bf C}+2\right)}\bigg(\bigg\{\left(\frac{k_{\bf INT}}{k_*}\right)^{\delta_{\bf C}+2}-\left(\frac{k_{\bf IR}}{k_*}\right)^{\delta_{\bf C}+2}\bigg\}\nonumber\\&&\quad\quad\quad\quad\quad\quad\quad\quad\quad\quad\quad\quad
+\left(\frac{k_{\bf INT}}{k_*}\right)^{\delta_{\bf C}+2}\bigg\{\left(\frac{\Lambda_{\bf UV}}{H}\right)^{\delta_{\bf C}+2}-1\bigg\}\bigg)\bigg]+{\bf C},\eea 
where we have utilized the fact that, $a(\tau)/c_s k_{\bf INT}=1/H$. Hence the counter term ${\bf C}$ in the present context at an arbitrary renormalization scale $\mu_{\bf REN}$ can be written as:
\bea {\bf C}\left(\mu_{\bf REN},\Lambda_{\bf UV}\right)&=&\bigg[\frac{{\bf A}}{\delta_{\bf C}}\left(\frac{k_{\bf INT}}{k_*}\right)^{\delta_{\bf C}}\bigg\{\left(\frac{\mu_{\bf REN}}{H}\right)^{\delta_{\bf C}}-\left(\frac{\Lambda_{\bf UV}}{H}\right)^{\delta_{\bf C}}\bigg\}\nonumber\\
&&\quad\quad\quad\quad\quad\quad+\frac{{\bf B}}{\left(\delta_{\bf C}+2\right)}\left(\frac{k_{\bf INT}}{k_*}\right)^{\delta_{\bf C}+2}\bigg\{\left(\frac{\mu_{\bf REN}}{H}\right)^{\delta_{\bf C}+2}-\left(\frac{\Lambda_{\bf UV}}{H}\right)^{\delta_{\bf C}+2}\bigg\}\bigg].\quad\quad\quad\eea
Hence the UV cut-off removed result for the generic one-loop momentum integration is described by the following expression:
\bea \label{yhxxx} {\bf E}_{1}(\mu_{\bf REN})&=&\bigg[\frac{{\bf A}}{\delta_{\bf C}}\bigg(\bigg\{\left(\frac{k_{\bf INT}}{k_*}\right)^{\delta_{\bf C}}-\left(\frac{k_{\bf IR}}{k_*}\right)^{\delta_{\bf C}}\bigg\}+\left(\frac{k_{\bf INT}}{k_*}\right)^{\delta_{\bf C}}\bigg\{\left(\frac{\Lambda_{\bf UV}}{H}\right)^{\delta_{\bf C}}-1\bigg\}\bigg)\nonumber\\
&&\quad\quad\quad\quad\quad\quad+\frac{{\bf B}}{\left(\delta_{\bf C}+2\right)}\bigg(\bigg\{\left(\frac{k_{\bf INT}}{k_*}\right)^{\delta_{\bf C}+2}-\left(\frac{k_{\bf IR}}{k_*}\right)^{\delta_{\bf C}+2}\bigg\}\nonumber\\&&\quad\quad\quad\quad\quad\quad\quad\quad\quad\quad\quad\quad
+\left(\frac{k_{\bf INT}}{k_*}\right)^{\delta_{\bf C}+2}\bigg\{\left(\frac{\Lambda_{\bf UV}}{H}\right)^{\delta_{\bf C}+2}-1\bigg\}\bigg)\bigg]\nonumber\\
&&+\bigg[\frac{{\bf A}}{\delta_{\bf C}}\left(\frac{k_{\bf INT}}{k_*}\right)^{\delta_{\bf C}}\bigg\{\left(\frac{\mu_{\bf REN}}{H}\right)^{\delta_{\bf C}}-\left(\frac{\Lambda_{\bf UV}}{H}\right)^{\delta_{\bf C}}\bigg\}\nonumber\\
&&\quad\quad\quad\quad\quad\quad+\frac{{\bf B}}{\left(\delta_{\bf C}+2\right)}\left(\frac{k_{\bf INT}}{k_*}\right)^{\delta_{\bf C}+2}\bigg\{\left(\frac{\mu_{\bf REN}}{H}\right)^{\delta_{\bf C}+2}-\left(\frac{\Lambda_{\bf UV}}{H}\right)^{\delta_{\bf C}+2}\bigg\}\bigg],\nonumber\eea\bea
&=&\bigg[\frac{{\bf A}}{\delta_{\bf C}}\bigg(\bigg\{\left(\frac{k_{\bf INT}}{k_*}\right)^{\delta_{\bf C}}-\left(\frac{k_{\bf IR}}{k_*}\right)^{\delta_{\bf C}}\bigg\}+\left(\frac{k_{\bf INT}}{k_*}\right)^{\delta_{\bf C}}\bigg\{\left(\frac{\mu_{\bf REN}}{H}\right)^{\delta_{\bf C}}-1\bigg\}\bigg)\nonumber\\
&&\quad\quad\quad\quad\quad\quad+\frac{{\bf B}}{\left(\delta_{\bf C}+2\right)}\bigg(\bigg\{\left(\frac{k_{\bf INT}}{k_*}\right)^{\delta_{\bf C}+2}-\left(\frac{k_{\bf IR}}{k_*}\right)^{\delta_{\bf C}+2}\bigg\}\nonumber\\&&\quad\quad\quad\quad\quad\quad\quad\quad\quad\quad\quad\quad
+\left(\frac{k_{\bf INT}}{k_*}\right)^{\delta_{\bf C}+2}\bigg\{\left(\frac{\mu_{\bf REN}}{H}\right)^{\delta_{\bf C}+2}-1\bigg\}\bigg)\bigg].\eea
Further, if we fix the scale of renormalization at the Hubble scale i.e. $\mu_{\bf REN}=H$ then we get the following simplified result for the loop integral:
\bea \label{yhxxxh} {\bf E}_{1}(\mu_{\bf REN}=H)
&=&\bigg[\frac{{\bf A}}{\delta_{\bf C}}\bigg\{\left(\frac{k_{\bf INT}}{k_*}\right)^{\delta_{\bf C}}-\left(\frac{k_{\bf IR}}{k_*}\right)^{\delta_{\bf C}}\bigg\}+\frac{{\bf B}}{\left(\delta_{\bf C}+2\right)}\bigg\{\left(\frac{k_{\bf INT}}{k_*}\right)^{\delta_{\bf C}+2}-\left(\frac{k_{\bf IR}}{k_*}\right)^{\delta_{\bf C}+2}\bigg\}\bigg].\quad\quad\quad\eea
Here the factors ${\bf A}$ and ${\bf B}$ are identified given by the following expressions in the contraction phase:
\bea {\bf A}={\bf B}=-\frac{4}{3}\bigg[\Delta^{2}_{\zeta,{\bf Tree}}(p)\bigg]^2_{\bf SRI}\times\Bigg(1+\frac{2}{15\pi^2}\frac{1}{c^2_{s}k^2_c}\bigg(-\left(1-\frac{1}{c^2_{s}}\right)\epsilon_c+6\frac{\bar{M}^3_1}{ HM^2_{ pl}}-\frac{4}{3}\frac{M^4_3}{H^2M^2_{ pl}}\bigg)\Bigg)\times \left(\frac{\epsilon_*}{\epsilon_c}\right).\quad\quad\eea
Also the counter term at $\mu_{\bf REN}=H$ scale is given by:
\bea {\bf C}\left(\mu_{\bf REN}=H,\Lambda_{\bf UV}\right)&=&{\cal K}_{\bf C}=\bigg[\frac{{\bf A}}{\delta_{\bf C}}\left(\frac{k_{\bf INT}}{k_*}\right)^{\delta_{\bf C}}\bigg\{1-\left(\frac{\Lambda_{\bf UV}}{H}\right)^{\delta_{\bf C}}\bigg\}\nonumber\\
&&\quad\quad\quad\quad\quad\quad+\frac{{\bf B}}{\left(\delta_{\bf C}+2\right)}\left(\frac{k_{\bf INT}}{k_*}\right)^{\delta_{\bf C}+2}\bigg\{1-\left(\frac{\Lambda_{\bf UV}}{H}\right)^{\delta_{\bf C}+2}\bigg\}\bigg].\quad\quad\quad\eea
Here connecting our findings with the standard Quantum Field Theory approach we found that:
\bea  \left(\delta_{{\cal Z}_{{\bf G}_1}}+\delta_{{\cal Z}_{{\bf G}_2}}+\delta_{{\cal Z}_{{\bf G}_3}}+\delta_{{\cal Z}_{{\bf G}_4}}+\delta_{{\cal Z}_{{\bf G}_5}}\right)={\bf C}(\mu_{\bf REN}=H,\Lambda_{\bf UV})={\cal K}_{\bf C}\quad{\rm with}\quad\delta_{{\cal Z}_{{\bf D}_6}}=0.\eea
For this reason, finally we get:
\bea  \left(\delta_{{\cal Z}_{{\bf G}_1}}+\delta_{{\cal Z}_{{\bf G}_2}}+\delta_{{\cal Z}_{{\bf G}_3}}+\delta_{{\cal Z}_{{\bf G}_4}}+\delta_{{\cal Z}_{{\bf G}_5}}\right)&=&-\frac{4}{3}\bigg[\Delta^{2}_{\zeta,{\bf Tree}}(p)\bigg]^2_{\bf SRI}\nonumber\\
&&\times\Bigg(1+\frac{2}{15\pi^2}\frac{1}{c^2_{s}k^2_c}\bigg(-\left(1-\frac{1}{c^2_{s}}\right)\epsilon_c+6\frac{\bar{M}^3_1}{ HM^2_{ pl}}-\frac{4}{3}\frac{M^4_3}{H^2M^2_{ pl}}\bigg)\Bigg)\times \left(\frac{\epsilon_*}{\epsilon_c}\right)\nonumber\\
&&\times\bigg[\frac{1}{\delta_{\bf C}}\left(\frac{k_{\bf INT}}{k_*}\right)^{\delta_{\bf C}}\bigg\{1-\left(\frac{\Lambda_{\bf UV}}{H}\right)^{\delta_{\bf C}}\bigg\}\nonumber\\
&&\quad\quad\quad\quad\quad\quad+\frac{1}{\left(\delta_{\bf C}+2\right)}\left(\frac{k_{\bf INT}}{k_*}\right)^{\delta_{\bf C}+2}\bigg\{1-\left(\frac{\Lambda_{\bf UV}}{H}\right)^{\delta_{\bf C}+2}\bigg\}\bigg].\eea
Then the regularized and renormalized expression for the one loop corrected contribution from the contraction phase is given by the following expression:
\bea \bigg[\Delta^{2}_{\zeta, {\bf One-loop}}(p)\bigg]_{\bf CONTRACTION}&=&-\frac{4}{3}\bigg[\Delta^{2}_{\zeta,{\bf Tree}}(p)\bigg]^2_{\bf SRI}\nonumber\\&&\quad\quad\times\Bigg(1+\frac{2}{15\pi^2}\frac{1}{c^2_{s}k^2_c}\bigg(-\left(1-\frac{1}{c^2_{s}}\right)\epsilon_c+6\frac{\bar{M}^3_1}{ HM^2_{ pl}}-\frac{4}{3}\frac{M^4_3}{H^2M^2_{ pl}}\bigg)\Bigg)\times \left(\frac{\epsilon_*}{\epsilon_c}\right)\nonumber\\
&&\quad\quad\times\bigg[\frac{1}{\delta_{\bf C}}\bigg\{\left(\frac{k_{b}}{k_*}\right)^{\delta_{\bf C}}-\left(\frac{k_{c}}{k_*}\right)^{\delta_{\bf C}}\bigg\}\nonumber\\
&&\quad\quad+\frac{1}{\left(\delta_{\bf C}+2\right)}\bigg\{\left(\frac{k_{b}}{k_*}\right)^{\delta_{\bf C}+2}-\left(\frac{k_{c}}{k_*}\right)^{\delta_{\bf C}+2}\bigg\}\bigg]\nonumber\\
&=&\bigg[\Delta^{2}_{\zeta,{\bf Tree}}(p)\bigg]_{\bf SRI}\times \overline{\bf W}_{\bf C},\eea
where $\overline{\bf W}_{\bf C}$ is given by:
\bea \label{t1}\overline{\bf W}_{\bf C}&=&-\frac{4}{3}\bigg[\Delta^{2}_{\zeta,{\bf Tree}}(p)\bigg]_{\bf SRI}\times\Bigg(1+\frac{2}{15\pi^2}\frac{1}{c^2_{s}k^2_c}\bigg(-\left(1-\frac{1}{c^2_{s}}\right)\epsilon_c+6\frac{\bar{M}^3_1}{ HM^2_{ pl}}-\frac{4}{3}\frac{M^4_3}{H^2M^2_{ pl}}\bigg)\Bigg)\times \left(\frac{\epsilon_*}{\epsilon_c}\right)\nonumber\\
&&\quad\quad\quad\quad\quad\quad\quad\quad\quad\quad\quad\quad\times\bigg[\frac{1}{\delta_{\bf C}}\bigg\{\left(\frac{k_{b}}{k_*}\right)^{\delta_{\bf C}}-\left(\frac{k_{c}}{k_*}\right)^{\delta_{\bf C}}\bigg\}\nonumber\\
&&\quad\quad\quad\quad\quad\quad\quad\quad\quad\quad\quad\quad\quad\quad+\frac{1}{\left(\delta_{\bf C}+2\right)}\bigg\{\left(\frac{k_{b}}{k_*}\right)^{\delta_{\bf C}+2}-\left(\frac{k_{c}}{k_*}\right)^{\delta_{\bf C}+2}\bigg\}\bigg].\eea
where we identify $k_{\bf INT}=k_b$ and $k_{\bf IR}=k_c$.
    \end{itemize}

    \item[\ding{43}] \underline{\textbf{Type-II: Bouncing phase integral}}
    \begin{itemize}
         \item[$\blacksquare$] \underline{\bf Method-I:}\\ \\ The representative one-loop momentum integral which describes the bouncing phase is given by the following expression:
\bea {\bf E}_2(\tau):&=&{\bf A}\int^{k_{\bf UV}}_{k_{\bf IR}}\frac{dk}{k}\;\left(\frac{k}{k_{\bf UV}}\right)^{2}\left(1+\left(\frac{k_{\bf UV}}{k}\right)^2\right)^{-\frac{1}{(\epsilon_b-1)}}\;\left(\frac{k}{k_{\bf UV}}\right)^{3-2\nu}\left(1+k^2c^2_s\tau^2\right)+{\bf C},\nonumber\\
 &=& {\bf A}\int^{1}_{k_{\bf IR}/k_{\bf UV}}d\left(\frac{k}{k_{\bf UV}}\right)\;\left(\frac{k}{k_{\bf UV}}\right)^{1+\delta_{\bf B}}\left(1+\left(\frac{k}{k_{\bf UV}}\right)^{2}\right)^{\frac{\epsilon_b-2}{\epsilon_b-1}}+{\bf C},\nonumber\\
 &=&\frac{{\bf A}}{\delta_{\bf B} +2}\bigg[\, _2F_1\left(\frac{\delta_{\bf B}+2}{2},\frac{1}{\epsilon_b-1}-1;\frac{\delta_{\bf B}+4}{2};-1\right)\nonumber\\
 &&\quad\quad\quad\quad\quad\quad\quad-\left(\frac{k_{\bf IR}}{k_{\bf UV}}\right)^{\delta_{\bf B}+2}\, _2F_1\left(\frac{\delta_{\bf B}+2}{2},\frac{1}{\epsilon_b-1}-1;\frac{\delta_{\bf B}+4}{2};-\left(\frac{k_{\bf IR}}{k_{\bf UV}}\right)^{2}\right)\bigg]+{\bf C}.\quad\quad\quad\eea
where we identify the UV and IR cut-offs of the contracting phase as, $k_{\bf UV}=k_*$ and $k_{\bf IR}=k_b$ and also the symbols ${\bf A}$, ${\bf B}$ are two constants and $\delta_{\bf B}$ is defined by, $\delta_{\bf B}:=\left(3-2\nu+\frac{2}{\epsilon_b-1}\right)$.
Here ${\bf C}$ is the counter term which we need to determine explicitly and this is directly associated with ${\cal K}_{\bf B}$ in the present context. In the limit $-k_*c_s\tau\rightarrow 1$ we have:
\bea {\bf E}_2(-k_*c_s\tau\rightarrow 1):
&=&\frac{{\bf A}}{\delta_{\bf B} +2}\bigg[\, _2F_1\left(\frac{\delta_{\bf B}+2}{2},\frac{1}{\epsilon_b-1}-1;\frac{\delta_{\bf B}+4}{2};-1\right)\nonumber\\
 &&\quad\quad\quad\quad\quad\quad\quad-\left(\frac{k_{\bf IR}}{k_{\bf UV}}\right)^{\delta_{\bf B}+2}\, _2F_1\left(\frac{\delta_{\bf B}+2}{2},\frac{1}{\epsilon_b-1}-1;\frac{\delta_{\bf B}+4}{2};-\left(\frac{k_{\bf IR}}{k_{\bf UV}}\right)^{2}\right)\bigg],\quad\quad\eea 
which further implies that here the counter term is fixed at:
\bea {\bf C}={\cal K}_{\bf B}=0.\eea
From the above-mentioned result, it is clearly visible that the IR and UV divergences do not appear in the logarithmic and quadratic forms. Since we know that $k_{\bf IR}/k_{\bf UV}\ll 1$, the term:
\bea \left(\frac{k_{\bf IR}}{k_{\bf UV}}\right)^{\delta_{\bf B}+2}\, _2F_1\left(\frac{\delta_{\bf B}+2}{2},\frac{1}{\epsilon_b-1}-1;\frac{\delta_{\bf B}+4}{2};-\left(\frac{k_{\bf IR}}{k_{\bf UV}}\right)^{2}\right)\ll 1, \eea is highly suppressed and for this reason it is clear that the bouncing phase is completely untouched by any sorts of IR and UV divergences. This further implies that in this context we obtain a completely finite result that can be trusted safely for further purposes.
 
          \item[$\blacksquare$] \underline{\bf Method-II:}\\ \\ 
          By following the same trick as mentioned before we decompose the integration limit for this purpose as:     
\bea \int^{k_{\bf UV}}_{k_{\bf IR}}:=\bigg(\int^{k_{\bf INT}}_{k_{\bf IR}}+\int^{k_{\bf UV}=\frac{\Lambda_{\bf UV} a(\tau)}{c_s}}_{k_{\bf INT}}\bigg),\eea
using which the loop integral can be further recast in the following simplified form:
\bea {\bf E}_2(\tau):&=&{\bf A}\bigg(\int^{k_{\bf INT}}_{k_{\bf IR}}+\int^{k_{\bf UV}=\frac{\Lambda_{\bf UV} a(\tau)}{c_s}}_{k_{\bf INT}}\bigg)\frac{dk}{k}\;\left(\frac{k}{k_{\bf INT}}\right)^{2}\left(1+\left(\frac{k_{\bf INT}}{k}\right)^2\right)^{-\frac{1}{(\epsilon_b-1)}}\nonumber\\
 &&\quad\quad\quad\quad\quad\quad\quad\quad\quad\quad\quad\quad\quad\times\;\left(\frac{k}{k_{\bf INT}}\right)^{3-2\nu}\left(1+k^2c^2_s\tau^2\right)+{\bf C},\nonumber\\
 &=& {\bf A}\bigg(\int^{1}_{k_{\bf IR}/k_{\bf INT}}+\int^{\frac{k_{\bf UV}}{k_{\bf INT}}=\frac{\Lambda_{\bf UV} a(\tau)}{c_sk_{\bf INT}}}_{1}\bigg)d\left(\frac{k}{k_{\bf INT}}\right)\;\left(\frac{k}{k_{\bf INT}}\right)^{1+\delta_{\bf B}}\left(1+\left(\frac{k}{k_{\bf INT}}\right)^{2}\right)^{\frac{\epsilon_b-2}{\epsilon_b-1}}+{\bf C},\nonumber\\
 &=&\frac{{\bf A}}{\delta_{\bf B} +2}\bigg[\, _2F_1\left(\frac{\delta_{\bf B}+2}{2},\frac{1}{\epsilon_b-1}-1;\frac{\delta_{\bf B}+4}{2};-1\right)\nonumber\\
 &&\quad\quad\quad\quad\quad\quad\quad-\left(\frac{k_{\bf IR}}{k_{\bf INT}}\right)^{\delta_{\bf B}+2}\, _2F_1\left(\frac{\delta_{\bf B}+2}{2},\frac{1}{\epsilon_b-1}-1;\frac{\delta_{\bf B}+4}{2};-\left(\frac{k_{\bf IR}}{k_{\bf INT}}\right)^{2}\right)\bigg]\nonumber\\
 &&+\frac{{\bf A}}{\delta_{\bf B} +2}\bigg[\left(\frac{\Lambda_{\bf UV} a(\tau)}{c_sk_{\bf INT}}\right)^{\delta_{\bf B}+2} \, _2F_1\left(\frac{\delta_{\bf B}+2}{2},\frac{1}{\epsilon_b-1}-1;\frac{\delta_{\bf B}+4}{2};-\left(\frac{\Lambda_{\bf UV} a(\tau)}{c_sk_{\bf INT}}\right)^2\right)\nonumber\\
 &&\quad\quad\quad\quad\quad\quad\quad\quad\quad\quad\quad\quad\quad\quad-\, _2F_1\left(\frac{\delta_{\bf B}+2}{2},\frac{1}{\epsilon_b-1}-1;\frac{\delta_{\bf B}+4}{2};-1\right)\bigg]+{\bf C},\nonumber\\
 &=&\frac{{\bf A}}{\delta_{\bf B} +2}\bigg[\left(\frac{\Lambda_{\bf UV} a(\tau)}{c_sk_{\bf INT}}\right)^{\delta_{\bf B}+2} \, _2F_1\left(\frac{\delta_{\bf B}+2}{2},\frac{1}{\epsilon_b-1}-1;\frac{\delta_{\bf B}+4}{2};-\left(\frac{\Lambda_{\bf UV} a(\tau)}{c_sk_{\bf INT}}\right)^2\right)\nonumber\\
 &&\quad\quad\quad\quad\quad\quad\quad\quad\quad\quad\quad\quad\quad\quad\nonumber\\
 &&\quad\quad\quad\quad\quad\quad\quad-\left(\frac{k_{\bf IR}}{k_{\bf INT}}\right)^{\delta_{\bf B}+2}\, _2F_1\left(\frac{\delta_{\bf B}+2}{2},\frac{1}{\epsilon_b-1}-1;\frac{\delta_{\bf B}+4}{2};-\left(\frac{k_{\bf IR}}{k_{\bf INT}}\right)^{2}\right)\bigg]+{\bf C},\nonumber\\
 &=&\frac{{\bf A}}{\delta_{\bf B} +2}\bigg[\left(\frac{\Lambda_{\bf UV}}{H}\right)^{\delta_{\bf B}+2} \, _2F_1\left(\frac{\delta_{\bf B}+2}{2},\frac{1}{\epsilon_b-1}-1;\frac{\delta_{\bf B}+4}{2};-\left(\frac{\Lambda_{\bf UV}}{H}\right)^2\right)\nonumber\\
 &&\quad\quad\quad\quad\quad\quad\quad\quad\quad\quad\quad\quad\quad\quad\nonumber\\
 &&\quad\quad\quad\quad\quad\quad\quad-\left(\frac{k_{\bf IR}}{k_{\bf INT}}\right)^{\delta_{\bf B}+2}\, _2F_1\left(\frac{\delta_{\bf B}+2}{2},\frac{1}{\epsilon_b-1}-1;\frac{\delta_{\bf B}+4}{2};-\left(\frac{k_{\bf IR}}{k_{\bf INT}}\right)^{2}\right)\bigg]+{\bf C},\quad\quad\quad\eea
where we have utilized the fact that, $a(\tau)/c_s k_{\bf INT}=1/H$. Hence the counter term ${\bf C}$ in the present context at an arbitrary renormalization scale $\mu_{\bf REN}$ can be written as:
\bea {\bf C}\left(\mu_{\bf REN},\Lambda_{\bf UV}\right)&=&\frac{{\bf A}}{\delta_{\bf B} +2}\bigg[\left(\frac{\mu_{\bf REN}}{H}\right)^{\delta_{\bf B}+2} \, _2F_1\left(\frac{\delta_{\bf B}+2}{2},\frac{1}{\epsilon_b-1}-1;\frac{\delta_{\bf B}+4}{2};-\left(\frac{\mu_{\bf REN}}{H}\right)^2\right)\nonumber\\
 &&\quad\quad\quad\quad\quad\quad\quad\quad\quad\quad\quad\quad\quad\quad\nonumber\\
 &&\quad\quad\quad\quad\quad\quad\quad-\left(\frac{\Lambda_{\bf UV}}{H}\right)^{\delta_{\bf B}+2} \, _2F_1\left(\frac{\delta_{\bf B}+2}{2},\frac{1}{\epsilon_b-1}-1;\frac{\delta_{\bf B}+4}{2};-\left(\frac{\Lambda_{\bf UV}}{H}\right)^2\right)\bigg].\quad\quad\quad\eea
 Hence the UV cut-off removed result for the generic one-loop momentum integration is described by the following expression:
 \bea {\bf E}_2(\mu_{\bf REN})
 &=&\frac{{\bf A}}{\delta_{\bf B} +2}\bigg[\left(\frac{\mu_{\bf REN}}{H}\right)^{\delta_{\bf B}+2} \, _2F_1\left(\frac{\delta_{\bf B}+2}{2},\frac{1}{\epsilon_b-1}-1;\frac{\delta_{\bf B}+4}{2};-\left(\frac{\mu_{\bf REN}}{H}\right)^2\right)\nonumber\\
 &&\quad\quad\quad\quad\quad\quad\quad\quad\quad\quad\quad\quad\quad\quad\nonumber\\
 &&\quad\quad\quad\quad\quad\quad\quad-\left(\frac{k_{\bf IR}}{k_{\bf INT}}\right)^{\delta_{\bf B}+2}\, _2F_1\left(\frac{\delta_{\bf B}+2}{2},\frac{1}{\epsilon_b-1}-1;\frac{\delta_{\bf B}+4}{2};-\left(\frac{k_{\bf IR}}{k_{\bf INT}}\right)^{2}\right)\bigg],\quad\quad\quad\eea
 Further, if we fix the scale of renormalization at the Hubble scale i.e. $\mu_{\bf REN}=H$ then we get the following simplified result for the loop integral:
 \bea {\bf E}_2(\mu_{\bf REN}=H)
 &=&\frac{{\bf A}}{\delta_{\bf B} +2}\bigg[\, _2F_1\left(\frac{\delta_{\bf B}+2}{2},\frac{1}{\epsilon_b-1}-1;\frac{\delta_{\bf B}+4}{2};-1\right)\nonumber\\
 &&\quad\quad\quad\quad\quad\quad\quad\quad\quad\quad\quad\quad\quad\quad\nonumber\\
 &&\quad\quad\quad\quad\quad\quad\quad-\left(\frac{k_{\bf IR}}{k_{\bf INT}}\right)^{\delta_{\bf B}+2}\, _2F_1\left(\frac{\delta_{\bf B}+2}{2},\frac{1}{\epsilon_b-1}-1;\frac{\delta_{\bf B}+4}{2};-\left(\frac{k_{\bf IR}}{k_{\bf INT}}\right)^{2}\right)\bigg].\quad\quad\quad\eea
 Here the factors ${\bf A}$ and ${\bf B}$ are identified given by the following expressions in the bouncing phase:
\bea {\bf A}=-\frac{4}{3}\bigg[\Delta^{2}_{\zeta,{\bf Tree}}(p)\bigg]^2_{\bf SRI}\times\Bigg(1+\frac{2}{15\pi^2}\frac{1}{c^2_{s}k^2_b}\bigg(-\left(1-\frac{1}{c^2_{s}}\right)\epsilon_b+6\frac{\bar{M}^3_1}{ HM^2_{ pl}}-\frac{4}{3}\frac{M^4_3}{H^2M^2_{ pl}}\bigg)\Bigg)\times \left(\frac{\epsilon_*}{\epsilon_b}\right).\quad\quad\eea
Also the counter term at $\mu_{\bf REN}=H$ scale is given by:
\bea {\bf C}\left(\mu_{\bf REN}=H,\Lambda_{\bf UV}\right)&=&\frac{{\bf A}}{\delta_{\bf B} +2}\bigg[\, _2F_1\left(\frac{\delta_{\bf B}+2}{2},\frac{1}{\epsilon_b-1}-1;\frac{\delta_{\bf B}+4}{2};-1\right)\nonumber\\
 &&\quad\quad\quad\quad\quad\quad\quad\quad\quad\quad\quad\quad\quad\quad\nonumber\\
 &&\quad\quad-\left(\frac{\Lambda_{\bf UV}}{H}\right)^{\delta_{\bf B}+2} \, _2F_1\left(\frac{\delta_{\bf B}+2}{2},\frac{1}{\epsilon_b-1}-1;\frac{\delta_{\bf B}+4}{2};-\left(\frac{\Lambda_{\bf UV}}{H}\right)^2\right)\bigg].\quad\quad\quad\eea
 Here connecting our findings with the standard Quantum Field Theory approach we found that:
\bea \left(\delta_{{\cal Z}_{{\bf G}_1}}+\delta_{{\cal Z}_{{\bf G}_2}}+\delta_{{\cal Z}_{{\bf G}_3}}+\delta_{{\cal Z}_{{\bf G}_4}}+\delta_{{\cal Z}_{{\bf G}_5}}\right)={\bf C}(\mu_{\bf REN}=H,\Lambda_{\bf UV})={\cal K}_{\bf B}\quad{\rm with}\quad\delta_{{\cal Z}_{{\bf G}_6}}=0.\eea
For this reason, finally we get:
\bea \left(\delta_{{\cal Z}_{{\bf G}_1}}+\delta_{{\cal Z}_{{\bf G}_2}}+\delta_{{\cal Z}_{{\bf G}_3}}+\delta_{{\cal Z}_{{\bf G}_4}}+\delta_{{\cal Z}_{{\bf G}_5}}\right)&=&-\frac{4}{3}\bigg[\Delta^{2}_{\zeta,{\bf Tree}}(p)\bigg]^2_{\bf SRI}\nonumber\\
&&\times\Bigg(1+\frac{2}{15\pi^2}\frac{1}{c^2_{s}k^2_b}\bigg(-\left(1-\frac{1}{c^2_{s}}\right)\epsilon_b+6\frac{\bar{M}^3_1}{ HM^2_{ pl}}-\frac{4}{3}\frac{M^4_3}{H^2M^2_{ pl}}\bigg)\Bigg)\times \left(\frac{\epsilon_*}{\epsilon_b}\right)\nonumber\\
&&\times\frac{1}{\delta_{\bf B} +2}\bigg[\, _2F_1\left(\frac{\delta_{\bf B}+2}{2},\frac{1}{\epsilon_b-1}-1;\frac{\delta_{\bf B}+4}{2};-1\right)\nonumber\\
 &&\quad\quad\quad\quad\quad\quad\quad\quad\quad\quad\quad\quad\quad\quad\nonumber\\
 &&-\left(\frac{\Lambda_{\bf UV}}{H}\right)^{\delta_{\bf B}+2} \, _2F_1\left(\frac{\delta_{\bf B}+2}{2},\frac{1}{\epsilon_b-1}-1;\frac{\delta_{\bf B}+4}{2};-\left(\frac{\Lambda_{\bf UV}}{H}\right)^2\right)\bigg].\quad\quad\quad\eea
 Then the regularized and renormalized expression for the one loop corrected contribution from the bouncing phase is given by the following expression:
\bea \bigg[\Delta^{2}_{\zeta, {\bf One-loop}}(p)\bigg]_{\bf BOUNCE}&=&-\frac{4}{3}\bigg[\Delta^{2}_{\zeta,{\bf Tree}}(p)\bigg]^2_{\bf SRI}\nonumber\\&&\quad\quad\times\Bigg(1+\frac{2}{15\pi^2}\frac{1}{c^2_{s}k^2_b}\bigg(-\left(1-\frac{1}{c^2_{s}}\right)\epsilon_b+6\frac{\bar{M}^3_1}{ HM^2_{ pl}}-\frac{4}{3}\frac{M^4_3}{H^2M^2_{ pl}}\bigg)\Bigg)\times \left(\frac{\epsilon_*}{\epsilon_b}\right)\nonumber\\
&&\quad\quad\times\frac{1}{\delta_{\bf B} +2}\bigg[\, _2F_1\left(\frac{\delta_{\bf B}+2}{2},\frac{1}{\epsilon_b-1}-1;\frac{\delta_{\bf B}+4}{2};-1\right)\nonumber\\
 &&\quad\quad\quad\quad\quad\quad\quad\quad\quad\quad\quad\quad\quad\quad\nonumber\\
 &&\quad\quad\quad\quad\quad\quad\quad-\left(\frac{k_{b}}{k_{*}}\right)^{\delta_{\bf B}+2}\, _2F_1\left(\frac{\delta_{\bf B}+2}{2},\frac{1}{\epsilon_b-1}-1;\frac{\delta_{\bf B}+4}{2};-\left(\frac{k_{b}}{k_{*}}\right)^{2}\right)\bigg]\nonumber\\
&=&\bigg[\Delta^{2}_{\zeta,{\bf Tree}}(p)\bigg]_{\bf SRI}\times \overline{\bf W}_{\bf B},\eea
where $\overline{\bf W}_{\bf B}$ is given by:
\bea \label{t2}\overline{\bf W}_{\bf B}&=&-\frac{4}{3}\bigg[\Delta^{2}_{\zeta,{\bf Tree}}(p)\bigg]_{\bf SRI}\times\Bigg(1+\frac{2}{15\pi^2}\frac{1}{c^2_{s}k^2_b}\bigg(-\left(1-\frac{1}{c^2_{s}}\right)\epsilon_b+6\frac{\bar{M}^3_1}{ HM^2_{ pl}}-\frac{4}{3}\frac{M^4_3}{H^2M^2_{ pl}}\bigg)\Bigg)\times \left(\frac{\epsilon_*}{\epsilon_b}\right)\nonumber\\
&&\quad\quad\times\frac{1}{\delta_{\bf B} +2}\bigg[\, _2F_1\left(\frac{\delta_{\bf B}+2}{2},\frac{1}{\epsilon_b-1}-1;\frac{\delta_{\bf B}+4}{2};-1\right)\nonumber\\
 &&\quad\quad\quad\quad\quad\quad\quad\quad\quad\quad\quad\quad\quad\quad\nonumber\\
 &&\quad\quad\quad\quad\quad\quad\quad-\left(\frac{k_{b}}{k_{*}}\right)^{\delta_{\bf B}+2}\, _2F_1\left(\frac{\delta_{\bf B}+2}{2},\frac{1}{\epsilon_b-1}-1;\frac{\delta_{\bf B}+4}{2};-\left(\frac{k_{b}}{k_{*}}\right)^{2}\right)\bigg].\eea
where we identify $k_{\bf INT}=k_*$ and $k_{\bf IR}=k_b$.
          \end{itemize}
\newpage
   \item[\ding{43}] \underline{\textbf{Type-III: SRI-USR-SRII phase integral}}
    \begin{itemize}
         \item[$\blacksquare$] \underline{\bf Method-I:}\\ \\The representative one-loop momentum integral which describes the SRI-USR-SRII phase is given by the following expression:
\bea {\bf E}_3(\tau):&=&\int^{k_{\bf UV}}_{k_{\bf IR}}\frac{dk}{k}\;\left(\frac{k}{k_{\bf IR}}\right)^{3-2\nu}\left({\bf A}+{\bf B}k^2c^2_s\tau^2\right)+{\bf C}\nonumber\\
 &=&\int^{k_{\bf UV}/k_{\bf IR}}_{1}d\left(\frac{k}{k_{\bf IR}}\right)\;\left(\frac{k}{k_{\bf IR}}\right)^{2\delta_{\bf X}-1}\left({\bf A}+{\bf B}k^2c^2_s\tau^2\right)+{\bf C},\nonumber\\
 &=&\bigg[\frac{{\bf A}}{2\delta_{\bf X}}\left\{\left(\frac{k_{\bf UV}}{k_{\bf IR}}\right)^{2\delta_{\bf X}}-1\right\}+\frac{{\bf B}}{2\left(\delta_{\bf X}+1\right)}k^2_{\bf IR}c^2_s\tau^2\left\{\left(\frac{k_{\bf UV}}{k_{\bf IR}}\right)^{2\left(\delta_{\bf X}+1\right)}-1\right\}\bigg]+{\bf C},\nonumber\\
 &=&\bigg[{\bf A}\bigg\{\frac{1}{2\delta_{\bf X}}+\ln\left(\frac{k_{\bf UV}}{k_{\bf IR}}\right)-\frac{1}{2\delta_{\bf X}}+\cdots\bigg\}+\frac{{\bf B}}{2\left(\delta_{\bf X}+1\right)}k^2_{\bf IR}c^2_s\tau^2\left\{\left(\frac{k_{\bf UV}}{k_{\bf IR}}\right)^{2\left(\delta_{\bf X}+1\right)}-1\right\}\bigg]+{\bf C},\nonumber\\
 &=&\bigg[{\bf A}\ln\left(\frac{k_{\bf UV}}{k_{\bf IR}}\right)+\frac{{\bf B}}{2\left(\delta_{\bf X}+1\right)}k^2_{\bf IR}c^2_s\tau^2\left\{\left(\frac{k_{\bf UV}}{k_{\bf IR}}\right)^{2\left(\delta_{\bf X}+1\right)}-1\right\}\bigg]+{\bf C}.\quad\quad\quad\eea
where we identify the UV and IR cut-offs of the contracting phase as, $k_{\bf UV}=k_s$ and $k_{\bf IR}=k_*$ and also the symbols ${\bf A}$, ${\bf B}$ are two constants and $\delta_{\bf B}$ is defined by, $\delta_{\bf X}:=\left(\frac{3}{2}-\nu\right)$ where {\bf X}:= 1({\bf SRI}), 2({\bf USR}), 3({\bf SRII}).
Here ${\bf C}$ is the counter term which we need to determine explicitly and this is directly associated with ${\cal K}_{\bf SRI}$, ${\cal K}_{\bf USR}$ and ${\cal K}_{\bf SRII}$ in the present context. In the super-horizon limit, we have:
\bea {\bf E}_3
&=&{\bf A}\ln\left(\frac{k_{\bf UV}}{k_{\bf IR}}\right),\quad\quad\eea 
which further implies that here the counter term is fixed at:
\bea {\bf C}=0.\eea
The above-mentioned analysis helps us to completely remove the UV divergence from the final result and sufficiently coarse grain the logarithmic IR divergent contribution. 
  
          \item[$\blacksquare$] \underline{\bf Method-II:}\\ \\
By following the same trick as mentioned before we decompose the integration limit for this purpose as:     
\bea \int^{k_{\bf UV}}_{k_{\bf IR}}:=\bigg(\int^{k_{\bf INT}}_{k_{\bf IR}}+\int^{k_{\bf UV}=\frac{\Lambda_{\bf UV} a(\tau)}{c_s}}_{k_{\bf INT}}\bigg),\eea
using which the loop integral can be further recast in the following simplified form:
\bea {\bf E}_3:&=&\bigg(\int^{k_{\bf INT}}_{k_{\bf IR}}+\int^{k_{\bf UV}=\frac{\Lambda_{\bf UV} a(\tau)}{c_s}}_{k_{\bf INT}}\bigg)\frac{dk}{k}\;\left(\frac{k}{k_{\bf IR}}\right)^{3-2\nu}\left({\bf A}+{\bf B}k^2c^2_s\tau^2\right)+{\bf C},\nonumber\\
 &=&\bigg[\frac{{\bf A}}{2\delta_{\bf X}}\left\{\left(\frac{k_{\bf INT}}{k_{\bf IR}}\right)^{2\delta_{\bf X}}-1\right\}+\frac{{\bf B}}{2\left(\delta_{\bf X}+1\right)}\left\{\left(\frac{k_{\bf INT}}{k_{\bf IR}}\right)^{2\left(\delta_{\bf X}+1\right)}-1\right\}\bigg]\nonumber\\
 &&+\bigg[\frac{{\bf A}}{2\delta_{\bf X}}\left\{\left(\frac{\Lambda_{\bf UV}a(\tau)}{c_sk_{\bf IR}}\right)^{2\delta_{\bf X}}-1\right\}+\frac{{\bf B}}{2\left(\delta_{\bf X}+1\right)}\left\{\left(\frac{\Lambda_{\bf UV}a(\tau)}{c_sk_{\bf IR}}\right)^{2(\delta_{\bf X}+1)}-\left(\frac{k_{\bf INT}}{k_{\bf IR}}\right)^{2\left(\delta_{\bf X}+1\right)}\right\}\bigg]+{\bf C},\nonumber\\
 &=&\bigg[{\bf A}\left\{\ln\left(\frac{k_{\bf INT}}{k_{\bf IR}}\right)+\ln\left(\frac{\Lambda_{\bf UV}}{H}\right)\right\}+\frac{{\bf B}}{2\left(\delta_{\bf X}+1\right)}\left\{\left(\frac{\Lambda_{\bf UV}}{H}\right)^{2\left(\delta_{\bf X}+1\right)}-1\right\}\bigg]+{\bf C}.\quad\quad\quad\eea
 where we have utilized the fact that, $a(\tau)/c_s k_{\bf IR}=1/H$. Hence the counter term ${\bf C}$ in the present context at an arbitrary renormalization scale $\mu_{\bf REN}$ can be written as:
\bea {\bf C}\left(\mu_{\bf REN},\Lambda_{\bf UV}\right)&=&\bigg[{\bf A}\left\{\ln\left(\frac{\mu_{\bf REN}}{H}\right)-\ln\left(\frac{\Lambda_{\bf UV}}{H}\right)\right\}+\frac{{\bf B}}{2\left(\delta_{\bf X}+1\right)}\left\{\left(\frac{\mu_{\bf REN}}{H}\right)^{2\left(\delta_{\bf X}+1\right)}-\left(\frac{\Lambda_{\bf UV}}{H}\right)^{2\left(\delta_{\bf X}+1\right)}\right\}\bigg].\quad\quad\quad\eea
Hence the UV cut-off removed result for the generic one-loop momentum integration is described by the following expression:
 \bea {\bf E}_3(\mu_{\bf REN})
 &=&\bigg[{\bf A}\left\{\ln\left(\frac{k_{\bf INT}}{k_{\bf IR}}\right)+\ln\left(\frac{\Lambda_{\bf UV}}{H}\right)\right\}+\frac{{\bf B}}{2\left(\delta_{\bf X}+1\right)}\left\{\left(\frac{\Lambda_{\bf UV}}{H}\right)^{2\left(\delta_{\bf X}+1\right)}-1\right\}\bigg]\nonumber\\
 &&+\bigg[{\bf A}\left\{\ln\left(\frac{\mu_{\bf REN}}{H}\right)-\ln\left(\frac{\Lambda_{\bf UV}}{H}\right)\right\}+\frac{{\bf B}}{2\left(\delta_{\bf X}+1\right)}\left\{\left(\frac{\mu_{\bf REN}}{H}\right)^{2\left(\delta_{\bf X}+1\right)}-\left(\frac{\Lambda_{\bf UV}}{H}\right)^{2\left(\delta_{\bf X}+1\right)}\right\}\bigg],\nonumber\\
 &=&\bigg[{\bf A}\left\{\ln\left(\frac{k_{\bf INT}}{k_{\bf IR}}\right)+\ln\left(\frac{\mu_{\bf REN}}{H}\right)\right\}+\frac{{\bf B}}{2\left(\delta_{\bf X}+1\right)}\left\{\left(\frac{\mu_{\bf REN}}{H}\right)^{2\left(\delta_{\bf X}+1\right)}-1\right\}\bigg].\eea
 Further, if we fix the scale of renormalization at the Hubble scale i.e. $\mu_{\bf REN}=H$ then we get the following simplified result for the loop integral:
 \bea {\bf E}_3(\mu_{\bf REN}=H)
 &=&{\bf A}\ln\left(\frac{k_{\bf INT}}{k_{\bf IR}}\right).\eea
  Here the factors ${\bf A}$ and ${\bf B}$ are identified given by the following expressions in the SRI, USR and SRII phases:
  \bea &&\underline{\bf SRI:}\quad\quad{\bf A}={\bf B}
  =-\frac{4}{3}\bigg[\Delta^{2}_{\zeta,{\bf Tree}}(p)\bigg]^2_{\bf SRI}\times\Bigg(1+\frac{2}{15\pi^2}\frac{1}{c^2_{s}k^2_*}\bigg(-\left(1-\frac{1}{c^2_{s}}\right)\epsilon_*+6\frac{\bar{M}^3_1}{ HM^2_{ pl}}-\frac{4}{3}\frac{M^4_3}{H^2M^2_{ pl}}\bigg)\Bigg),\quad\quad\quad\\
  &&\underline{\bf USR:}\quad\quad{\bf A}={\bf B}
  =\frac{1}{4}\bigg[\Delta^{2}_{\zeta,{\bf Tree}}(p)\bigg]^2_{\bf SRI}\times\bigg[\bigg(\frac{\Delta\eta(\tau_{e})}{\tilde{c}^{4}_{s}}\bigg)^{2}{\bigg(\frac{k_{e}}{k_{s}}\bigg)^6} - \left(\frac{\Delta\eta(\tau_{s})}{\tilde{c}^{4}_{s}}\right)^{2}\bigg],\quad\quad\quad\\
  &&\underline{\bf SRII:}\quad\quad{\bf A}={\bf B}
  =\bigg[\Delta^{2}_{\zeta,{\bf Tree}}(p)\bigg]^2_{\bf SRI}\times\Bigg(1+\frac{2}{15\pi^2}\frac{1}{c^2_{s}k^2_*}\bigg(-\left(1-\frac{1}{c^2_{s}}\right)\epsilon_*+6\frac{\bar{M}^3_1}{ HM^2_{ pl}}-\frac{4}{3}\frac{M^4_3}{H^2M^2_{ pl}}\bigg)\Bigg).\quad\quad\quad\eea
Also the counter term at $\mu_{\bf REN}=H$ scale is given by:
\bea {\bf C}\left(\mu_{\bf REN}=H,\Lambda_{\bf UV}\right)&=&-\bigg[{\bf A}\ln\left(\frac{\Lambda_{\bf UV}}{H}\right)+\frac{{\bf B}}{2\left(\delta_{\bf X}+1\right)}\left\{\left(\frac{\Lambda_{\bf UV}}{H}\right)^{2\left(\delta_{\bf X}+1\right)}-1\right\}\bigg].\quad\quad\quad\eea
Here connecting our findings with the standard Quantum Field Theory approach we found that:
\bea &&\underline{\bf SRI:}\quad\quad\left(\delta_{{\cal Z}_{{\bf G}_1}}+\delta_{{\cal Z}_{{\bf G}_2}}+\delta_{{\cal Z}_{{\bf G}_3}}+\delta_{{\cal Z}_{{\bf G}_4}}+\delta_{{\cal Z}_{{\bf G}_5}}\right)={\bf C}(\mu_{\bf REN}=H,\Lambda_{\bf UV})={\cal K}_{\bf SRI}\quad{\rm with}\quad\delta_{{\cal Z}_{{\bf G}_6}}=0,\quad\quad\\
&&\underline{\bf USR:}\quad\quad\delta_{{\cal Z}_{{\bf G}_6}}={\bf C}(\mu_{\bf REN}=H,\Lambda_{\bf UV})={\cal K}_{\bf USR}\quad{\rm with}\quad\left(\delta_{{\cal Z}_{{\bf G}_1}}+\delta_{{\cal Z}_{{\bf G}_2}}+\delta_{{\cal Z}_{{\bf G}_3}}+\delta_{{\cal Z}_{{\bf G}_4}}+\delta_{{\cal Z}_{{\bf G}_5}}\right)=0,\quad\quad\\
&&\underline{\bf SRII:}\quad\quad\left(\delta_{{\cal Z}_{{\bf G}_1}}+\delta_{{\cal Z}_{{\bf G}_2}}+\delta_{{\cal Z}_{{\bf G}_3}}+\delta_{{\cal Z}_{{\bf G}_4}}+\delta_{{\cal Z}_{{\bf G}_5}}\right)={\bf C}(\mu_{\bf REN}=H,\Lambda_{\bf UV})={\cal K}_{\bf SRII}\quad{\rm with}\quad\delta_{{\cal Z}_{{\bf G}_6}}=0.
\eea
For this reason, finally we get:
\bea &&\underline{\bf SRI:}\quad\quad\left(\delta_{{\cal Z}_{{\bf G}_1}}+\delta_{{\cal Z}_{{\bf G}_2}}+\delta_{{\cal Z}_{{\bf G}_3}}+\delta_{{\cal Z}_{{\bf G}_4}}+\delta_{{\cal Z}_{{\bf G}_5}}\right)=\frac{4}{3}\bigg[\Delta^{2}_{\zeta,{\bf Tree}}(p)\bigg]^2_{\bf SRI}\nonumber\\
&&\quad\quad\quad\quad\quad\quad\quad\quad\quad\quad\quad\quad\quad\quad\quad\quad\quad\quad\quad\quad\times\Bigg(1+\frac{2}{15\pi^2}\frac{1}{c^2_{s}k^2_*}\bigg(-\left(1-\frac{1}{c^2_{s}}\right)\epsilon_*+6\frac{\bar{M}^3_1}{ HM^2_{ pl}}-\frac{4}{3}\frac{M^4_3}{H^2M^2_{ pl}}\bigg)\Bigg)\nonumber\\
&&\quad\quad\quad\quad\quad\quad\quad\quad\quad\quad\quad\quad\quad\quad\quad\quad\quad\quad\quad\quad\times\bigg[\ln\left(\frac{\Lambda_{\bf UV}}{H}\right)+\frac{1}{2\left(\delta_{\bf SRI}+1\right)}\left\{\left(\frac{\Lambda_{\bf UV}}{H}\right)^{2\left(\delta_{\bf SRI}+1\right)}-1\right\}\bigg],\quad\quad\\
&&\underline{\bf USR:}\quad\quad\delta_{{\cal Z}_{{\bf G}_6}}=-\frac{1}{4}\bigg[\Delta^{2}_{\zeta,{\bf Tree}}(p)\bigg]^2_{\bf SRI}\times\frac{1}{4}\bigg[\bigg(\frac{\Delta\eta(\tau_{e})}{\tilde{c}^{4}_{s}}\bigg)^{2}{\bigg(\frac{k_{e}}{k_{s}}\bigg)^6} - \left(\frac{\Delta\eta(\tau_{s})}{\tilde{c}^{4}_{s}}\right)^{2}\bigg]\nonumber\\
&&\quad\quad\quad\quad\quad\quad\quad\quad\quad\quad\quad\quad\quad\quad\times\bigg[\ln\left(\frac{\Lambda_{\bf UV}}{H}\right)+\frac{1}{2\left(\delta_{\bf USR}+1\right)}\left\{\left(\frac{\Lambda_{\bf UV}}{H}\right)^{2\left(\delta_{\bf USR}+1\right)}-1\right\}\bigg],\quad\quad\eea
\bea
&&\underline{\bf SRII:}\quad\quad\left(\delta_{{\cal Z}_{{\bf G}_1}}+\delta_{{\cal Z}_{{\bf G}_2}}+\delta_{{\cal Z}_{{\bf G}_3}}+\delta_{{\cal Z}_{{\bf G}_4}}+\delta_{{\cal Z}_{{\bf G}_5}}\right)=-\bigg[\Delta^{2}_{\zeta,{\bf Tree}}(p)\bigg]^2_{\bf SRI}\nonumber\\
&&\quad\quad\quad\quad\quad\quad\quad\quad\quad\quad\quad\quad\quad\quad\quad\quad\quad\quad\quad\quad\times\Bigg(1+\frac{2}{15\pi^2}\frac{1}{c^2_{s}k^2_*}\bigg(-\left(1-\frac{1}{c^2_{s}}\right)\epsilon_*+6\frac{\bar{M}^3_1}{ HM^2_{ pl}}-\frac{4}{3}\frac{M^4_3}{H^2M^2_{ pl}}\bigg)\Bigg)\nonumber\\
&&\quad\quad\quad\quad\quad\quad\quad\quad\quad\quad\quad\quad\quad\quad\quad\quad\quad\times\bigg[\ln\left(\frac{\Lambda_{\bf UV}}{H}\right)+\frac{1}{2\left(\delta_{\bf SRII}+1\right)}\left\{\left(\frac{\Lambda_{\bf UV}}{H}\right)^{2\left(\delta_{\bf SRII}+1\right)}-1\right\}\bigg].
\eea
Then the regularized and renormalized expression for the one loop corrected contribution from the SRI, USR, and SRII phases are given by the following expression:
\bea \bigg[\Delta^{2}_{\zeta, {\bf One-loop}}(p)\bigg]_{\bf SRI}&=&-\frac{4}{3}\bigg[\Delta^{2}_{\zeta,{\bf Tree}}(p)\bigg]^2_{\bf SRI}\nonumber\\&&\quad\quad\times\Bigg(1+\frac{2}{15\pi^2}\frac{1}{c^2_{s}k^2_*}\bigg(-\left(1-\frac{1}{c^2_{s}}\right)\epsilon_*+6\frac{\bar{M}^3_1}{ HM^2_{ pl}}-\frac{4}{3}\frac{M^4_3}{H^2M^2_{ pl}}\bigg)\Bigg)\times\ln\left(\frac{k_s}{k_*}\right),\nonumber\\
&=&\bigg[\Delta^{2}_{\zeta,{\bf Tree}}(p)\bigg]_{\bf SRI}\times \overline{\bf W}_{\bf SRI},\\
\bigg[\Delta^{2}_{\zeta, {\bf One-loop}}(p)\bigg]_{\bf USR}&=&\frac{1}{4}\bigg[\Delta^{2}_{\zeta,{\bf Tree}}(p)\bigg]^2_{\bf SRI}\times\bigg[\bigg(\frac{\Delta\eta(\tau_{e})}{\tilde{c}^{4}_{s}}\bigg)^{2}{\bigg(\frac{k_{e}}{k_{s}}\bigg)^6} - \left(\frac{\Delta\eta(\tau_{s})}{\tilde{c}^{4}_{s}}\right)^{2}\bigg]\times\ln\left(\frac{k_e}{k_s}\right),\nonumber\\
&=&\bigg[\Delta^{2}_{\zeta,{\bf Tree}}(p)\bigg]_{\bf SRI}\times \overline{\bf W}_{\bf USR},\\
\bigg[\Delta^{2}_{\zeta, {\bf One-loop}}(p)\bigg]_{\bf SRII}&=&\bigg[\Delta^{2}_{\zeta,{\bf Tree}}(p)\bigg]^2_{\bf SRI}\nonumber\\&&\quad\quad\times\Bigg(1+\frac{2}{15\pi^2}\frac{1}{c^2_{s}k^2_*}\bigg(-\left(1-\frac{1}{c^2_{s}}\right)\epsilon_*+6\frac{\bar{M}^3_1}{ HM^2_{ pl}}-\frac{4}{3}\frac{M^4_3}{H^2M^2_{ pl}}\bigg),\Bigg)\times\ln\left(\frac{k_{\rm end}}{k_e}\right)\nonumber\\
&=&\bigg[\Delta^{2}_{\zeta,{\bf Tree}}(p)\bigg]_{\bf SRI}\times \overline{\bf W}_{\bf SRII},\eea
where $\overline{\bf W}_{\bf SRI}$, $\overline{\bf W}_{\bf USR}$, and $\overline{\bf W}_{\bf SRII}$ are given by:
\bea \label{t3}\overline{\bf W}_{\bf SRI}&=&-\frac{4}{3}\bigg[\Delta^{2}_{\zeta,{\bf Tree}}(p)\bigg]_{\bf SRI}\nonumber\\
&&\quad\quad\quad\quad\times\Bigg(1+\frac{2}{15\pi^2}\frac{1}{c^2_{s}k^2_*}\bigg(-\left(1-\frac{1}{c^2_{s}}\right)\epsilon_*+6\frac{\bar{M}^3_1}{ HM^2_{ pl}}-\frac{4}{3}\frac{M^4_3}{H^2M^2_{ pl}}\bigg)\Bigg)\times\ln\left(\frac{k_s}{k_*}\right),\nonumber\\
&=&-\frac{4}{3}\bigg[\Delta^{2}_{\zeta,{\bf Tree}}(N)\bigg]_{\bf SRI}\nonumber\\
&&\quad\quad\quad\quad\times\Bigg(1+\frac{2}{15\pi^2}\frac{1}{c^2_{s}k^2_*}\bigg(-\left(1-\frac{1}{c^2_{s}}\right)\epsilon_*+6\frac{\bar{M}^3_1}{ HM^2_{ pl}}-\frac{4}{3}\frac{M^4_3}{H^2M^2_{ pl}}\bigg)\Bigg)\times\Delta N_{\bf SRI},\quad\quad\\
\label{t4}\overline{\bf W}_{\bf USR}&=&\frac{1}{4}\bigg[\Delta^{2}_{\zeta,{\bf Tree}}(p)\bigg]_{\bf SRI}\times\bigg[\bigg(\frac{\Delta\eta(\tau_{e})}{\tilde{c}^{4}_{s}}\bigg)^{2}{\bigg(\frac{k_{e}}{k_{s}}\bigg)^6} - \left(\frac{\Delta\eta(\tau_{s})}{\tilde{c}^{4}_{s}}\right)^{2}\bigg]\times\ln\left(\frac{k_e}{k_s}\right),\nonumber\\
&=&\frac{1}{4}\bigg[\Delta^{2}_{\zeta,{\bf Tree}}(N)\bigg]_{\bf SRI}\times\bigg[\bigg(\frac{\Delta\eta(\tau_{e})}{\tilde{c}^{4}_{s}}\bigg)^{2}\exp(6\Delta N_{\bf USR}) - \left(\frac{\Delta\eta(\tau_{s})}{\tilde{c}^{4}_{s}}\right)^{2}\bigg]\times\Delta N_{\bf USR},\quad\quad\quad\\
\label{t5}\overline{\bf W}_{\bf SRII}&=&\bigg[\Delta^{2}_{\zeta,{\bf Tree}}(p)\bigg]^2_{\bf SRI}\times\Bigg(1+\frac{2}{15\pi^2}\frac{1}{c^2_{s}k^2_*}\bigg(-\left(1-\frac{1}{c^2_{s}}\right)\epsilon_*+6\frac{\bar{M}^3_1}{ HM^2_{ pl}}-\frac{4}{3}\frac{M^4_3}{H^2M^2_{ pl}}\bigg)\Bigg)\times\ln\left(\frac{k_{\rm end}}{k_e}\right),\nonumber\\
&=&\bigg[\Delta^{2}_{\zeta,{\bf Tree}}(N)\bigg]^2_{\bf SRI}\times\Bigg(1+\frac{2}{15\pi^2}\frac{1}{c^2_{s}k^2_*}\bigg(-\left(1-\frac{1}{c^2_{s}}\right)\epsilon_*+6\frac{\bar{M}^3_1}{ HM^2_{ pl}}-\frac{4}{3}\frac{M^4_3}{H^2M^2_{ pl}}\bigg)\Bigg)\times\Delta N_{\bf SRII}.\quad\quad\quad\eea
Here UV and IR cut-off scales in the SRI, USR, and SRII phases are identified with the following values for the present computation:
\bea &&\underline{\bf SRI:}\quad\quad k_{\rm UV}=k_s,\quad k_{\bf IR}=k_*,\\
&&\underline{\bf USR:}\quad\quad k_{\rm UV}=k_e,\quad k_{\bf IR}=k_s,\\
&&\underline{\bf SRII:}\quad\quad k_{\rm UV}=k_{\rm end},\quad k_{\bf IR}=k_e.\eea
\end{itemize}

\end{itemize}
  Hence, the total regularized and renormalized one-loop corrected power spectrum for scalar modes can be expressed by the following expression:
   \bea \label{one-loopRR} \overline{\Delta^{2}_{\zeta, {\bf EFT}}(p)}&=&\Delta^{2}_{\zeta, {\bf RR}}(p)\nonumber\\  
   &=&\bigg[\Delta^{2}_{\zeta,{\bf Tree}}(p)\bigg]_{\bf SRI}\times\bigg(1+\underbrace{\overline{{\bf W}}_{\bf C}+\overline{{\bf W}}_{\bf B}+\overline{{\bf W}}_{\bf SRI}+\overline{{\bf W}}_{\bf USR}+\overline{{\bf W}}_{\bf SRII}}_{\textbf{Regularized and Renormalized one-loop correction}}\bigg)\nonumber\\
   &=&\begin{tikzpicture}[baseline={([yshift=-3.5ex]current bounding box.center)},very thick]
  
    % Loop
  \def\radius{1}
  \scalebox{1}{\draw[red,very thick] (0,\radius) circle (\radius);
  \draw[red,very thick] (4.5*\radius,0) circle (\radius);}

  % External lines
  %\filldraw;
  \draw[black, very thick] (-4*\radius,0) -- 
  (-2.5*\radius,0);
  \node at (-2*\radius,0) {+};
  \draw[black, very thick] (-1.5*\radius,0) -- (0,0);
  \draw[blue,fill=blue] (0,0) circle (.5ex);
  \draw[black, very thick] (0,0)  -- (1.5*\radius,0);
  \node at (2*\radius,0) {+};
  \draw[black, very thick] (2.5*\radius,0) -- (3.5*\radius,0); 
  \draw[blue,fill=blue] (3.5*\radius,0) circle (.5ex);
  \draw[blue,fill=blue] (5.5*\radius,0) circle (.5ex);
  \draw[black, very thick] (5.5*\radius,0) -- (6.5*\radius,0);
\end{tikzpicture},\eea
   where $\overline{{\bf W}}_{\bf C}$, $\overline{{\bf W}}_{\bf B}$, $\overline{{\bf W}}_{\bf SRI}$, $\overline{{\bf W}}_{\bf USR}$, and $\overline{{\bf W}}_{\bf SRII}$ are already computed in equation(\ref{t1}), equation(\ref{t2}), equation(\ref{t3}), equation(\ref{t4}) and equation(\ref{t5}) respectively.

\subsection{Step II: Adiabatic/Wave-function renormalization and removal of UV/power law divergence}

Our main goal in this paragraph is to eliminate the contributions from quadratic UV divergences that show up in the primordial power spectrum computation for the comoving curvature perturbation from the contraction, bounce, SRI, USR, and SRII periods in a one-loop manner. Technically speaking, the origin of these UV divergences is the sub-horizon area ($-kc_s\tau\gg 1$), where quantum mechanical disturbances have a major influence. It is clear that, at the late time limit where the scalar modes leave the horizon and enter the super horizon area, only logarithmic divergent contributions remain in the non-trivial FLRW background which describes the previously mentioned five phases. As a result, one may ignore the quadratic UV divergence later on. But in the current environment, it is critical to methodically offer a technical framework that enables us to do so. The influence of quadratic UV divergence at the late time limit has not been eliminated in this work. This section focuses on providing the technical calculation of a renormalization strategy that enables us to include a suitable counter term to automatically negate the contribution from the quadratic divergence. Furthermore, since we have quantized the comoving curvature perturbation and are treating it at the same footage of the scalar quantum field, we must implement a smoothing scheme in order to remove the UV divergences of the underlying cosmological perturbation theory. This is because adiabatic renormalization precisely serves to smooth the effect of quantum fluctuations at the sub-horizon region where the short-range UV modes manifest.

In this instance, we are working with the Quantum Field Theory of FLRW space-time, where the well-known adiabatic regularization and the associated renormalization method play a crucial role in eliminating the contributions from the UV divergences evident at various powers. Refer to references \cite{Durrer:2009ii,Wang:2015zfa,PhysRevD.9.341,Finelli:2007fr,Marozzi:2011da,Boyanovsky:2005sh,PhysRevD.10.3905,PhysRevD.35.2955} for a more thorough discussion of the physical ramifications and possible applications. The sub-horizon quantum fluctuations are captured by applying the minimum subtraction rule to the appropriate UV modes, which makes it simple to eliminate the UV divergence contributions from the underlying Quantum Field Theoretic setup. To eliminate the UV divergent contribution from the short-range UV modes that are prevalent in the sub-horizon area, the minimum subtraction in this case technically equates to inserting a counter term in the underlying theory. The effects of quadratic UV divergences, which come from both the SR and USR periods, are sufficiently eliminated in this computation by using a second-order subtraction. It becomes significant to remove the counter terms beyond the second-order minimal subtraction only when addressing the contributions of the UV divergent components that arise at higher powers. Fortunately, the adiabatic renormalization approach applied in this calculation has enough power to eliminate the quadratic divergence contribution, thus we don't need to consider counterterms higher than the second order.
In the context of curved space-time within Quantum Field Theory, specifically in FLRW backgrounds, the adiabatic renormalization approach was developed with the specific goal of eliminating UV divergences alone. IR divergences cannot be addressed with this strategy. See refs. \cite{Durrer:2009ii,Wang:2015zfa,PhysRevD.9.341,Finelli:2007fr,Marozzi:2011da,Boyanovsky:2005sh,PhysRevD.10.3905,PhysRevD.35.2955} for more details. It is noteworthy that in this case, the comoving curvature perturbation modes are renormalized directly in the adiabatic limit of the cosmological perturbation using the adiabatic renormalization procedure.  This is precisely comparable to the wave function renormalization that we often carry out in the context of quantum field theory.  The one-loop contribution to the cosmic power spectrum in the current context is further renormalized by such renormalized modes or the wave function.  It is also crucial to note that, in the current context, the WKB approximation approach is useful in helping to create a generalized version of the mode function, or more specifically, the regularised wave function in the adiabatic limit. As such, the contributions of UV divergences from the short-range modes may be entirely eliminated from the regularised form of the mode function.  The generalized structure of the regularised modes is meant to ensure that UV divergences may be correctly handled in any order while maintaining the generality of the underlying Quantum Field Theoretical setup.

By expanding the field modes to $m$th order, a WKB-like adiabatic expansion provides the foundation for the regularization in adiabatic renormalization. 
\bea 
 v_{k}^{(m)} &\equiv& \frac{1}{\sqrt{2 {\cal W}_{k}^{(m)}(\tau)}} \exp{\bigg(-i \int_{\tau_0}^{\tau}{\cal W}_{k}^{(m)}(\tau')d\tau \bigg)}, \\
 v_{k}^{(m)^{*}} &\equiv& \frac{1}{\sqrt{2 {\cal W}_{k}^{(m)}(\tau)}} \exp{\bigg(i \int_{\tau_0}^{\tau}{\cal W}_{k}^{(m)}(\tau')d\tau \bigg)},
\eea
where the general form of the function ${\cal W}_{k}^{m}$ is given by the following expression:
\bea
{\cal W}_{k}^{(m)} \equiv \bigg(\omega_{k}^{(0)} + \omega_{k}^{(1)}+\omega_{k}^{(2)}+ \cdots + \omega_{k}^{(m)}\bigg).
\eea
The amount of time derivatives in $\omega_{k}$ indicates its adiabatic order, which is shown by the superscripts. The WKB approximation aids in the construction of a regularized wave function across the adiabatic limit in the current situation, which is an essential point to notice. This, in turn, helps to eliminate the UV divergence contributions from modes that are at short range. Thus, we take into account that in the adiabatic limit, the UV divergent contributions show up as $m$-th power in the mode functions. Crucially, the comoving curvature perturbation modes inside the adiabatic limit of cosmic perturbations are directly renormalized by the adiabatic renormalization procedure. Initially, we need to take into account that the UV divergences in the adiabatic limiting scenario are occurring at the $m$-th power, which essentially signifies the $m$-th order UV divergences originating from short-range scalar modes.  For each of the five subsequent phases, the renormalized mode function for the comoving curvature perturbation may be represented as follows using the well-known WKB approximation approach that was previously mentioned:
\bea
\label{genWKBcv}
\zeta_{\bf k, {\bf C}}^{(m)}(\tau) &=&-
2^{\nu-\frac{3}{2}} (-k c_s \tau )^{\frac{3}{2}-\nu}\left(\frac{\tau}{\tau_0}\right)^{-\frac{1}{(\epsilon-1)}}\sqrt{\left(\frac{\epsilon_*}{\epsilon_c}\right)}\Bigg(\frac{c_s H \tau}{2 M_{pl} \sqrt{\epsilon_*} \sqrt{{\cal W}_{k}^{(m)}(\tau)}}\Bigg) \Bigg|\frac{\Gamma(\nu)}{\Gamma(\frac{3}{2})}\Bigg|\nonumber\\
&&\quad\quad\quad\quad\quad\quad\quad\times\bigg[\alpha_1 \exp{\bigg(-i \int_{\tau_{0}}^{\tau} d \tau^{'} {\cal W}_{k}^{(m)}(\tau ^{'})}\bigg) + \beta_1 \exp{\bigg(i \int_{\tau_{0}}^{\tau} d \tau^{'} {\cal W}_{k}^{(m)}(\tau ^{'})}\bigg)\bigg], \\
\zeta_{\bf k, {\bf B}}^{(m)}(\tau) &=&-
2^{\nu-\frac{3}{2}} (-k c_s \tau )^{\frac{3}{2}-\nu}\bigg[1+\left(\frac{\tau}{\tau_0}\right)^2\bigg]^{-\frac{1}{2(\epsilon-1)}}\sqrt{\left(\frac{\epsilon_*}{\epsilon_b}\right)}\Bigg(\frac{c_s H \tau}{2 M_{pl} \sqrt{\epsilon_*} \sqrt{{\cal W}_{k}^{(m)}(\tau)}}\Bigg) \Bigg|\frac{\Gamma(\nu)}{\Gamma(\frac{3}{2})}\Bigg|\nonumber\\
&&\quad\quad\quad\quad\quad\quad\quad\times\bigg[\alpha_1 \exp{\bigg(-i \int_{\tau_{0}}^{\tau} d \tau^{'} {\cal W}_{k}^{(m)}(\tau ^{'})}\bigg) + \beta_1 \exp{\bigg(i \int_{\tau_{0}}^{\tau} d \tau^{'} {\cal W}_{k}^{(m)}(\tau ^{'})}\bigg)\bigg], \\
\zeta_{\bf k, \bf SRI}^{(m)}(\tau) &=& -
2^{\nu-\frac{3}{2}} (-k c_s \tau )^{\frac{3}{2}-\nu}\Bigg(\frac{c_s H \tau}{2 M_{pl} \sqrt{\epsilon_*} \sqrt{{\cal W}_{k}^{(m)}(\tau)}}\Bigg)\Bigg|\frac{\Gamma(\nu)}{\Gamma(\frac{3}{2})}\Bigg|\nonumber\\
&&\quad\quad\quad\quad\quad\quad\quad\times \bigg[\alpha_1 \exp{\bigg(-i \int_{\tau_{0}}^{\tau} d \tau^{'} {\cal W}_{k}^{(m)}(\tau ^{'})}\bigg) + \beta_1 \exp{\bigg(i \int_{\tau_{0}}^{\tau} d \tau^{'} {\cal W}_{k}^{(m)}(\tau ^{'})}\bigg)\bigg], \\
\zeta_{\bf k, \bf USR}^{(m)}(\tau) &=& -2^{\nu-\frac{3}{2}} (-k c_s \tau )^{\frac{3}{2}-\nu}\Bigg(\frac{c_s H \tau}{2 M_{pl} \sqrt{\epsilon_*} \sqrt{{\cal W}_{k}^{(m)}(\tau)}}\Bigg)\Bigg|\frac{\Gamma(\nu)}{\Gamma(\frac{3}{2})}\Bigg| \bigg(\frac{\tau_{0}}{\tau}\bigg)^{3} \nonumber\\
&&\quad\quad\quad\quad\quad\quad\quad\times \bigg[\alpha_2 \exp{\bigg(-i \int_{\tau_{0}}^{\tau}d \tau^{'} {\cal W}_{k}^{(m)}(\tau ^{'})}\bigg) + \beta_2 \exp{\bigg(i \int_{\tau_{0}}^{\tau}d \tau^{'} {\cal W}_{k}^{(m)}(\tau ^{'})}\bigg)\bigg], \eea\bea
\zeta_{\bf k, \bf SRII}^{(m)}(\tau) &=& -2^{\nu-\frac{3}{2}} (-k c_s \tau )^{\frac{3}{2}-\nu}\Bigg(\frac{c_s H \tau}{2 M_{pl} \sqrt{\epsilon_*} \sqrt{{\cal W}_{k}^{(m)}(\tau)}}\Bigg)\Bigg|\frac{\Gamma(\nu)}{\Gamma(\frac{3}{2})}\Bigg| \bigg(\frac{\tau_{0}}{\tau_e}\bigg)^{3} \nonumber\\
&&\quad\quad\quad\quad\quad\quad\quad\times \bigg[\alpha_3 \exp{\bigg(-i \int_{\tau_{0}}^{\tau}d \tau^{'} {\cal W}_{k}^{(m)}(\tau ^{'})}\bigg) + \beta_3 \exp{\bigg(i \int_{\tau_{0}}^{\tau}d \tau^{'} {\cal W}_{k}^{(m)}(\tau ^{'})}\bigg)\bigg]. 
\eea
In this case, the features function ${\cal W}_{k}^{(m)}$ stands for the conformal time dependence factor inside adiabatic regularisation, which is defined as follows for the $m$-th order:
\bea
{\cal W}^{(m)}(\tau) = \sqrt{\bigg(c_s^2 k^2 - \frac{z^{''}}{z}\bigg) - \frac{1}{2}\bigg[\frac{{\cal W}_{k}^{(m-2)''}(\tau)}{{\cal W}_{k}^{m-2}(\tau)} - \frac{3}{2}\bigg(\frac{{\cal W}_{k}^{(m-2)''}(\tau)}{{\cal W}_{k}^{m-2}(\tau)}\bigg)^2\bigg]}, \quad \quad {\rm where} \quad \frac{z''}{z}\approx \frac{1}{\tau ^2}\left(\nu^2-\frac{1}{4}\right)
\eea
In this computation, the initial choice of the vacuum is maintained to Bunch Davies whose structure is preserved during the contraction, bouncing, and SRI phases. In terms of the Bogoliubov coefficients, it is characterized by $\alpha_1=1$ and $\beta_1=0$. However, due to having sharp transitions at the boundaries of the USR phase, which is attached to the SRI and SRII phases, the corresponding vacuum structure changes in both of these mentioned phases. This fact is captured in the new Bogoliubov coefficients $(\alpha_2,\beta_2)$ and $(\alpha_3,\beta_3)$ which describes the USR and SRII phases. It is possible to argue that the formulations for the Bogolibov coefficients differ from the findings explicitly obtained in the USR and SRII periods because of the adiabaticity in the scalar modes. Nonetheless, it is anticipated that these Bogoliubov coefficients in the USR and SRII period would not deviate considerably from the values that we have previously computed because of the extremely slight shift in the adiabatic limit. This is perfectly acceptable in terms of the validity of the adiabatic regularisation in the corresponding mode at the $m$-th order and well justified from a physical point of view. Before moving on to the remainder of the computation, it is preferable to make this fact clear even if we won't be using the explicit equation for the Bogoliubov coefficients in the USR and SRII periods. In the process of computing the counter term that appears in the renormalized version of the one-loop power spectrum in the USR and SRII periods, we will clearly demonstrate that the explicit structure of the Bogoliubov coefficients $(\alpha_2,\beta_2)$ and $(\alpha_3,\beta_3)$ will not have a significant impact on the final result, nor on the short-range UV modes in the corresponding computation. In the remaining computations for the USR and SRII phases in this section, we shall elaborate on this topic. One could argue here, for a more physical justification, that in the USR and SRII periods, the quantum initial condition and the corresponding quantum vacuum state shift in comparison to the Bunch Davies initial state that exists in the contraction, bouncing, and SRI phases.  It will be contrary to common sense and the usefulness of the adiabatic regularisation scheme itself if the shifted initial vacuum as it appears in the USR and SRII periods differs noticeably in the adiabatic limiting situation from that which we initially obtained by precisely solving the Mukhanov Sasaki equation. By enforcing the restriction that adiabaticity will not change the structure of the underlying quantum vacuum state in the USR and SRII periods, the adiabatic limiting approximation in terms of the WKB regularised modes is realized. In the current calculation, the strong adiabaticity limit validates the uniqueness of the quantum vacuum state and the underlying quantum starting condition in the USR and SRII periods. Because of this, it is possible to use the same structure for the Bogoliubov coefficients $(\alpha_2,\beta_2)$ and $(\alpha_3,\beta_3)$ as they were previously determined without encountering any further ambiguity from a physical perspective.  We are making full use of this intrinsic power of the Quantum Field Theory of curved space-time, which is written in the backdrop of FLRW space-time, in the current computation. We also need to emphasize that this possibility does not arise in the context of the contraction, bouncing, and SRI phases because we have fixed the quantum initial condition to be Bunch Davies. As a result of this fixed choice, the structure of the Bogolibov coefficient is also fixed, and it will remain unchanged whether we take into account the adiabatically regularised mode or the precise mode that we have computed by solving the Mukhanov-Sasaki equation.

Let us first discuss a few key points that will be very helpful for the remaining analyses conducted in this part and throughout the remainder of the work, before delving deeper into the technical calculation. The five renormalization scheme dependent parameters, ${\cal K}_{\bf C}$, ${\cal K}_{\bf B}$, ${\cal K}_{\bf SRI}$, ${\cal K}_{\bf USR}$, and ${\cal K}_{\bf SRII}$, were introduced in the previous section during the computation of on-loop effects in the primordial power spectrum for the comoving curvature perturbation. Since the renormalization scheme was not fixed previously, we have not been able to fix any of these parameters. In the current discussion context, once the renormalization method is fixed to be adiabatic, the equation for the counter terms ${\cal K}_{\bf C}$, ${\cal K}_{\bf B}$, ${\cal K}_{\bf SRI}$, ${\cal K}_{\bf USR}$, and ${\cal K}_{\bf SRII}$ in the five phases may be explicitly computed automatically. In this part, we will explicitly carry out this computation, which will allow us to demonstrate the total elimination of the short-range modes to the quadratic UV divergence. We shall show in the calculation carried out in this part that IR logarithmic divergences cannot be eliminated from the underlying theory using the current renormalization approach. First, we exploited the fact that the adiabatic limit holds for the $n$-th order and can be used to subtract the $n$-th power of UV divergences from the contributions of the short-range modes, in order to illustrate the generalized structure of the scalar modes in the five phases. Nevertheless, we are already aware that UV divergences occur in the different phases with quadratic power for $\nu=3/2$ or for $\nu\neq 3/2$ with having other power, as shown by the computation carried out in the preceding section to illustrate the entire one-loop adjusted primordial power spectrum for scalar modes. With the help of the adiabatic regularisation technical structure, we can eliminate the quadratic divergences from the one-loop correction terms in all five of the previously mentioned phases by limiting the remainder of the computation to the second order, or fixing $n=2$. Within the previously specified limitations, the following formulas may be used to further simplify the adiabatically regularised WKB approximated modes for comoving scalar curvature perturbation in each of the five phases:
\bea
\label{genWKBcv2}
\zeta_{\bf k, {\bf C}}^{(2)}(\tau) &=&-
2^{\nu-\frac{3}{2}} (-k c_s \tau )^{\frac{3}{2}-\nu}\left(\frac{\tau}{\tau_0}\right)^{-\frac{1}{(\epsilon-1)}}\sqrt{\left(\frac{\epsilon_*}{\epsilon_c}\right)}\Bigg(\frac{c_s H \tau}{2 M_{pl} \sqrt{\epsilon_*} \sqrt{{\cal W}_{k}^{(2)}(\tau)}}\Bigg) \Bigg|\frac{\Gamma(\nu)}{\Gamma(\frac{3}{2})}\Bigg|\nonumber\\
&&\quad\quad\quad\quad\quad\quad\quad\times\bigg[\alpha_1 \exp{\bigg(-i \int_{\tau_{0}}^{\tau} d \tau^{'} {\cal W}_{k}^{(2)}(\tau ^{'})}\bigg) + \beta_1 \exp{\bigg(i \int_{\tau_{0}}^{\tau} d \tau^{'} {\cal W}_{k}^{(2)}(\tau ^{'})}\bigg)\bigg], \\
\zeta_{\bf k, {\bf B}}^{(2)}(\tau) &=&-
2^{\nu-\frac{3}{2}} (-k c_s \tau )^{\frac{3}{2}-\nu}\bigg[1+\left(\frac{\tau}{\tau_0}\right)^2\bigg]^{-\frac{1}{2(\epsilon-1)}}\sqrt{\left(\frac{\epsilon_*}{\epsilon_b}\right)}\Bigg(\frac{c_s H \tau}{2 M_{pl} \sqrt{\epsilon_*} \sqrt{{\cal W}_{k}^{(2)}(\tau)}}\Bigg) \Bigg|\frac{\Gamma(\nu)}{\Gamma(\frac{3}{2})}\Bigg|\nonumber\\
&&\quad\quad\quad\quad\quad\quad\quad\times\bigg[\alpha_1 \exp{\bigg(-i \int_{\tau_{0}}^{\tau} d \tau^{'} {\cal W}_{k}^{(2)}(\tau ^{'})}\bigg) + \beta_1 \exp{\bigg(i \int_{\tau_{0}}^{\tau} d \tau^{'} {\cal W}_{k}^{(2)}(\tau ^{'})}\bigg)\bigg], \\
\zeta_{\bf k, \bf SRI}^{(2)}(\tau) &=& -
2^{\nu-\frac{3}{2}} (-k c_s \tau )^{\frac{3}{2}-\nu}\Bigg(\frac{c_s H \tau}{2 M_{pl} \sqrt{\epsilon_*} \sqrt{{\cal W}_{k}^{(2)}(\tau)}}\Bigg)\Bigg|\frac{\Gamma(\nu)}{\Gamma(\frac{3}{2})}\Bigg|\nonumber\\
&&\quad\quad\quad\quad\quad\quad\quad\times \bigg[\alpha_1 \exp{\bigg(-i \int_{\tau_{0}}^{\tau} d \tau^{'} {\cal W}_{k}^{(2)}(\tau ^{'})}\bigg) + \beta_1 \exp{\bigg(i \int_{\tau_{0}}^{\tau} d \tau^{'} {\cal W}_{k}^{(2)}(\tau ^{'})}\bigg)\bigg], \\
\zeta_{\bf k, \bf USR}^{(2)}(\tau) &=& -2^{\nu-\frac{3}{2}} (-k c_s \tau )^{\frac{3}{2}-\nu}\Bigg(\frac{c_s H \tau}{2 M_{pl} \sqrt{\epsilon_*} \sqrt{{\cal W}_{k}^{(2)}(\tau)}}\Bigg)\Bigg|\frac{\Gamma(\nu)}{\Gamma(\frac{3}{2})}\Bigg| \bigg(\frac{\tau_{0}}{\tau}\bigg)^{3} \nonumber\\
&&\quad\quad\quad\quad\quad\quad\quad\times \bigg[\alpha_2 \exp{\bigg(-i \int_{\tau_{0}}^{\tau}d \tau^{'} {\cal W}_{k}^{(2)}(\tau ^{'})}\bigg) + \beta_2 \exp{\bigg(i \int_{\tau_{0}}^{\tau}d \tau^{'} {\cal W}_{k}^{(2)}(\tau ^{'})}\bigg)\bigg], \\
\zeta_{\bf k, \bf SRII}^{(2)}(\tau) &=& -2^{\nu-\frac{3}{2}} (-k c_s \tau )^{\frac{3}{2}-\nu}\Bigg(\frac{c_s H \tau}{2 M_{pl} \sqrt{\epsilon_*} \sqrt{{\cal W}_{k}^{(2)}(\tau)}}\Bigg)\Bigg|\frac{\Gamma(\nu)}{\Gamma(\frac{3}{2})}\Bigg| \bigg(\frac{\tau_{0}}{\tau_e}\bigg)^{3} \nonumber\\
&&\quad\quad\quad\quad\quad\quad\quad\times \bigg[\alpha_3 \exp{\bigg(-i \int_{\tau_{0}}^{\tau}d \tau^{'} {\cal W}_{k}^{(2)}(\tau ^{'})}\bigg) + \beta_3 \exp{\bigg(i \int_{\tau_{0}}^{\tau}d \tau^{'} {\cal W}_{k}^{(2)}(\tau ^{'})}\bigg)\bigg]. 
\eea
Here the typical conformal time-dependent frequency function, denoted as ${\cal W}^{(2)}_{k}(\tau)$, is defined as follows for the adiabatic $2$-nd order:
\bea
{\cal W}_{k}^{(2)} (\tau) = \sqrt{c_s ^2k^2 - \frac{1}{\tau ^2}\left(\nu^2-\frac{1}{4}\right)} \approx c_s k.
\eea
In the previous stage, we took the restriction $-kc_s\tau\rightarrow \infty$ to account for the contribution from the short-range UV mode frequency in the formula above. In the case when adiabatic regularisation holds perfectly, this approximation is fairly justified. The equation for the counter terms for the contraction and bounce phases may be further computed in the presence of the previously given adiabatically regularised scalar mode function for the comoving curvature perturbation. specifically SRI, USR, and SRII. Prior to delving into the technical computation, let us at last record the equation for the scalar modes of comoving curvature perturbation in the five phases that were previously described. This expression will be utilized to calculate the counter terms:
\bea
\label{genWKBcv22}
\zeta_{\bf k, {\bf C}}^{(2)}(\tau) &\approx&-
2^{\nu-\frac{3}{2}} (-k c_s \tau )^{\frac{3}{2}-\nu}\left(\frac{\tau}{\tau_0}\right)^{-\frac{1}{(\epsilon-1)}}\sqrt{\left(\frac{\epsilon_*}{\epsilon_c}\right)}\Bigg(\frac{c_s H \tau}{2 M_{pl} \sqrt{\epsilon_*} \sqrt{c_s k}}\Bigg) \Bigg|\frac{\Gamma(\nu)}{\Gamma(\frac{3}{2})}\Bigg|\nonumber\\
&&\quad\quad\quad\quad\quad\quad\quad\times\bigg[\alpha_1 \exp{\bigg(-ic_sk(\tau - \tau_{0})}\bigg) + \beta_1 \exp{\bigg(ic_sk(\tau - \tau_{0})}\bigg)\bigg], \eea\bea
\zeta_{\bf k, {\bf B}}^{(2)}(\tau) &\approx&-
2^{\nu-\frac{3}{2}} (-k c_s \tau )^{\frac{3}{2}-\nu}\bigg[1+\left(\frac{\tau}{\tau_0}\right)^2\bigg]^{-\frac{1}{2(\epsilon-1)}}\sqrt{\left(\frac{\epsilon_*}{\epsilon_b}\right)}\Bigg(\frac{c_s H \tau}{2 M_{pl} \sqrt{\epsilon_*} \sqrt{c_s k}}\Bigg) \Bigg|\frac{\Gamma(\nu)}{\Gamma(\frac{3}{2})}\Bigg|\nonumber\\
&&\quad\quad\quad\quad\quad\quad\quad\times\bigg[\alpha_1 \exp{\bigg(-ic_sk(\tau - \tau_{0})}\bigg) + \beta_1 \exp{\bigg(ic_sk(\tau - \tau_{0})}\bigg)\bigg], \\
\zeta_{\bf k, \bf SRI}^{(2)}(\tau) &\approx& -
2^{\nu-\frac{3}{2}} (-k c_s \tau )^{\frac{3}{2}-\nu}\Bigg(\frac{c_s H \tau}{2 M_{pl} \sqrt{\epsilon_*} \sqrt{c_s k}}\Bigg)\Bigg|\frac{\Gamma(\nu)}{\Gamma(\frac{3}{2})}\Bigg|\nonumber\\
&&\quad\quad\quad\quad\quad\quad\quad\times \bigg[\alpha_1 \exp{\bigg(-ic_sk(\tau - \tau_{0})}\bigg) + \beta_1 \exp{\bigg(ic_sk(\tau - \tau_{0})}\bigg)\bigg], \\
\zeta_{\bf k, \bf USR}^{(2)}(\tau) &\approx& -2^{\nu-\frac{3}{2}} (-k c_s \tau )^{\frac{3}{2}-\nu}\Bigg(\frac{c_s H \tau}{2 M_{pl} \sqrt{\epsilon_*} \sqrt{c_s k}}\Bigg)\Bigg|\frac{\Gamma(\nu)}{\Gamma(\frac{3}{2})}\Bigg| \bigg(\frac{\tau_{0}}{\tau}\bigg)^{3} \nonumber\\
&&\quad\quad\quad\quad\quad\quad\quad\times \bigg[\alpha_2 \exp{\bigg(-ic_sk(\tau - \tau_{0})}\bigg) + \beta_2 \exp{\bigg(ic_sk(\tau - \tau_{0})}\bigg)\bigg], \\
\zeta_{\bf k, \bf SRII}^{(2)}(\tau) &\approx& -2^{\nu-\frac{3}{2}} (-k c_s \tau )^{\frac{3}{2}-\nu}\Bigg(\frac{c_s H \tau}{2 M_{pl} \sqrt{\epsilon_*} \sqrt{c_s k}}\Bigg)\Bigg|\frac{\Gamma(\nu)}{\Gamma(\frac{3}{2})}\Bigg| \bigg(\frac{\tau_{0}}{\tau_e}\bigg)^{3} \nonumber\\
&&\quad\quad\quad\quad\quad\quad\quad\times \bigg[\alpha_3 \exp{\bigg(-ic_sk(\tau - \tau_{0})}\bigg) + \beta_3 \exp{\bigg(ic_sk(\tau - \tau_{0})}\bigg)\bigg]. 
\eea
Currently, we are calculating the expression for the counter terms of these five periods, which is the primary task of this system. The counter terms that follow are introduced; they are dependent on the parameters that determine the adiabatic renormalization scheme, as previously seen. Here ${\cal K}_{\bf C}(\mu, \mu_0)$, ${\cal K}_{\bf B}(\mu, \mu_0)$, ${\cal K}_{\bf SRI}(\mu, \mu_0)$, ${\cal K}_{\bf USR}(\mu, \mu_0)$, and ${\cal K}_{\bf SRII}(\mu, \mu_0)$ for the corresponding five phases contraction, bounce, SRI, USR, and SRII.
\bea
\label{count1}
{\bf Z}_{\bf \zeta, \bf C}^{\bf UV} (\mu, \mu_0) &=& \bigg[\Delta^{2}_{\zeta,\textbf{Tree}}(k)\bigg]_{\textbf{SRI}}^{2} \times {\cal K}_{\bf C}(\mu, \mu_0)\times\left(\frac{\epsilon_*}{\epsilon_c}\right), \\
\label{count2}
{\bf Z}_{\bf \zeta, \bf B}^{\bf UV} (\mu, \mu_0) &=& \bigg[\Delta^{2}_{\zeta,\textbf{Tree}}(k)\bigg]_{\textbf{SRI}}^{2} \times {\cal K}_{\bf B}(\mu, \mu_0)\times\left(\frac{\epsilon_*}{\epsilon_b}\right), \\
\label{count3}
{\bf Z}_{\bf \zeta, \bf SRI}^{\bf UV} (\mu, \mu_0) &=& \bigg[\Delta^{2}_{\zeta,\textbf{Tree}}(k)\bigg]_{\textbf{SRI}}^{2} \times {\cal K}_{\bf SRI}(\mu, \mu_0), \\
\label{count4}
{\bf Z}_{\bf \zeta, \bf USR}^{\bf UV} (\mu, \mu_0) &=& \frac{1}{4}\bigg[\Delta^{2}_{\zeta,\textbf{Tree}}(k)\bigg]_{\textbf{SRi}}^{2} \times {\cal K}_{\bf USR}(\mu, \mu_0), \\
\label{count5}
{\bf Z}_{\bf \zeta, \bf SRII}^{\bf UV} (\mu, \mu_0) &=& \bigg[\Delta^{2}_{\zeta,\textbf{Tree}}(k)\bigg]_{\textbf{SRI}}^{2} \times {\cal K}_{\bf SRII}(\mu, \mu_0). 
\eea
It is also important to note that the dependent parameters of the adiabatic renormalization scheme and the counter terms are functions of a new mass scale $\mu$. This can be interpreted as the renormalization scale of the underlying quantum field theory, against which all of the computations are being done. In order to carry out the adiabatic regularisation procedure, $\mu_0$ also denotes the renormalization scale at the conformal time $\tau_0$. As long as it makes sense in the given situation, one may use these scales as convenient. Now let us examine the parameters that are reliant on the renormalization technique explicitly:
\bea
{\cal K}_{\bf C}(\mu, \mu_0) &=& \int_{\mu_0}^{\mu}\frac{dk}{k}\;\left(\frac{k}{\mu_0}\right)^{\delta_{\bf C}}\;(1+k^2c^2_s\tau^2) =\Bigg[\frac{1}{\delta_{\bf C}}\bigg\{\left(\frac{\mu}{\mu_0}\right)^{\delta_{\bf C}}-1\bigg\}+\frac{1}{\left(\delta_{\bf C}+2\right)}\bigg\{\left(\frac{\mu}{\mu_0}\right)^{\delta_{\bf C}+2}-1\bigg\}\Bigg], \\
{\cal K}_{\bf B}(\mu, \mu_0) &=& \int_{\mu_0}^{\mu}\frac{dk}{k}\;\left(\frac{k}{\mu_{0}}\right)^{2}\left(1+\left(\frac{\mu_{0}}{k}\right)^2\right)^{-\frac{1}{(\epsilon_b-1)}}\;\left(\frac{k}{\mu_{0}}\right)^{3-2\nu}\left(1+k^2c^2_s\tau^2\right)\nonumber\\
&=&\frac{1}{\left(\delta_{\bf B} +2\right)}\bigg[\, _2F_1\left(\frac{\delta_{\bf B}+2}{2},\frac{1}{\epsilon_b-1}-1;\frac{\delta_{\bf B}+4}{2};-1\right)\nonumber\\
 &&\quad\quad\quad\quad\quad\quad\quad-\left(\frac{\mu_0}{\mu}\right)^{\delta_{\bf B}+2}\, _2F_1\left(\frac{\delta_{\bf B}+2}{2},\frac{1}{\epsilon_b-1}-1;\frac{\delta_{\bf B}+4}{2};-\left(\frac{\mu_0}{\mu}\right)^{2}\right)\bigg], \eea\bea
 {\cal K}_{\bf SRI}(\mu, \mu_0) &=& \int_{\mu_0}^{\mu}\frac{dk}{k}\;\left(\frac{k}{\mu_{0}}\right)^{3-2\nu}k^2c^2_s\tau^2 =\frac{1}{2\left(\delta_{\bf SRI}+1\right)}\left\{\left(\frac{\mu}{\mu_0}\right)^{2\left(\delta_{\bf SRI}+1\right)}-1\right\}, \\
{\cal K}_{\bf USR}(\mu, \mu_0) &=& \bigg[\left(\frac{\Delta\eta(\tau)}{\tilde{c}^{4}_{s}}\right)^{2}\bigg(\frac{\mu}{\mu_0}\bigg)^6  - \left(\frac{\Delta\eta(\tau_{0})}{\tilde{c}^{4}_{s}}\right)^{2}\bigg] \int_{\mu_0}^{\mu}\frac{dk}{k}\;\left(\frac{k}{\mu_{0}}\right)^{3-2\nu}k^2c^2_s\tau^2 \nonumber \\
&=& \frac{1}{2\left(\delta_{\bf USR}+1\right)}\bigg[\left(\frac{\Delta\eta(\tau)}{\tilde{c}^{4}_{s}}\right)^{2}\bigg(\frac{\mu}{\mu_0}\bigg)^6  - \left(\frac{\Delta\eta(\tau_{0})}{\tilde{c}^{4}_{s}}\right)^{2} \bigg]\left\{\left(\frac{\mu}{\mu_0}\right)^{2\left(\delta_{\bf USR}+1\right)}-1\right\},\\
{\cal K}_{\bf SRII}(\mu, \mu_0) &=& \int_{\mu_0}^{\mu}\frac{dk}{k}\;\left(\frac{k}{\mu_{0}}\right)^{3-2\nu}k^2c^2_s\tau^2 =\frac{1}{2\left(\delta_{\bf SRII}+1\right)}\left\{\left(\frac{\mu}{\mu_0}\right)^{2\left(\delta_{\bf SRII}+1\right)}-1\right\}.
\eea
Upon replacing the counter terms at any renormalization scale $\mu$ with the previously calculated formulas, the corresponding counter terms of the one-loop primordial power spectrum for the scalar modes may be further represented as follows:
\bea
\label{count1x}
{\bf Z}_{\bf \zeta, \bf C}^{\bf UV} (\mu, \mu_0) &=& \bigg[\Delta^{2}_{\zeta,\textbf{Tree}}(k)\bigg]_{\textbf{SRI}}^{2} \times\left(\frac{\epsilon_*}{\epsilon_c}\right)\nonumber\\
&&\quad\quad\quad\quad\quad\quad\times\Bigg[\frac{1}{\delta_{\bf C}}\bigg\{\left(\frac{\mu}{\mu_0}\right)^{\delta_{\bf C}}-1\bigg\}+\frac{1}{\left(\delta_{\bf C}+2\right)}\bigg\{\left(\frac{\mu}{\mu_0}\right)^{\delta_{\bf C}+2}-1\bigg\}\Bigg], \\
\label{count2x}
{\bf Z}_{\bf \zeta, \bf B}^{\bf UV} (\mu, \mu_0) &=&\frac{1}{\left(\delta_{\bf B} +2\right)}\times \bigg[\Delta^{2}_{\zeta,\textbf{Tree}}(k)\bigg]_{\textbf{SRI}}^{2} \times\left(\frac{\epsilon_*}{\epsilon_b}\right)\nonumber\\
&&\quad\quad\quad\quad\quad\quad\times\bigg[\, _2F_1\left(\frac{\delta_{\bf B}+2}{2},\frac{1}{\epsilon_b-1}-1;\frac{\delta_{\bf B}+4}{2};-1\right)\nonumber\\
 &&\quad\quad\quad\quad\quad\quad\quad-\left(\frac{\mu_0}{\mu}\right)^{\delta_{\bf B}+2}\, _2F_1\left(\frac{\delta_{\bf B}+2}{2},\frac{1}{\epsilon_b-1}-1;\frac{\delta_{\bf B}+4}{2};-\left(\frac{\mu_0}{\mu}\right)^{2}\right)\bigg], \\
\label{count3x}
{\bf Z}_{\bf \zeta, \bf SRI}^{\bf UV} (\mu, \mu_0) &=&\frac{1}{2\left(\delta_{\bf SRI}+1\right)}\times \bigg[\Delta^{2}_{\zeta,\textbf{Tree}}(k)\bigg]_{\textbf{SRI}}^{2} \times \left\{\left(\frac{\mu}{\mu_0}\right)^{2\left(\delta_{\bf SRI}+1\right)}-1\right\}, \\
\label{count4x}
{\bf Z}_{\bf \zeta, \bf USR}^{\bf UV} (\mu, \mu_0) &=& \frac{1}{8\left(\delta_{\bf USR}+1\right)}\times\bigg[\Delta^{2}_{\zeta,\textbf{Tree}}(k)\bigg]_{\textbf{SRI}}^{2} \nonumber\\
&&\quad\quad\quad\quad\quad\times \bigg[\left(\frac{\Delta\eta(\tau)}{\tilde{c}^{4}_{s}}\right)^{2}\bigg(\frac{\mu}{\mu_0}\bigg)^6  - \left(\frac{\Delta\eta(\tau_{0})}{\tilde{c}^{4}_{s}}\right)^{2} \bigg]\times\left\{\left(\frac{\mu}{\mu_0}\right)^{2\left(\delta_{\bf USR}+1\right)}-1\right\}, \\
\label{count5x}
{\bf Z}_{\bf \zeta, \bf SRII}^{\bf UV} (\mu, \mu_0) &=&\frac{1}{2\left(\delta_{\bf SRII}+1\right)}\times \bigg[\Delta^{2}_{\zeta,\textbf{Tree}}(k)\bigg]_{\textbf{SRI}}^{2} \times \left\{\left(\frac{\mu}{\mu_0}\right)^{2\left(\delta_{\bf SRII}+1\right)}-1\right\}. 
\eea
Consequently, the following may be used to express the distinct contributions from the contraction, bounce, SRI, USR, and SRII phases of the one-loop corrected adiabatically renormalized scalar power spectrum:
\bea
\label{p1}
\bigg[\Delta^{2}_{\zeta,\textbf{One-loop}}(k,\mu,\mu_0)\bigg]_{\textbf{C}} &=& \bigg[\Delta^{2}_{\zeta,\textbf{Tree}}(k)\bigg]_{\textbf{SRI}}^{2}
\times \Bigg(1+\frac{2}{15\pi^2}\frac{1}{c^2_{s}k^2_c}\bigg(-\left(1-\frac{1}{c^2_{s}}\right)\epsilon_c+6\frac{\bar{M}^3_1}{ HM^2_{ pl}}-\frac{4}{3}\frac{M^4_3}{H^2M^2_{ pl}}\bigg)\Bigg)\times \left(\frac{\epsilon_*}{\epsilon_c}\right)\nonumber \\
&& \quad \quad \times\bigg(\Bigg[\frac{1}{\delta_{\bf C}}\bigg\{\left(\frac{\mu}{\mu_0}\right)^{\delta_{\bf C}}-1\bigg\}+\frac{1}{\left(\delta_{\bf C}+2\right)}\bigg\{\left(\frac{\mu}{\mu_0}\right)^{\delta_{\bf C}+2}-1\bigg\}\Bigg] \nonumber\\
&&- \frac{4}{3}\bigg[\frac{1}{\delta_{\bf C}}\bigg\{\left(\frac{k_{b}}{k_*}\right)^{\delta_{\bf C}}-\left(\frac{k_{c}}{k_*}\right)^{\delta_{\bf C}}\bigg\}+\frac{1}{\left(\delta_{\bf C}+2\right)}\bigg\{\left(\frac{k_{b}}{k_*}\right)^{\delta_{\bf C}+2}-\left(\frac{k_{c}}{k_*}\right)^{\delta_{\bf C}+2}\bigg\}\bigg]\bigg), \\
\label{p2}
\bigg[\Delta^{2}_{\zeta,\textbf{One-loop}}(k,\mu,\mu_0)\bigg]_{\textbf{B}} &=& \bigg[\Delta^{2}_{\zeta,\textbf{Tree}}(k)\bigg]_{\textbf{SRI}}^{2}
\times \Bigg(1+\frac{2}{15\pi^2}\frac{1}{c^2_{s}k^2_b}\bigg(-\left(1-\frac{1}{c^2_{s}}\right)\epsilon_b+6\frac{\bar{M}^3_1}{ HM^2_{ pl}}-\frac{4}{3}\frac{M^4_3}{H^2M^2_{ pl}}\bigg)\Bigg)\times \left(\frac{\epsilon_*}{\epsilon_b}\right)\nonumber \\
&& \quad \quad \times\frac{1}{(\delta_{\bf B} +2)}\times\bigg(\bigg[\, _2F_1\left(\frac{\delta_{\bf B}+2}{2},\frac{1}{\epsilon_b-1}-1;\frac{\delta_{\bf B}+4}{2};-1\right)\nonumber\\
 &&\quad\quad\quad\quad\quad\quad\quad-\left(\frac{\mu_0}{\mu}\right)^{\delta_{\bf B}+2}\, _2F_1\left(\frac{\delta_{\bf B}+2}{2},\frac{1}{\epsilon_b-1}-1;\frac{\delta_{\bf B}+4}{2};-\left(\frac{\mu_0}{\mu}\right)^{2}\right)\bigg] \nonumber\eea\bea
&&- \frac{4}{3}\bigg[\, _2F_1\left(\frac{\delta_{\bf B}+2}{2},\frac{1}{\epsilon_b-1}-1;\frac{\delta_{\bf B}+4}{2};-1\right)\nonumber\\
 &&\quad\quad\quad\quad\quad\quad\quad\quad\quad\quad\quad\quad\quad\quad\nonumber\\
 &&\quad\quad\quad\quad\quad\quad\quad-\left(\frac{k_{b}}{k_{*}}\right)^{\delta_{\bf B}+2}\, _2F_1\left(\frac{\delta_{\bf B}+2}{2},\frac{1}{\epsilon_b-1}-1;\frac{\delta_{\bf B}+4}{2};-\left(\frac{k_{b}}{k_{*}}\right)^{2}\right)\bigg] \bigg), \quad\quad\\
\label{p3}
\bigg[\Delta^{2}_{\zeta,\textbf{One-loop}}(k,\mu,\mu_0)\bigg]_{\textbf{SRI}} &=& \bigg[\Delta^{2}_{\zeta,\textbf{Tree}}(k)\bigg]_{\textbf{SRI}}^{2}
\times \Bigg(1+\frac{2}{15\pi^2}\frac{1}{c^2_{s}k^2_*}\bigg(-\left(1-\frac{1}{c^2_{s}}\right)\epsilon_*+6\frac{\bar{M}^3_1}{ HM^2_{ pl}}-\frac{4}{3}\frac{M^4_3}{H^2M^2_{ pl}}\bigg)\Bigg)\nonumber \\
&& \quad \quad \times\bigg(\frac{1}{2\left(\delta_{\bf SRI}+1\right)}\bigg\{ \bigg(\frac{\mu}{\mu_0}\bigg)^{2\left(\delta_{\bf SRI}+1\right)} - \bigg(\frac{k_{s}}{k_{*}}\bigg)^{2\left(\delta_{\bf SRI}+1\right)} \bigg\} - \frac{4}{3} \ln\bigg({\frac{k_{s}}{k_{*}}}\bigg) \bigg), \\
\label{p4}
\bigg[\Delta^{2}_{\zeta,\textbf{One-loop}}(k,\mu,\mu_0)\bigg]_{\textbf{USR}} &=& \frac{1}{4}\bigg[\Delta^{2}_{\zeta,\textbf{Tree}}(k)\bigg]_{\textbf{SRI}}^{2} \nonumber \\
&& \quad \times \bigg\{\bigg[\bigg(\frac{\Delta\eta(\tau_{e})}{\tilde{c}^{4}_{s}}\bigg)^{2}\bigg(\frac{k_{e}}{k_{s}}\bigg)^6  - \left(\frac{\Delta\eta(\tau_{s})}{\tilde{c}^{4}_{s}}\right)^{2}\bigg]\nonumber \\
&& \quad \quad\quad\quad\times 
\bigg[\ln\bigg({\frac{k_{e}}{k_{s}}}\bigg) + \frac{1}{2\left(\delta_{\bf USR}+1\right)}\bigg\{ \bigg(\frac{k_{e}}{k_{s}}\bigg)^{2\left(\delta_{\bf USR}+1\right)} - 1 \bigg\}\bigg] \nonumber \\
&& \quad \quad \quad  - \frac{1}{2\left(\delta_{\bf USR}+1\right)}\bigg[\bigg(\frac{\Delta\eta(\tau_{e})}{\tilde{c}^{4}_{s}}\bigg)^{2} \bigg(\frac{\mu}{\mu_0}\bigg)^6 - \left(\frac{\Delta\eta(\tau_{s})}{\tilde{c}^{4}_{s}}\right)^{2}\bigg]\nonumber\\
&&\quad \quad\quad\quad \times\bigg[\bigg(\frac{\mu}{\mu_0}\bigg)^{2\left(\delta_{\bf USR}+1\right)} -1\bigg] \bigg\}, \\
\label{p5}
\bigg[\Delta^{2}_{\zeta,\textbf{One-loop}}(k,\mu,\mu_0)\bigg]_{\textbf{SRII}} &=& \bigg[\Delta^{2}_{\zeta,\textbf{Tree}}(k)\bigg]_{\textbf{SRI}}^{2} \times \Bigg(1+\frac{2}{15\pi^2}\frac{1}{c^2_{s}k^2_*}\bigg(-\left(1-\frac{1}{c^2_{s}}\right)\epsilon_*+6\frac{\bar{M}^3_1}{ HM^2_{ pl}}-\frac{4}{3}\frac{M^4_3}{H^2M^2_{ pl}}\bigg)\Bigg) \nonumber \\
&& \quad \quad\quad\quad\quad \quad\quad\quad\times  \bigg(\frac{1}{2\left(\delta_{\bf SRII}+1\right)}\bigg[\bigg(\frac{\mu}{\mu_0}\bigg)^{2\left(\delta_{\bf SRII}+1\right)} - 1 \bigg]+ \ln{\bigg(\frac{k_{end}}{k_{e}}}\bigg)\bigg).
\eea
Given a fixed renormalization scale denoted by $\mu$ and a matching reference scale $\mu_0$, the UV divergences will be eliminated, as can be shown from the preceding formulas for the power spectrum of the comoving curvature perturbation. Still unknown, though, is what will happen to the IR divergences. It leads, then, to a quantum field theory of curved space-time that is UV-protected but IR-sensitive. It is evident from this kind of IR nature that perturbative approximations are legitimate and should always be followed. While an arbitrary renormalization scale may always be assumed, if the scale is fixed in the vicinity of the UV cut-off, the perturbativity is preserved over time. The representative catalogue of the UV and IR cut-off scales, renormalization and references scales for the contraction, bouncing, SRI, USR and SRII phases are appended below:
\begin{itemize}
 \item[\ding{43}] \underline{\textbf{Scales for contracting phase:}}
\bea \Lambda_{\rm UV} = k_{b}, \quad  \Lambda_{\rm IR} = k_{c},\quad \mu=\mu_0,\quad\eea
\item[\ding{43}] \underline{\textbf{Scales for bouncing phase:}}
\bea \Lambda_{\rm UV} = k_{*}, \quad  \Lambda_{\rm IR} = k_{b},\quad \mu=\mu_0,\quad\eea

\item[\ding{43}] \underline{\textbf{Scales for SRI phase:}}
\bea \Lambda_{\rm UV} = k_{s}=\mu, \quad  \Lambda_{\rm IR} = k_{*}=\mu_0,\quad\eea

\item[\ding{43}] \underline{\textbf{Scales for USR phase:}}
\bea \Lambda_{\rm UV} = k_{e}=\mu, \quad  \Lambda_{\rm IR} = k_{s}=\mu_0,\quad\eea

\item[\ding{43}] \underline{\textbf{Scales for SRII phase:}}
\bea \Lambda_{\rm UV} = k_{end}=\mu=\mu_0, \quad  \Lambda_{\rm IR} = k_{e},\quad\eea

\end{itemize}

   Here connecting our findings with the standard Quantum Field Theory approach we found that:
\bea &&\underline{\bf Contraction:}\quad\quad\left(\delta_{{\cal Z}_{{\bf G}_1}}+\delta_{{\cal Z}_{{\bf G}_2}}+\delta_{{\cal Z}_{{\bf G}_3}}+\delta_{{\cal Z}_{{\bf G}_4}}+\delta_{{\cal Z}_{{\bf G}_5}}\right)={\bf Z}_{\bf \zeta, \bf C}^{\bf UV}\quad{\rm with}\quad\delta_{{\cal Z}_{{\bf D}_6}}=0,\\
&&\underline{\bf Bouncing:}\quad\quad\left(\delta_{{\cal Z}_{{\bf G}_1}}+\delta_{{\cal Z}_{{\bf G}_2}}+\delta_{{\cal Z}_{{\bf G}_3}}+\delta_{{\cal Z}_{{\bf G}_4}}+\delta_{{\cal Z}_{{\bf G}_5}}\right)={\bf Z}_{\bf \zeta, \bf B}^{\bf UV}\quad{\rm with}\quad\delta_{{\cal Z}_{{\bf D}_6}}=0,\\
&&\underline{\bf SRI:}\quad\quad\left(\delta_{{\cal Z}_{{\bf G}_1}}+\delta_{{\cal Z}_{{\bf G}_2}}+\delta_{{\cal Z}_{{\bf G}_3}}+\delta_{{\cal Z}_{{\bf G}_4}}+\delta_{{\cal Z}_{{\bf G}_5}}\right)={\bf Z}_{\bf \zeta, \bf SRI}^{\bf UV}\quad{\rm with}\quad\delta_{{\cal Z}_{{\bf G}_6}}=0,\quad\quad\\
&&\underline{\bf USR:}\quad\quad\delta_{{\cal Z}_{{\bf G}_6}}={\bf Z}_{\bf \zeta, \bf USR}^{\bf UV}\quad{\rm with}\quad\left(\delta_{{\cal Z}_{{\bf G}_1}}+\delta_{{\cal Z}_{{\bf G}_2}}+\delta_{{\cal Z}_{{\bf G}_3}}+\delta_{{\cal Z}_{{\bf G}_4}}+\delta_{{\cal Z}_{{\bf G}_5}}\right)=0,\quad\quad\\
&&\underline{\bf SRII:}\quad\quad\left(\delta_{{\cal Z}_{{\bf G}_1}}+\delta_{{\cal Z}_{{\bf G}_2}}+\delta_{{\cal Z}_{{\bf G}_3}}+\delta_{{\cal Z}_{{\bf G}_4}}+\delta_{{\cal Z}_{{\bf G}_5}}\right)=={\bf Z}_{\bf \zeta, \bf SRII}^{\bf UV}\quad{\rm with}\quad\delta_{{\cal Z}_{{\bf G}_6}}=0.
\eea
Upon fixing the renormalization scale and the reference scale at the previously mentioned values, we get:
\bea
\label{count1xa}
{\bf Z}_{\bf \zeta, \bf C}^{\bf UV}&=&0, \\
\label{count2xa}
{\bf Z}_{\bf \zeta, \bf B}^{\bf UV} &=&0, \\
\label{count3x}
{\bf Z}_{\bf \zeta, \bf SRI}^{\bf UV}&=&\frac{1}{2\left(\delta_{\bf SRI}+1\right)}\times \bigg[\Delta^{2}_{\zeta,\textbf{Tree}}(k)\bigg]_{\textbf{SRI}}^{2} \times \left\{\left(\frac{k_s}{k_*}\right)^{2\left(\delta_{\bf SRI}+1\right)}-1\right\}, \\
\label{count4x}
{\bf Z}_{\bf \zeta, \bf USR}^{\bf UV}&=& \frac{1}{8\left(\delta_{\bf USR}+1\right)}\times\bigg[\Delta^{2}_{\zeta,\textbf{Tree}}(k)\bigg]_{\textbf{SRI}}^{2} \nonumber\\
&&\quad\quad\quad\quad\quad\times \bigg[\left(\frac{\Delta\eta(\tau)}{\tilde{c}^{4}_{s}}\right)^{2}\bigg(\frac{\mu}{\mu_0}\bigg)^6  - \left(\frac{\Delta\eta(\tau_{0})}{\tilde{c}^{4}_{s}}\right)^{2} \bigg]\times\left\{\left(\frac{k_e}{k_s}\right)^{2\left(\delta_{\bf USR}+1\right)}-1\right\}, \\
\label{count5x}
{\bf Z}_{\bf \zeta, \bf SRII}^{\bf UV}&=&0. 
\eea
Hence the one-loop corrections to the power spectrum for the scalar modes derived for the five consecutive phases are fixed as: 

\bea
\label{p1w}
\bigg[\Delta^{2}_{\zeta,\textbf{One-loop}}(k)\bigg]_{\textbf{C}} &=&- \frac{4}{3} \bigg[\Delta^{2}_{\zeta,\textbf{Tree}}(k)\bigg]_{\textbf{SRI}}^{2}
\times \Bigg(1+\frac{2}{15\pi^2}\frac{1}{c^2_{s}k^2_c}\bigg(-\left(1-\frac{1}{c^2_{s}}\right)\epsilon_c+6\frac{\bar{M}^3_1}{ HM^2_{ pl}}-\frac{4}{3}\frac{M^4_3}{H^2M^2_{ pl}}\bigg)\Bigg)\times \left(\frac{\epsilon_*}{\epsilon_c}\right)\nonumber \\
&& \quad \quad \times\bigg[\frac{1}{\delta_{\bf C}}\bigg\{\left(\frac{k_{b}}{k_*}\right)^{\delta_{\bf C}}-\left(\frac{k_{c}}{k_*}\right)^{\delta_{\bf C}}\bigg\}+\frac{1}{\left(\delta_{\bf C}+2\right)}\bigg\{\left(\frac{k_{b}}{k_*}\right)^{\delta_{\bf C}+2}-\left(\frac{k_{c}}{k_*}\right)^{\delta_{\bf C}+2}\bigg\}\bigg],\nonumber\\
&=&\bigg[\Delta^{2}_{\zeta,{\bf Tree}}(k)\bigg]_{\bf SRI}\times \overline{\bf W}_{\bf C}, \\
\label{p2w}
\bigg[\Delta^{2}_{\zeta,\textbf{One-loop}}(k)\bigg]_{\textbf{B}} &=&- \frac{4}{3} \bigg[\Delta^{2}_{\zeta,\textbf{Tree}}(k)\bigg]_{\textbf{SRI}}^{2}
\times \Bigg(1+\frac{2}{15\pi^2}\frac{1}{c^2_{s}k^2_b}\bigg(-\left(1-\frac{1}{c^2_{s}}\right)\epsilon_b+6\frac{\bar{M}^3_1}{ HM^2_{ pl}}-\frac{4}{3}\frac{M^4_3}{H^2M^2_{ pl}}\bigg)\Bigg)\times \left(\frac{\epsilon_*}{\epsilon_b}\right)\nonumber \\
&& \quad \quad \times\frac{1}{(\delta_{\bf B} +2)}\times\bigg[\, _2F_1\left(\frac{\delta_{\bf B}+2}{2},\frac{1}{\epsilon_b-1}-1;\frac{\delta_{\bf B}+4}{2};-1\right)\nonumber\\
 &&\quad\quad\quad\quad\quad\quad\quad\quad\quad\quad\quad\quad\quad\quad\nonumber\\
 &&\quad\quad\quad\quad\quad\quad\quad-\left(\frac{k_{b}}{k_{*}}\right)^{\delta_{\bf B}+2}\, _2F_1\left(\frac{\delta_{\bf B}+2}{2},\frac{1}{\epsilon_b-1}-1;\frac{\delta_{\bf B}+4}{2};-\left(\frac{k_{b}}{k_{*}}\right)^{2}\right)\bigg],\nonumber\\
&=&\bigg[\Delta^{2}_{\zeta,{\bf Tree}}(k)\bigg]_{\bf SRI}\times \overline{\bf W}_{\bf B}, \\
\label{p3w}
\bigg[\Delta^{2}_{\zeta,\textbf{One-loop}}(k)\bigg]_{\textbf{SRI}} &=&-\frac{4}{3} \bigg[\Delta^{2}_{\zeta,\textbf{Tree}}(k)\bigg]_{\textbf{SRI}}^{2}\nonumber \\
&& \quad\times  \Bigg(1+\frac{2}{15\pi^2}\frac{1}{c^2_{s}k^2_*}\bigg(-\left(1-\frac{1}{c^2_{s}}\right)\epsilon_*+6\frac{\bar{M}^3_1}{ HM^2_{ pl}}-\frac{4}{3}\frac{M^4_3}{H^2M^2_{ pl}}\bigg)\Bigg)\times\ln\bigg({\frac{k_{s}}{k_{*}}}\bigg),\nonumber\\
&=&\bigg[\Delta^{2}_{\zeta,{\bf Tree}}(k)\bigg]_{\bf SRI}\times \overline{\bf W}_{\bf SRI}, \eea\bea
\label{p4w}
\bigg[\Delta^{2}_{\zeta,\textbf{One-loop}}(k)\bigg]_{\textbf{USR}} &=& \frac{1}{4}\bigg[\Delta^{2}_{\zeta,\textbf{Tree}}(k)\bigg]_{\textbf{SRI}}^{2} \nonumber \\
&& \quad \times\bigg[\bigg(\frac{\Delta\eta(\tau_{e})}{\tilde{c}^{4}_{s}}\bigg)^{2}\bigg(\frac{k_{e}}{k_{s}}\bigg)^6  - \left(\frac{\Delta\eta(\tau_{s})}{\tilde{c}^{4}_{s}}\right)^{2}\bigg]\times 
\ln\bigg({\frac{k_{e}}{k_{s}}}\bigg),\nonumber\\
&=&\bigg[\Delta^{2}_{\zeta,{\bf Tree}}(k)\bigg]_{\bf SRI}\times \overline{\bf W}_{\bf USR}, \\
\label{p5w}
\bigg[\Delta^{2}_{\zeta,\textbf{One-loop}}(k)\bigg]_{\textbf{SRII}} &=& \bigg[\Delta^{2}_{\zeta,\textbf{Tree}}(k)\bigg]_{\textbf{SRI}}^{2} \nonumber \\
&& \quad\times \Bigg(1+\frac{2}{15\pi^2}\frac{1}{c^2_{s}k^2_*}\bigg(-\left(1-\frac{1}{c^2_{s}}\right)\epsilon_*+6\frac{\bar{M}^3_1}{ HM^2_{ pl}}-\frac{4}{3}\frac{M^4_3}{H^2M^2_{ pl}}\bigg)\Bigg) \times  \bigg(\ln{\bigg(\frac{k_{end}}{k_{e}}}\bigg)\bigg),\nonumber\\
&=&\bigg[\Delta^{2}_{\zeta,{\bf Tree}}(k)\bigg]_{\bf SRI}\times \overline{\bf W}_{\bf SRII}.
\eea
Hence, the total regularized and renormalized one-loop corrected power spectrum for scalar modes can be expressed by the following expression:
   \bea \label{one-loopRR} \overline{\Delta^{2}_{\zeta, {\bf EFT}}(k)}&=&\Delta^{2}_{\zeta, {\bf RR}}(p)\nonumber\\  
   &=&\bigg[\Delta^{2}_{\zeta,{\bf Tree}}(p)\bigg]_{\bf SRI}\times\bigg(1+\underbrace{\overline{{\bf W}}_{\bf C}+\overline{{\bf W}}_{\bf B}+\overline{{\bf W}}_{\bf SRI}+\overline{{\bf W}}_{\bf USR}+\overline{{\bf W}}_{\bf SRII}}_{\textbf{Regularized and Renormalized one-loop correction}}\bigg),\nonumber\\
   &=&\begin{tikzpicture}[baseline={([yshift=-3.5ex]current bounding box.center)},very thick]
  
    % Loop
  \def\radius{1}
  \scalebox{1}{\draw[green,very thick] (0,\radius) circle (\radius);
  \draw[green,very thick] (4.5*\radius,0) circle (\radius);}

  % External lines
  %\filldraw;
  \draw[black, very thick] (-4*\radius,0) -- 
  (-2.5*\radius,0);
  \node at (-2*\radius,0) {+};
  \draw[black, very thick] (-1.5*\radius,0) -- (0,0);
  \draw[blue,fill=blue] (0,0) circle (.5ex);
  \draw[black, very thick] (0,0)  -- (1.5*\radius,0);
  \node at (2*\radius,0) {+};
  \draw[black, very thick] (2.5*\radius,0) -- (3.5*\radius,0); 
  \draw[blue,fill=blue] (3.5*\radius,0) circle (.5ex);
  \draw[blue,fill=blue] (5.5*\radius,0) circle (.5ex);
  \draw[black, very thick] (5.5*\radius,0) -- (6.5*\radius,0);
\end{tikzpicture},\eea
   where $\overline{{\bf W}}_{\bf C}$, $\overline{{\bf W}}_{\bf B}$, $\overline{{\bf W}}_{\bf SRI}$, $\overline{{\bf W}}_{\bf USR}$, and $\overline{{\bf W}}_{\bf SRII}$ are already computed in equation(\ref{t1}), equation(\ref{t2}), equation(\ref{t3}), equation(\ref{t4}) and equation(\ref{t5}) respectively.

We also found from our analysis that the renormalized two-point amplitude of the power spectrum for the scalar modes obtained from the late time and adiabatic renormalization schemes turns out be exactly same and our obtained result is scheme independent. This further implies in terms of diagramatics that:
\bea
&&\quad\quad\quad\quad\quad\quad\bigg[\overline{\Delta^{2}_{\zeta, {\bf EFT}}(k)}\bigg]_{\textbf{Late-time renormalization}}=\bigg[\overline{\Delta^{2}_{\zeta, {\bf EFT}}(k)}\bigg]_{\textbf{Adiabatic renormalization}}\nonumber\\
&&
\begin{tikzpicture}[baseline={([yshift=-3.5ex]current bounding box.center)},very thick]
  
    % Loop
  \def\radius{1}
  \scalebox{1}{\draw[red,very thick] (-0.6*\radius,0.75*\radius) circle (0.75*\radius);
  \draw[red,very thick] (2.45*\radius,0) circle (0.75*\radius);\draw[green,very thick](8.4*\radius,0.75*\radius) circle(0.75*\radius);\draw[green,very thick](11.55*\radius,0) circle(0.75*\radius);}

  % External lines
  %\filldraw;
   \draw[black, very thick] (-3*\radius,0) -- (-2.2*\radius,0);
   \node at (-1.8*\radius,0) {+};
  \draw[black, very thick] (-1.4*\radius,0) -- (-0.6*\radius,0);
  \draw[blue,fill=blue] (-0.6*\radius,0) circle (.5ex);
  \draw[black, very thick] (-0.6*\radius,0)  -- (0.2*\radius,0);
  \node at (0.6*\radius,0) {+};
  \draw[black, very thick] (0.9*\radius,0) -- (1.7*\radius,0); 
  \draw[blue,fill=blue] (1.7*\radius,0) circle (.5ex);
  \draw[blue,fill=blue] (3.2*\radius,0) circle (.5ex);
  \draw[black, very thick] (3.2*\radius,0) -- (4*\radius,0);
  \node at (5*\radius,0) {=};
  %RHS
   \draw[black, very thick] (6*\radius,0) -- (6.8*\radius,0);
   \node at (7.2*\radius,0) {+};
   \draw[black, very thick] (7.6*\radius,0) -- (8.4*\radius,0);
  \draw[blue,fill=blue] (8.4*\radius,0) circle (.5ex);
  \draw[black, very thick] (8.4*\radius,0)  -- (9.2*\radius,0);
  \node at (9.6*\radius,0) {+};
  \draw[black, very thick] (10*\radius,0) -- (10.8*\radius,0); 
  \draw[blue,fill=blue] (10.8*\radius,0) circle (.5ex);
  \draw[blue,fill=blue] (12.3*\radius,0) circle (.5ex);
  \draw[black, very thick] (12.3*\radius,0) -- (13.1*\radius,0);
   
\end{tikzpicture}  %\quad = \quad \text{One-loop contributions}
\eea
This identification not only reveals that the quadratic UV divergence and other UV-like power-law divergences can be eliminated entirely from the expression for ${\bf Z}^{\bf UV}$, but it also helps us comprehend the underlying relationship between the counter-terms appearing in the bare action (UV sensitive part) and current contexts. This identification will be useful in building a bridge between the current adiabatic/wave function scheme and the conventional renormalization method that is accessible inside the framework of Quantum Field Theory of FLRW space-time, now that the link has been established. Primarily, the adiabatic/wave function approach aids in precisely fixing the mathematical structure of the number ${\bf Z}^{\bf UV}$, which is used to compute the total power spectrum of the scalar modes following renormalizations. The difficulty lies only in the fact that the structure of ${\bf Z}^{\bf UV}$ may be used to compute the IR-counter term ${\bf Z}^{\bf IR}$, utilizing the constraint that appears at the CMB pivot scale beforehand. Furthermore, our computations reveal that the unique structure of the term ${\bf Z}^{\bf UV}$ arises from late-time and adiabatic renormalization techniques, after the elimination of the quadratic divergence and other power-law-like contributions. As a result, the ultimate outcome is found to be independent of schemes, and we are left with the renormalized one-loop spectrum, in which the logarithmic IR divergent contributions represent the one-loop effect. Given that the structure of the term ${\bf Z}^{\bf UV}$ has been uniquely established, the IR-counter term ${\bf Z}^{\bf IR}$ may be further determined using the constraint that was previously calculated at the CMB pivot scale. We will be able to correct the exact structure of the IR counter-term by using the calculation carried out in the next section, which involves applying power spectrum renormalization. The findings obtained in the next section are essentially the continuation of the unknown outcomes we calculated using common renormalization techniques as they appear in the settings of Quantum Field Theory.

\subsection{Step IV: Power spectrum renormalization and softening of IR divergence}

From this point on, we use the power spectrum renormalization approach, which suppresses and improves controllability of logarithmic infrared divergences, so saving us from them. In order to get the renormalized form of the scalar power spectrum, this approach employs a counter-term that is found by invoking a renormalization condition at the pivot scale $k_{*}$. In addition to having abrupt transitions that arise at the USR limits and are connected to the SRI and SRII phases, the operation is carried out in the presence of contraction and bounce phases. The last phase is to use the resummation technique to produce a physical output that is finite. Within the current discourse, the renormalized 1PI correlation function for any given $m$-point amplitude, calculated using the EFT of Single Field framework, can characterize phases such as contraction, bounce, SRI, USR, and SRII. Additionally, it can be simplified into the following expression:
\bea \overline{\Gamma[\zeta]}=\sum^{\infty}_{m=2}\frac{i}{m!}\int\prod^{m}_{j=1}d^4x_j\,\overline{\Gamma^{(m)}_{\zeta}(x_i)}\,\zeta(x_j),\eea
where it is significant to note that $\overline{\Gamma^{(m)}_{\zeta,{\bf EFT}}(x_j)}$ may be further written as follows in the Fourier space for such a $m$-point renormalized amplitude:
\bea \overline{\Gamma^{(m)}_{\zeta}(x_j)}:=\int \frac{d^4k_j}{(2\pi)^4}\, e^{ik_j. x_j}\,\overline{\Gamma^{(m)}_{\zeta}(k_j,\mu,\mu_0)}\times (2\pi)^4\delta^4\left(\sum^{m}_{j=1}k_j\right)\quad\quad\forall\quad j=1,2,\cdots,m.\eea
where the renormalization scale and reference scale are denoted, respectively, by $\mu$ and $\mu_0$.
Furthermore, any generic $m$-point renormalized amplitude in the Fourier space may be described as follows in terms of the 1PI effective action:
\bea \overline{\Gamma^{(m)}_{\zeta}(k_1,k_2,k_3,\cdots,k_m,\mu,\mu_0)}=\left({\bf Z}^{\bf IR}\right)^{\frac{m}{2}}\Gamma^{(m)}_{\zeta}(k_1,k_2,k_3,\cdots,k_m)\eea
Now that we have a counter-term for each of the contributions from the five phases—contraction, bounce, SRI, USR, and SRII—as well as a sum-total version of them all, we can carry out the renormalization. The renormalization factor, or counter-term, is represented going forward as ${\bf Z}^{\bf IR}$. The following simple formulation may be used to convert this assertion into terms of the $m$-point renormalized cosmological correlation function, which can then be expressed in terms of the unrenormalized/bare contribution:
\bea \overline{\langle \zeta_{\bf k_1}\zeta_{\bf k_2}\zeta_{\bf k_3}\cdots\cdots\zeta_{\bf k_m}\rangle}=\left({\bf Z}^{\bf IR}\right)^{\frac{m}{2}}\langle \zeta_{\bf k_1}\zeta_{\bf k_2}\zeta_{\bf k_3}\cdots\cdots\zeta_{\bf k_m}\rangle.\eea
We now limit our study by limiting $m=2$, which characterizes the two-point amplitude of the cosmic correlation function in the Fourier space, as we are interested in the renormalization of the scalar power spectrum. Once this is fixed, the two-point amplitude-related 1PI effective action may be further stated as follows:
\bea \overline{\Gamma^{(2)}_{\zeta}(k_1,k_2,\mu,\mu_0)}={\bf Z}^{\bf IR}\times\Gamma^{(2)}_{\zeta}(k_1,k_2).\eea
This assertion is made at the $2$-point correlation function level, and it can be readily understood in terms of the relationship that connects the renormalized, unrenormalized/bare, and counter-term contributions in the expression for the gauge invariant comoving scalar curvature perturbation. This expression is as follows:
\bea \zeta^{\bf R}_{\bf k}=\sqrt{{\bf Z}^{\bf IR}}\times \zeta^{\bf B}_{\bf k},\eea
using which the renormalized version of the cosmological two-point function can be expressed by the following simplified expression:
\bea \overline{\langle \zeta_{\bf k_1}\zeta_{\bf k_2}\rangle}=\langle \zeta^{\bf R}_{\bf k_1}\zeta^{\bf R}_{\bf k_2}\rangle&=&{\bf Z}^{\bf IR}\times \langle \zeta^{\bf B}_{\bf k_1}\zeta^{\bf B}_{\bf k_2}\rangle \quad\quad\quad{\rm where}\quad\quad\langle \zeta^{\bf B}_{\bf k_1}\zeta^{\bf B}_{\bf k_2}\rangle=(2\pi)^3\delta^3\left({\bf k_1}+{\bf k_2}\right)\frac{2\pi^2}{k^3_1}\small[\Delta_{\zeta,\textbf{EFT}}^{2}(k)\small].\eea
Consequently, we get the following result in terms of the two-point amplitude of the power spectrum for the scale modes after performing renormalization:
\bea\overline{\langle \zeta_{\bf k_1}\zeta_{\bf k_2}\rangle}&=&{\bf Z}^{\bf IR}\times \langle \zeta^{\bf B}_{\bf k_1}\zeta^{\bf B}_{\bf k_2}\rangle=(2\pi)^3\delta^3\left({\bf k_1}+{\bf k_2}\right)\frac{2\pi^2}{k^3_1}\small[\overline{\Delta_{\zeta,\textbf{EFT}}^{2}(k)}\small]={\bf Z}^{\bf IR}\times (2\pi)^3\delta^3\left({\bf k_1}+{\bf k_2}\right)\frac{2\pi^2}{k^3_1}\small[\Delta_{\zeta,\textbf{EFT}}^{2}(k)\small],\quad\quad\eea
which further implies the following fact:
\bea \label{zeftdd}
\overline{\Delta_{\zeta,\textbf{EFT}}^{2}(k)}= {\bf Z}^{\bf IR}\times\;\Delta_{\zeta,\textbf{EFT}}^{2}(k).
\eea
This is very crucial information in the present computational purpose as it serves the purpose of softening or smoothening which is ensured via the IR counter term multiplication in the two-point amplitude. Now we implement the renormalization condition I as quoted in equation(\ref{recon1}), whose immediate consequence is the following:
\bea {\bf Z}^{\bf IR}=  \frac{\bigg[\overline{\Delta_{\zeta,\textbf{EFT}}^{2}(k_{*})}\bigg]}{\bigg[\Delta_{\zeta,\textbf{EFT}}^{2}(k_{*})\bigg]} = \frac{\bigg[  \Delta_{\zeta,\textbf{Tree}}^{2}(k_{*})\bigg]_{\textbf{SRI}}}{\bigg[\Delta_{\zeta,\textbf{EFT}}^{2}(k_{*})\bigg]}\quad\Longrightarrow\quad {\bf Z}^{\bf IR}(k_*){\bf Z}^{\bf UV}(k_*)=1.\eea
Here the symbol ${\bf Z}^{\bf UV}(k_*)$ signifies the the total contribution of the UV counter-term computed from the five consecutive phases - contraction, bounce, SRI, USR and SRII and evaluated at the CMB pivot scale $k=k_*$ and in the present context it is given by the following expression:
\bea {\bf Z}^{\bf UV}(k_*):=1+\bigg({\bf Z}_{\bf \zeta, \bf C}^{\bf UV}(k)+{\bf Z}_{\bf \zeta, \bf B}^{\bf UV}(k)+{\bf Z}_{\bf \zeta, \bf SRI}^{\bf UV}(k)+{\bf Z}_{\bf \zeta, \bf USR}^{\bf UV}(k)+{\bf Z}_{\bf \zeta, \bf SRII}^{\bf UV}(k)\bigg)_{k=k_*},\eea
using which the IR counter-term in the present context is computed as:
\bea {\bf Z}^{\bf IR}&=&\frac{1}{\left(1+\left[\underbrace{\overline{{\bf W}}_{\bf C}+\overline{{\bf W}}_{\bf B}+\overline{{\bf W}}_{\bf SRI}+\overline{{\bf W}}_{\bf USR}+\overline{{\bf W}}_{\bf SRII}}_{\textbf{Regularized and Renormalized one-loop correction}}\right]_{k=k*}\right)}\nonumber\\
&=&\frac{1}{\left(1+\overline{{\bf W}}_{\bf C,*}+\overline{{\bf W}}_{\bf B,*}+\overline{{\bf W}}_{\bf SRI,*}+\overline{{\bf W}}_{\bf USR,*}+\overline{{\bf W}}_{\bf SRII,*}\right)}\nonumber\\
&\approx&\left(1-\overline{{\bf W}}_{\bf C,*}-\overline{{\bf W}}_{\bf B,*}-\overline{{\bf W}}_{\bf SRI,*}-\overline{{\bf W}}_{\bf USR,*}-\overline{{\bf W}}_{\bf SRII,*}\right).\eea
Further utilizing the result for the IR counter-term we derive the following expression for a fully renormalized power spectrum for the scalar modes:
\bea \label{renormpowerspectrum}
\overline{\Delta_{\zeta,\textbf{EFT}}^{2}(k)}&=&\bigg[\Delta_{\zeta,\textbf{Tree}}^{2}(k_{*})\bigg]_{\textbf{SRI}}\times\left(1-\overline{{\bf W}}_{\bf C,*}-\overline{{\bf W}}_{\bf B,*}-\overline{{\bf W}}_{\bf SRI,*}-\overline{{\bf W}}_{\bf USR,*}-\overline{{\bf W}}_{\bf SRII,*}\right)\nonumber\\
&&\quad\quad\quad\quad\quad\quad\quad\quad\quad\quad\quad\quad\quad\quad\quad\quad\times\left(1+\overline{{\bf W}}_{\bf C}+\overline{{\bf W}}_{\bf B}+\overline{{\bf W}}_{\bf SRI}+\overline{{\bf W}}_{\bf USR}+\overline{{\bf W}}_{\bf SRII}\right)\nonumber\\
&=&\bigg[\Delta_{\zeta,\textbf{Tree}}^{2}(k_{*})\bigg]_{\textbf{SRI}}\times \bigg(1+\overline{{\bf X}}_{\bf Loop}\bigg),\eea
where the quantity $\overline{{\bf X}}_{\bf Loop}$ encompasses the renormalized loop contribution

\textcolor{black}{At this stage, it is important to note that the above calculation can be simplified after introducing some convenient expressions which will prove helpful in the present context of the computation and its details are elaborated in the appendix \ref{appC1}. }
In principle, during the computation of the IR counter-term, one can consider higher order terms in the series expansion of the result, which gives rise to higher even order Feynman graphs. This is because of the fact that all odd-order graphs cancel at that order itself in the super-horizon scale constraint which makes them vanish. \textcolor{black}{Consequently, following the manipulations in the appendix \ref{appC1}, we have the following result for the renormalized power spectrum for scalar modes}:
\bea \overline{\Delta_{\zeta,\textbf{EFT}}^{2}(k)}
&=&\bigg[\Delta_{\zeta,\textbf{Tree}}^{2}(k_{*})\bigg]_{\textbf{SRI}}\times\left(1-\overline{{\bf W}}_{\bf C,*}-\overline{{\bf W}}_{\bf B,*}-\overline{{\bf W}}_{\bf SRI,*}-\overline{{\bf W}}_{\bf USR,*}-\overline{{\bf W}}_{\bf SRII,*}-\cdots\right)\nonumber\\
&&\quad\quad\quad\quad\quad\quad\quad\quad\quad\quad\quad\quad\quad\quad\quad\quad\times\left(1+\overline{{\bf W}}_{\bf C}+\overline{{\bf W}}_{\bf B}+\overline{{\bf W}}_{\bf SRI}+\overline{{\bf W}}_{\bf USR}+\overline{{\bf W}}_{\bf SRII}\right)\nonumber\\
&=&\bigg[\Delta_{\zeta,\textbf{Tree}}^{2}(k_{*})\bigg]_{\textbf{SRI}}\times \Bigg(1-\left\{\frac{\bigg[  \Delta_{\zeta,\textbf{Tree}}^{2}(k)\bigg]_{\textbf{SRI}}}{\bigg[  \Delta_{\zeta,\textbf{Tree}}^{2}(k_{*})\bigg]_{\textbf{SRI}}}\right\}\times\bigg[\bigg(\overline{{\bf W}}^2_{\bf C,*}+\overline{{\bf W}}^2_{\bf B,*}+\overline{{\bf W}}^2_{\bf SRI,*}+\overline{{\bf W}}^2_{\bf USR,*}+\overline{{\bf W}}^2_{\bf SRII,*}\bigg)\nonumber\\
&&\quad\quad\quad\quad\quad+2
\bigg(\overline{{\bf W}}_{\bf B,*}\overline{{\bf W}}_{\bf C,*}+\overline{{\bf W}}_{\bf SRI,*}\overline{{\bf W}}_{\bf C,*}+\overline{{\bf W}}_{\bf USR,*}\overline{{\bf W}}_{\bf C,*}+\overline{{\bf W}}_{\bf SRII,*}\overline{{\bf W}}_{\bf C,*}
\nonumber\\
&&\quad\quad\quad\quad\quad\quad\quad\quad\quad+\overline{{\bf W}}_{\bf SRI,*}\overline{{\bf W}}_{\bf B,*}+2\overline{{\bf W}}_{\bf USR,*}\overline{{\bf W}}_{\bf B,*}+\overline{{\bf W}}_{\bf SRII,*}\overline{{\bf W}}_{\bf B,*}
\nonumber\\
&&\quad\quad\quad\quad\quad\quad\quad\quad\quad+\overline{{\bf W}}_{\bf USR,*}\overline{{\bf W}}_{\bf SRI,*}+\overline{{\bf W}}_{\bf SRII,*}\overline{{\bf W}}_{\bf SRI,*}+\overline{{\bf W}}_{\bf SRII,*}\overline{{\bf W}}_{\bf USR,*}\bigg)\bigg]+\cdots\Bigg).\eea
In the diagrammatic representation the above-mentioned statement further can be written as:
\bea  \overline{\Delta_{\zeta,\textbf{EFT}}^{2}(k)}&=&\bigg[\Delta_{\zeta,\textbf{Tree}}^{2}(k_{*})\bigg]_{\textbf{SRI}}\times \bigg(1+\sum_{\textbf{All even graphs G}}{\cal F}_{\bf G}\bigg),\eea
where we have:
\bea \sum_{\textbf{All even graphs G}}{\cal F}_{\bf G}:={\cal F}_{\bf 2}+{\cal F}_{\bf 4}+{\cal F}_{\bf 6}+\cdots\eea
Here the corresponding two-loop and four-loop graphs are given by:
\bea &&\label{twoloopx}
\begin{tikzpicture}[baseline={([yshift=-3.5ex]current bounding box.center)},very thick]
  
   % Loop
  \def\radius{0.76}
  \scalebox{0.5}{
  \draw[red,very thick] (3*\radius,0) circle (\radius);
  \draw[red,very thick] (5*\radius,0) circle (\radius);
  \draw[red,very thick] (13*\radius,\radius) circle (\radius);
  \draw[red,very thick] (13*\radius,3*\radius) circle (\radius);
  \draw[red,very thick] (21*\radius,0) circle (\radius);}
  % External lines
  %\filldraw
  \draw[black, very thick] (0,0) -- (\radius,0); 
  \draw[blue,fill=blue] (\radius,0) circle (.3ex);
  \draw[blue,fill=blue] (2*\radius,0) circle (.3ex);
  \draw[blue,fill=blue] (3*\radius,0) circle (.3ex);
  \draw[black, very thick] (3*\radius,0) -- (4*\radius,0);
  \node at (4.5*\radius,0) {+};
  \draw[black, very thick] (5*\radius,0) -- (6.5*\radius,0);
  \draw[blue,fill=blue] (6.5*\radius,0) circle (.3ex);
  \draw[blue,fill=blue] (6.5*\radius,\radius) circle (.3ex);
  \draw[black, very thick] (6.5*\radius,0) -- (8*\radius,0);
  \node at (8.5*\radius,0) {+};
  \draw[black, very thick] (9*\radius,0) -- (10*\radius,0);
  \draw[blue,fill=blue] (10*\radius,0) circle (.3ex);
  \draw[red, very thick] (10*\radius,0) -- (11*\radius,0);
  \draw[blue,fill=blue] (11*\radius,0) circle (.3ex);
  \draw[black, very thick] (11*\radius,0) -- (12*\radius,0);
  \node at (15*\radius,0) {=${\cal F}_{\bf 2}$,};,
\end{tikzpicture}
\\
&&\label{fourloopx}
    \begin{tikzpicture}[baseline={([yshift=-.5ex]current bounding box.center)},very thick]
    \draw [line width=1pt] (-14.5,0)-- (-14,0);
    \draw[blue,fill=blue] (-14,0) circle (.3ex);
    \draw [red,line width=0.8pt] (-14,0)-- (-13.5,0.5);
    \draw[blue,fill=blue] (-13.5,0.5) circle (.3ex);
    \draw [red,line width=0.8pt] (-14,0)-- (-13.5,-0.5);
    \draw[blue,fill=blue] (-13.5,-0.5) circle (.3ex);
    \draw [red,line width=0.8pt] (-13.5,0.5)-- (-13.5,-0.5);
    \draw [red,line width=0.8pt] (-13.5,0.5)-- (-12.5,0.5);
    \draw [red,line width=0.8pt] (-12.5,0.5)-- (-12.5,-0.5);
    \draw [red,line width=0.8pt] (-13.5,0)-- (-12.5,0);
    \draw[blue,fill=blue] (-13.5,0) circle (.3ex);
    \draw[blue,fill=blue] (-12.5,0) circle (.3ex);
    \draw[blue,fill=blue] (-12.5,0.5) circle (.3ex);
    \draw [red,line width=0.8pt] (-13.5,-0.5)-- (-12.5,-0.5);
    \draw[blue,fill=blue] (-12.5,-0.5) circle (.3ex);
    \draw [red,line width=0.8pt] (-12.5,0.5)-- (-12,0);
    \draw[blue,fill=blue] (-12,0) circle (.3ex);
    \draw[blue,fill=blue] (-12.5,-0.5) circle (.3ex);
    \draw [red,line width=0.8pt] (-12.5,-0.5)-- (-12,0);
    \draw [black,line width=0.8pt] (-12,0)-- (-11.5,0);

    \node at (-11,0) {+};

    \draw [line width=1pt] (-10.5,0)-- (-10,0);
    \draw[blue,fill=blue] (-10,0) circle (.3ex);
    \draw [red,line width=0.8pt] (-10,0)-- (-9.5,0.5);
    \draw [red,line width=0.8pt] (-10,0)-- (-9.5,-0.5);
    \draw [red,line width=0.8pt] (-9.5,0.5)-- (-8.5,0.5);
    \draw [red,line width=0.8pt] (-9.5,-0.5)-- (-8.5,-0.5);
    \draw [red,line width=0.8pt] (-8.5,0.5)-- (-8,0);
    \draw [red,line width=0.8pt] (-8.5,-0.5)-- (-8,0);
    \draw [line width=1pt] (-8,0)-- (-7.5,0);
    \draw [red,line width=0.8pt] (-9.5,0.5)-- (-9,0);
    \draw [red,line width=0.8pt] (-9.5,-0.5)-- (-9,0);
    \draw [red,line width=0.8pt] (-8.5,0.5)-- (-9,0);
    \draw [red,line width=0.8pt] (-8.5,-0.5)-- (-9,0);
    \draw[blue,fill=blue] (-9.5,0.5) circle (.3ex);
    \draw[blue,fill=blue] (-9.5,-0.5) circle (.3ex);
    \draw[blue,fill=blue] (-8.5,0.5) circle (.3ex);
    \draw[blue,fill=blue] (-8.5,-0.5) circle (.3ex);
    \draw[blue,fill=blue] (-9,0) circle (.3ex);
    \draw[blue,fill=blue] (-8,0) circle (.3ex);

    \node at (-7,0) {+};

    \draw [line width=1pt] (-6.5,0)-- (-6,0);
    \draw[blue,fill=blue] (-6,0) circle (.3ex);
    \draw [red,line width=0.8pt] (-6,0)-- (-5.5,0.5);
    \draw[blue,fill=blue] (-5.5,0.5) circle (.3ex);
    \draw [red,line width=0.8pt] (-6,0)-- (-5.5,-0.5);
    \draw[blue,fill=blue] (-5.5,-0.5) circle (.3ex);
    \draw [red,line width=0.8pt] (-5.5,0.5)-- (-5.5,-0.5);
    \draw [red,line width=0.8pt] (-5.5,0.5)-- (-4.5,0.5);
    \draw [red,line width=0.8pt] (-4.5,0.5)-- (-4.5,-0.5);
    \draw [red,line width=0.8pt] (-5,0.5)-- (-5,-0.5);
    \draw [red,line width=0.8pt] (-5.5,-0.5)-- (-4.5,-0.5);
    \draw [red,line width=0.8pt] (-4.5,0.5)-- (-4,0);
    \draw [red,line width=0.8pt] (-4.5,-0.5)-- (-4,0);
    \draw [line width=0.8pt] (-4,-0)-- (-3.5,0);
    \draw[blue,fill=blue] (-5,0.5) circle (.3ex);
    \draw[blue,fill=blue] (-5,-0.5) circle (.3ex);
    \draw[blue,fill=blue] (-4.5,0.5) circle (.3ex);
    \draw[blue,fill=blue] (-4.5,-0.5) circle (.3ex);
    \draw[blue,fill=blue] (-4,0) circle (.3ex);

    \node at (-3,0) {\quad\quad+ $\cdots$=${\cal F}_{\bf 4}$};
    %\node at (-3,0) { = \quad \text{Four-loop contributions}};
\end{tikzpicture}.
\eea
After substituting the explicit form of the IR and UV counter-terms within the current context using the renormalization scale, the cosmological flow equations, which are commonly known as the cosmological beta functions are computed as:
\begin{itemize}[label={\checkmark}]
\item The first flow equation which describes the renormalized version of the spectral tilt for the scalar modes reads:
 \bea \label{renormflowtilt}
\left[\overline{n_{\zeta,\textbf{EFT}}(k)-1}\right]
&=&{\bf Z}^{\bf IR}\times\Bigg[{\bf Z}^{\bf UV}\bigg(\bigg[n_{\zeta,\textbf{Tree}}(k)\bigg]_{{\bf SRI}}-1\bigg)+\bigg(\frac{d{\bf Z}^{\bf UV}}{d\ln k}\bigg)\bigg(\ln \bigg[\Delta^{2}_{\zeta,\textbf{Tree}}(k)\bigg]_{\textbf{SRI}}\bigg)\Bigg]\nonumber\\
&=&\bigg(1-\alpha_{\bf C}{\cal A}_{\bf C}-\alpha_{\bf B}{\cal A}_{\bf B}-\alpha_{\bf SRI}{\cal A}_{\bf SRI}-\alpha_{\bf USR}{\cal A}_{\bf USR}-\alpha_{\bf SRII}{\cal A}_{\bf SRII}\bigg)\nonumber\\
&&\quad\quad\quad\quad\quad\quad\quad\quad\quad\quad\quad\quad\quad\quad\times\Bigg(\alpha_{\bf C}{\cal A}_{\bf C}+\alpha_{\bf B}{\cal A}_{\bf B}+\alpha_{\bf SRI}{\cal A}_{\bf SRI}+\alpha_{\bf USR}{\cal A}_{\bf USR}+\alpha_{\bf SRII}{\cal A}_{\bf SRII}\Bigg)\nonumber\\
&&\quad\quad\quad\quad\times\Bigg[\bigg[\Delta^{2}_{\zeta,\textbf{Tree}}(k)\bigg]^2_{\textbf{SRI}}+2\ln \bigg[\Delta^{2}_{\zeta,\textbf{Tree}}(k)\bigg]_{\textbf{SRI}}\Bigg]\times\bigg(\bigg[n_{\zeta,\textbf{Tree}}(k)\bigg]_{{\bf SRI}}-1\bigg),\eea
\textcolor{black}{where the various symbols like, $\alpha_{\bf C}$, $\alpha_{\bf B}$, $\alpha_{\bf SRI}$, $\alpha_{\bf USR}$, $\alpha_{\bf SRII}$ and, ${\cal A}_{\bf C}$, ${\cal A}_{\bf B}$, ${\cal A}_{\bf SRI}$, ${\cal A}_{\bf USR}$, ${\cal A}_{\bf SRII}$, are provided in the appendix \ref{appC2}. }
\item The second flow equation which describes the renormalized version of the running of the spectral tilt for the scalar modes is given by:
 \bea \label{renormflowalpha}
\left[\overline{\alpha_{\zeta,\textbf{EFT}}(k)}\right] 
&=&{\bf Z}^{\bf IR}\times\Bigg[{\bf Z}^{\bf UV}\bigg(\bigg[\alpha_{\zeta,\textbf{Tree}}(k)\bigg]_{{\bf SRI}}\bigg)+2\bigg(\frac{d{\bf Z}^{\bf UV}}{d\ln k}\bigg)\bigg(\bigg[n_{\zeta,\textbf{Tree}}(k)\bigg]_{{\bf SRI}}-1\bigg)\nonumber\\
&&\quad\quad\quad\quad\quad\quad\quad\quad\quad\quad\quad\quad\quad\quad\quad\quad+\bigg(\frac{d^2{\bf Z}^{\bf UV}}{d\ln k^2}\bigg)\bigg(\ln \bigg[\Delta^{2}_{\zeta,\textbf{Tree}}(k)\bigg]_{\textbf{SRI}}\bigg)\Bigg],\nonumber\\
&=&\bigg(1-\alpha_{\bf C}{\cal A}_{\bf C}-\alpha_{\bf B}{\cal A}_{\bf B}-\alpha_{\bf SRI}{\cal A}_{\bf SRI}-\alpha_{\bf USR}{\cal A}_{\bf USR}-\alpha_{\bf SRII}{\cal A}_{\bf SRII}\bigg)\nonumber\\
&&\quad\quad\quad\quad\quad\quad\quad\quad\quad\quad\quad\quad\quad\quad\times\Bigg(\alpha_{\bf C}{\cal A}_{\bf C}+\alpha_{\bf B}{\cal A}_{\bf B}+\alpha_{\bf SRI}{\cal A}_{\bf SRI}+\alpha_{\bf USR}{\cal A}_{\bf USR}+\alpha_{\bf SRII}{\cal A}_{\bf SRII}\Bigg)\nonumber\\
&&\quad\times\Bigg\{\Bigg[\bigg[\Delta^{2}_{\zeta,\textbf{Tree}}(k)\bigg]^2_{\textbf{SRI}}+2\ln \bigg[\Delta^{2}_{\zeta,\textbf{Tree}}(k)\bigg]_{\textbf{SRI}}\Bigg]\times\bigg[\alpha_{\zeta,\textbf{Tree}}(k)\bigg]_{{\bf SRI}}\nonumber\\
&&\quad\quad\quad\quad\quad\quad\quad\quad\quad\quad\quad\quad\quad\quad\quad\quad\quad\quad\quad +6\bigg(\bigg[n_{\zeta,\textbf{Tree}}(k)\bigg]_{{\bf SRI}}-1\bigg)^2\Bigg\}.\eea
\item  The third flow equation which describes the renormalized version of the running of the running of spectral tilt for the scalar modes is given by:
\bea \label{renormflowbeta}
\left[\overline{\beta_{\zeta,\textbf{EFT}}(k)}\right]
&=&{\bf Z}^{\bf IR}\times\Bigg[{\bf Z}^{\bf UV}\bigg(\bigg[\beta_{\zeta,\textbf{Tree}}(k)\bigg]_{{\bf SRI}}\bigg)+2\bigg(\frac{d{\bf Z}^{\bf UV}}{d\ln k}\bigg)\bigg(\bigg[\alpha_{\zeta,\textbf{Tree}}(k)\bigg]_{{\bf SRI}}\bigg)\nonumber\\
&&\quad\quad\quad\quad\quad+3\bigg(\frac{d^2{\bf Z}^{\bf UV}}{d\ln k^2}\bigg)\bigg(\bigg[n_{\zeta,\textbf{Tree}}(k)\bigg]_{{\bf SRI}}-1\bigg)+\bigg(\frac{d^3{\bf Z}^{\bf UV}}{d\ln k^3}\bigg)\bigg(\ln \bigg[\Delta^{2}_{\zeta,\textbf{Tree}}(k)\bigg]_{\textbf{SRI}}\bigg)\Bigg],\nonumber\\
&=&\bigg(1-\alpha_{\bf C}{\cal A}_{\bf C}-\alpha_{\bf B}{\cal A}_{\bf B}-\alpha_{\bf SRI}{\cal A}_{\bf SRI}-\alpha_{\bf USR}{\cal A}_{\bf USR}-\alpha_{\bf SRII}{\cal A}_{\bf SRII}\bigg)\nonumber\\
&&\quad\quad\quad\quad\quad\quad\quad\quad\quad\quad\quad\quad\quad\quad\times\Bigg(\alpha_{\bf C}{\cal A}_{\bf C}+\alpha_{\bf B}{\cal A}_{\bf B}+\alpha_{\bf SRI}{\cal A}_{\bf SRI}+\alpha_{\bf USR}{\cal A}_{\bf USR}+\alpha_{\bf SRII}{\cal A}_{\bf SRII}\Bigg)\nonumber\\
&&\quad\times\Bigg\{\Bigg[\bigg[\Delta^{2}_{\zeta,\textbf{Tree}}(k)\bigg]^2_{\textbf{SRI}}+2\ln \bigg[\Delta^{2}_{\zeta,\textbf{Tree}}(k)\bigg]_{\textbf{SRI}}\Bigg]\times\bigg[\beta_{\zeta,\textbf{Tree}}(k)\bigg]_{{\bf SRI}}\nonumber\\
&&\quad\quad\quad\quad\quad\quad\quad\quad\quad\quad\quad\quad\quad\quad\quad\quad +16\bigg(\bigg[n_{\zeta,\textbf{Tree}}(k)\bigg]_{{\bf SRI}}-1\bigg)\times\bigg[\alpha_{\zeta,\textbf{Tree}}(k)\bigg]_{{\bf SRI}}\Bigg\}.\quad\quad\eea
\end{itemize}
\section{Resummation in loop corrected scalar power spectrum}
\label{s11}

Let us state up front that the prime component of the one-loop correction, denoted as:
\bea \sum_{\textbf{All even graphs G}}{\cal F}_{\bf G}={\cal F}_{\bf 2}+{\cal F}_{\bf 4}+{\cal F}_{\bf 6}+\cdots=\overline{{\bf X}}_{\bf Loop},\eea is not the basis for the computation that is being done in this section. The overall size of this component must be kept inside the perturbative limit in order to execute the resummation and ultimately obtain a finite result considering the contributions up to all even loop order. For our computational purposes, this strategy works well, as we quote the output in such a way. Here, our main goal is to introduce the Dynamical Renormalization Group (DRG) technique \cite{Chen:2016nrs,Baumann:2019ghk,Boyanovsky:1998aa,Boyanovsky:2001ty,Boyanovsky:2003ui,Burgess:2015ajz,Burgess:2014eoa,Burgess:2009bs,Dias:2012qy,Chaykov:2022zro,Chaykov:2022pwd}, which allows us to resum across all of the logarithmically divergent contributions in the current calculation. More specifically, this resummed conclusion holds true for perturbative computations in all loop orders where the quantum effects can be accurately captured. Nevertheless, this is only possible if the resummed infinite series satisfies the tight convergence requirements at super-horizon and horizon-crossing values. The aforementioned components in the convergent series are all results of the hypothesis of cosmic perturbation of scalar modes in all possible loop orders. The DRG process is commonly understood as the inherent mechanism that permits the justification of secular momentum-dependent contributions to the convergent infinite series at horizon crossing and super-horizon scales. This process is utilized in the context of primordial EFT-driven cosmological setup. Upon completing the resummation process, this approach enables one to accurately determine the behavior at the horizon crossing and super-horizon scales, rather than obtaining the whole behavior from the series expansion term by term. At first, momentum-dependent contributions to the equations for scale-dependent running couplings of the underlying theory in terms of beta functions were introduced using the notion of the Renomrmalization Group (RG) resummation approach. The improved upon RG resummation method is known as the DRG resummation methodology. This finding is applicable in the minuscule coupling domain where perturbative approximations hold precisely within the underlying EFT framework of the set-up and in a specific observationally feasible larger interval of running of momentum scales. Rather than examining the running couplings behavior in relation to the underlying scale, we examine the features and shape of the primordial power spectrum in the context of Cosmological DGR resummation at the late time scale. These features are defined by the spectral tilt, running, and running of the running of the spectral tilt in relation to the momentum scale. The presence of these cosmic beta functions, which are all distinctive physical variables, validates the small deviation of the primordial power spectrum from its precise scale invariant feature. DRG resummation is frequently referred to as the resummation under the impact of exponentiation in the related context of computation and discussion. Finally, the resummed form of the scalar modes power spectrum is described by the following relations:
\bea \label{DRG1} \overline{\overline{\Delta^{2}_{\zeta,{\bf EFT}}(k)}}
&\approx&\bigg[\Delta^{2}_{\zeta,{\bf Tree}}(k)\bigg]_{\bf SRI}\times \Bigg(\sum^{\infty}_{n=0}\frac{\overline{{\bf X}}_{\bf Loop}^n}{n!}\Bigg)\times\bigg\{1+{\cal O}\bigg(\bigg[\Delta^{2}_{\zeta,{\bf Tree}}(k_*)\bigg]^2_{\bf SRI}\bigg)\bigg\},\nonumber\\
&=&\bigg[\Delta^{2}_{\zeta,{\bf Tree}}(k)\bigg]_{\bf SRI}\exp\bigg(\overline{{\bf X}}_{\bf Loop}\bigg)\times\bigg\{1+{\cal O}\bigg(\bigg[\Delta^{2}_{\zeta,{\bf Tree}}(k_*)\bigg]^2_{\bf SRI}\bigg)\bigg\},\nonumber\\
&=&\bigg[\Delta^{2}_{\zeta,{\bf Tree}}(k)\bigg]_{\bf SRI}\exp\bigg(\sum_{\textbf{All even graphs G}}{\cal F}_{\bf G}\bigg)\times\bigg\{1+{\cal O}\bigg(\bigg[\Delta^{2}_{\zeta,{\bf Tree}}(k_*)\bigg]^2_{\bf SRI}\bigg)\bigg\},\nonumber\\
&=&\bigg[\Delta^{2}_{\zeta,{\bf Tree}}(k)\bigg]_{\bf SRI}\nonumber\\
&&\times\exp\left(\begin{tikzpicture}[baseline={([yshift=-5.5ex]current bounding box.center)},very thick]

      % Loop
  \def\radius{0.5}
  \scalebox{0.5}{
  \draw[red, ultra thick] (3*\radius,0) circle (\radius);
  \draw[red,ultra thick] (5*\radius,0) circle (\radius);
  \draw[red,ultra thick] (13*\radius,\radius) circle (\radius);
  \draw[red,ultra thick] (13*\radius,3*\radius) circle (\radius);
  \draw[red,ultra thick] (21*\radius,0) circle (\radius);}

  % External lines
  %\filldraw
  %\draw[black, very thick] (-2*\radius,0) -- (-1*\radius,0); 
  %\node at (-0.5*\radius,0) {+};
  \draw[black, very thick] (0,0) -- (\radius,0); 
  \draw[blue,fill=blue] (\radius,0) circle (.3ex);
  \draw[blue,fill=blue] (2*\radius,0) circle (.3ex);
  \draw[blue,fill=blue] (3*\radius,0) circle (.3ex);
  \draw[black, very thick] (3*\radius,0) -- (4*\radius,0);
  \node at (4.5*\radius,0) {+};
  \draw[black, very thick] (5*\radius,0) -- (6.5*\radius,0);
  \draw[blue,fill=blue] (6.5*\radius,0,0) circle (.3ex);
  \draw[blue,fill=blue] (6.5*\radius,\radius) circle (.3ex);
  \draw[black, very thick] (6.5*\radius,0,0) -- (8*\radius,0);
  \node at (8.5*\radius,0) {+};
  \draw[black, very thick] (9*\radius,0) -- (10*\radius,0);
  \draw[blue,fill=blue] (10*\radius,0,0) circle (.3ex);
  \draw[red, ultra thick] (10*\radius,0) -- (11*\radius,0);
  \draw[blue,fill=blue] (11*\radius,0,0) circle (.3ex);
  \draw[black, very thick] (11*\radius,0) -- (12*\radius,0);
  \node at (12.5*\radius,0) {+};
  \draw[black, very thick](13*\radius,0) --(14*\radius,0);
  \draw[blue,fill=blue](14*\radius,0,0) circle (.3ex);
  \draw[red,  thick](14*\radius,0) --(14.5*\radius,\radius);
  \draw[blue,fill=blue](14.5*\radius,\radius) circle (.3ex);
\draw[red, thick](14.5*\radius,\radius) --(15.5*\radius,\radius);
\draw[blue,fill=blue](15.5*\radius,\radius) circle (.3ex);
\draw[red, thick](14*\radius,0) -- (14.5*\radius,-\radius);\draw[blue,fill=blue](14.5*\radius,-\radius) circle (.3ex);
\draw[red,  thick](14.5*\radius,\radius) -- (14.5*\radius,-\radius);
\draw[red, thick](14.5*\radius,-\radius) -- (15.5*\radius,-\radius);
\draw[blue,fill=blue](15.5*\radius,-\radius) circle (.3ex);
\draw[blue,fill=blue](14.5*\radius,0) circle (.3ex);
\draw[red, thick](14.5*\radius,0) -- (15.5*\radius,0);
\draw[red,  thick](15.5*\radius,\radius) -- (15.5*\radius,-\radius);
\draw[blue,fill=blue](15.5*\radius,0) circle (.3ex);
\draw[red, thick](15.5*\radius,\radius) -- (16*\radius,0);
\draw[blue,fill=blue](16*\radius,0) circle (.3ex);
\draw[red, thick](16*\radius,0) -- (15.5*\radius,-\radius);
\draw[black, very thick](16*\radius,0) -- (17*\radius,0);
\node at (17.5*\radius,0) {+};
\draw[black, very thick](18*\radius,0) -- (19*\radius,0);
\draw[blue,fill=blue](19*\radius,0) circle (.3ex);
\draw[red, thick](19*\radius,0) -- (19.5*\radius,\radius);
\draw[red, thick](19*\radius,0) -- (19.5*\radius,-\radius);
\draw[blue,fill=blue](19.5*\radius,\radius) circle (.3ex);
 \draw[blue,fill=blue](19.5*\radius,-\radius) circle (.3ex); 
 \draw[red, thick](19.5*\radius,\radius) -- (20.5*\radius,\radius);
 \draw[red, thick](19.5*\radius,-\radius) -- (20.5*\radius,-\radius);
 \draw[blue,fill=blue](20.5*\radius,\radius) circle (.3ex);
 \draw[blue,fill=blue](20.5*\radius,-\radius) circle (.3ex);
 \draw[red, thick](20.5*\radius,\radius) -- (21*\radius,0);
 \draw[red, thick](20.5*\radius,-\radius) -- (21*\radius,0);
\draw[blue,fill=blue](21*\radius,0) circle (.3ex);
\draw[black,  thick](21*\radius,0) -- (22*\radius,0); 
\draw[red, thick](19.5*\radius,\radius) -- (20.5*\radius,-\radius);
\draw[red, thick](20.5*\radius,\radius) -- (19.5*\radius,-\radius);
\node at (22.5*\radius,0) {+};
\draw[black, thick](23*\radius,0) -- (24*\radius,0);
\draw[red, thick](24*\radius,0) -- (24.5*\radius,\radius);
\draw[red, thick](24*\radius,0) -- (24.5*\radius,-\radius);
\draw[blue,fill=blue](24*\radius,0) circle (.3ex);
\draw[blue,fill=blue](24.5*\radius,\radius) circle (.3ex);
\draw[blue,fill=blue](24.5*\radius,-\radius) circle (.3ex);
\draw[red, thick](24.5*\radius,\radius) -- (25.5*\radius,\radius);
\draw[red,  thick](24.5*\radius,-\radius) -- (25.5*\radius,-\radius);
\draw[red,  thick](25.5*\radius,\radius) -- (26*\radius,0);
\draw[red, thick](26*\radius,0) -- (25.5*\radius,-\radius);
\draw[blue,fill=blue](25.5*\radius,\radius) circle (.3ex);
\draw[blue,fill=blue](26*\radius,0) circle (.3ex);
\draw[blue,fill=blue](25.5*\radius,-\radius) circle (.3ex);
\draw[red,  thick](24.5*\radius,\radius) -- (24.5*\radius,-\radius);
\draw[red, thick](25*\radius,\radius) -- (25*\radius,-\radius);
\draw[red, thick](25.5*\radius,\radius) -- (25.5*\radius,-\radius);
\draw[blue,fill=blue](25*\radius,\radius) circle (.3ex);
\draw[blue,fill=blue](25*\radius,-\radius) circle (.3ex);
\draw[black, thick](26*\radius,0) -- (27*\radius,0); 
\node at (28*\radius,0) {+};
\node at (29*\radius,0) {...};

\end{tikzpicture}\right)\nonumber\\
&&\quad\quad\quad\quad\quad\quad\quad\quad\quad\quad\quad\quad\quad\quad\quad\quad\quad\quad\quad\quad\quad\quad\times\bigg\{1+{\cal O}\bigg(\bigg[\Delta^{2}_{\zeta,{\bf Tree}}(k_*)\bigg]^2_{\bf SRI}\bigg)\bigg\},\eea
In this case, the result in the DRG resummed version is valid for all orders of the function $\overline{{\bf X}}_{\bf Loop}$. It should be emphasized that $|\overline{{\bf X}}_{\bf Loop}|\ll 1$ is strictly required by the convergence criteria and is fully met in the present discussion. The most important result of the DRG resummed version of the one-loop corrected primordial scalar power spectrum is that, in contrast to the renormalized one-loop power spectrum derived in the preceding section, it yields a controlled version of the two-point amplitude after summing over all even loop graphs where the behaviour of the logarithmic divergences are softened enough and that makes the computation trustable. It is crucial to note that, in the current context, chain diagrams that continuously add cubic self-energy are the source of the leading order logarithmically divergent contributions, even though the explicit details of the Feynman diagrams and the subgraphs are not required to perform the DRG resummation method. The dominance of all chain diagrams over other potential diagrams in the computation is not necessary for the application of the DRG resummation, but it will undoubtedly add to the leading contributions from these logarithmic-dependant components. The higher-order convergent components in the infinite series in the aforementioned result precisely replicate the function of upper-loop contributions in the situation under consideration. It is truly remarkable that one can study the behaviour of each correction term in all-loop order without explicitly calculating higher-loop corrections to the primordial power spectrum for the scalar modes. This means that we can study the non-perturbative but convergent behaviour of the spectrum as the sum over all-loop contribution becomes finite and can be expressed in terms of an exponential function in this particular context. To get a more realistic outcome by substituting the individual contributions from contracting, bouncing, SRI, USR, and SRII phases in the loop-dominated graphs we get the following simplified result for the regularized-renormalized-resummed (RRR) version of the primordial power spectrum for the scalar modes in the present context of the computation:
\bea \label{DRG2} \overline{\overline{\Delta^{2}_{\zeta,{\bf EFT}}(k)}}
&\approx&\bigg[\Delta^{2}_{\zeta,{\bf Tree}}(k)\bigg]_{\bf SRI}\times \underbrace{\exp\bigg[\alpha^2_{\bf C}{\cal A}^2_{\bf C}(k_c,k_b)\bigg]}_{\textbf{Contraction phase}}\times\underbrace{\exp\bigg[\alpha^2_{\bf B}{\cal A}^2_{\bf B}(k_b,k_*)\bigg]}_{\textbf{Bouncing phase}}\nonumber\\
&&\quad\quad\quad\times\underbrace{\left(\frac{k_s}{k_*}\right)^{2.3\alpha^2_{\bf SRI}}}_{\textbf{SRI phase}}\times\underbrace{\left(\frac{k_e}{k_s}\right)^{-2.3\alpha^2_{\bf USR}}}_{\textbf{USR phase}}\times \underbrace{\left(\frac{k_{\rm end}}{k_e}\right)^{-2.3\alpha^2_{\bf SRII}}}_{\textbf{SRII phase}}\times
\bigg\{1+\underbrace{{\cal O}\bigg(\bigg[\Delta^{2}_{\zeta,{\bf Tree}}(k_*)\bigg]^2_{\bf SRI}}_{\textbf{Correction term}}\bigg)\bigg\}.\eea
In terms of the number of e-foldings, the above expression can be adjusted in the following form:
\bea \label{DRG2a} \overline{\overline{\Delta^{2}_{\zeta,{\bf EFT}}(N)}}
&\approx&\bigg[\Delta^{2}_{\zeta,{\bf Tree}}(N)\bigg]_{\bf SRI}\times \underbrace{\exp\bigg[\alpha^2_{\bf C}{\cal A}^2_{\bf C}(\Delta N_{\bf C})\bigg]}_{\textbf{Contraction phase}}\times\underbrace{\exp\bigg[\alpha^2_{\bf B}{\cal A}^2_{\bf B}(\Delta N_{\bf B})\bigg]}_{\textbf{Bouncing phase}}\nonumber\\
&&\quad\quad\quad\times\underbrace{\exp\left(2.3\Delta N_{\bf SRI}\alpha^2_{\bf SRI}\right)}_{\textbf{SRI phase}}\times\underbrace{\exp\left(-2.3\Delta N_{\bf USR}\alpha^2_{\bf USR}\right)}_{\textbf{USR phase}}\times \underbrace{\exp\left(-2.3\Delta N_{\bf SRII}\alpha^2_{\bf SRII}\right)}_{\textbf{SRII phase}}\nonumber\\
&&\quad\quad\quad\quad\quad\quad\quad\quad\quad\quad\quad\quad\quad\quad\quad\quad\quad\quad\quad\quad\quad\quad\times
\bigg\{1+\underbrace{{\cal O}\bigg(\bigg[\Delta^{2}_{\zeta,{\bf Tree}}N_*)\bigg]^2_{\bf SRI}}_{\textbf{Correction term}}\bigg)\bigg\}.\eea
The temporal dependency of the cosmological $n$-point correlation functions may be effectively ascertained by using the separate universe technique, often known as the $\delta {\cal N}$ formalism to perturbation theory \cite{Sugiyama:2012tj,Dias:2012qy,Naruko:2012fe,Takamizu:2013gy,Abolhasani:2013zya,Clesse:2013jra,Chen:2013eea,Choudhury:2014uxa,vandeBruck:2014ata,Dias:2014msa,Garriga:2015tea,Choudhury:2015hvr,Choudhury:2016wlj,Choudhury:2017cos,Choudhury:2018glz, Starobinsky:1985ibc,Sasaki:1995aw,Sasaki:1998ug,Lyth:2005fi,Lyth:2004gb,Abolhasani:2018gyz,Passaglia:2018ixg,Dias:2012qy, Burgess:2009bs, Burgess:2014eoa, Burgess:2015ajz, Chaykov:2022zro, Chaykov:2022pwd, Jackson:2023obv}. Our observables are these correlations, which are assessed later in time. In order to precisely calculate these observables, one needs a set of beginning circumstances that need understanding the sub-horizon quantum fluctuations, which cross the horizon and eventually become classical. The distinct universe technique is limited to vast sizes, making it challenging to collect this information. It is highlighted that such correlators can be defined in terms of their value at an earlier horizon-crossing time and entail specified coefficients along with it. This technique is applicable to the cosmological correlators assessed at late times, $c_sk/aH \ll 1$. As demonstrated subsequently in \cite{Dias:2012qy}, the coefficients in question provide information on the late-time divergent IR adjustments to the required correlators at the lowest order. Using the framework of quantum field theory, namely the Callan-Symanzik equations, is the most reliable method to calculate the dynamics of such coefficients. It is important to note that the coefficients that were previously described and carried the IR divergences are expressed as form factors that may be found in calculations using quantum chromodynamics \cite{Dias:2012qy}. In order to provide a finite result and account for the infinite divergent contributions from the late time restriction in the correlators, resummation is thus required. The effects from large scales inside the aforesaid coefficients may be packaged using the renormalization group equations (RGE), and the initial conditions to solve such equations are created at horizon-crossing time by the same correlators. In the end, using just quantum field theoretic techniques, the dynamical renormalization group (DRG) analysis results in an all-order reconstruction of the correlation functions, which were previously necessary from the separate universe approach. We are able to refocus our attention on the cosmological beta functions, which are the real physical quantities—the spectral tilt, running, and running of the running of the running of spectral tilt with the momentum scales—after power-spectrum renormalization, freeing them from the one-loop IR divergences at pivot scale. According to our findings, this equivalency leads to an enhancement in the trustworthiness of the $\delta{\cal N}$ method in the superhorizon scales when the results produced from DRG resummation match the observational data. A comparable approach to the logarithmic IR divergences covered in the previous paragraph is demonstrated by Eqn.(\ref{DRG1}) and Eqn.(\ref{DRG2}). Prior to engaging in the final exponentiated form, each term in the series expansion represents the contribution from all even-order loop correction terms. The terms that represent the contributions from each even-order loop correction term are as follows: $\overline{{\bf X}}_{\bf Loop}\sim {\cal F}_{\bf 2}$, which resembles a two-loop contribution; $\overline{{\bf X}}_{\bf Loop}^{2}\sim {\cal F}_{\bf 4}$, which represents a four-loop contribution; and so forth. This behavior demonstrates how the knowledge of the lowest-order terms in the perturbative expansion provided by the DRG resummation enables an all-order reconstruction of the correlations. 

Further using the resummed spectra one can compute the cosmological beta functions which are given by the following expressions:
\begin{itemize}[label={\checkmark}]
\item The first flow equation which describes the resummed version of the spectral tilt for the scalar modes is given by:
 \bea \label{resumflowtilt}
\left[\overline{\overline{n_{\zeta,\textbf{EFT}}(k)-1}}\right]
&=&\bigg(\bigg[n_{\zeta,\textbf{Tree}}(k)\bigg]_{{\bf SRI}}-1\bigg)+\bigg(\frac{d\overline{\bf X}_{\bf Loop}}{d\ln k}\bigg),\nonumber\\
&=&\bigg(\bigg[n_{\zeta,\textbf{Tree}}(k)\bigg]_{{\bf SRI}}-1\bigg)+\frac{d}{d\ln k}\bigg(\sum_{\textbf{All even graphs G}}{\cal F}_{\bf G}\bigg),\nonumber\\
&=&\bigg(1-\alpha_{\bf C}{\cal A}_{\bf C}-\alpha_{\bf B}{\cal A}_{\bf B}-\alpha_{\bf SRI}{\cal A}_{\bf SRI}-\alpha_{\bf USR}{\cal A}_{\bf USR}-\alpha_{\bf SRII}{\cal A}_{\bf SRII}\bigg)\nonumber\\
&&\quad\times\Bigg(\alpha_{\bf C}{\cal A}_{\bf C}+\alpha_{\bf B}{\cal A}_{\bf B}+\alpha_{\bf SRI}{\cal A}_{\bf SRI}+\alpha_{\bf USR}{\cal A}_{\bf USR}+\alpha_{\bf SRII}{\cal A}_{\bf SRII}\Bigg) \nonumber\\
&&\quad\quad\quad\quad\quad\quad\quad\quad\quad\times\bigg(\bigg[n_{\zeta,\textbf{Tree}}(k)\bigg]_{{\bf SRI}}-1\bigg).\quad\quad\quad\eea

\item The second flow equation which describes the resummed version of the running of the spectral tilt for the scalar modes is given by:
 \bea \label{resumflowalpha}
\left[\overline{\overline{\alpha_{\zeta,\textbf{EFT}}(k)}}\right] 
&=&\bigg[\alpha_{\zeta,\textbf{Tree}}(k)\bigg]_{{\bf SRI}}+\bigg(\frac{d^2\overline{\bf X}_{\bf Loop}}{d\ln k^2}\bigg),\nonumber\\
&=&\bigg[\alpha_{\zeta,\textbf{Tree}}(k)\bigg]_{{\bf SRI}}+\frac{d^2}{d\ln k^2}\bigg(\sum_{\textbf{All even graphs G}}{\cal F}_{\bf G}\bigg),\nonumber\\
&=&\bigg(1-\alpha_{\bf C}{\cal A}_{\bf C}-\alpha_{\bf B}{\cal A}_{\bf B}-\alpha_{\bf SRI}{\cal A}_{\bf SRI}-\alpha_{\bf USR}{\cal A}_{\bf USR}-\alpha_{\bf SRII}{\cal A}_{\bf SRII}\bigg)\nonumber\\
&&\quad\times\Bigg(\alpha_{\bf C}{\cal A}_{\bf C}+\alpha_{\bf B}{\cal A}_{\bf B}+\alpha_{\bf SRI}{\cal A}_{\bf SRI}+\alpha_{\bf USR}{\cal A}_{\bf USR}+\alpha_{\bf SRII}{\cal A}_{\bf SRII}\Bigg)\times\bigg[\alpha_{\zeta,\textbf{Tree}}(k)\bigg]_{{\bf SRI}}.\quad\quad\quad\eea
\item  The third flow equation which describes the resummed version of the running of the running of spectral tilt for the scalar modes is given by:
\bea \label{resumflowbeta}
\left[\overline{\overline{\beta_{\zeta,\textbf{EFT}}(k)}}\right]
&=&\bigg[\beta_{\zeta,\textbf{Tree}}(k)\bigg]_{{\bf SRI}}+\bigg(\frac{d^3\overline{\bf X}_{\bf Loop}}{d\ln k^3}\bigg),\nonumber\\
&=&\bigg[\beta_{\zeta,\textbf{Tree}}(k)\bigg]_{{\bf SRI}}+\frac{d^3}{d\ln k^3}\bigg(\sum_{\textbf{All even graphs G}}{\cal F}_{\bf G}\bigg),\nonumber\\
&=&\bigg(1-\alpha_{\bf C}{\cal A}_{\bf C}-\alpha_{\bf B}{\cal A}_{\bf B}-\alpha_{\bf SRI}{\cal A}_{\bf SRI}-\alpha_{\bf USR}{\cal A}_{\bf USR}-\alpha_{\bf SRII}{\cal A}_{\bf SRII}\bigg)\nonumber\\
&&\quad\times\Bigg(\alpha_{\bf C}{\cal A}_{\bf C}+\alpha_{\bf B}{\cal A}_{\bf B}+\alpha_{\bf SRI}{\cal A}_{\bf SRI}+\alpha_{\bf USR}{\cal A}_{\bf USR}+\alpha_{\bf SRII}{\cal A}_{\bf SRII}\Bigg)\times\bigg[\beta_{\zeta,\textbf{Tree}}(k)\bigg]_{{\bf SRI}}.\quad\quad\eea
\end{itemize}

\section{Comment on the scheme dependence of the renormalization and trustworthiness}
\label{s12}

It is crucial to note, in keeping with the discussion from earlier portions of this work, that we have discovered that the renormalization scheme that is being considered affects the explicit mathematical form of the counter term. It is clearly demonstrated by us that the final determined form of the counter terms differs in the late-time and adiabatic/wavefunction renormalization techniques. It was discovered, nevertheless, that the final computed results for the one-loop momentum integrals, at least for the two renormalization schemes mentioned, exhibit exact equivalency. In both cases, quadratic UV or other similar types of power law divergence can be entirely eliminated, and the result is dependent on a coarse-grained smooth version of logarithmic IR divergence. After performing late-time or adiabatic/wave function renormalization, that is, after the UV divergent quadratic detrimental contribution has been completely eliminated, we must execute the power spectrum renormalization strategy to further smooth out this IR dependency. Our final conclusion is entirely predicated on the particular renormalization strategies used in this linked debate. In addition to the previously listed schemes, which are:
\begin{itemize}[label={\checkmark}]
\item the Late-time (LT) scheme \cite{Choudhury:2023jlt}, 
\item the Adiabatic-Wave function (AWF) scheme \cite{Choudhury:2023vuj}, and 
\item the Power Spectrum (PS) scheme \cite{Choudhury:2023vuj,Choudhury:2023jlt}, \end{itemize}
there are a number of other effective renormalization schemes that can be found in the literature on the quantum field theory of curved space-time:
\begin{itemize}
    \item[\ding{43}] The schemes for minimal subtraction (MS) and modified minimal subtraction ($\overline{\rm MS}$) are presented in \cite{tHooft:1973mfk,Weinberg:1973xwm,Collins:1984xc} and \cite{tHooft:1973mfk,Weinberg:1973xwm,Collins:1984xc},

 \item[\ding{43}] The on-shell approach \cite{Peskin:1995ev}, 
 
 \item[\ding{43}] Bogoliubov-Parasiuk-Hepp-Zimmermann (BPHZ) scheme \cite{Dyson:1949ha, Kraus:1997bi, Piguet:1986ug, Zimmermann:1968mu, Zimmermann:1969jj},

 \item[\ding{43}] Bogoliubov-Parasiuk-Hepp-Zimmermann-Lowenstein (BPHZL) scheme \cite{Lowenstein:1975rg, Lowenstein:1975ps}, 

 \item[\ding{43}] Dimensional Renormalization (DR) scheme \cite{Binetruy:1980xn, Coquereaux:1979eq, Belusca-Maito:2020ala}, 

 \item[\ding{43}] Algebraic Renormalization (AR) scheme \cite{Adler:1969er,Batalin:1981jr,Becchi:1973gu,tHooft:1972tcz,Piguet:1995er}, and many more are included in the list. 

 \end{itemize}
 The analysis for the possibilities indicated in the above-mentioned cases has not been completed yet, and the validity, application, and accuracy of the result reached in this study regarding the proof of a strict no-go theorem on the PBH mass have not been cross-checked. Doing this kind of study in the near future might be intriguing. Therefore, at this point, we can say that, at least in the case of the first three schemes mentioned, one can eliminate the quadratic or similar types of UV divergence, smoothen the logarithmic IR divergence, and ultimately provide a justification to evade the no-go theorem on PBH mass by placing a constraint on the span of USR phase along with the insertion of contracting, bouncing, SRI and SRII phases in order to preserve the perturbative approximations within the currently examined framework. As a result, we, therefore, disclaim any further strong claims about the validity and application of the suggested evading the previously proposed no-go theorem on the PBH mass, as these can only be established following analysis for every class of the aforementioned renormalization schemes. Last but not least, before we wrap up the following part, there is one more significant issue that we must specifically address. Using the DRG resummation approach, we have supplied the resummed version of the one-loop adjusted power spectrum in our study. With this enhanced version of the popular RG resummation method, one can construct a finite and controllable amplitude of the scalar power spectrum, perfectly consistent with cosmological beta functions and the associated slow-roll hierarchy at the CMB pivot scale of the computation, by summing over every secular contribution possible, i.e., using the repetitive structure in the higher order loop diagrams. Consequently, the PBH creation phenomena derived from the DRG resummed spectrum are in full agreement with the Renormalization Group (RG) flow. This is taken into account by the evading of the suggested no-go theorem when determining the formation of PBH mass within the current context. It would be great to include the above-described schemes of renormalization followed by DRG resummation to make further strong remarks concerning the generation of PBHs connected to the current one-loop correction. We are intending to work on these schemes in great detail in the near future.

\section{Numerical outcomes III: Studying the constraints on PBH mass}
\label{s13}

In this section, we study the impact of the primordial scalar power spectrum on generating PBHs with masses that remain of significant interest from a cosmological perspective. We will incorporate the analytical results developed in the previous sections for both the tree-level and the regularized-renormalized-resummed power spectrum and start by demonstrating the differences we observe coming from both types of power spectrum across the five consecutive phases in our setup.

\subsection{Constraints from tree-level power spectrum}

  %%%%%%%%%%%%%%%%%%%%%%%%%%%%%%%%%%%%%%%%%%%%%
\begin{figure*}[htb!]
    	\centering
    \subfigure[]{
      	\includegraphics[width=8.5cm,height=7.5cm]{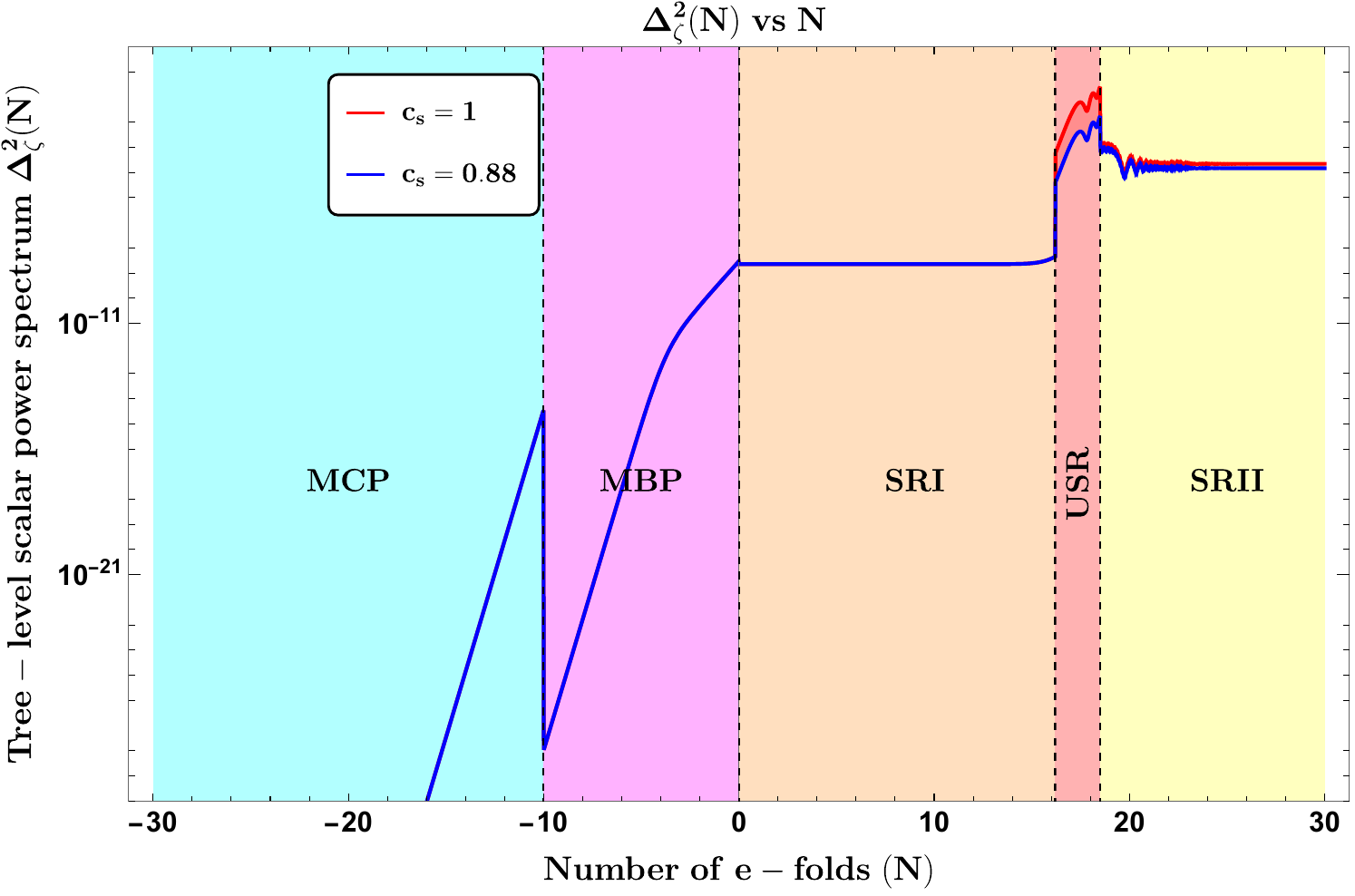}
        \label{treematter}
    }
    \subfigure[]{
       \includegraphics[width=8.5cm,height=7.5cm]{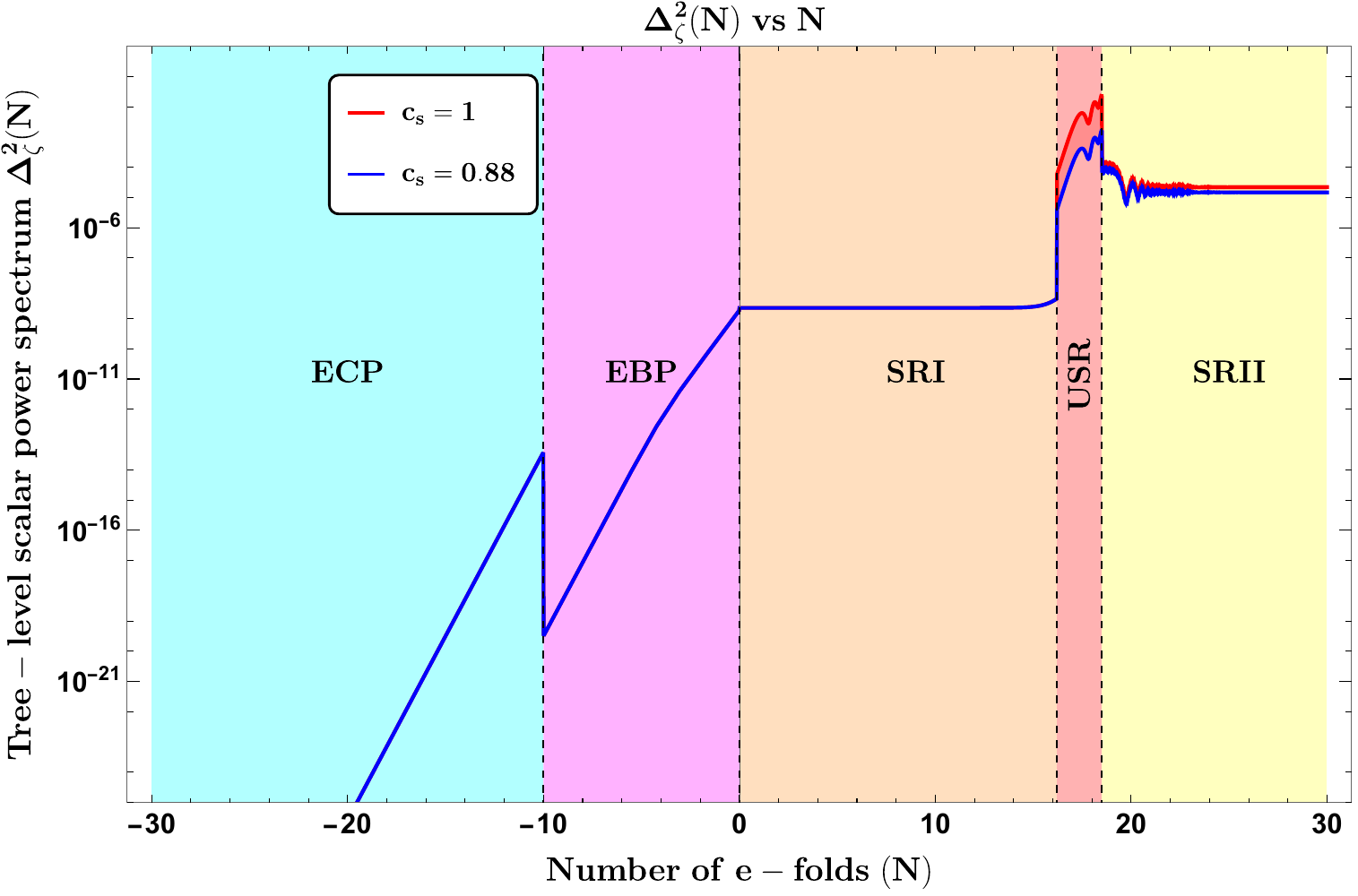}
        \label{treeekpy}
    }
    	\caption[Optional caption for list of figures]{Plots of tree-level scalar power spectrum as a function of the e-foldings ${\rm N}$. In the left we observe the matter contraction (MCP) and bounce (MBP) phases where the slow-roll parameter $\epsilon=3/2$ is fixed. In the right we observe the ekpyrotic contraction (ECP) and bounce (EBP) phases where the slow-roll parameter $\epsilon=7/2$ is fixed. The stages after the bouncing phase remain the same in their behaviour with ${\rm N}$. The red and blue lines correspond to the effective sound speed values $c_{s}=1,0.88$, respectively. } 
    	\label{treelevelpspec }
    \end{figure*}
%%%%%%%%%%%%%%%%%%%%%%%%%%%%%%%%%%%%%%%%%%%%%

We begin with analyzing the tree-level scalar power spectrum from our setup. In fig.(\ref{treematter},\ref{treeekpy}), this behaviour is shown as a function of the number of e-foldings $({\rm N})$. The first thing we notice is that for the phases before the slow-roll conditions, the contraction, and the bounce, the e-foldings are written with a negative signature. During these two phases, the strength of the tree-level power spectrum is extremely sub-dominant relative to its standard result for the SRI amplitude $\sim 2.2\times 10^{-9}$. In the case of the matter-type contraction and bounce scenarios, the amplitude falls drastically with an increase in the duration of both phases. We have shown the contracting phase with a considerable duration of ${\cal O}(20)$ and the bouncing phase with its e-foldings ${\cal O}(10)$. For the same interval length, the tree-level power spectrum on the right with the ekpyrotic contraction and bounce phases show significant enhancements relative to the corresponding matter scenarios on the left. However, the overall strength remains sub-dominant, and both power spectra ultimately join at ${\rm N}=0$ with the amplitude observable from SRI. 

The SRI amplitude stays constant throughout until we encounter a sharp transition at the instant of ${\rm N}\sim {\cal O}(16)$ from where the USR commences. This USR phase lasts for the shortest duration of e-folds $\sim {\cal O}(2)$ and here, the tree-level power spectrum observes the most significant enhancement in its amplitude, leading to values of ${\cal O}(10^{-2})$. The position of the USR phase sets the conditions on the mass of PBH produced, and such magnitudes of amplitude are necessary to generate large enough fluctuations that facilitate PBH formation after Horizon re-entry. Following the exit from the USR via another sharp transition and into the SRII phase, the scalar power spectrum drops in amplitude ${\cal O}(10^{-5})$ and continues till we reach the end of inflation. 

The figures also highlight the role of the effective sound speed in changing the total amplitude. The particular choice of the sound speed values will be made clear in the next subsection where we discuss the one-loop level results of the scalar power spectrum. With the value of $c_{s}=1$ we observe from the spectrum in red the maximum amplitude of order ${\cal O}(10^{-2})$ from the USR while with $c_{s}=0.88$ value the spectrum in blue shows a magnitude of order ${\cal O}(10^{-3})$ in the USR. The impact of $c_{s}$ is much suppressed in the SRII phase as can be seen and no noticeable changes are brought to the SRI phase and the bounce and contraction phases preceding before owning to their highly suppressed amplitudes.

\subsection{Constraints from regularized-renormalized-resummed power spectrum}

  %%%%%%%%%%%%%%%%%%%%%%%%%%%%%%%%%%%%%%%%%%%%%
\begin{figure*}[htb!]
    	\centering
    \subfigure[]{
      	\includegraphics[width=8.5cm,height=7.5cm]{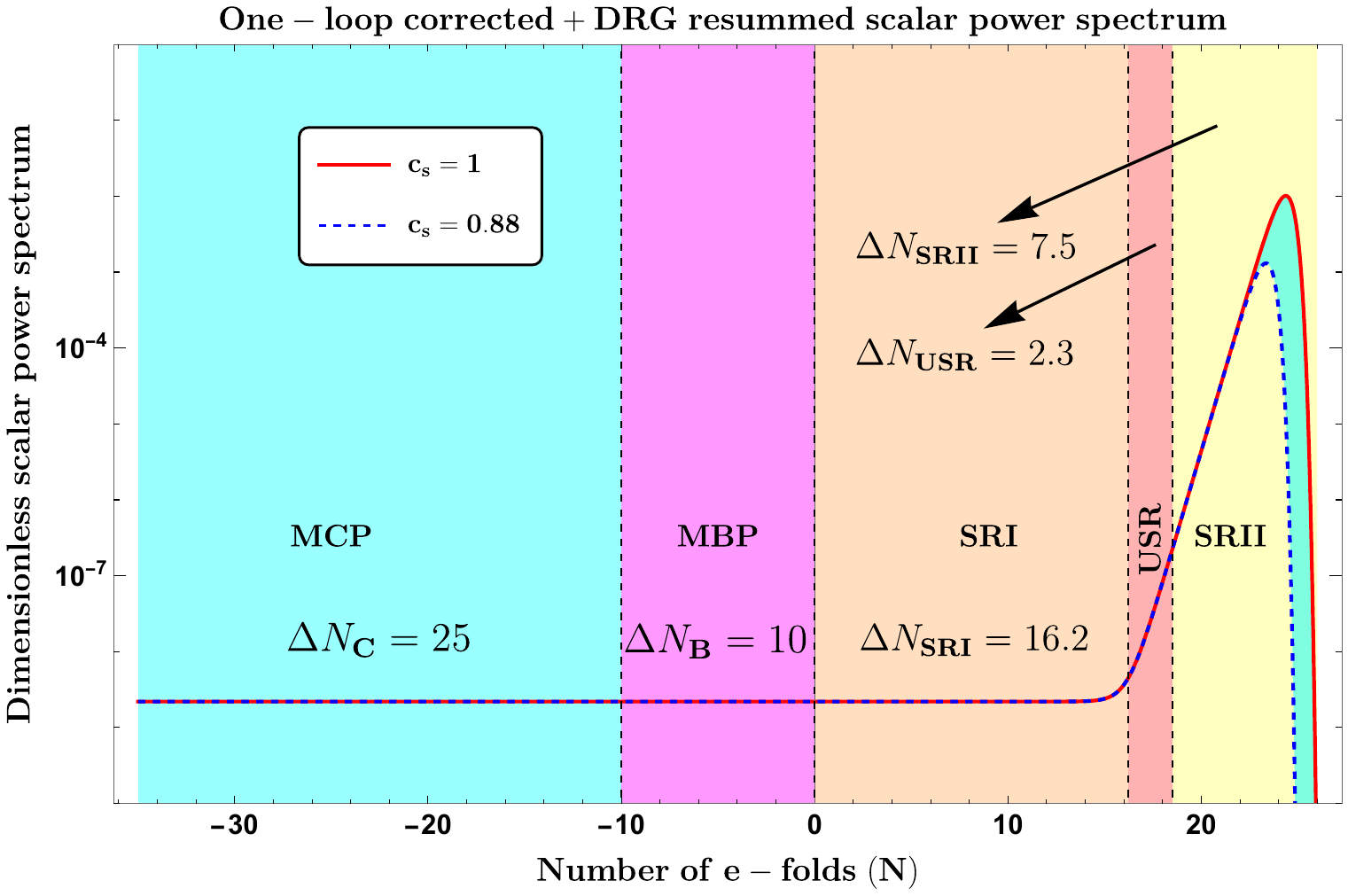}
        \label{drgmatter}
    }
    \subfigure[]{
       \includegraphics[width=8.5cm,height=7.5cm]{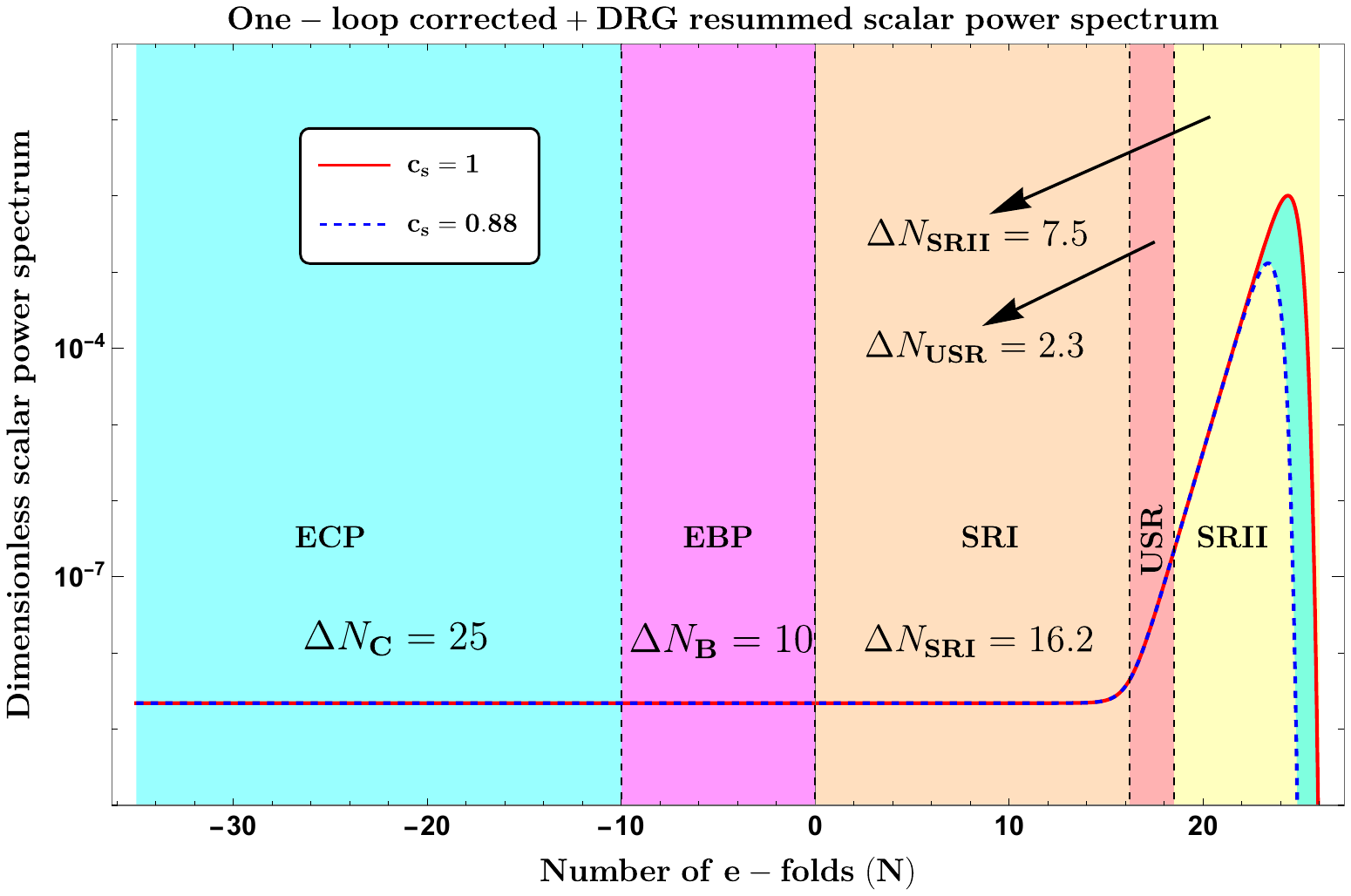}
        \label{drgekpy}
    }
    	\caption[Optional caption for list of figures]{Plots of regularized-renormalized-resummed scalar power spectrum as a function of the e-foldings ${\rm N}$. The left power spectrum contains the matter contraction (MCP) and bounce (MBP) phases with the slow-roll parameter $\epsilon=3/2$ fixed. The right power spectrum contains the ekpyrotic contraction (ECP) and bounce (EBP) phases with the slow-roll parameter $\epsilon=7/2$ fixed. The stages after the bouncing phase remain the same in their behaviour with ${\rm N}$. The red and blue lines correspond to the effective sound speed values $c_{s}=1,0.88$, respectively. } 
    	\label{drgpspec }
    \end{figure*}
%%%%%%%%%%%%%%%%%%%%%%%%%%%%%%%%%%%%%%%%%%%%%

Here, we discuss the nature of the scalar power spectrum that results from the analytic treatment developed thoroughly in sections \ref{s9},\ref{s10}, and \ref{s11}. Fig. (\ref{drgmatter},\ref{drgekpy}) depicts the final version of the regularized-renormalized-resummed scalar power spectrum. The left figure considers the matter-type contraction and bounce scenarios, and the figure on the right focuses on the ekpyrotic-type contraction and bounce. The most important feature to focus on is the exponential increase in the amplitude after the beginning of the USR phase, where the amplitude reaches its value of order ${\cal O}(10^{-2})$ and after which the power spectrum drops, a sudden exponential decrease in its amplitude at around $ N\sim 26$ marking the end of inflation. The total duration of the five consecutive phases sums up to give $\Delta N_{\rm Total} = \Delta N_{\rm C}+\Delta N_{\rm B}+\Delta N_{\rm SRI}+\Delta N_{\rm USR}+\Delta N_{\rm SRII}=  61$. Both the figures show similar behaviour with the e-foldings number ${\rm N}$ and reason for that is the suppressed contributions from the one-loop results in all the phases and inclusion of the tree-level results for bounce and contraction does not make any difference due to them being also extremely small as shown before. 

The red and dashed blue curves highlight the spectra for the values of $c_{s}=1,0.88$, respectively. An immediate change of lowering $c_{s}$ is found to lower the peak amplitude of both spectra to the amplitude $10^{-3}$ with the gray color in between highlights the allowed region of interest. Below this amplitude, the conditions to form PBH reduce significantly and thus are of no interest to us here. This particular response to the choice of having $c_{s}\leq 1$ suggests that the setup preserves the causality and unitarity conditions overall and as going above $c_{s}>1$ would mean raising amplitude to higher than ${\cal O}(10^{-1})$ thus putting perturbativity arguments in danger. We also point our that one can notice similar features in the regularized-renormalized-resummed power spectrum by extending the moment of sharp transition into the USR to a higher value of e-folding and still observe the power spectrum to fall in the upcoming SRII phase in the same amount of duration. By changing the transition point one can address the cases of PBHs in a variety of mass ranging from the high solar mass, ${\cal O}(M_{\odot})$, to the extremely low, ${\cal O}(10^2{\rm gm})$, sub-solar mass scales. In the upcoming sections we will utilise the knowledge gained from studying the tree-level and one-loop corrected and resummed scalar power spectrum to understand their impact on PBH formation across the above mentioned ranges of mass scales.  

\section{Numerical outcomes IV: Studying the cosmological flow equations and corresponding beta functions}
\label{s14}

In this section we understand the behaviour of the three different cosmological flow equations as a function of the e-foldings. 

%%%%%%%%%%%%%%%%%%%%%%%%%%%%%%%%%%%%%%%%%%%%%
\begin{figure*}[ht!]
    	\centering
    \subfigure[]{
      	\includegraphics[width=8.5cm,height=7.5cm]{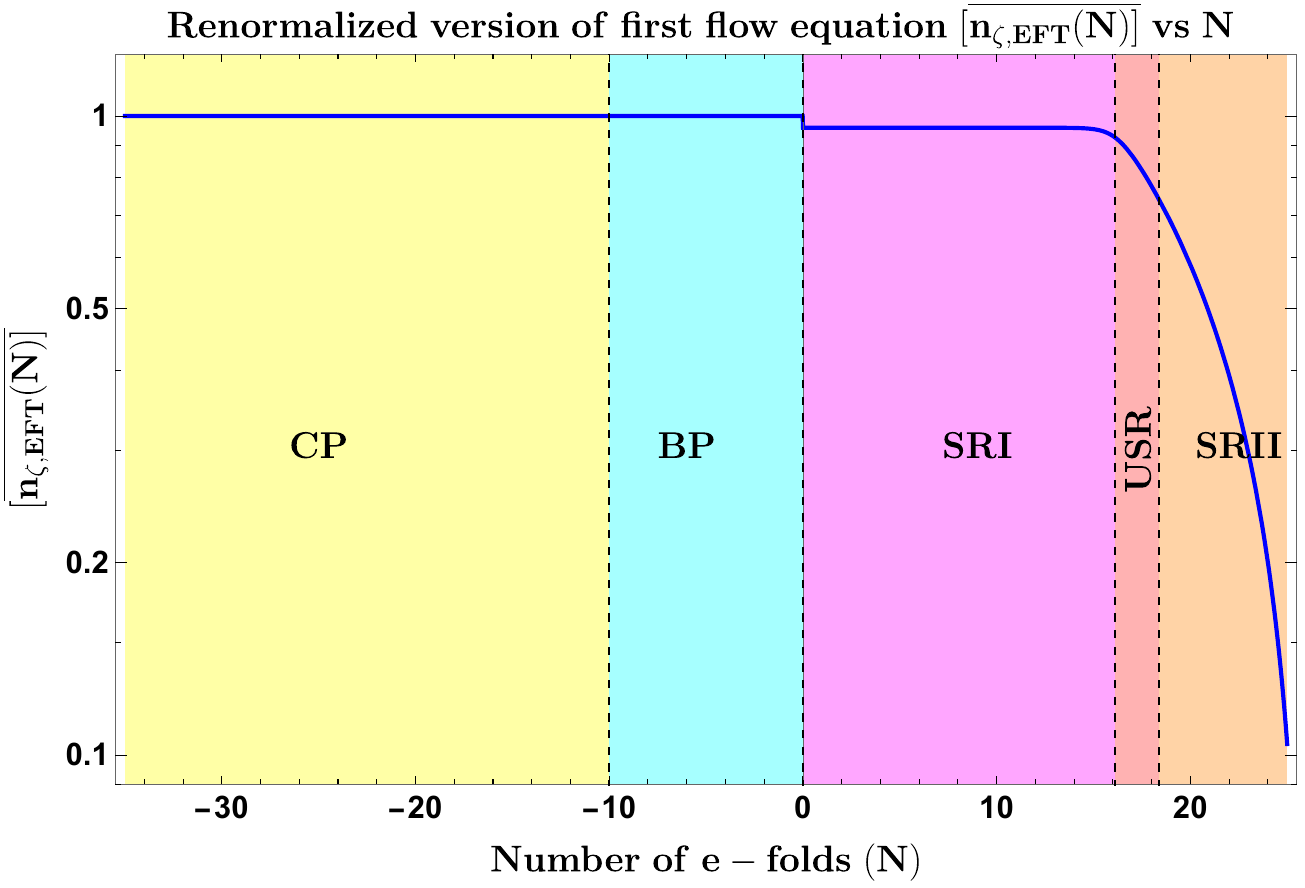}
        \label{renormtilt}
    }
    \subfigure[]{
        \includegraphics[width=8.5cm,height=7.5cm]{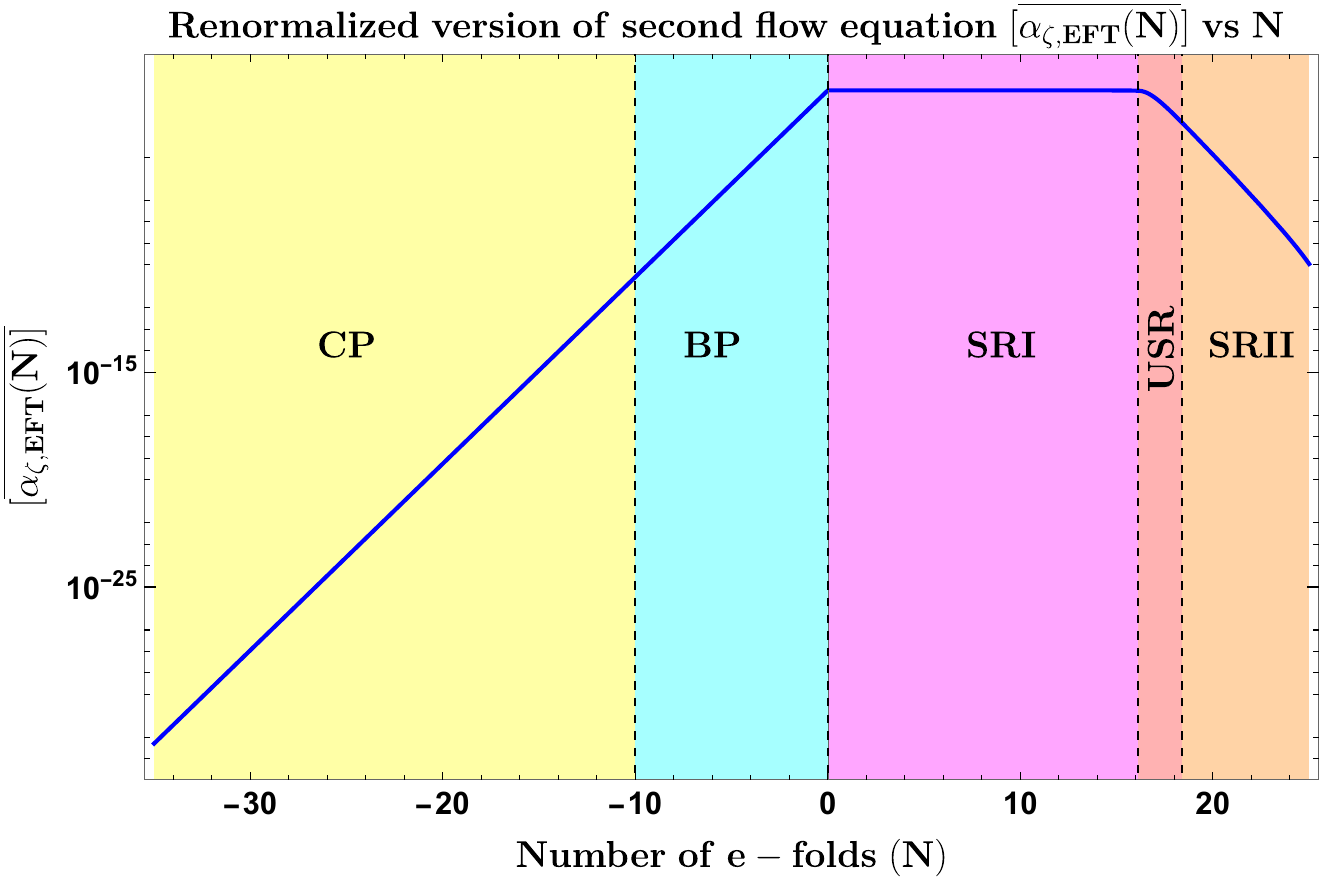}
        \label{renormalpha}
    }
       \subfigure[]{
        \includegraphics[width=9cm,height=7.5cm]{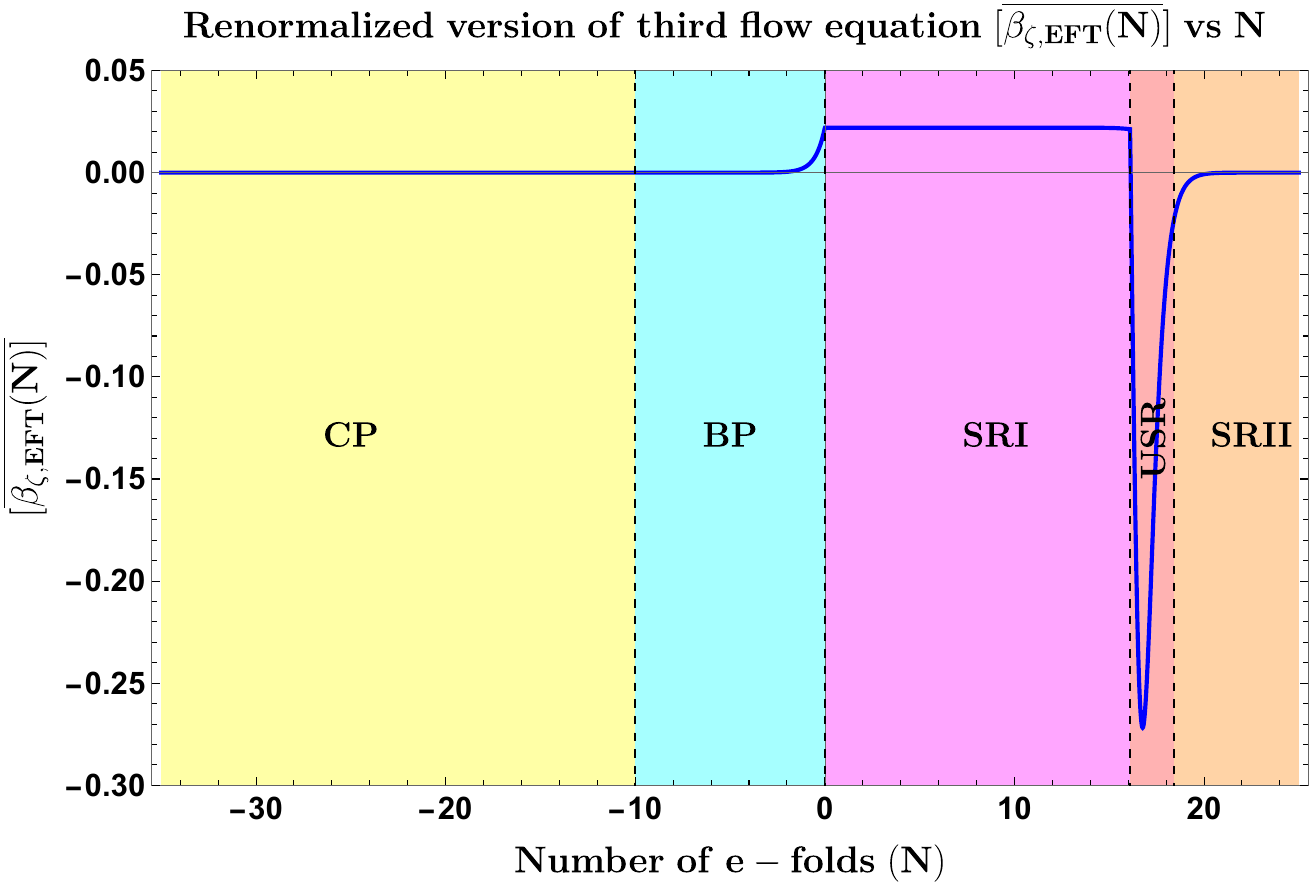}
        \label{renormbeta}
    }
    	\caption[Optional caption for list of figures]{Behaviour of the renormalized versions of the cosmological flow equations as a function of the e-folds ${\rm N}$. The top-row depicts the first and second flow equations and the bottom row the third flow equation. The solid blue curve shows the behaviour for each parameter and the different phases are distinguished by shaded regions. Here CP stands for contracting phase and BP stands for the bouncing phase. } 
    	\label{renormflow}
    \end{figure*}
%%%%%%%%%%%%%%%%%%%%%%%%%%%%%%%%%%%%%%%%%%%%

%%%%%%%%%%%%%%%%%%%%%%%%%%%%%%%%%%%%%%%%%%%%%
\begin{figure*}[ht!]
    	\centering
    \subfigure[]{
      	\includegraphics[width=8.5cm,height=7.5cm]{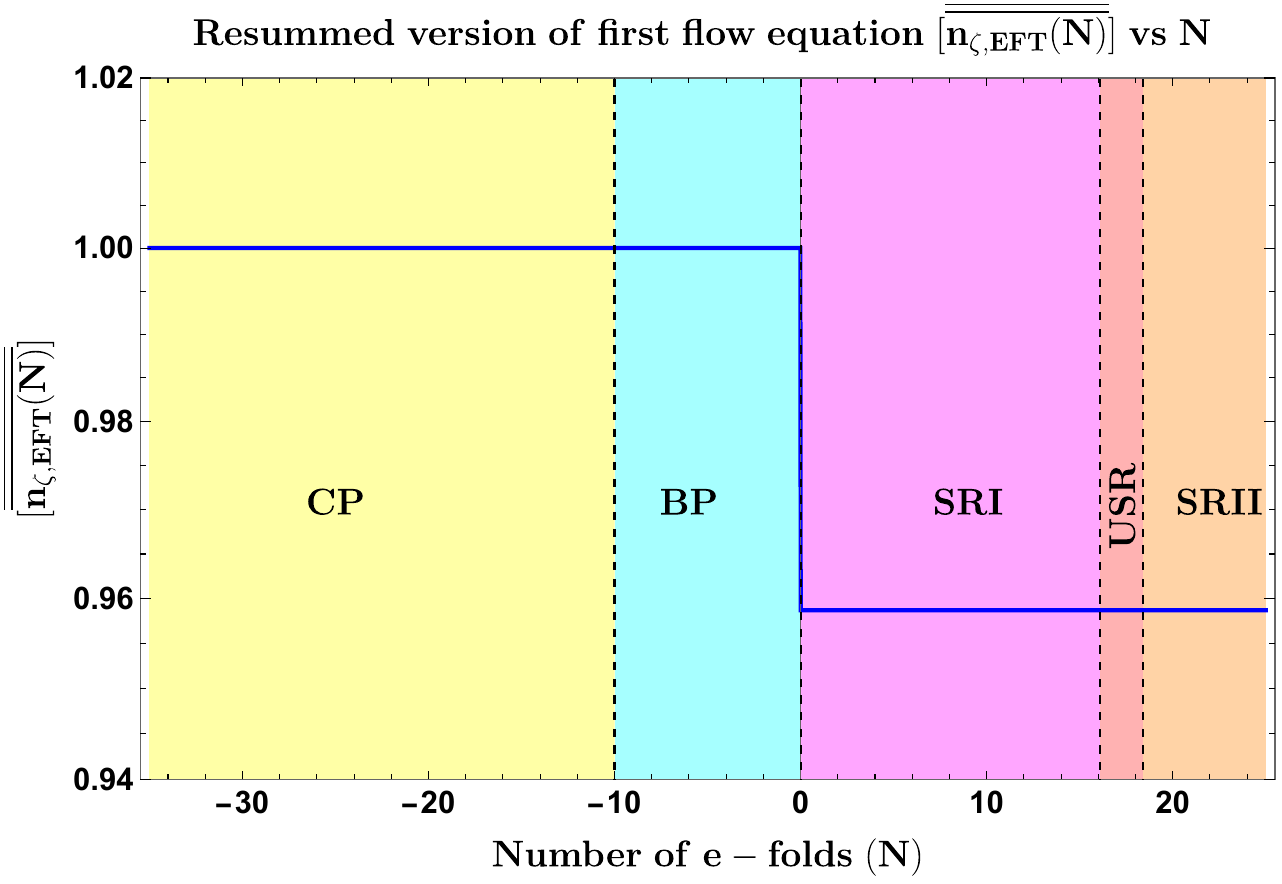}
        \label{resummedtilt}
    }
    \subfigure[]{
        \includegraphics[width=8.5cm,height=7.5cm]{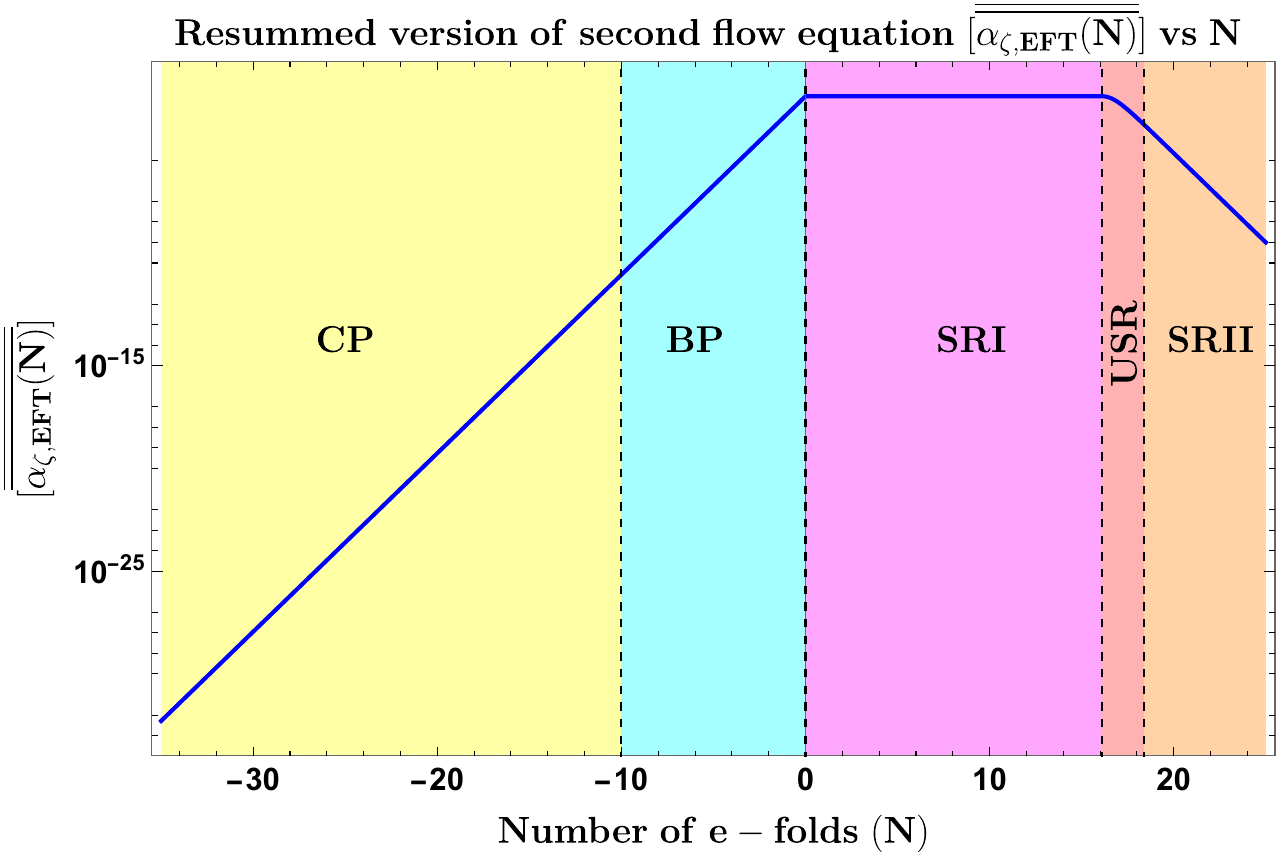}
        \label{resummedalpha}
    }
       \subfigure[]{
        \includegraphics[width=9cm,height=7.5cm]{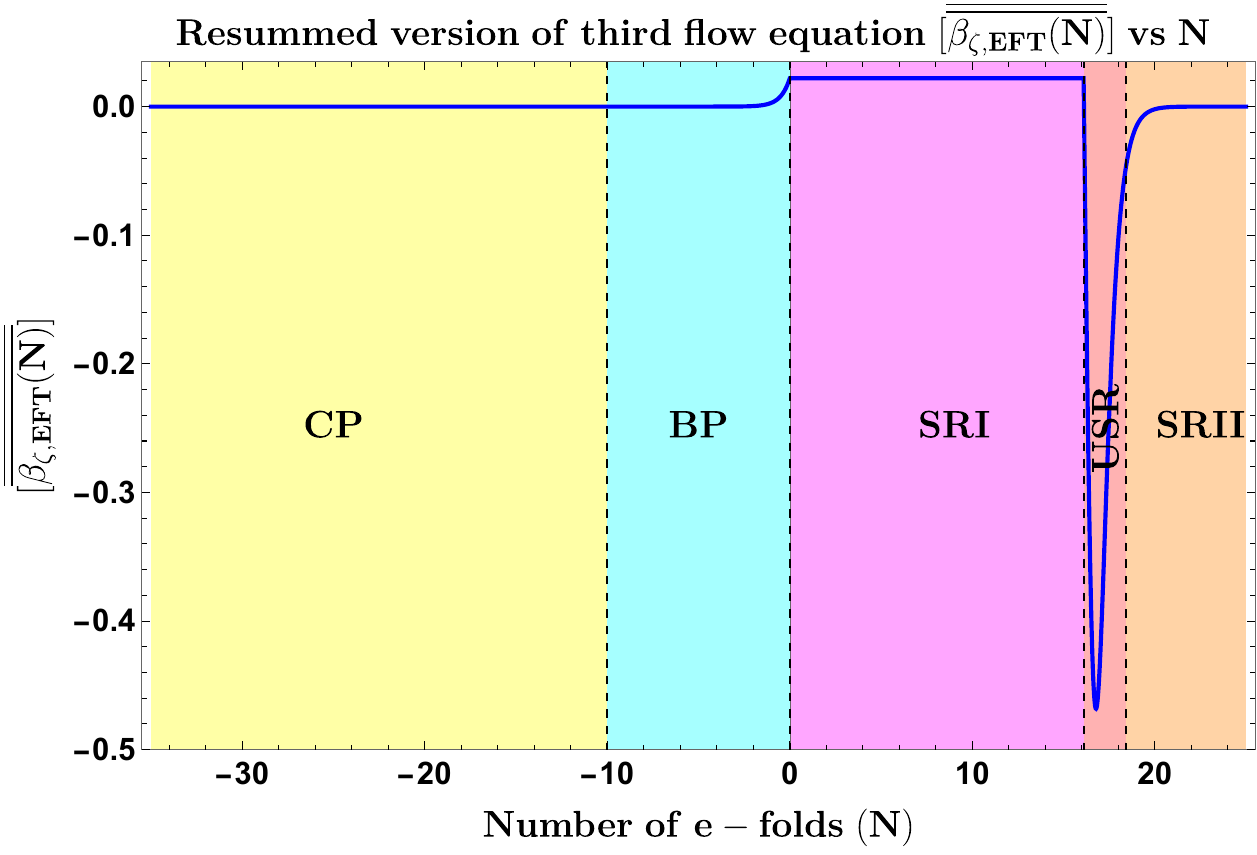}
        \label{resummedbeta}
    }
    	\caption[Optional caption for list of figures]{Behaviour of the resummed versions of the cosmological flow equations as a function of the e-folds ${\rm N}$. The top-row depicts the first and second flow equations and the bottom row the third flow equation. The solid blue curve shows the behaviour for each parameter and the different phases are distinguished by shaded regions. Here CP stands for contracting phase and BP stands for bouncing phase. } 
    	\label{resumflow}
    \end{figure*}
%%%%%%%%%%%%%%%%%%%%%%%%%%%%%%%%%%%%%%%%%%%%

We want to stress the importance of the analysis conducted in this section. The cosmological beta functions are a consequence of maintaining a slow-roll hierarchy in our theory, and without solving the Callan-Symanzik equation, we can capture crucial information about the variation of the underlying couplings of the theory with energy from these flow equations. In a cosmological study, it is impossible to extract information regarding the couplings directly from some observations, in contrast to the studies conducted in high-energy physics, where one can measure the strength of the couplings from experiments. The above cosmological flow equations thus help perform a similar task as observed in high-energy physics but in a different language. These equations capture all the effects of the actual couplings underlying our theory, which can later be used to extract knowledge about physics at the relevant scales instead of trying to go for their direct measurements. In the present context, a qualitative analysis of the flow equations in the presence of quantum loop corrections, followed by the regularization-renormalization-resummation (RRR) procedure, is being done for the first time. Here, we show explicitly how these beta functions behave with their dependence on the scales in each phase after subjecting them to the RRR procedure, such that the boundary conditions for each function are satisfied at the pivot scale that matches with the observational values.

The above fig.\ref{renormflow} depicts the behaviour of the three cosmological flow equations introduced earlier in equations.(\ref{renormflowtilt},\ref{renormflowalpha},\ref{renormflowbeta}). Starting with the top-left fig.\ref{renormtilt}, we observe that during the contracting and bouncing phases, the renormalized version of the spectral tilt, $\overline{n_{\zeta,{\bf EFT}}}$, is almost close to, but slightly larger, than one. It is not less than one before it crosses into the SRI phase at ${\rm N}=0$. During the complete duration of SRI, the value does not deviate from its corresponding pivot scale normalization value \cite{Planck:2018jri}, $n_{s,*}\sim 0.9587$. After exiting from the SRI into the USR, the quantity falls quickly and continues similarly into the SRII till the end of inflation at ${\rm N}\sim 25$, where it reaches close to $\sim 0.105$. 

Next, we consider the renormalized version of the second flow equation in fig.\ref{renormalpha}. This quantity represents the derivative of the spectral tilt with respect to the logarithm, $\ln{k}$. We notice that starting from the contracting phase and moving into the bounce, the quantity $\overline{\alpha_{\zeta,{\bf EFT}}}$, increases monotonically from extremely small values till it encounters the beginning of SRI. Inside the duration of SRI, the value stays at its pivot scale normalization result \cite{Planck:2018jri}, $\alpha_{s,*}\sim 0.013$. The values fall as we continue from the USR, showing a negative slope feature. We can see that each phase, except for the SRI, has prominent scale-dependent features. 

Lastly, we look into the renormalized version of the third flow equation in fig.\ref{renormbeta}. In the contracting and bouncing phases, the quantity starts with an extremely small value but keeps rising till its nature becomes noticeable as the bounce phase ends. After entering the SRI phase, its value equals the normalized estimate at the pivot scale \cite{Planck:2018jri}, $\beta_{s,*}\sim 0.022$. Then, as soon as we cross into the USR phase, the quantity $\overline{\beta_{\zeta,{\bf EFT}}}$ drops rapidly to become negative, stopping close to $\sim -0.27$ and from there it starts rising while still in the USR and end into the SRII as an extremely small but negative value still.

The results for the resummed versions of the cosmological flow equations are displayed in fig.\ref{resumflow}. These quantities were introduced before in the equations.(\ref{resumflowtilt},\ref{resumflowalpha},\ref{resumflowbeta}). In fig.\ref{resummedtilt}, we show how the resummed version of the spectral tilt, $\overline{\overline{n_{\zeta,{\bf EFT}}}}$ changes with the e-folds. Similar to the renormalized version, this quantity is close to one in the contracting and bouncing phases. As we transition into the SRI phase and continue forward, the value remains close to the pivot scale renormalized value of $n_{s,*}\sim 0.9587$. This is in contrast to the renormalized version behaviour after the USR phase. 

In fig.\ref{resummedalpha}, we have shown the resummed version of the second flow equation, $\overline{\overline{\alpha_{\zeta,{\bf EFT}}}}$. The behaviour shown is almost similar to the renormalized version except that towards the end of inflation the renormalized version of this equation falls relatively faster, than its resummed version. In the contracting and bouncing phase it rises in the same monotonic manner till it reaches the pivot scale normalization value in the SRI, $\alpha_{s,*}\sim 0.013$, where it stays the same.

Finally, in fig.\ref{resummedbeta}, we have shown the resummed version of the third flow equation, $\overline{\overline{\beta_{\zeta,{\bf EFT}}}}$. The contracting and bouncing phase features are the same as in the renormalized version, with its value being close to zero but positive. During the SRI, the quantity stays at the pivot scale normalized result, $\beta_{s,*}\sim 0.022$, till it drops rapidly but this time falls to a maximum negative value of $\sim -0.46$. After this, it rises when in the USR and joins with the SRII while still being negative but close to zero till the end of inflation.

\section{PBH formation from EFT of bounce}
\label{s15}

We outline here the production details of PBHs in context of the EFT of bounce and discuss the necessary variables required in the study of PBHs which make up for being possible dark matter candidates. 

 \subsection{Mechanism behind PBH formation}

The formation of PBHs is closely associated with study of large enhancements in the primordial density perturbations. The inclusion of a USR phase in our setup is one example of a means to give the necessary large amount of excitement to the perturbations in the early universe. Upon their re-entry into the horizon, these large curvature perturbations suffer gravitational instability leading to their collapse and formation of PBH after exceeding certain conditions on their threshold of formation. We plan to carry out our present analysis based on the Press-Schechter formalism (threshold statistics) for the PBH formation. 
%Other useful mechanisms to study PBHs have seen increased applications in the recent past like the peak theory \cite{}, or the compaction function approaches \cite{}, but we do not involve ourselves into the complicacies of these for our current work.  

We highlight the relation between properties of the curvature perturbation and the notion of PBH formation mentioned before through use of the following linear relation between these two in the Fourier space:
\bea \label{lineardelta}
\delta_{k} \simeq -\frac{4}{9}\bigg(\frac{k}{aH}\bigg)^{2}\zeta_{k},
\eea
which also translates in to the scalar power spectrum definition that will be important to characterise the other necessary variable related to PBH formation. In this linear regime approximation, the PBH formation threshold is investigated rigorously and gives the interval of $2/5\leq \delta_{\rm th}\leq 2/3$ \cite{Musco:2020jjb} which the most favorable under such conditions.

Before moving to the statistical features of the PBH collapse, we mention the useful relation for relating the masses of PBHs with the scale of sharp transition into the USR $(k_{s})$ from where the large fluctuations are generated:
\bea \label{MPBHks}
\frac{M_{\rm PBH}}{M_{\odot}} = 1.13\times 10^{15}\bigg(\frac{\gamma}{0.2}\bigg)\bigg(\frac{g_{*}}{106.75}\bigg)^{-1/6}\bigg(\frac{k_{*}}{k_{s}}\bigg)^{2},
\eea
here $\gamma\sim 0.2$ is the critical collapse factor, $g_{*}=106.75$ is the total number of standard model relativistic degrees of freedom, and $k_{*}$ is the standard pivot scale.

 \subsection{Mass fraction of produced PBHs}

The consequence of relation in equation(\ref{lineardelta}) can be found when estimating the variance of the density perturbation distribution that is given by:
\bea \label{pbhvariance}
\sigma^{2} = \frac{16}{81}\int\;d\ln{k}\;(kR)^{4}\;W^{2}(kR)\;\overline{\overline{\Delta^{2}_{\zeta,{\bf EFT}}(k)}},
\eea
where $R=1/(c_{s}k_{s})$ is the scale associated with the size of horizon collapsing to form PBHs and $k_{s}$ is the transition scale related to the PBH mass from equation(\ref{MPBHks}). The function $W(kR)$ here acts as a smoothing function for the large perturbations and we choose it to be of the Gaussian form, $\exp{(-k^{2}R^{2}/4)}$. The variance is a quantity highly sensitive to the amplitude of the power spectrum and using this one can determine the mass fraction of PBHs. 

In the current use of threshold statistics, the mass fraction at PBH formation for small variances $(\sigma\ll \delta_{\rm th})$ reads as follows:
\bea \label{pbhmfrac}
\beta(M_{\rm PBH}) \simeq \gamma\times\frac{\sigma}{\sqrt{2\pi}\delta_{\rm th}}\exp{\bigg(\frac{-\delta^{2}_{\rm th}}{2\sigma^{2}}\bigg)},
\eea
where the mass dependence comes from the variance that implicitly depends on $M_{\rm PBH}$ through the scale $k_{s}$.

 \subsection{Calculation of PBH abundance}

With the knowledge of the mass fraction $\beta$ one can determine the abundance of PBH, or the present-day fraction of dark matter (DM) contained in form of PBH through use of the formula:
\bea \label{fpbhformula}
f_{\rm PBH} = 1.68\times 10^{8}\;\bigg(\frac{\gamma}{0.2}\bigg)^{1/2} \bigg(\frac{g_{*}}{106.75}\bigg)^{-1/4} \bigg(\frac{M_{\rm PBH}}{M_{\odot}}\bigg)^{-1/2}\times \beta(M_{\rm PBH})
\eea
where the fraction approaches $f_{\rm PBH}\rightarrow 1$ in the case where the total DM is made up of PBHs. In the case we observe the fraction to go $f_{\rm PBH}\geq 1$, it signals the overproduction of PBHs and one should watch out for conditions that result in such estimates. 

In the next section we consider the numerical results on the PBH abundance obtained from our use of the equations(\ref{pbhvariance},\ref{pbhmfrac},\ref{fpbhformula}) and definitions of the tree-level scalar power spectrum and the regularized-renormalized-resummed power spectrum.

  %%%%%%%%%%%%%%%%%%%%%%%%%%%%%%%%%%%%%%%%%%%%%
\begin{figure*}[htb!]
    	\centering
    \subfigure[]{
      	\includegraphics[width=8.5cm,height=7.5cm]{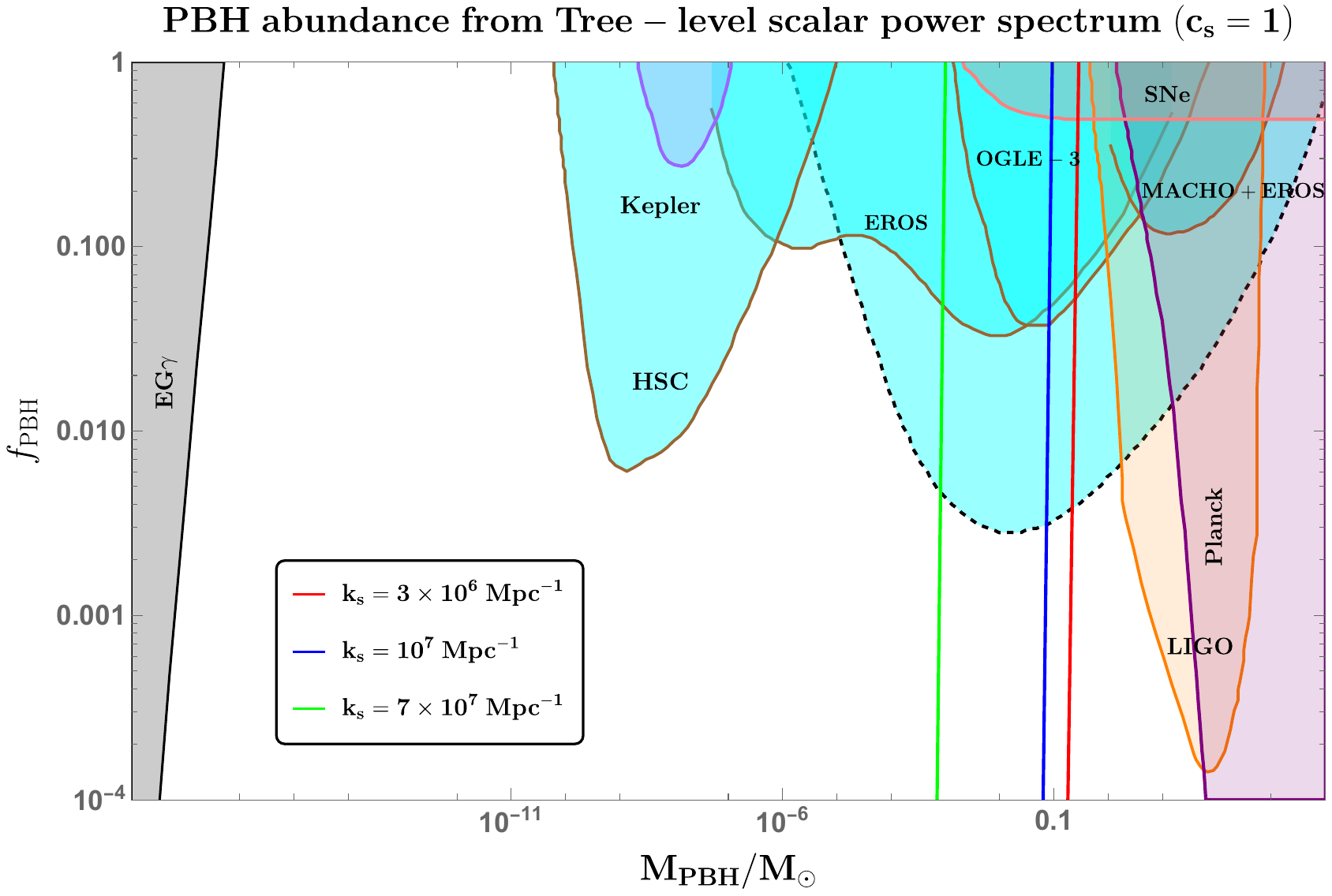}
        \label{fpbhtreecs1HM}
    }
    \subfigure[]{
       \includegraphics[width=8.5cm,height=7.5cm]{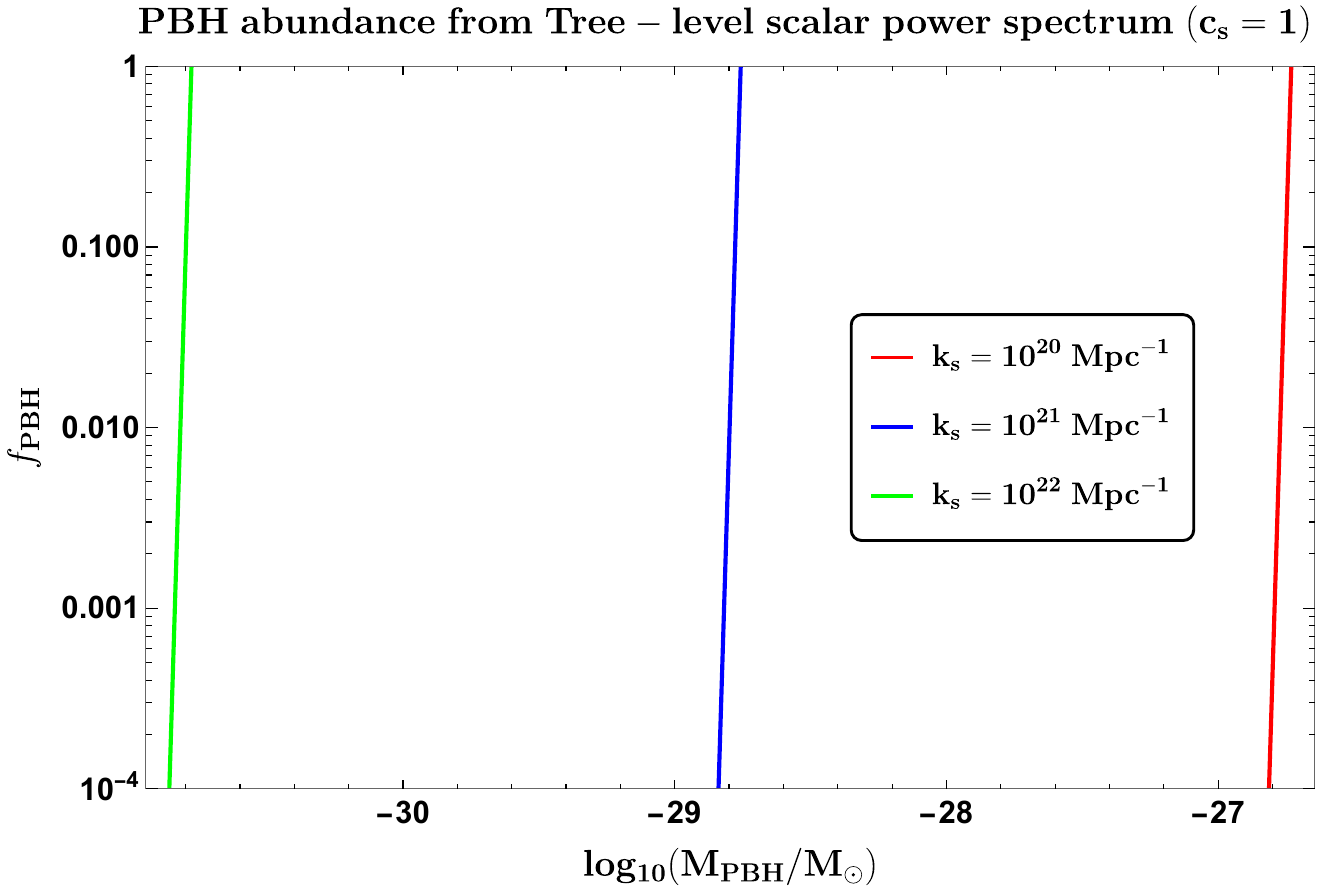}
        \label{fpbhtreecs1LM}
    }
    \subfigure[]{
      	\includegraphics[width=8.5cm,height=7.5cm]{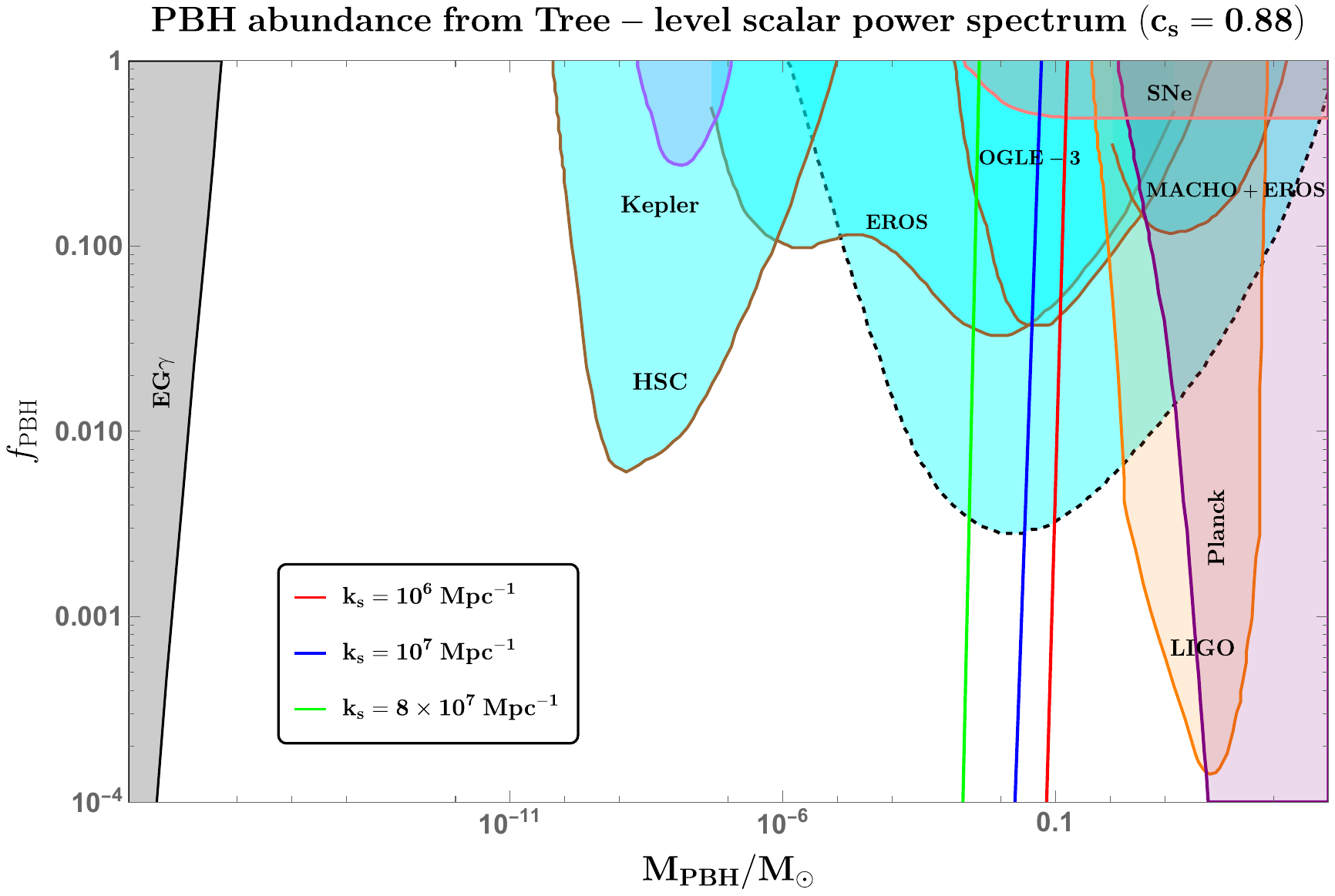}
        \label{fpbhtreecs2HM}
    }
    \subfigure[]{
      	\includegraphics[width=8.5cm,height=7.5cm]{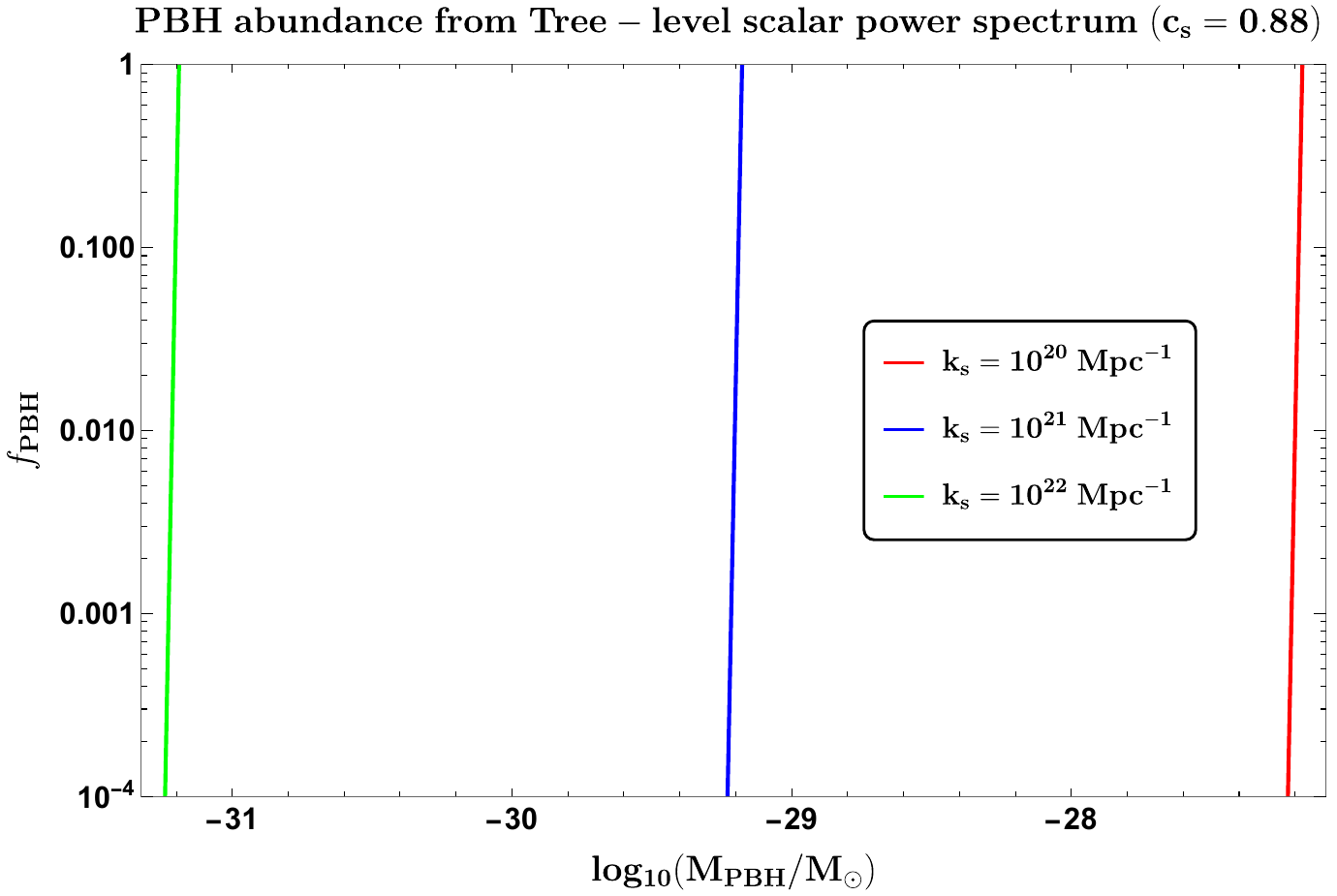}
        \label{fpbhtreecs2LM}
    }
    	\caption[Optional caption for list of figures]{Plots showing the fraction of PBH energy density, $f_{\rm PBH}$ as a function of their masses in solar mass units, $M_{\rm PBH}/M_{\odot}$ calculated using the tree-level scalar power spectrum. The top row focuses on the high (top-left) and extremely low (top-right) mass limits when we fix $c_{s}=1$. The bottom row focuses on the high (bottom-left) and extremely low (bottom-right) mass limits when we fix $c_{s}=0.88$. The different colours, red, blue, and green, denote the various choices of the transition wavenumber $k_{s}$. \textcolor{black}{For $M_{\rm PBH}\sim {\cal O}(10^{-10}-10^{4})M_{\odot}$, we include the latest constraints coming from various microlensing experiments: cyan-coloured regions highlighting the recently obtained $95\%$ upper limits on PBH abundance with the dashed black boundary marking strict limits on $f_{\rm PBH}$ \cite{Mroz:2024mse} and also includes limits from other dark matter surveys: EROS \cite{EROS-2:2006ryy}, OGLE-III \cite{wyrzykowski2011ogle}, MACHO+EROS \cite{Blaineau:2022nhy}, and Subaru Hyper Suprime-Cam (HSC) \cite{Niikura:2017zjd} all with brown borders. There are also constraints in purple, red, and pink for larger than solar mass PBHs coming from the Planck data \cite{Ali-Haimoud:2016mbv}, LIGO \cite{Vaskonen:2019jpv}, and supernovae (SNe) \cite{Zumalacarregui:2017qqd}. The Magenta region inside HSC includes constraints from the Kepler mission \cite{Griest:2013aaa}, and the far black constraints come from evaporation effects \cite{Carr:2009jm} on the extra-galactic photon (EG$\gamma$) background. See \cite{Carr:2020gox} for more details on PBH constraints.} } 
    	\label{treefpbhHMLM}
    \end{figure*}
%%%%%%%%%%%%%%%%%%%%%%%%%%%%%%%%%%%%%%%%%%%%%

 \section{Numerical outcomes V: Constraints from PBH formation}
 \label{s16}

This section focuses on the numerical outcomes concerning PBH formation, and for this, we use the help of the standard PBH formation mechanism and the statistical variables involved in this process. We later confront the values with recent observational constraints coming out of detailed microlensing experiments data analysis. As stated at the end of the previous section, we will use the different forms of the scalar power spectrum to visualize and compare their impact on the abundance of PBH.

% \subsection{Constraints from Mass fraction}

% \subsection{Constraints from PBH abundance}
  %%%%%%%%%%%%%%%%%%%%%%%%%%%%%%%%%%%%%%%%%%%%%
\begin{figure*}[htb!]
    	\centering
    \subfigure[]{
      	\includegraphics[width=8.5cm,height=7.5cm]{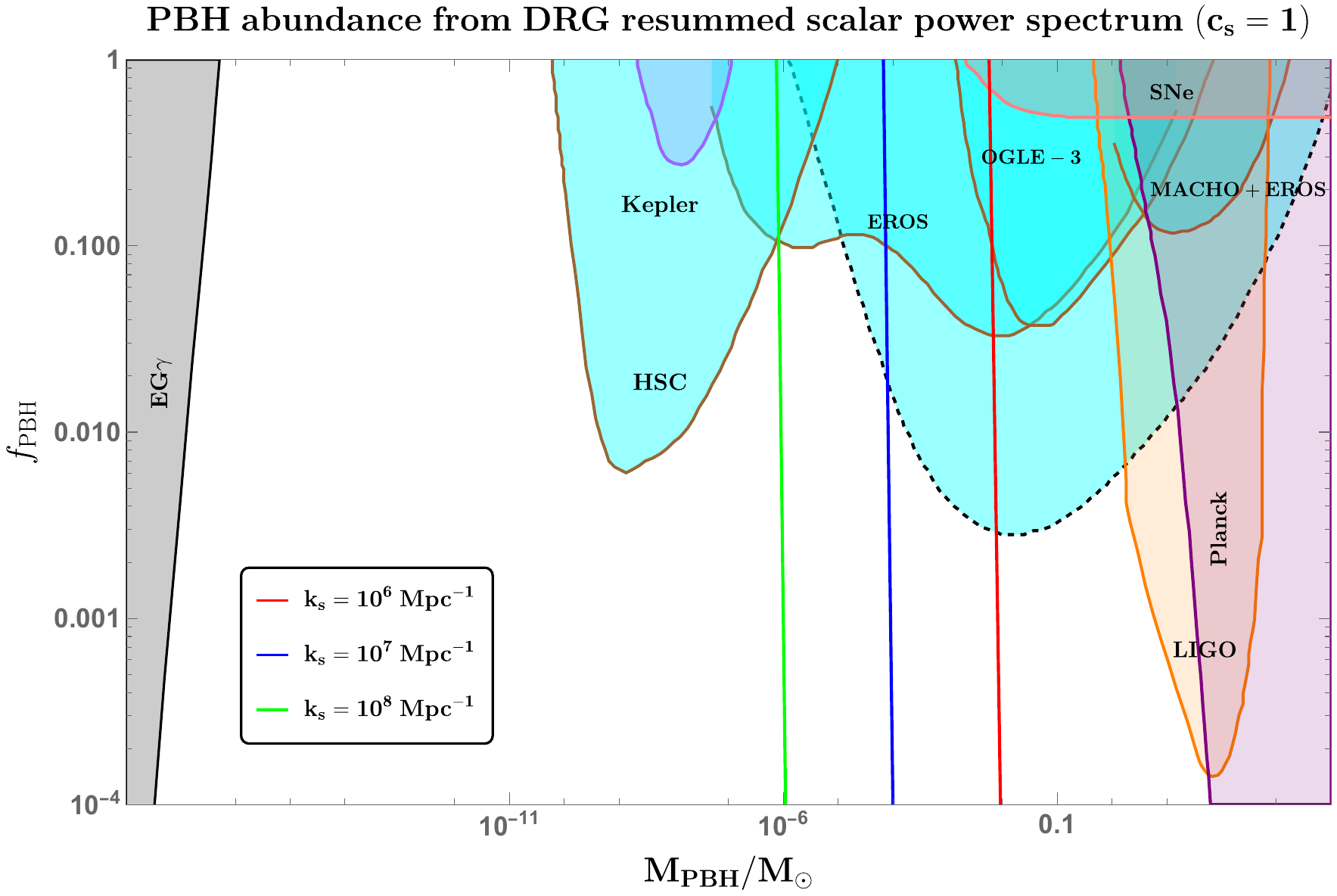}
        \label{fpbhdrgcs1HM}
    }
    \subfigure[]{
      	\includegraphics[width=8.5cm,height=7.5cm]{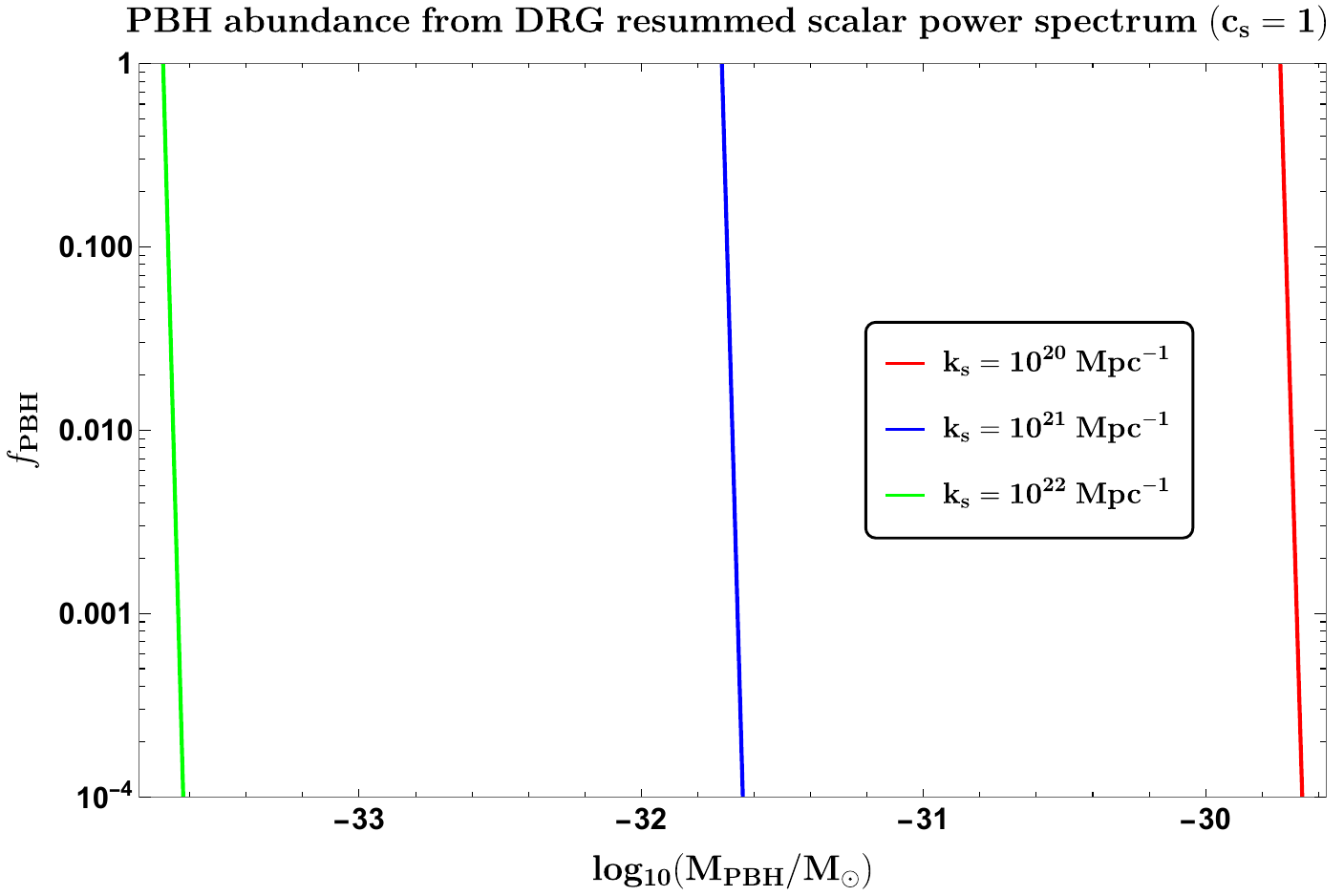}
        \label{fpbhdrgcs1LM}
    }
    \subfigure[]{
      	\includegraphics[width=8.5cm,height=7.5cm]{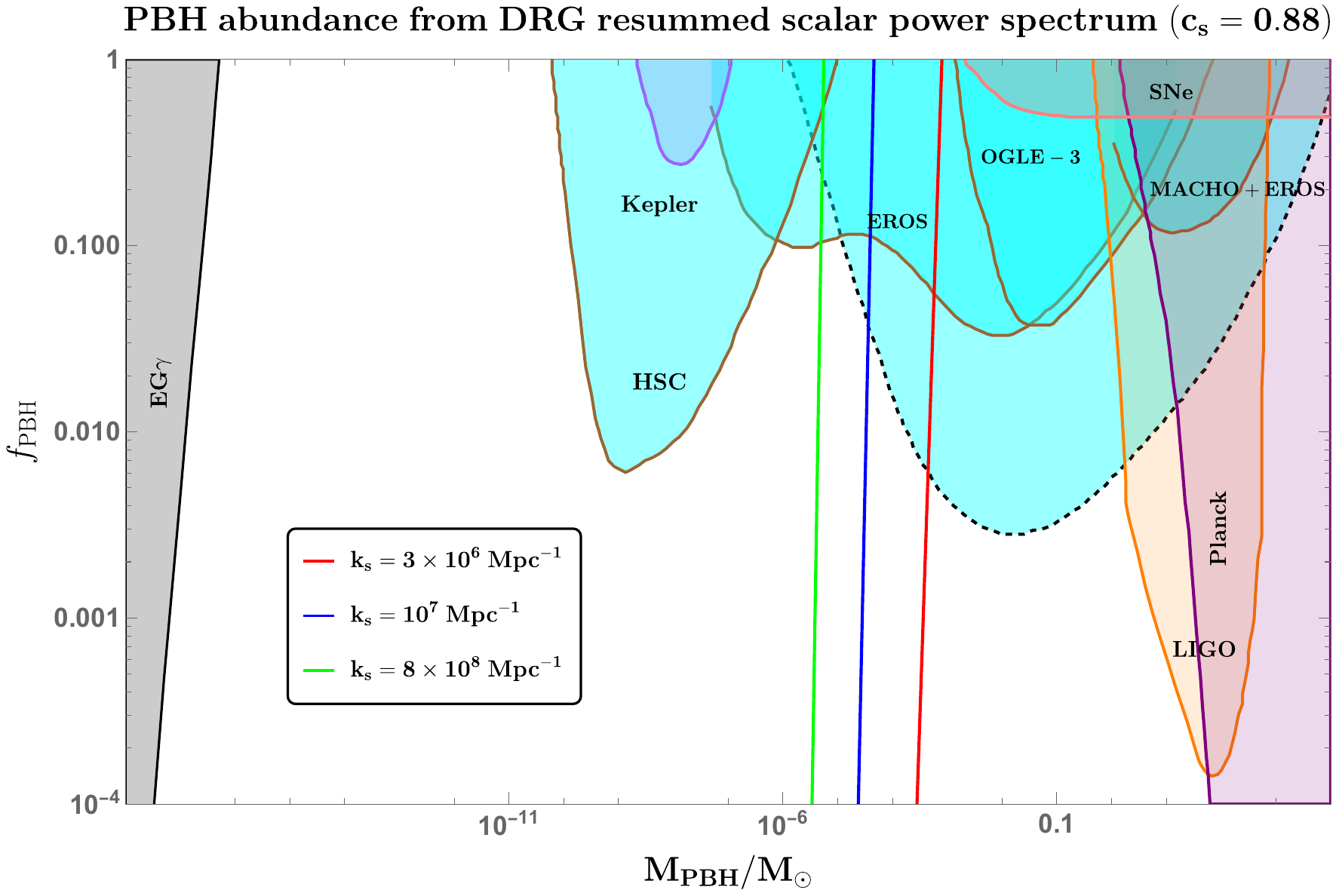}
        \label{fpbhdrgcs2HM}
    }
    \subfigure[]{
      	\includegraphics[width=8.5cm,height=7.5cm]{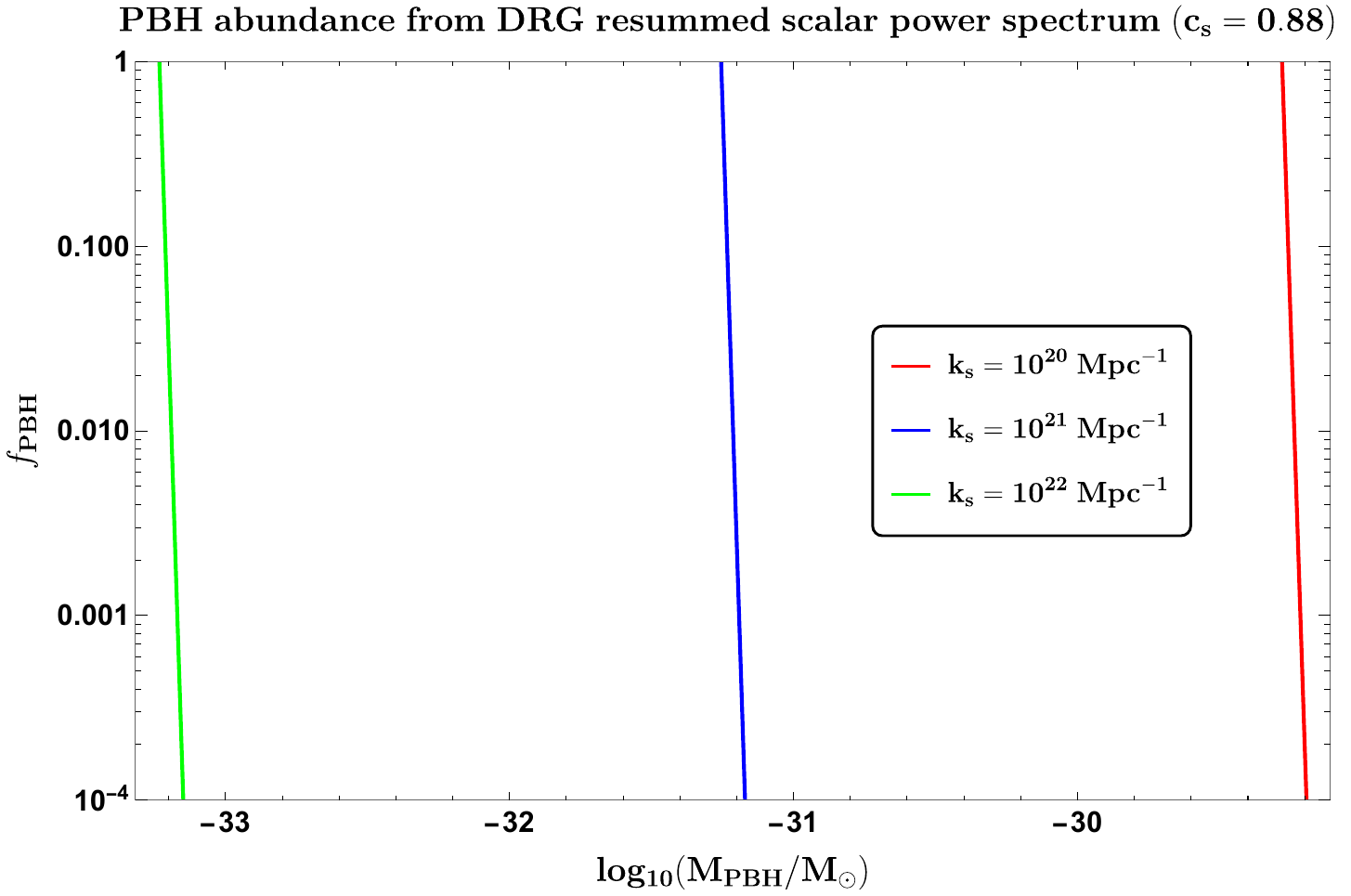}
        \label{fpbhdrgcs2LM}
    }
    	\caption[Optional caption for list of figures]{Plots show the fraction of PBH energy density, $f_{\rm PBH}$, as a function of their masses in solar mass units, $M_{\rm PBH}/M_{\odot}$ calculated using the regularized-renormalized-resummed scalar power spectrum. The top row focuses on the high (top-left) and low (top-right) mass limits when we fix $c_{s}=1$. The bottom row focuses on the high (bottom-left) and low (bottom-right) mass limits when we fix $c_{s}=0.88$. The different colours red, blue, and green, denote the various choices of the transition wavenumber $k_{s}$. \textcolor{black}{Equivalent constraints are displayed as in the case for the high-mass limit with the tree-level power spectrum in subfigures \ref{fpbhtreecs1HM} and \ref{fpbhtreecs2HM}. } } 
    	\label{DRGfpbhHMLM}
    \end{figure*}
%%%%%%%%%%%%%%%%%%%%%%%%%%%%%%%%%%%%%%%%%%%%%

From the two figures \ref{treefpbhHMLM},\ref{DRGfpbhHMLM}, we can observe the variation of abundance of PBH as a function of their masses present in the low and the high mass ranges. We also consider the impact that a different value of $c_{s}$ brings in the fraction of PBHs formed. Here, we start with results from using the tree-level scalar power spectrum and then move on to using the regularized-renormalized-resummed scalar power spectrum. 

In fig.(\ref{treefpbhHMLM}), we highlight this behaviour considering different $c_{s}$ and $k_{s}$ values. From the top row, figs.(\ref{fpbhtreecs1HM},\ref{fpbhtreecs1LM}), we observe the abundance values decrease quickly to $f_{\rm PBH}\lesssim 10^{-4}$ as the mass increases in presence of $c_{s}=1$. We see that the relationship of the fraction $f_{\rm PBH}$ with the mass $M_{\rm PBH}$ in both the high and small mass limits falls off to become negligible in a short interval where the different masses are considered on basis of the choice for the transition wavenumber $k_{s}$. The range of interest significantly decreases when we aim for the sub-solar mass values on the right. The tree-level power spectrum allows for higher abundance for fairly large mass PBHs and we will see this fact change in the fig.(\ref{DRGfpbhHMLM}).
When we start to consider the lower value of $c_{s}=0.88$, see figs.(\ref{fpbhtreecs2HM},\ref{fpbhtreecs2LM}), it gives large abundance estimates when looking into the PBH masses lower relative to the $c_{s}=1$ case. The impact of variable $c_{s}$ is less visible when we are concerned with the sub-solar masses on the right. Another effect of changing $c_{s}$ is seen with the possible mass values allowed till abundance becomes negligible. In the case of lower $c_{s}$, the masses allowed are smaller than $c_{s}=1$, but the range of non-negligible abundance values is spread out larger than $c_{s}=1$.

We now compare the above scenarios with the PBH abundance estimated using the regularized-renormalized-resummed scalar power spectrum.
From the case of $c_{s}=1$ in the top row, we again observe relationship of the fraction $f_{\rm PBH}$ with the mass $M_{\rm PBH}$ in both the high (near solar \ref{fpbhdrgcs1HM}) and small (sub-solar \ref{fpbhdrgcs1LM}) mass limits falling to become negligible in a short interval. The low-mass PBHs fall rapidly in mass ranges showing any significant abundance. In the case of the large mass PBHs, we consider up to $M_{\rm PBH}\sim {\cal O}(10^{-6}-10^{-2})$ where in a small gap of the mass values, the abundance falls quickly to become $f_{\rm PBH}\lesssim 10^{-4}$. An important distinction to note when comparing with the similar scenario of $c_{s}=1$ in fig.(\ref{treefpbhHMLM}) is that the tree-level power spectrum only allowed for higher abundance for PBH masses larger relative to the ones allowed with the regularized-renormalized-resummed scalar power spectrum. The pattern persists when we study the $c_{s}=0.88$ case in the bottom row with figs.\ref{fpbhtreecs2HM},\ref{fpbhtreecs2LM}.

In contrast, when using the other value of $c_{s}=0.88$, the abundance $f_{\rm PBH}$ falls again in a short interval of PBH masses but for the case of high PBH mass, fig.\ref{fpbhdrgcs2HM}, the values shown are smaller compared to the previous case of $c_{s}=1$. Since the peak amplitude of the scalar power spectrum for $c_{s}=0.88$ is smaller by an order of magnitude with value $10^{-3}$, we can see why we get greater abundance estimates for the similar transition wavenumber with masses slightly smaller than when $c_{s}$ was unity with power spectrum amplitude $10^{-2}$. The situation for the sub-solar masses when $c_{s}=0.88$, fig.\ref{fpbhdrgcs2LM}, is almost the same as for $c_{s}=1$. So changing $c_{s}$ does not affect much the abundances of the extremely small PBHs. \textcolor{black}{In tables \ref{tab1tree} and \ref{tab2drg} we summarize the results for the PBH masses and their abundance as plotted inside the figures \ref{treefpbhHMLM}-\ref{DRGfpbhHMLM} after accounting for the constraints coming from various experiments. In both tables we mention the entries corresponding to the two separate scenarios of the sound speed $c_{s}=1,0.88$. }

\begin{table} 
  \begin{tabular}{cccc|ccc}
    \toprule
    \multirow{2}{*}{Effective sound speed} &
      \multicolumn{3}{c|}{High mass} &
      \multicolumn{3}{c}{Low mass} \\
      $c_{s}$ & $M_{\rm PBH}$ & $f_{\rm PBH}$ & $k_{s}/{\rm Mpc^{-1}}$ & $M_{\rm PBH}$ & $f_{\rm PBH}$ & $k_{s}/{\rm Mpc^{-1}}$ \\
      \midrule\midrule
     & $(8.1\times 10^{-4},7\times 10^{-4})$ & $(0.0046,10^{-4})$ &  $7\times 10^{7}$ & $(10^{-28.76},10^{-28.84})$ & $(1,10^{-4})$  & $10^{22}$\\
    \cmidrule{2-7}
    1 & $(0.072,0.063)$ & $(0.003,10^{-4})$ & $10^{7}$ & $(10^{-30.78},10^{-30.86})$ & $(1,10^{-4})$ & $10^{21}$ \\
    \cmidrule{2-7}
     & $(0.215,0.18)$ & $(0.0037,10^{-4})$ & $3\times 10^{6}$ & $(10^{-26.73},10^{-26.81})$ & $(1,10^{-4})$ & $10^{20}$ \\
    \hline
     & $(2.6\times 10^{-3},2\times 10^{-3})$ & $(0.0033,10^{-4})$ & $8\times 10^{7}$ & $(10^{-31.19},10^{-31.24})$ & $(1,10^{-4})$ & $10^{22}$  \\
    \cmidrule{2-7}
    0.88 & $(0.027,0.018)$ & $(0.0028,10^{-4})$ & $10^{7}$ & $(10^{-29.18},10^{-29.23})$ & $(1,10^{-4})$ & $10^{21}$  \\
    \cmidrule{2-7}
     & $(0.095,0.068)$ & $(0.0031,10^{-4})$ & $10^{6}$ & $(10^{-27.17},10^{-27.23})$ & $(1,10^{-4})$ & $10^{20}$ \\
    \bottomrule
  \end{tabular}
  \label{tab1tree}
  \caption{ \textcolor{black}{Summary of the values taken by various parameters such as PBH masses $M_{\rm PBH}$, in the high (near solar-mass) and low (extremely smaller than solar-mass) mass regimes, their respective abundances $f_{\rm PBH}$, and the peak wavenumber $k_{s}$ when using the tree-level scalar power spectrum. The set of parameters are listed for two separate values of $c_{s}=1,0.88$.} }
\end{table}

\begin{table} 
  \begin{tabular}{cccc|ccc}
    \toprule
    \multirow{2}{*}{Effective sound speed} &
      \multicolumn{3}{c|}{High mass} &
      \multicolumn{3}{c}{Low mass} \\
      $c_{s}$ & $M_{\rm PBH}/M_{\odot}$ & $f_{\rm PBH}$ & $k_{s}/{\rm Mpc^{-1}}$ & $M_{\rm PBH}/M_{\odot}$ & $f_{\rm PBH}$ & $k_{s}/{\rm Mpc^{-1}}$ \\
      \midrule\midrule
     & $(7.5\times 10^{-7}, 1.1 \times 10^{-6})$ & $(1,10^{-4})$ & $10^{8}$ & $(10^{-33.69},10^{-33.62})$ & $(1,10^{-4})$ & $10^{22}$ \\
    \cmidrule{2-7}
    1 & $(7.8\times 10^{-5},1\times 10^{-4})$ & $(0.018,10^{-4})$ & $10^{7}$ & $(10^{-31.71},10^{-31.64})$ & $(1,10^{-4})$ & $10^{21}$ \\
    \cmidrule{2-7}
     & $(7.7\times 10^{-3},9.2\times 10^{-3})$ & $(0.0028,10^{-4})$ & $10^{6}$ & $(10^{-29.74},10^{-29.66})$ & $(1,10^{-4})$ & $10^{20}$ \\
    \hline
     & $(5.1\times 10^{-6}, 3.4 \times 10^{-6})$ & $(0.246,10^{-4})$ & $8\times 10^{8}$ & $(10^{-33.23},10^{-33.15})$ & $(1,10^{-4})$ & $10^{22}$ \\
    \cmidrule{2-7}
    0.88 & $(3.6\times 10^{-5}, 2.4 \times 10^{-5})$ & $(0.035,10^{-4})$ & $10^{7}$ & $(10^{-31.25},10^{-31.17})$ & $(1,10^{-4})$ & $10^{21}$  \\
    \cmidrule{2-7}
     & $(4.4\times 10^{-4}, 2.8 \times 10^{-4})$ & $(0.006,10^{-4})$ & $3\times 10^{6}$ & $(10^{-29.28},10^{-29.19})$ & $(1,10^{-4})$ & $10^{20}$ \\
    \bottomrule
  \end{tabular}
  \label{tab2drg}
  \caption{ \textcolor{black}{Summary of the values taken by various parameters such as PBH masses $M_{\rm PBH}$, in the high (near solar-mass) and low (extremely smaller than solar-mass) mass regimes, their respective abundances $f_{\rm PBH}$, and the peak wavenumber $k_{s}$ when using the DRG resummed scalar power spectrum. The set of parameters are listed for two separate values of $c_{s}=1,0.88$.} }
\end{table}

% \begin{table}[H]
% \vspace{-0.6cm}
%   \centering
%   \caption{}
% \begin{tabular}{ccc}
%   \hline
%   % after \\: \hline or \cline{col1-col2} \cline{col3-col4} ...
% $\#$ & \quad $n_{s}$\quad &\quad $M_{\rm PBH}$\quad & \quad $f_{\rm PBH}$ \quad \\ \hline\hline
% val1 &\quad val1b \quad  &\quad val1c\quad & \quad val1d \quad \\ \hline
% val2 &\quad val2b\quad  & \quad val1c\quad & \quad val2d\quad\\ \hline
% val3 &\quad val3b\quad  & \quad val3c\quad & \quad val3d\quad\\ \hline
% \end{tabular}
% \label{tab1}
% \end{table}

\section{Spectral distortion and PBH overproduction}
\label{s17}

In this section we examine another effect of observational importance which are the spectral distortions in the CMB background and analyse the results that we find from the use of our scalar power spectrum and also comment on the overproduction issue.

The CMB spectral features are very close to that predicted by the theory of a perfect black-body. However, its frequency spectrum can suffer small departures from equilibrium when we consider the history of the early universe and the variety of processes that can occur at larger redshift $(z)$ values. Detection of these distortions thus directly allows us to probe the very early universe history, making such effects increasingly significant. The thermal history of the early universe can be understood much better by examining the CMB spectral distortions as accurately as possible. Different distortion effects are generated and classified based on the redshift, which ultimately results from the radiation and matter content that goes out of thermal equilibrium, see refs.\cite{zeldovich1969interaction,Sunyaev:1970eu,Illarionov:1975ei,Hu:1992dc,Chluba:2012we,Khatri:2012tw,Chluba:2003exb,Stebbins:2007ve,Chluba:2011hw,Pitrou:2014ota,Hooper:2023nnl,Deng:2021edw} for detailed study of on such effects.

The primordial density perturbations exhibit dissipation effects in the early universe, which generates the distortions that are observable in the CMB spectrum. These important signals are of two major types, the $\mu-$ and $y-$ distortions, and depend on the strength of energy exchanges between the photons and the surrounding electrons. At large redshifts, $z> 2\times 10^{6}$, the era of well-maintained black-body spectrum continues with almost negligible spectral distortions. In the range of $2\times 10^{5}< z< 2\times 10^{6}$, the Compton scattering processes remain efficient, leading to a Bose-Einstein distribution for the photon energies that suggests a non-zero chemical potential $(\mu)$ value and defines the $\mu$ distortion era. With redshifts lower than $z\lesssim 2\times 10^{5}$, the scattering processes become inefficient, with the photon energy distribution suffering departures from equilibrium and leading to spectral distortions labeled as the $y$ distortions.

The two distortion effects can be estimated using the scalar power spectrum through the following relations \cite{Hooper:2023nnl}:
\bea \label{mudistortion}
\mu &\simeq& 2.2\int_{k_{i}}^{\infty}\frac{dk}{k}\Delta^{2}_{\zeta}(k)\bigg\{\exp{\bigg[-\frac{k}{5400\;{\rm Mpc^{-1}}}\bigg]}-\exp{\bigg[-\bigg(\frac{k}{31.6\;{\rm Mpc^{-1}}}\bigg)^{2}\bigg]}\bigg\},\\
\label{ydistortion}
y &\simeq& 0.4\int_{k_{i}}^{\infty}\frac{dk}{k}\Delta^{2}_{\zeta}(k)\exp{\bigg[-\bigg(\frac{k}{31.6\;{\rm Mpc^{-1}}}\bigg)^{2}\bigg]},
\eea
where we keep $k_{i}=1\;{\rm Mpc^{-1}}$. In the above expressions, the required scalar power spectrum comes from either the tree-level version in equation(\ref{treetotalpspec}), $\Delta^{2}_{\zeta,{\bf Tree-Total}}(k)$, or the regularized-renormalized-resummed version in equation(\ref{DRG2}), $\overline{\overline{\Delta^{2}_{\zeta,{\bf EFT}}(k)}}$. We show the resulting values and related analysis through use of each of them in the next section. Also related to this is the fact that accurate and detailed constraints exist from various observations on the scalar power spectrum amplitude at different scales involved during inflation. The distortion effects help to provide a large set of such strong constraints on the amplitude, and in the upcoming section, we also visualise the behaviour of the current amplitude with changing wavenumbers and how it compares when contrasted with the space of constraints. For this purpose, we mention the formula for the peak amplitude of the power spectrum:
\bea \label{amplitude}
A(k_{s}) = \bigg(\frac{H^2}{8\pi^2\epsilon  c_s M_p^2}\bigg)_{*}\bigg(\frac{k_{e}}{k_{s}}\bigg)^{6},
\eea
where $k_{s}$ is the scale associated with the sharp transition into the USR phase and the notation $*$ is used to indicate value estimated at the pivot scale. 

The relation in equation(\ref{amplitude}) will be required in the next section to comment on another highly crucial problem closely related to the PBHs getting overproduced from the significant enhancement of the primordial fluctuations. This overproduction issue was recently brought to attention and quickly incited interest in the community \cite{Inomata:2023zup,Balaji:2023ehk,Franciolini:2023pbf,Gorji:2023sil,DeLuca:2023tun,Gow:2023zzp,Firouzjahi:2023xke} due to its direct connections with the latest observations of a stochastic gravitational wave background signal from the pulsar timing array collaborations (NANOGrav and EPTA). Thus, in the next section we also make some interesting remarks and show the resolution of the overproduction problem.
 
\section{Numerical outcomes VI: Constraints from Spectral distortion and solving PBH overproduction issue}
\label{s18}

  %%%%%%%%%%%%%%%%%%%%%%%%%%%%%%%%%%%%%%%%%%%%%
\begin{figure*}[htb!]
    	\centering
    \subfigure[]{
      	\includegraphics[width=8.5cm,height=7.5cm]{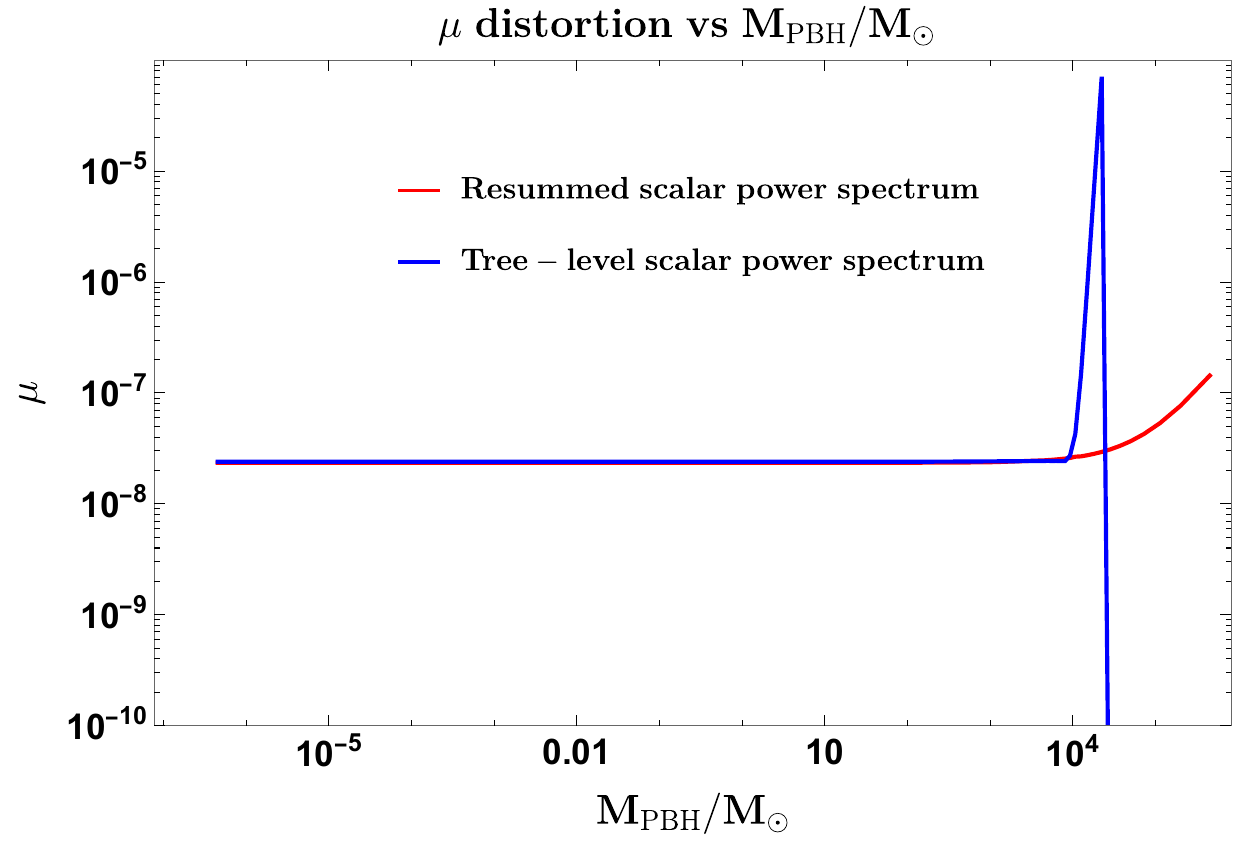}
        \label{mudistort}
    }
    \subfigure[]{
       \includegraphics[width=8.5cm,height=7.5cm]{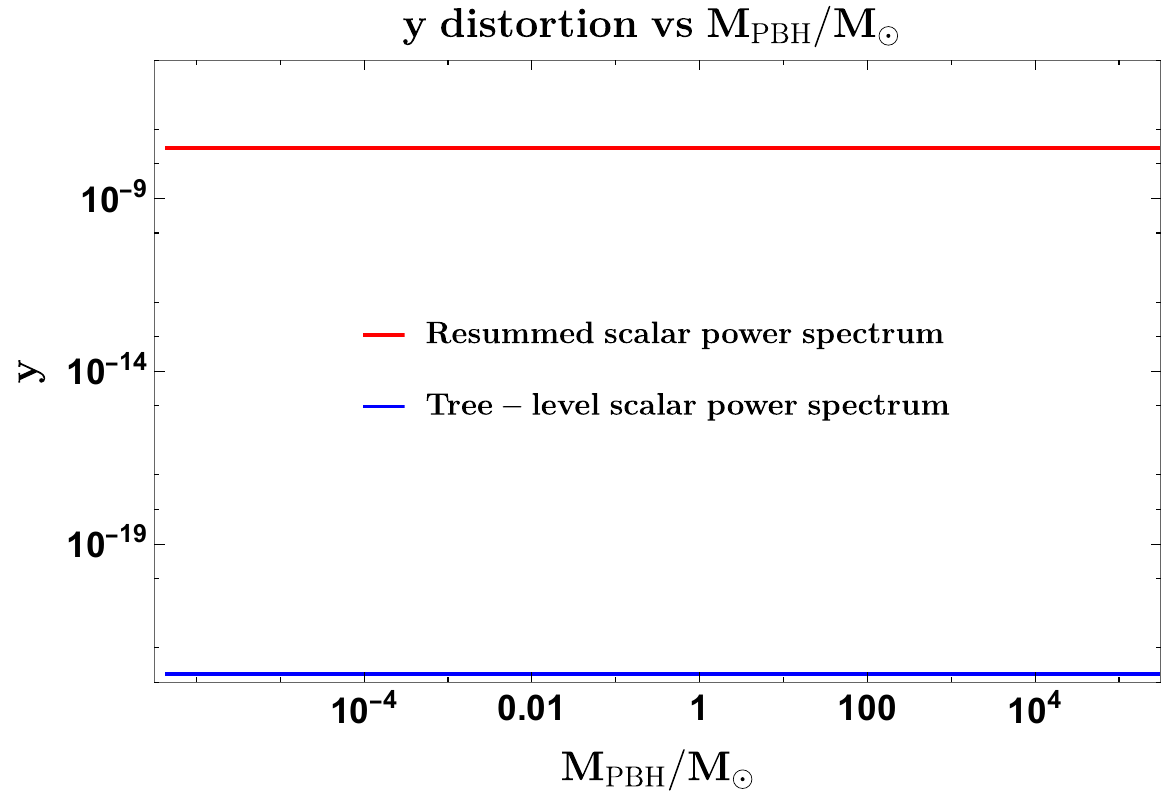}
        \label{ydistort}
    }
    \caption[Optional caption for list of figures]{ Plot shows the different $\mu-$ and $y-$ distortion effects in the left and right as a function of the PBH mass. The red color denotes use of the regularized-renormalized-resummed scalar power spectrum and the blue color highlights use of the tree-level scalar power spectrum. 
    }
    	\label{distortioneffects }
    \end{figure*}
%%%%%%%%%%%%%%%%%%%%%%%%%%%%%%%%%%%%%%%%%%%%%

%%%%%%%%%%%%%%%%%%%%%%%%%%%%%%%%%%%%%%%%%%%%%
\begin{figure*}[ht!]
    	\centering
    {
       \includegraphics[width=17cm,height=11cm]{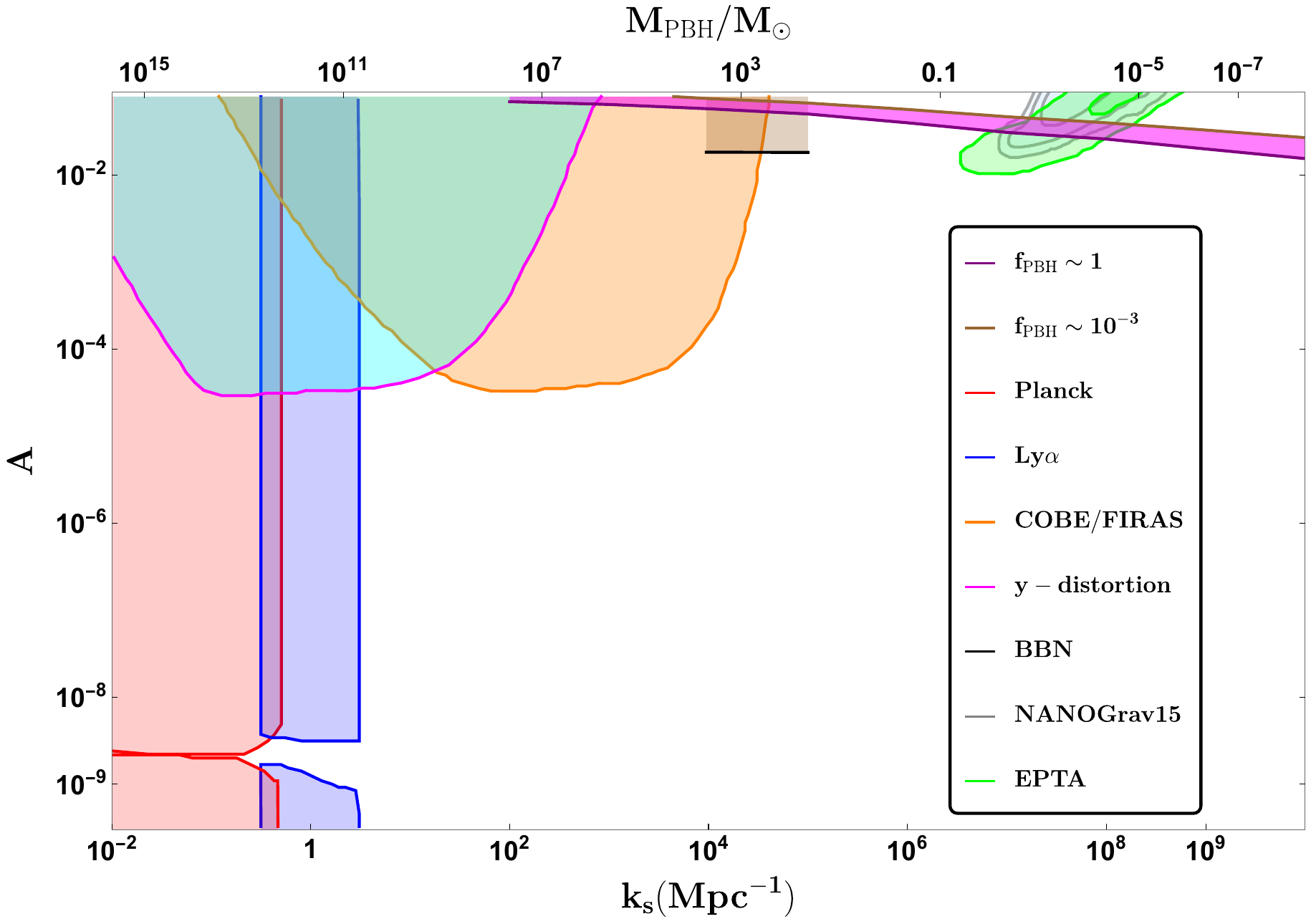}
        \label{distortionbounce}
    } 
    \caption[Optional caption for list of figures]{Scalar power spectrum amplitude ($A$) necessary to achieve significant PBH abundance as a function of the transition wavenumber $k_{s}$ associated with the PBH mass, $M_{\rm PBH}/M_{\odot}$. The background involves constraints on the amplitude coming from the CMB temperature anisotropies (red) \cite{Planck:2018jri} at the large scales, Lyman-$\alpha$ forest (blue) \cite{bird2011minimally}, COBE/FIRAS (orange) and the y-distortion (cyan) effects \cite{Cyr:2023pgw}, BBN (black)\cite{Jeong:2014gna}, the $1\sigma$ and $2\sigma$ contours reported by the pulsar timing array (EPTA) collaboration (in green) \cite{EPTA:2023xxk}, along with the posteriors corresponding to NANOGrav15 data (in gray) \cite{Franciolini:2023pbf}. The magenta-coloured band, bounded by the brown and purple lines, highlights the amplitude region where PBH abundance lies within $f_{\rm PBH}\sim {\cal O}(10^{-3}-1)$. }
\label{amplitudedistortion}
    \end{figure*}
%%%%%%%%%%%%%%%%%%%%%%%%%%%%%%%%%%%%%%%%%%%%%

In this section we present the outcomes for the spectral distortion estimates using the relations mentioned in the previous section and see the results coming from use of both the tree-level and the regularized-renormalized-resummed scalar power spectrum.

In fig.\ref{distortioneffects }, we show how the distortion estimates change with variation in the PBH mass across a wide range of values. Fig.\ref{mudistort} contains the $\mu-$distortion values obtained from using both versions of the scalar power spectrum. For most of the mass range starting with $M_{\rm PBH}\sim 10^{-6}M_{\odot}$ to $M_{\rm PBH}\sim 10^{3}M_{\odot}$, its value remains the same under use of both power spectra. However, near $M_{\rm PBH}\sim 10^{4}M_{\odot}$ curve changes its nature and shoots quickly to give $\mu\sim 7\times 10^{-5}$ and it also drops quickly soon for masses, $M_{\rm PBH}\geq 2\times 10^{4}M_{\odot}.$ Such drastic changes are unobserved when using the RRR version of the spectrum. The turning point remains, however, the same, that is when $M_{\rm PBH}\sim 10^{4}M_{\odot}$, after which the red curve starts to increase gradually but ultimately is still pretty low in its estimates. On the other side in fig.\ref{ydistort}, we notice the estimates remain flat throughout the entire mass range with not showing any interesting behaviour. We show here the results from using the RRR power spectrum version which stays at $y\sim 2\times 10^{-8}$ everywhere while the tree-level estimates remain extremely suppressed, at ${\cal O}(10^{-23})$ as indicated by the blue line at the bottom. 

The distortion effects have found use to help constrain the amplitudes of the scalar power spectrum at large scales; see ref.\cite{Chluba:2012gq,Chluba:2012we} for the earliest steps towards this. In light of the previous discussions on the distortion effects from different power spectra, we now look into how the required amplitude behaves with the transition wavenumber such that one can obtain significant PBH abundance and what constraints the observations at different scales can put on this amplitude. In fig.\ref{amplitudedistortion}, we show this behaviour and the restricted parameter space from different experiments. The orange coloured contour highlights $\mu-$distortion constraints from the COBE/FIRAS observations and the magenta contour for the $y-$distortion constraints. The wavenumber region between, $k\simeq 10^{4}-10^5\;{\rm Mpc^{-1}}$, shows a black shaded region putting constraints from BBN, and at the larger scales or smaller wavenumbers, we have the strict CMB anisotropies constraints. Beyond $k\simeq 10^{7}{\rm Mpc^{-1}}$ lies the recent EPTA $1\sigma$ and $2\sigma$ contours in green along with the gray posteriors corresponding to the NANOGrav15 data, and we notice the magenta-shaded band of significant abundance passing outside the $1\sigma$ region of EPTA, steadily dropping in amplitude for even higher wavenumbers. The same band avoiding overproduction observes less tension coming from the NANOGrav15 data as it crosses through the $1\sigma$ region in gray. It is important to stress this point in the context of resolving overproduction. The magenta band indicates that on going further inside the $1\sigma$ regime in the best-fit region of the latest SGWB signal analysis from NANOGrav15 and EPTA data, the power spectrum amplitude $(A)$ must lie higher inside this regime, leading to PBH overproduction, which is undesirable for the theory. In this way, the current power spectrum does not exhibit overproduction related issues from the scales of interest. We also notice that it requires increasingly larger amplitudes for $M_{\rm PBH}\geq 10 M_{\odot}$ to have $f_{\rm PBH}\rightarrow 1$, and all this region is strictly ruled out by the constraints from distortion effects.

 \section{Comparing with previous works and evading no-go theorem on PBH mass}
 \label{s21}

We now present some comparative discussions on other studies conducted in the same vein of PBH formation. The question of PBH formation in single-field inflation has suffered a long and ongoing debate that has involved different theoretical approaches toward the possibility of their formation and on the conditions for observing cosmologically relevant PBHs. One such condition introduced by \cite{Choudhury:2023vuj,Choudhury:2023jlt,Choudhury:2023rks} in the form of a no-go theorem allows for the total number of e-foldings $N$ elapsed in a setup with USR to be only roughly $\Delta N_{\rm total}\sim {\cal O}(25)$ if one wishes to form $M_{\rm PBH}\sim {\cal O}(M_{\odot})$. An immediate implication of this is an incomplete inflationary setup for which the standard is to achieve at least $\Delta N_{\rm Total}\sim 60$. Following this, in refs.\cite{Choudhury:2023hvf,Choudhury:2023kdb,Choudhury:2023hfm,Choudhury:2023fwk,Choudhury:2024one,Bhattacharya:2023ysp,Choudhury:2023fjs}, some other possibilities are later investigated to evade this no-go theorem. The crux of the analysis done in \cite{Choudhury:2023vuj,Choudhury:2023jlt,Choudhury:2023rks} is the complete regularization, renormalization, and resummation (RRR) procedure, and in the present paper, we tie this rigorous approach with the effective field theory of non-singular bounce. This current setup allows us with the means to complete the minimum e-foldings requirement and make generation of large, near solar-mass PBHs possible in an EFT framework of single-field inflation. Apart from the above mentioned authors, none of the other studies in PBH formation has yet talked about properly incorporating the RRR procedure in a detailed manner. By extending this complete procedure to include the bounce and pre-bounce scenarios we show a robust and improved version of the simple three phase, SRI-USR-SRII, scenario evading the no-go theorem. With our current EFT description, we have shown, after incorporating non-singular bouncing features, that the quadratic UV divergences get entirely removed at all loop orders of calculation, and the less harmful IR divergences can be further resummed owning to their similar structure getting repeated at every loop order. Thus, in the limit of validity of the perturbative arguments, the resummation procedure remains applicable for a large number of momentum modes, giving, in the end, a finite result in the Horizon crossing and super-Horizon regimes with encapsulating the quantum effects.

 \section{Smooth vs Sharp transition in the light of EFT of bounce}
 \label{s19}

The exact nature of the transition into the USR phase is also a vital part of the recent debate concerning PBHs. Multiple studies have backed the use of either a sharp transition or doing away with a smooth transition. In papers utilising the sharp transition, refs.\cite{Kristiano:2022maq,Kristiano:2023scm} claims that in presence of such a transition formation of any PBHs is directly ruled out, while with the use of a smooth transition, others have claimed \cite{Riotto:2023hoz,Riotto:2023gpm,Firouzjahi:2023aum,Firouzjahi:2023ahg,Firouzjahi:2023bkt} that large one-loop corrections remain suppressed enough to allow PBH formation. Few of the present authors and those in \cite{Choudhury:2023vuj,Choudhury:2023jlt,Choudhury:2023rks} have pursued the sharp transition route to arrive at the result of PBH formation, but with a bound on their mass, $M_{\rm PBH}\simeq 100{\rm gm}$, if inflation is to occur successfully. In the scenario of a smooth transition, the authors of refs.\cite{Firouzjahi:2023aum,Firouzjahi:2023ahg,Firouzjahi:2023bkt} have shown that tuning the transition properties can lead to removing the quantum loop corrections that hinder PBH generation. This tuning gets performed via a parameter, $h$, that controls the strength of the Heaviside Theta function, as in $h\Theta(k-k_{t})$ for $t=s,e$ when going into SRI to USR or USR to SRII, respectively. By this method, the authors successfully got away with the quantum loop contributions, but only at the lowest order in the perturbation theory. Using the factor $h$ in this calculation is akin to performing the regularization procedure with a cut-off parameter. However, the most essential condition here remains the complete removal of this cut-off at all orders in the perturbation theory. From our detailed exposition of the RRR program, we have shown explicitly that after successful renormalization, one notices shifting of the logarithmic IR divergences to higher loop orders getting, in turn, softened.
In contrast, the quadratic UV divergences have been removed entirely from the calculations. The similar structure of the contributions from loop diagrams appearing at all orders of perturbation theory allows one to perform resummation further to give a finite result. Thus, it is crucial not just to remove the cut-off from the lowest order calculation since it will get promoted to the higher orders, and if we require a different $h$ at each order to remove the divergences, then the original introduction of $h$ becomes obsolete. An excellent example of this issue is the four Fermi theory, where one encounters adding an infinite number of counter-terms to the theory in order to quantize it. However, since it turned out to be an EFT description, it does not require any quantization. In the present work, we have elaborated further on the choice of the transition and show that the manner of transitioning here is irrelevant to PBH formation while giving details on evading the above-mentioned no-go bound on PBH mass. The procedure of RRR removes the need to consider any specific nature of transition for its conclusion, which we have shown explicitly in this paper.

\section{Conclusion}
\label{s20}

In this study, we have introduced the construction of an effective field theory of bounce and through this addressed the issue of producing large mass PBHs in single-field inflation. After developing the EFT of bounce, our goal has been to study the proper theoretical procedure to handle quantum loop corrections from higher-order interaction terms and their impact on the tree-level power spectrum of the comoving curvature perturbations. We consider a setup of five consecutive phases, starting with the contraction and bounce followed by the sequence of SRI-USR-SRII phases after which inflation is supposed to end. In this setup, we have highlighted the differences due to having a matter or ekpyrotic type contraction and bouncing scenario at each stage in the rest of our work. We continued with deriving the scalar power spectrum contribution from each phase and after that added further discussions were we establish connection between the smooth and sharp nature of transitions encountered within the USR phase. We then turned our focus towards analysing the one-loop contributions to the scalar power spectrum using the Schwinger-Keldysh (in-in) formalism from all phases separately. This concluded our discussion on the regularization and we began to discuss in detail how to successfully remove the quadratic UV divergences and take care of the troubling logarithmic IR divergences present in the underlying theory. We performed the next procedure of renormalization using both the late time and adiabatic/wave-function renormalization techniques where the idea was to completely eradicate the UV/power law divergences in the theory and by arriving at similar results from the two different schemes we illustrated the scheme independence of our approach. Following this with the power spectrum renormalization allowed us to soften the impact of the remaining logarithmic IR divergences which would later undergo through the procedure of Dynamical Renormalization Group resummation technique to finally provide a finite result for the scalar power spectrum that effectively captures the quantum effects from all loop orders. 

%%%%%%%%%%%%%%%%%%%%%%%%%%%%%%%%%%%%%%%%%%%%%
\begin{figure*}[htb!]
    	\centering
    {
       \includegraphics[width=19cm,height=14.5cm]{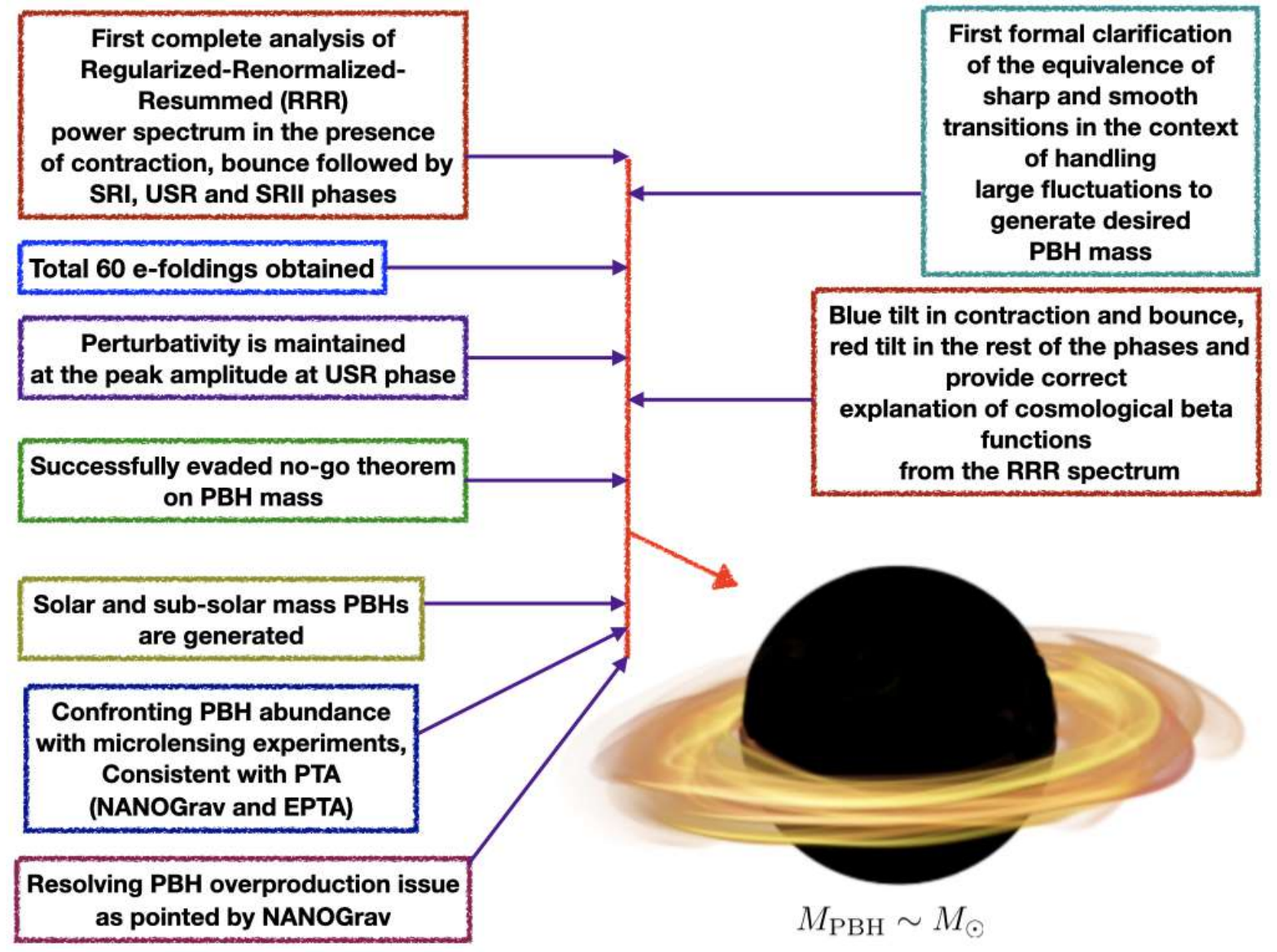}
        \label{conclusionQB}
    } 
    \caption[Optional caption for list of figures]{Schematic diagram describing the main highlighting results from this work.  }
\label{bouncesummary}
    \end{figure*}
%%%%%%%%%%%%%%%%%%%%%%%%%%%%%%%%%%%%%%%%%%%%%

After development of the regularization-renormalization-resummation procedure within the EFT of bounce framework our task was to study the versions of the scalar power spectrum before and after adding the quantum loop corrections. We found that the final version of the scalar power spectrum after resummation showed a smooth behaviour in joining each successive phase, where in the USR the power spectrum rises quickly to achieve the maximum desirable amplitude for PBH formation, ${\cal O}(10^{-2})$ and then sharply decreases in the SRII region to end inflation. In this resummed version of the power spectrum we noticed fulfilment of the minimum criteria on the number of e-folds during inflation, that is $\Delta N_{\rm Total}\sim {\cal O}(60)$. We also examined here the impact of having a changing effective sound speed $c_{s}$ on the power spectrum amplitude and found that lower values of $c_{s}$ lowers the peak amplitude of the final spectrum, with $c_{s}=0.88$ giving us an amplitude of order ${\cal O}(10^{-3})$, and with $c_{s}=1$ we would observe the maximum amplitude of order ${\cal O}(10^{-2})$. This indicates that causality and unitarity is preserved in our overall setup and perturbativity arguments remain intact. 

We next utilise the scalar power spectrum constructed through the above procedures to study PBH formation from our setup. Both the tree-level and resummed version of the power spectrum allowed for large mass PBH formation depending on the wavenumber considered to transition into the USR. We observed that from the tree-level spectrum the PBH masses that gave significant abundance, $f_{\rm PBH}\sim (10^{-3},1)$, were at least an order of magnitude larger relative to the ones produced using the resummed power spectrum. This result was found when looking into PBHs in windows of extremely small, $M_{\rm PBH}\sim {\cal O}(10^{-33}-10^{-27})M_{\odot}$, and the near solar-mass, $M_{\rm PBH}\sim {\cal O}(10^{-6}-10^{-1})M_{\odot}$. The impact of changing sound speed values between $c_{s}=1$ to $c_{s}=0.88$ was also analyzed and we found that the mass estimates giving significant abundance got noticeably smaller in the near solar-mass regime while the extremely small masses did not show much change. This result was obtained regardless of the power spectrum being used and came as a consequence of having reduced peak amplitude with lower values of $c_{s}$. The abundance estimates for the near solar-mass PBHs were further confronted with the recent microlensing constraints showing mass windows where the abundance still qualifies in the significant regime mentioned above. We also take into account another observationally important feature of estimating the spectral distortion effects using both tree-level and resummed power spectra. From this we found that with both spectra the $\mu-$distortion estimates remained constant at $\sim2\times 10^{-8}$, and beyond $M_{\rm PBH}\geq 2\times 10^{4}M_{\odot}$, the results from the resummed power spectrum changed gradually and more smoothly compared with the use of tree-level spectrum. The $y-$distortion magnitude was found to be highly suppressed with the tree-level spectrum and with a constant value of $\sim 5\times 10^{-8}$ with the resummed power spectrum. Lastly, we also addressed the PBH overproduction issue using the resummed power spectrum and observed the corresponding peak amplitude to decrease until the region of significant abundance, $f_{\rm PBH}\sim (10^{-3},1)$ fell below the $1\sigma$ contour reported via the latest EPTA analysis but crossing the $1\sigma$ region coming from analysis of the NANOGrav15 data posteriors. In the future we aim to address the overproduction issue in greater detail when we study the SIGWs produced from this framework and show the resolution of PBH overproduction with proper justification. The schematic diagram in fig.\ref{bouncesummary} summarizes our overall findings from this work.

\subsection*{Acknowledgement}

SC would like to thank The North American Nanohertz Observatory for Gravitational Waves (NANOGrav) collaboration and the National Academy of Sciences (NASI), Prayagraj, India, for being elected as an associate member and the member of the academy respectively. SC would like to acknowledge the inputs from The North American Nanohertz Observatory for Gravitational Waves (NANOGrav) collaboration members, especially Stephen Taylor and Andrea Mitridate for useful comments and discussions, which helped the improvements of the presentation of the article. Also SC would like to especially thank Soumitra SenGupta and Supratik Pal for inviting to IACS, Kolkata, and ISI, Kolkata, during the work. Additionally, SC thanks Supratik Pal and his students for inviting to give an inaugural plenary talk at the discussion meeting titled, {\it Cosmo Mingle}, where part of the work was presented. SC would also like to thank all the members of Quantum Aspects of the Space-Time \& Matter
(QASTM) for elaborative discussions. SP is supported by the INSA Senior scientist position at NISER, Bhubaneswar through the Grant number INSA/SP/SS/2023. Last but not least, we would like to acknowledge our debt to the people
belonging to the various parts of the world for their generous and steady support for research in natural sciences.

\newpage
\section*{Appendix}
\appendix

\section{Regularized scalar power spectrum formulas}
\label{appA}

\textcolor{black}{In this section, we present the resulting expressions for the scalar power spectrum from each phase after implementing the cut-off regularization procedure for the momentum integrals and taking the late-time limit for the temporal integrals in section \ref{s9c}. Specifically, we introduce the UV and IR cut-offs to regulate the various momentum integrals for $k_{\rm UV}< k< k_{\rm IR}$ which are the intervals in between which the phase of interest operates. After the said procedure we are left with this result for the one-loop regularized and unrenormalized power spectrum: }
\bea \Delta^{2}_{\zeta, {\bf EFT}}(p)
   &=&\bigg[\Delta^{2}_{\zeta,{\bf Tree}}(p)\bigg]_{\bf SRI}\times\bigg(1+\underbrace{{\bf W}_{\bf C}+{\bf W}_{\bf B}+{\bf W}_{\bf SRI}+{\bf W}_{\bf USR}+{\bf W}_{\bf SRII}}_{\textbf{Regularized one-loop correction}}\bigg)
\eea
\textcolor{black}{where the different symbols ${\bf W}_{\bf C}$ (for contraction), ${\bf W}_{\bf B}$ (for bounce), ${\bf W}_{\bf SRI}$ (for SRI), ${\bf W}_{\bf USR}$ (for USR) and ${\bf W}_{\bf SRII}$ (for SRII) are defined by the following expressions as a result of the cut-off regularization:}

 \bea  {\bf W}_{\bf C}:&=&\bigg[\Delta^{2}_{\zeta,{\bf Tree}}(p)\bigg]_{\bf SRI}\times\Bigg(1+\frac{2}{15\pi^2}\frac{1}{c^2_{s}k^2_c}\bigg(-\left(1-\frac{1}{c^2_{s}}\right)\epsilon_c+6\frac{\bar{M}^3_1}{ HM^2_{ pl}}-\frac{4}{3}\frac{M^4_3}{H^2M^2_{ pl}}\bigg)\Bigg)\nonumber\\
     &&\quad\quad\quad\quad\quad\quad\quad\quad\quad\quad\quad\quad\quad\quad\quad\quad\quad\quad\quad\quad\quad\quad\quad\quad\quad\quad\quad\quad\quad\quad\times\Bigg({\cal K}_{\bf C}-\frac{4}{3}{\bf I}_{\bf C}\Bigg),\\
     {\bf W}_{\bf B}:&=&\bigg[\Delta^{2}_{\zeta,{\bf Tree}}(p)\bigg]_{\bf SRI}\times\Bigg(1+\frac{2}{15\pi^2}\frac{1}{c^2_{s}k^2_b}\bigg(-\left(1-\frac{1}{c^2_{s}}\right)\epsilon_b+6\frac{\bar{M}^3_1}{ HM^2_{ pl}}-\frac{4}{3}\frac{M^4_3}{H^2M^2_{ pl}}\bigg)\Bigg)\nonumber\\
     &&\quad\quad\quad\quad\quad\quad\quad\quad\quad\quad\quad\quad\quad\quad\quad\quad\quad\quad\quad\quad\quad\quad\quad\quad\quad\quad\quad\quad\quad\quad\times\Bigg({\cal K}_{\bf B}-\frac{4}{3}{\bf I}_{\bf B}\Bigg),\\
     {\bf W}_{\bf SRI}:&=&\bigg[\Delta^{2}_{\zeta,{\bf Tree}}(p)\bigg]_{\bf SRI}\times\Bigg(1+\frac{2}{15\pi^2}\frac{1}{c^2_{s}k^2_*}\bigg(-\left(1-\frac{1}{c^2_{s}}\right)\epsilon_*+6\frac{\bar{M}^3_1}{ HM^2_{ pl}}-\frac{4}{3}\frac{M^4_3}{H^2M^2_{ pl}}\bigg)\Bigg)\nonumber\\
     &&\quad\quad\quad\quad\quad\quad\quad\quad\quad\quad\quad\quad\quad\quad\quad\quad\quad\quad\quad\quad\quad\quad\quad\quad\quad\quad\quad\quad\quad\quad\times\Bigg({\cal K}_{\bf SRI}-\frac{4}{3}{\bf I}_{\bf SRI}\Bigg),\\
     {\bf W}_{\bf USR}:&=&\frac{1}{4}\bigg[\Delta^{2}_{\zeta,{\bf Tree}}(p)\bigg]_{\bf SRI}\times\bigg\{\bigg(\frac{\left(\Delta\eta(\tau_e)\right)^2}{c^8_s} \left(\frac{k_e}{k_s}\right)^{6}- \frac{\left(\Delta\eta(\tau_s)\right)^2}{c^8_s}\bigg){\bf I}_{\bf USR}-{\cal K}_{\bf USR}\bigg\},\eea\bea
    {\bf W}_{\bf SRII}:&=&\bigg[\Delta^{2}_{\zeta,{\bf Tree}}(p)\bigg]_{\bf SRI}\times\Bigg(1+\frac{2}{15\pi^2}\frac{1}{c^2_{s}k^2_*}\bigg(-\left(1-\frac{1}{c^2_{s}}\right)\epsilon_*+6\frac{\bar{M}^3_1}{ HM^2_{ pl}}-\frac{4}{3}\frac{M^4_3}{H^2M^2_{ pl}}\bigg)\Bigg)\nonumber\\
     &&\quad\quad\quad\quad\quad\quad\quad\quad\quad\quad\quad\quad\quad\quad\quad\quad\quad\quad\quad\quad\quad\quad\quad\quad\quad\quad\quad\quad\quad\quad\times\Bigg({\cal K}_{\bf SRII}+{\bf I}_{\bf SRII}\Bigg).\eea

\textcolor{black}{ Here the one-loop integrals ${\bf I}_{\bf C}$ for contraction, ${\bf I}_{\bf B}$ for bounce, ${\bf I}_{\bf SRI}$ for SRI, ${\bf I}_{\bf USR}$ for USR and ${\bf I}_{\bf SRII}$ for SRII phases are explicitly computed in equations (\ref{eqx1}), (\ref{eqx2}), (\ref{eqx3}), (\ref{eqx4}) and (\ref{eqx5}) respectively. The remaining terms namely ${\cal K}_{\bf C}$, ${\cal K}_{\bf B}$, ${\cal K}_{\bf SRI}$, ${\cal K}_{\bf USR}$ and, ${\cal K}_{\bf SRII}$, play the role of counter-terms for each phase and which have to be later fixed by the renormalization procedure as detailed later in section \ref{s10}. }

\section{Connection with Quantum Field Theory of curved space-time}
\label{appB}

Our major objective here is to translate the computation into a more intelligible language so that it may be more easily related to the standard renormalization methods used in quantum field theory. Rather of using many approaches of this type in this section, we will concentrate solely on the way in which divergences at the level of the unrenormalized/bare action are removed by inserting counter-terms. Finally, this will result in the renormalized version of the action, in which any potentially detrimental divergences, particularly quadratic UV divergence, may be completely eradicated and logarithmic IR divergences smoothed when the method is completed successfully.

Let us now begin by putting out the expression for the third-order perturbed bare action for the comoving curvature disturbance:
\bea
\label{actionU}
         S_{\zeta,{\bf B}}^{(3)} &=& M^{2}_{p}\int d\tau\;d^3x\;\bigg [\left({\bf G}_1\right)_{\bf B}\; \zeta^{\prime} {^2}_{\bf B} \zeta_{\bf B} + \left({\bf G}_2\right)_{\bf B}\;(\partial_i \zeta_{\bf B})^2 \zeta_{\bf B}  -  \left({\bf G}_3\right)_{\bf B}\;\zeta^{\prime}_{\bf B} (\partial_i \zeta_{\bf B}) \bigg (\partial_i \partial ^{-2}\bigg(\frac{\epsilon \zeta^{\prime}_{\bf B}}{c_s ^2}\bigg)\bigg) \nonumber \\ 
        && \quad \quad \quad \quad \quad \quad \quad  - \left({\bf G}_4\right)_{\bf B}\;\left(\zeta^{\prime} {^3}_{\bf B}+\zeta^{\prime}_{\bf B}(\partial_i \zeta_{\bf B})^2 \right) +  \left({\bf G}_5\right)_{\bf B}\;\zeta_{\bf B} \bigg(\partial_i \partial_j \partial^{-2}\bigg (\frac{\epsilon \zeta^{\prime}_{\bf B}}{c_s ^2}\bigg)\bigg)^2 + \underbrace{ \left({\bf G}_6\right)_{\bf B}\zeta^{\prime}_{\bf B} \zeta^2_{\bf B}}_{\textbf{Dominant term in USR}}+.....\bigg],\quad\quad\quad
   \eea 
where the bare coupling parameters $\left({\bf G}_i\right)_{\bf B}\forall i=1,2,\cdots,6$ are given by:
\bea  &&\left({\bf G}_1\right)_{\bf B}=\bigg(3(c_s ^2 -1)\epsilon + \epsilon ^2 - \frac{\epsilon ^3}{2}\bigg )a^2,\\
    &&\left({\bf G}_2\right)_{\bf B}=\frac{\epsilon}{c_s ^2}\bigg(\epsilon - 2s +1 -c_s ^2 \bigg)a^2, \\
    &&\left({\bf G}_3\right)_{\bf B}=\frac{2 \epsilon}{c_s ^2}a^2,\\
    &&\left({\bf G}_4\right)_{\bf B}=\frac{\epsilon}{aH}\bigg (1-\frac{1}{c_s ^2}\bigg)a^2,\\
    &&\left({\bf G}_5\right)_{\bf B}=\frac{\epsilon}{2}a^2,\\
    &&\left({\bf G}_6\right)_{\bf B}=\frac{\epsilon}{2c_s ^2}\bigg(\frac{\eta}{c_s ^2}\bigg)^{'}a^2.\eea
    To avoid confusion, the subscript {\bf B} specifically indicates bare contributions.

    The rescaling ansatz of the gauge invariant modes, which is extremely useful in determining the relationship between the renormalized, unrenormalized/bare, and counter-term contributions, can be used to create the renormalized version of the third order action for the comoving scalar curvature perturbation. The following expression describes the renormalized form of the third-order action:
\bea \zeta_{\bf R}=\zeta_{\bf B}-\zeta_{\bf C}=\sqrt{{\bf Z}^{\rm IR}}\times \zeta_{\bf B}\quad\quad\quad{\rm where}\quad\quad\quad {\bf Z}^{\rm IR}:=\left(1+\delta_{{\bf Z}^{\rm IR}}\right).\eea
The superscripts {\bf R}, {\bf B}, and {\bf C} represent renormalized, bare, and counter-term contributions, respectively. It is worth noting that the quantity ${\bf Z}^{\rm IR}$, or more precisely $\delta_{{\bf Z}^{\rm IR}}$, is commonly referred to as the counter-term. In this computation, we must use the physical renormalization condition to explicitly estimate this quantity.

Next, we must utilise the recently released rescaled renormalized form of the gauge invariant scalar curvature perturbation to translate the formula for the second- and third-order unrenormalized/bare action into terms of the renormalized version. One may easily achieve this by performing the following actions:
\begin{itemize}[label={\checkmark}]
    \item Using the previously provided ansatz, the renormalized coupling parameters contained in the third-order perturbed action may be described as follows in terms of the contributions of the bare and counter-terms:
 \bea \left({\bf G}_i\right)_{\bf R}&=&\left({\bf G}_i\right)_{\bf B}-\left({\bf G}_i\right)_{\bf C}={\bf Z}_{{\bf G}_i}\times \left({\bf G}_i\right)_{\bf B}\quad\quad\quad{\rm where}\quad\quad\quad {\bf Z}_{{\bf G}_i}:=\left(1+\delta_{{\bf Z}_{{\bf G}_i}}\right)\forall~i=1,2,\cdots,6,\eea

 \item Here, the separate operator contributions and couplings may be written as:
\bea &&\left({\bf G}_1\right)_{\bf R} \zeta^{\prime} {^2}_{\bf R} \zeta_{\bf R}
     = \left(1+\delta_{{\bf Z}_{{\bf G}_1}}+\frac{3}{2}\delta_{{\bf Z}^{\rm IR}}+\cdots\right)\times \left({\bf G}_1\right)_{\bf B}\zeta^{\prime} {^2}_{\bf B} \zeta_{\bf B},\\
&&\left({\bf G}_2\right)_{\bf R} (\partial_i \zeta_{\bf R})^2 \zeta_{\bf R}
     = \left(1+\delta_{{\bf Z}_{{\bf G}_2}}+\frac{3}{2}\delta_{{\bf Z}^{\rm IR}}+\cdots\right)\times \left({\bf G}_2\right)_{\bf B}(\partial_i \zeta_{\bf B})^2 \zeta_{\bf B},\\
    && \left({\bf G}_3\right)_{\bf R} \zeta^{\prime}_{\bf R} (\partial_i \zeta_{\bf R}) \bigg (\partial_i \partial ^{-2}\bigg(\frac{\epsilon \zeta^{\prime}_{\bf R}}{c_s ^2}\bigg)\bigg)= \left(1+\delta_{{\bf Z}_{{\bf G}_3}}+\frac{3}{2}\delta_{{\bf Z}^{\rm IR}}+\cdots\right)\times \left({\bf G}_3\right)_{\bf B}(\partial_i \zeta_{\bf B}) \bigg (\partial_i \partial ^{-2}\bigg(\frac{\epsilon \zeta^{\prime}_{\bf B}}{c_s ^2}\bigg)\bigg),\\
     &&\left({\bf G}_4\right)_{\bf R} \left(\zeta^{\prime} {^3}_{\bf R}+\zeta^{\prime}_{\bf R}(\partial_i \zeta_{\bf R})^2 \right)= \left(1+\delta_{{\bf Z}_{{\bf G}_4}}+\frac{3}{2}\delta_{{\bf Z}^{\rm IR}}+\cdots\right)\times \left({\bf G}_4\right)_{\bf B}\left(\zeta^{\prime} {^3}_{\bf B}+\zeta^{\prime}_{\bf B}(\partial_i \zeta_{\bf B})^2 \right),\\
&&\left({\bf G}_5\right)_{\bf R} \zeta_{\bf R} \bigg (\partial_i\partial_j \partial ^{-2}\bigg(\frac{\epsilon \zeta^{\prime}_{\bf R}}{c_s ^2}\bigg)\bigg)^2= \left(1+\delta_{{\bf Z}_{{\bf G}_5}}+\frac{3}{2}\delta_{{\bf Z}^{\rm IR}}+\cdots\right)\times \left({\bf G}_5\right)_{\bf B}\zeta_{\bf B}\bigg (\partial_i\partial_j \partial ^{-2}\bigg(\frac{\epsilon \zeta^{\prime}_{\bf B}}{c_s ^2}\bigg)\bigg)^2,\\
     &&\left({\bf G}_6\right)_{\bf R} \zeta^{\prime}_{\bf R} \zeta^2_{\bf R}
     = \left(1+\delta_{{\bf Z}_{{\bf G}_6}}+\frac{3}{2}\delta_{{\bf Z}^{\rm IR}}+\cdots\right)\times \left({\bf G}_6\right)_{\bf B}\zeta^{\prime}_{\bf B} \zeta^2_{\bf B},
   \eea
    where we have utilized the following universal scaling relationships:
    \bea &&{\bf Z}_{{\bf G}_i}\left({\bf Z}^{\rm IR}\right)^{\frac{3}{2}}\approx \left(1+\delta_{{\bf Z}_{{\bf G}_i}}+\frac{3}{2}\delta_{{\bf Z}^{\rm IR}}+\cdots\right)\quad\quad\forall \quad i=1,2,\cdots,6.\eea
    The dotted contributions $\cdots$ in this instance reflect the higher-order components in the corresponding power-series expansion. We have ignored any higher-order minor effects and restricted our analysis to first-order variables. This implies, then, that we have finished the remaining computations to determine the precise contributions of the counter-terms inside the linear domain of the parallel expansion.
\end{itemize}
Thus, by following all the steps mentioned above, one may finally obtain the third-ordered renormalized version of the perturbed actions for the gauge invariant comoving curvature perturbation:
\bea
\label{ractionx}
         S_{\zeta,{\bf R}}^{(3)} &=& M^{2}_{pl}\int d\tau\;d^3x\;\bigg [\left({\bf G}_1\right)_{\bf R}\; \zeta^{\prime} {^2}_{\bf R} \zeta_{\bf R} + \left({\bf G}_2\right)_{\bf R}\;(\partial_i \zeta_{\bf R})^2 \zeta_{\bf R}  -  \left({\bf G}_3\right)_{\bf R}\;\zeta^{\prime}_{\bf R} (\partial_i \zeta_{\bf R}) \bigg (\partial_i \partial ^{-2}\bigg(\frac{\epsilon \zeta^{\prime}_{\bf R}}{c_s ^2}\bigg)\bigg) \nonumber \\ 
        && \quad \quad \quad \quad \quad \quad \quad  - \left({\bf G}_4\right)_{\bf R}\;\left(\zeta^{\prime} {^3}_{\bf R}+\zeta^{\prime}_{\bf R}(\partial_i \zeta_{\bf R})^2 \right) +  \left({\bf G}_5\right)_{\bf R}\;\zeta_{\bf R} \bigg(\partial_i \partial_j \partial^{-2}\bigg (\frac{\epsilon \zeta^{\prime}_{\bf R}}{c_s ^2}\bigg)\bigg)^2 + \underbrace{ \left({\bf G}_6\right)_{\bf R}\zeta^{\prime}_{\bf R} \zeta^2_{\bf R}}_{\textbf{Dominant term in USR}}+.....\bigg]\nonumber\\
        &=&M^{2}_{pl}\int d\tau\;d^3x\;\bigg [\left(1+\delta_{{\bf Z}_{{\bf G}_1}}+\frac{3}{2}\delta_{{\bf Z}^{\rm IR}}+\cdots\right)\left({\bf G}_1\right)_{\bf B}\; \zeta^{\prime} {^2}_{\bf B} \zeta_{\bf B} +\left(1+\delta_{{\bf Z}_{{\bf G}_2}}+\frac{3}{2}\delta_{{\bf Z}^{\rm IR}}+\cdots\right)\left({\bf G}_2\right)_{\bf B}\;(\partial_i \zeta_{\bf B})^2 \zeta_{\bf B}\nonumber \\ 
        && \quad \quad \quad \quad \quad \quad \quad    -  \left(1+\delta_{{\bf Z}_{{\bf G}_3}}+\frac{3}{2}\delta_{{\bf Z}^{\rm IR}}+\cdots\right)\left({\bf G}_3\right)_{\bf B}\;\zeta^{\prime}_{\bf B} (\partial_i \zeta_{\bf B}) \bigg (\partial_i \partial ^{-2}\bigg(\frac{\epsilon \zeta^{\prime}_{\bf B}}{c_s ^2}\bigg)\bigg) \nonumber \\
        && \quad \quad \quad \quad \quad \quad \quad  - \left(1+\delta_{{\bf Z}_{{\bf G}_4}}+\frac{3}{2}\delta_{{\bf Z}^{\rm IR}}+\cdots\right)\left({\bf G}_4\right)_{\bf B}\;\left(\zeta^{\prime} {^3}_{\bf B}+\zeta^{\prime}_{\bf B}(\partial_i \zeta_{\bf B})^2 \right) \nonumber \\
        && \quad \quad \quad \quad \quad \quad \quad  +  \left(1+\delta_{{\bf Z}_{{\bf G}_5}}+\frac{3}{2}\delta_{{\bf Z}^{\rm IR}}+\cdots\right)\left({\bf G}_5\right)_{\bf B}\;\zeta_{\bf B} \bigg(\partial_i \partial_j \partial^{-2}\bigg (\frac{\epsilon \zeta^{\prime}_{\bf B}}{c_s ^2}\bigg)\bigg)^2\nonumber \\ 
        && \quad \quad \quad \quad \quad \quad \quad  + \underbrace{\left(1+\delta_{{\bf Z}_{{\bf G}_6}}+\frac{3}{2}\delta_{{\bf Z}^{\rm IR}}+\cdots\right) \left({\bf G}_6\right)_{\bf B}\zeta^{\prime}_{\bf B} \zeta^2_{\bf B}}_{\textbf{Dominant term in USR}}+.....\bigg],
   \eea 
With the derived form of the renormalized form of the third-order perturbed action for the gauge invariant curvature perturbation in equation (\ref{ractionx}), we now aim to compute the explicit expression for the one-loop corrected power spectrum in the presence of all the counter-terms previously introduced. In the description given by Quantum Field Theory, this is nothing more than the calculation of the renormalized one-loop 1PI effective action corresponding to the two-point amplitude. In the present computer context, we use the in-in formalism that was previously presented to do this. For each of the various stages at the one-loop level, we were able to obtain the following simplified result for the regularized and renormalized version of the power spectrum:
 \bea  \overline{\bigg[\Delta^{2}_{\zeta, {\bf One-loop}}(p)\bigg]}_{\bf Total}&=&\bigg[\Delta^{2}_{\zeta, {\bf One-loop,RR}}(p)\bigg]_{\bf Total}\nonumber\\
 &=&\bigg[\Delta^{2}_{\zeta,{\bf Tree}}(p)\bigg]_{\bf SRI}\times\bigg(\overline{{\bf W}}_{\bf C}+\overline{{\bf W}}_{\bf B}+\overline{{\bf W}}_{\bf SRI}+\overline{{\bf W}}_{\bf USR}+\overline{{\bf W}}_{\bf SRII}\bigg),\eea
 where the symbols $\overline{{\bf W}}_{\bf C}$ (for contraction), $\overline{{\bf W}}_{\bf B}$ (for bounce), $\overline{{\bf W}}_{\bf SRI}$ (for SRI), $\overline{{\bf W}}_{\bf USR}$ (for USR) and $\overline{{\bf W}}_{\bf SRII}$ (for SRII) after renormalization are defined by the following expressions:
 \bea  \overline{{\bf W}}_{\bf C}:&=&\bigg[\Delta^{2}_{\zeta,{\bf Tree}}(p)\bigg]_{\bf SRI}\times\Bigg(1+\frac{2}{15\pi^2}\frac{1}{c^2_{s}k^2_c}\bigg(-\left(1-\frac{1}{c^2_{s}}\right)\epsilon_c+6\frac{\bar{M}^3_1}{ HM^2_{ pl}}-\frac{4}{3}\frac{M^4_3}{H^2M^2_{ pl}}\bigg)\Bigg)\nonumber\\
     &&\quad\quad\quad\quad\quad\quad\quad\quad\quad\quad\quad\quad\quad\quad\quad\quad\times\Bigg(\left(\delta_{{\bf Z}_{{\bf G}_1}}+\delta_{{\bf Z}_{{\bf G}_2}}+\delta_{{\bf Z}_{{\bf G}_3}}+\delta_{{\bf Z}_{{\bf G}_4}}+\delta_{{\bf Z}_{{\bf G}_5}}\right)_{\textbf{C}}-\frac{4}{3}{\bf I}_{\bf C}\Bigg),\\
     \overline{{\bf W}}_{\bf B}:&=&\bigg[\Delta^{2}_{\zeta,{\bf Tree}}(p)\bigg]_{\bf SRI}\times\Bigg(1+\frac{2}{15\pi^2}\frac{1}{c^2_{s}k^2_b}\bigg(-\left(1-\frac{1}{c^2_{s}}\right)\epsilon_b+6\frac{\bar{M}^3_1}{ HM^2_{ pl}}-\frac{4}{3}\frac{M^4_3}{H^2M^2_{ pl}}\bigg)\Bigg)\nonumber\\
     &&\quad\quad\quad\quad\quad\quad\quad\quad\quad\quad\quad\quad\quad\quad\quad\quad\times\Bigg(\left(\delta_{{\bf Z}_{{\bf G}_1}}+\delta_{{\bf Z}_{{\bf G}_2}}+\delta_{{\bf Z}_{{\bf G}_3}}+\delta_{{\bf Z}_{{\bf G}_4}}+\delta_{{\bf Z}_{{\bf G}_5}}\right)_{\textbf{B}}-\frac{4}{3}{\bf I}_{\bf B}\Bigg),\\
     \overline{{\bf W}}_{\bf SRI}:&=&\bigg[\Delta^{2}_{\zeta,{\bf Tree}}(p)\bigg]_{\bf SRI}\times\Bigg(1+\frac{2}{15\pi^2}\frac{1}{c^2_{s}k^2_*}\bigg(-\left(1-\frac{1}{c^2_{s}}\right)\epsilon_*+6\frac{\bar{M}^3_1}{ HM^2_{ pl}}-\frac{4}{3}\frac{M^4_3}{H^2M^2_{ pl}}\bigg)\Bigg)\nonumber\\
     &&\quad\quad\quad\quad\quad\quad\quad\quad\quad\quad\quad\quad\quad\quad\quad\quad\times\Bigg(\left(\delta_{{\bf Z}_{{\bf G}_1}}+\delta_{{\bf Z}_{{\bf G}_2}}+\delta_{{\bf Z}_{{\bf G}_3}}+\delta_{{\bf Z}_{{\bf G}_4}}+\delta_{{\bf Z}_{{\bf G}_5}}\right)_{\bf SRI}-\frac{4}{3}{\bf I}_{\bf SRI}\Bigg),\eea\bea
     \overline{{\bf W}}_{\bf USR}:&=&\frac{1}{4}\bigg[\Delta^{2}_{\zeta,{\bf Tree}}(p)\bigg]_{\bf SRI}\times\bigg\{\bigg(\frac{\left(\Delta\eta(\tau_e)\right)^2}{c^8_s} \left(\frac{k_e}{k_s}\right)^{6}- \frac{\left(\Delta\eta(\tau_s)\right)^2}{c^8_s}\bigg){\bf I}_{\bf USR}-\left(\delta_{{\bf Z}_{{\bf G}_6}}\right)_{\textbf{USR}}\bigg\},\\
    \overline{{\bf W}}_{\bf SRII}:&=&\bigg[\Delta^{2}_{\zeta,{\bf Tree}}(p)\bigg]_{\bf SRI}\times\Bigg(1+\frac{2}{15\pi^2}\frac{1}{c^2_{s}k^2_*}\bigg(-\left(1-\frac{1}{c^2_{s}}\right)\epsilon_*+6\frac{\bar{M}^3_1}{ HM^2_{ pl}}-\frac{4}{3}\frac{M^4_3}{H^2M^2_{ pl}}\bigg)\Bigg)\nonumber\\
     &&\quad\quad\quad\quad\quad\quad\quad\quad\quad\quad\quad\quad\quad\quad\quad\quad\times\Bigg(\left(\delta_{{\bf Z}_{{\bf G}_1}}+\delta_{{\bf Z}_{{\bf G}_2}}+\delta_{{\bf Z}_{{\bf G}_3}}+\delta_{{\bf Z}_{{\bf G}_4}}+\delta_{{\bf Z}_{{\bf G}_5}}\right)_{\bf SRII}+{\bf I}_{\bf SRII}\Bigg).\eea
     Here the one-loop integrals ${\bf I}_{\bf C}$ for contraction, ${\bf I}_{\bf B}$ for bounce, ${\bf I}_{\bf SRI}$ for SRI, ${\bf I}_{\bf USR}$ for USR and ${\bf I}_{\bf SRII}$ for SRII phases are explicitly computed in equations (\ref{eqx1}), (\ref{eqx2}), (\ref{eqx3}), (\ref{eqx4}) and (\ref{eqx5}) respectively.
     Hence, the total regularized and renormalized one-loop corrected power spectrum for scalar modes can be expressed by the following expression:
   \bea \label{one-loopRR} \overline{\Delta^{2}_{\zeta, {\bf EFT}}(p)}&=&\Delta^{2}_{\zeta, {\bf RR}}(p)\nonumber\\
   &=& {\bf Z}^{\rm IR}\times \left[\Delta_{\zeta,\textbf{EFT}}^{2}(p)\right]\nonumber\\
   &=&{\bf Z}^{\rm IR}{\bf Z}^{\rm UV}\times \bigg[\Delta^{2}_{\zeta,\textbf{Tree}}(p)\bigg]_{\textbf{SRI}}\nonumber\\   
   &=&\bigg[\Delta^{2}_{\zeta,{\bf Tree}}(p)\bigg]_{\bf SRI}\times\bigg(1+\underbrace{\overline{{\bf W}}_{\bf C}+\overline{{\bf W}}_{\bf B}+\overline{{\bf W}}_{\bf SRI}+\overline{{\bf W}}_{\bf USR}+\overline{{\bf W}}_{\bf SRII}}_{\textbf{Regularized and Renormalized one-loop correction}}\bigg),\eea
where IR and UV counter-terms are defined as:
\bea && {\bf Z}^{\rm IR}=\left(1+\delta_{{\bf Z}^{\rm IR}}\right),\\
 && {\bf Z}^{\rm UV}=\left(1+\delta_{{\bf Z}^{\rm UV}}\right).\eea
 We next apply the renormalization condition to obtain the expression for the counter-terms as they appear in the above-mentioned derived result. Renormalization Group (RG) flow may be explicitly used to understand the present issue, and it will effectively correct the structure of any counter-terms that emerge in the computation that are sensitive to UV and IR. This is technically possible with the help of the flow equation and matching beta functions written for the corresponding 1PI one-loop corrected renormalized two-point amplitude in the Fourier space, which is essentially a representation of the renormalized scalar power spectrum in this specific context. Here, we may write the Callan–Symanzik equation as follows for this cosmological configuration:
\bea \frac{d}{d\ln \mu}\Bigg\{\bigg[\Delta^{2}_{\zeta,\textbf{Tree}}(k)\bigg]_{\textbf{SRI}}\Bigg\}=\frac{d}{d\ln \mu}\Bigg\{\frac{\left[\overline{\Delta_{\zeta,\textbf{EFT}}^{2}(k)}\right]}{{\bf Z}^{\rm IR}{\bf Z}^{\rm UV}}\Bigg\}=0.\eea
It is now essential to consider that the following formula may be used to further minimize the corresponding total differential operator in this instance:
\bea \frac{d}{d\ln \mu}=\bigg(\frac{\partial}{\partial\ln \mu}+\sum^{6}_{i=1}\beta_{{\bf G}_i}\frac{\partial}{\partial {\bf G}_i}-\gamma_{\bf IR}-\gamma_{\bf UV}\Bigg)\quad{\rm where}\quad\gamma_{\rm IR}:=\left(\frac{\partial\ln {\bf Z}^{\rm IR}}{\partial\ln \mu}\right),
 \gamma_{\bf UV}:=\left(\frac{\partial\ln {\bf Z}^{\rm UV}}{\partial\ln \mu}\right),\quad\eea
 which further implies:
 \bea \bigg(\frac{\partial}{\partial\ln \mu}+\sum^{6}_{i=1}\beta_{{\bf G}_i}\frac{\partial}{\partial {\bf G}_i}-\gamma_{\bf IR}-\gamma_{\bf UV}\Bigg)\overline{\Delta_{\zeta,\textbf{EFT}}^{2}(p)}=0.\eea
where the beta functions are described as:
\bea \beta_{{\bf G}_1}&=&\left(\frac{\partial {\bf G}_1}{\partial\ln \mu}\right)= 2\epsilon a^2\Bigg[(\epsilon-\eta)\bigg(3(c^2_s-1)+2\epsilon-\frac{3\epsilon^2}{2}\bigg)+\bigg(3(c^2_s-1)+\epsilon-\frac{\epsilon^2}{2}+6\frac{ s c^2_s}{H}\bigg)\Bigg]\Bigg[1+\left(\epsilon+\frac{s}{H}\right)\Bigg],\\
\beta_{{\bf G}_2}
&=&\left(\frac{\partial {\bf G}_2}{\partial\ln \mu}\right)=\frac{2\epsilon a^2}{c^2_s}\Bigg[(\epsilon-\eta)\bigg(2\epsilon-2s+1-c^2_s\bigg)+\bigg(\epsilon-2s+1-c^2_s\bigg)\left(1-\frac{s}{H}\right)-\left(\frac{\dot{s}}{H}+\frac{sc^2_s}{H}\right)\Bigg]\Bigg[1+\left(\epsilon+\frac{s}{H}\right)\Bigg],\quad\quad\quad\\
\beta_{{\bf G}_3}
&=&\left(\frac{\partial {\bf G}_3}{\partial\ln \mu}\right)=\frac{4\epsilon a^2}{c^2_s}\Bigg(\epsilon-\eta+1-\frac{s}{H}\Bigg)\Bigg[1+\left(\epsilon+\frac{s}{H}\right)\Bigg],\\
\beta_{{\bf G}_4}
&=&\left(\frac{\partial {\bf G}_4}{\partial\ln \mu}\right)=\frac{2a\epsilon}{H}\Bigg[(\epsilon-\eta)\left(1-\frac{1}{c^2_s}\right)+\frac{s}{H c^2_s}+\frac{1}{2}(1+\epsilon H)\left(1-\frac{1}{c^2_s}\right)\Bigg]\Bigg[1+\left(\epsilon+\frac{s}{H}\right)\Bigg],\\
\beta_{{\bf G}_5}
&=&\left(\frac{\partial {G}_5}{\partial\ln \mu}\right)=\epsilon a^2 \left(\epsilon-\eta+1\right)\Bigg[1+\left(\epsilon+\frac{s}{H}\right)\Bigg],\\
\beta_{{\bf G}_6}
&=&\left(\frac{\partial {\bf G}_6}{\partial\ln \mu}\right)=\frac{a^2\epsilon}{c^2_s}\left(\frac{\eta}{c^2_s}\right)^{'}\Bigg[\bigg(\epsilon-\eta+1-\frac{s}{H}\bigg)+\frac{1}{2}\frac{d}{d\ln \mu}\ln \left(\frac{\eta}{c^2_s}\right)^{'}\Bigg]\Bigg[1+\left(\epsilon+\frac{s}{H}\right)\Bigg].
\eea
To aid in the identification of the IR and UV counter-terms within the current context using the renormalization scale, the flow equations shown below can be calculated:
\begin{itemize}[label={\checkmark}]
\item The first flow equation which describes the renormalized version of the spectral tilt for the scalar modes is given by:
 \bea 
\left[\overline{n_{\zeta,\textbf{EFT}}(p)-1}\right]&=&\frac{d}{d\ln p}\bigg(\ln \left[\overline{\Delta_{\zeta,\textbf{EFT}}^{2}(p)}\right]\bigg)\nonumber\\
&=&{\bf Z}^{\bf IR}\times\Bigg[{\bf Z}^{\bf UV}\bigg(\bigg[n_{\zeta,\textbf{Tree}}(p)\bigg]_{{\bf SRI}}-1\bigg)+\bigg(\frac{d{\bf Z}^{\bf UV}}{d\ln p}\bigg)\bigg(\ln \bigg[\Delta^{2}_{\zeta,\textbf{Tree}}(p)\bigg]_{\textbf{SRI}}\bigg)\Bigg].\eea
\item The second flow equation which describes the renormalized version of the running of the spectral tilt for the scalar modes is given by:
 \bea 
\left[\overline{\alpha_{\zeta,\textbf{EFT}}(p)}\right] &=&\frac{d}{d\ln p}\bigg(\left[\overline{n_{\zeta,\textbf{EFT}}(p)}\right]\bigg)\nonumber\\
&=&{\bf Z}^{\bf IR}\times\Bigg[{\bf Z}^{\bf UV}\bigg(\bigg[\alpha_{\zeta,\textbf{Tree}}(p)\bigg]_{{\bf SRI}}\bigg)+2\bigg(\frac{d{\bf Z}^{\bf UV}}{d\ln p}\bigg)\bigg(\bigg[n_{\zeta,\textbf{Tree}}(p)\bigg]_{{\bf SRI}}-1\bigg)\nonumber\\
&&\quad\quad\quad\quad\quad\quad\quad\quad\quad\quad\quad\quad\quad\quad\quad\quad+\bigg(\frac{d^2{\bf Z}^{\bf UV}}{d\ln p^2}\bigg)\bigg(\ln \bigg[\Delta^{2}_{\zeta,\textbf{Tree}}(p)\bigg]_{\textbf{SRI}}\bigg)\Bigg].\eea
\item  The third flow equation which describes the renormalized version of the running of the running of spectral tilt for the scalar modes is given by:
\bea 
\left[\overline{\beta_{\zeta,\textbf{EFT}}(k)}\right]&=&\frac{d}{d\ln p}\bigg(\left[\overline{\alpha_{\zeta,\textbf{EFT}}(p)}\right]\bigg)\nonumber\\
&=&{\bf Z}^{\bf IR}\times\Bigg[{\bf Z}^{\bf UV}\bigg(\bigg[\beta_{\zeta,\textbf{Tree}}(p)\bigg]_{{\bf SRI}}\bigg)+2\bigg(\frac{d{\bf Z}^{\bf UV}}{d\ln p}\bigg)\bigg(\bigg[\alpha_{\zeta,\textbf{Tree}}(p)\bigg]_{{\bf SRI}}\bigg)\nonumber\\
&&\quad\quad\quad\quad\quad+3\bigg(\frac{d^2{\bf Z}^{\bf UV}}{d\ln p^2}\bigg)\bigg(\bigg[n_{\zeta,\textbf{Tree}}(p)\bigg]_{{\bf SRI}}-1\bigg)+\bigg(\frac{d^3{\bf Z}^{\bf UV}}{d\ln p^3}\bigg)\bigg(\ln \bigg[\Delta^{2}_{\zeta,\textbf{Tree}}(p)\bigg]_{\textbf{SRI}}\bigg)\Bigg].\quad\quad\eea
\end{itemize}
The preceding flow equations show that when the IR and UV counter-term effects are present in each individual expression, there is scale dependency in the two-point amplitude of the primordial scalar modes power spectrum.

The structure of the counter-terms that are sensitive to both UV and IR will be fixed when we apply the renormalization criteria. This will now be done using the known facts at the CMB pivot scale $p_*$, which prohibits us from considering the following crucial restrictions, which are described in terms of renormalization conditions:
\begin{itemize}[label={\checkmark}]
\item The first renormalization condition is that the two-point amplitude of the scalar power spectrum after renormalization at the CMB pivot scale $k_*$ must exactly match the tree-level contribution measured during the initial SR phase. In technical terms, the following might be used to represent this assertion:
\bea \label{recon1}\underline{\textbf{Renormalization Condition I:}}\quad\quad\quad\left[\overline{\Delta_{\zeta,\textbf{EFT}}^{2}(p_*)}\right] &=&\left[\Delta_{\zeta,\textbf{RR}}^{2}(p_*)\right]= \bigg[\Delta^{2}_{\zeta,\textbf{Tree}}(p_*)\bigg]_{\textbf{SRI}}.\eea
This is by no means an ad hoc condition; rather, it is fully warranted from a physical perspective. It is fair to assume that such quantum corrections are absent at the pivot scale, the analogous momentum scale, as no observable quantum effects have yet been discovered at the CMB map level. Any a-causal features outside the horizon (that is, in the super-Hubble scale) have no bearing on the reliable cosmic data. The distribution of hot and cold regions on the maps is how the CMB shows the outcome of causative events.

\item The second requirement for renormalization is that the tree-level contribution computed in the first SR phase at the CMB pivot scale $p_*$ must precisely match the logarithmic derivative of the amplitude of the scalar power spectrum with respect to the momentum scale after renormalization. This logarithmic derivative with respect to the momentum scale, in the context of primordial cosmology, effectively computes the scale dependence of the two-point amplitude, which is alternatively known as the {\it spectral-tilt} or {\it spectral-index} of the scalar power spectrum. Technically, the following might be used to state this claim:
 \bea 
\underline{\textbf{Renormalization Condition II:}}\quad\quad\quad\left[\overline{n_{\zeta,\textbf{EFT}}(p_*)-1}\right]&=&\Bigg[\frac{d}{d\ln p}\bigg(\ln \left[\overline{\Delta_{\zeta,\textbf{EFT}}^{2}(p)}\right]\bigg)\Bigg]_{p=p_*}\nonumber\\
&=&\Bigg[\frac{d}{d\ln p}\bigg(\ln \left[\Delta_{\zeta,\textbf{RR}}^{2}(p)\right]\bigg)\Bigg]_{p=p_*}\nonumber\\
&=&\bigg(\bigg[n_{\zeta,\textbf{Tree}}(p_*)\bigg]_{{\bf SRI}}-1\bigg).\eea
The aforementioned conditions are completely compatible. In fact, it sheds light on the structure of the scalar power spectrum in the current context. Despite the tilt being calculated in the CMB pivot scale $p_*$, the non-zero value allows us to further fix the shape of the primordial scalar modes power spectrum.

\item In our computational setup, the tree-level contribution computed during the first slow-roll phase must perfectly fulfill the third renormalization criterion, which is the second logarithmic derivative of the two-point amplitude of the scalar power spectrum with respect to the momentum scale. This is effectively the execution of the scalar spectral tilt following renormalization. In primordial cosmology, the {\it running of the spectral-tilt} or {\it running of the spectral-index} of the scalar power spectrum is computed as the scale dependence of the two-point amplitude using a logarithmic double derivative with respect to the momentum scale. Alternatively, this assertion can be technically phrased as: 
\bea 
\underline{\textbf{Renormalization Condition III:}}\quad\quad\quad\left[\overline{\alpha_{\zeta,\textbf{EFT}}(p_*)}\right]&=&\Bigg[\frac{d}{d\ln p}\bigg(\left[\overline{n_{\zeta,\textbf{EFT}}(p)}\right]\bigg)\Bigg]_{p=p_*}\nonumber\\
&=&\bigg[\alpha_{\zeta,\textbf{Tree}}(p_*)\bigg]_{{\bf SRI}}.\eea
The scenario mentioned above is completely consistent with the two previously mentioned requirements. In fact, it provides more information on the structure of the scalar power spectrum in this case, rather than having a tilt. Even if the running of the tilt is computed in the CMB pivot scale $k_*$, we can fix the shape of the primordial power spectrum in terms of concavity or convexity in the original form of the underlying effective potential or Hubble parameter of the underlying EFT setup. A saddle point, inflection point, bump/dip, or other features in the Hubble parameter or the mathematical structure of the effective potential as observed in the current EFT computation are some of the additional features that might be present depending on the existence of such running farther.

\item The fourth renormalization condition states that the third logarithmic derivative of the two-point amplitude of the scalar power spectrum with respect to the momentum scale must precisely match the tree-level contribution that was computed in the first slow-roll phase at the CMB pivot scale $k_*$. This is the scalar spectral tilt running following renormalization. This logarithmic triple derivative with respect to the momentum scale effectively calculates the further minute scale dependence of the two-point amplitude in the context of primordial cosmology. This is also known as the {\it running of the running of spectral-tilt} or {\it running of the running of spectral-index} of the scalar power spectrum. Put another way, the technical formulation of this assertion is as follows:
\bea 
\underline{\textbf{Renormalization Condition IV:}}\quad\quad\quad\left[\overline{\beta_{\zeta,\textbf{EFT}}(p_*)}\right]&=&\Bigg[\frac{d}{d\ln p}\bigg(\left[\overline{\alpha_{\zeta,\textbf{EFT}}(p)}\right]\bigg)\Bigg]_{p=p_*}\nonumber\\
&=&\bigg(\bigg[\beta_{\zeta,\textbf{Tree}}(p_*)\bigg]_{{\bf SRI}}\bigg).\eea
The above-described situation is fully compatible with the three previously mentioned conditions. In the current situation, it truly offers more information about the structure of the scalar power spectrum rather than a running of the spectral tilt. As previously noted in the current version of the EFT computation, the non-zero value of the primordial power spectrum allows us to further fix the shape of the spectrum very minutely computed for the scalar modes in terms of concavity or convexity, while the running of the spectral tilt is calculated in the CMB pivot scale $p_*$.

\end{itemize}
The four renormalization criteria stated before directly lead to the following additional constraints on the properties of the UV and IR-sensitive counter-terms. The following is a point-by-point list of these restrictions:
\begin{itemize}
    \item[$\blacksquare$] The constraint condition that conveys the immediate outcome of the first renormalization condition is as follows:
    \bea \underline{\textbf{Consequence I:}}\quad\quad\quad{\bf Z}^{\bf IR}=  \frac{\small[\overline{\Delta_{\zeta,\textbf{EFT}}^{2}(p_{*})}\small]}{\small[\Delta_{\zeta,\textbf{EFT}}^{2}(p_{*})\small]} = \frac{\small[  \Delta_{\zeta,\textbf{Tree}}^{2}(p_{*})\small]_{\textbf{SRI}}}{\small[\Delta_{\zeta,\textbf{EFT}}^{2}(p_{*})\small]}\quad\Longrightarrow\quad {\bf Z}^{\bf IR}(p_*){\bf Z}^{\bf UV}(p_*)=1.\eea
This link will be quite helpful in this situation to explicitly compute the IR counter-term. It is important to keep in mind that the UV counter-term ${\bf Z}^{\bf UV}$ in the current computation is the opposite of the IR counter-term ${\bf Z}^{\bf IR}$, which we receive from this relation. Consequently, without determining the form of the UV counter-term, it is likewise impossible to determine the IR counter-term. 

    \item[$\blacksquare$] The constraint condition that conveys the immediate outcome of the second renormalization condition is as follows:
    \bea \underline{\textbf{Consequence II:}}\quad\quad\quad{\bf Z}^{\bf IR}(p_*){\bf Z}^{\bf UV}(p_*)=1\quad {\rm and}\quad \bigg(\frac{d{\bf Z}^{\rm UV}}{d\ln p}\bigg)_{p=p_*}=0.\eea
    Even while condition II already takes care of the immediate outcome of condition I, a closer look reveals that it is more limited than condition I. This may help in deciphering the mathematical structure of the UV counter-term because of an additional limitation developing in terms of the vanishing logarithmic momentum scale dependant derivative computed at the CMB pivot scale $p_*$. The contribution of such words is shockingly not easily apparent or recognizable on that scale, which totally justifies this finding physically in the current computing setup. Direct determination of the restricted structure of the UV counter-term allows one to promptly compute the contribution of the IR counter-term and fix the structure of the renormalized scalar spectrum derived from this EFT configuration.

    \item[$\blacksquare$] The constraint condition that conveys the immediate outcome of the third renormalization condition is as follows:
    \bea \underline{\textbf{Consequence III:}}\quad\quad\quad{\bf Z}^{\bf IR}(p_*){\bf Z}^{\bf UV}(p_*)=1,\quad \bigg(\frac{d{\bf Z}^{\bf UV}}{d\ln p}\bigg)_{p=p_*}=0\quad{\rm and}\quad\bigg(\frac{d^2{\bf Z}^{bf UV}}{d\ln p^2}\bigg)_{p=p_*}=0.\eea
In contrast to the previous two, it offers stricter constraints that further restrict the scale-dependent behaviour of the UV-counter term at the pivot scale.

\item[$\blacksquare$] The constraint condition that conveys the immediate outcome of the fourth renormalization condition is as follows:
 \bea \underline{\textbf{Consequence IV:}}\quad{\bf Z}^{\bf IR}(p_*){\bf Z}^{\bf UV}(p_*)=1,\bigg(\frac{d{\bf Z}^{\bf UV}}{d\ln p}\bigg)_{p=p_*}=0,\bigg(\frac{d^2{\bf Z}^{\bf UV}}{d\ln p^2}\bigg)_{p=p_*}=0\quad{\rm and}\quad \bigg(\frac{d^3{\bf Z}^{\bf UV}}{d\ln p^3}\bigg)_{p=p_*}=0.\quad\quad\quad\eea
 It offers even stricter constraints than the previous three, significantly reducing the UV-counter term at the pivot scale with respect to size.

\end{itemize}
Further investigation of the problem led us to conclude that careful determination of the UV counter-term ${\bf Z}^{\bf UV}$ is necessary for it to satisfy the previously obtained sets of constraint requirements. This can only be accomplished by precisely removing the quadratic divergence contribution. After this, the shape of the IR counter-term ${\bf Z}^{\bf IR}$ will be automatically determined. However, at this level, determining the exact UV counter-term ${\bf Z}^{\bf UV}$ at the CMB pivot scale with just the aforementioned constraints is quite difficult. The main technical problem is that counter-terms in the five successive phases must each eliminate the quadratic divergence contribution. As simple as it appears, completing the computation we have done so far is technically difficult. The significance of the next three sections lies here: the late time renormalization scheme (or equivalently, the adiabatic renormalization scheme) completely eliminates the power law type and, in particular, the $\nu=3/2$ quadratic UV divergences from the five consecutive phases in each case. Once this is done, we can easily determine the explicit IR counter-term by using the condition ${\bf Z}^{\bf IR}(p_*){\bf Z}^{\bf UV}(p_*)=1$. The adiabatic renormalization scheme for ${\bf Z}^{\bf UV}$, or the quadratic divergence free result obtained from the late time renormalization scheme, have been used in the context of power spectrum renormalization together with this condition.

The conventional method of adding a counter-term at the level of action, which is used in this article, and the underlying relationship between the power spectrum, adiabatic/wave function, and late-time renormalization methods will be covered in the following subsections. This will assist in dispelling any doubt that could exist about the renormalization schemes utilized or the relationships between the many instruments and methods applied in this work. We are sure that this type of explanation will help readers understand the significance of the findings drawn from this effort. Let us examine each of the following points in detail using the explanations that are attached:
\begin{itemize}

\item[\ding{43}]  In the present calculation, the complete elimination of the destructive quadratic UV divergence and other power-law types of divergences are closely correlated with the counter-term contribution of the third-order perturbed action. As a direct consequence of the calculation performed in this section, it is possible to show that the cumulative factor $\delta_{{\bf Z}_{{\bf G}_i}}\forall i=1,2,\cdots,6$ can be used to express the sum of the counter-terms for the six previously described operators. This component is what we refer to as the counter-term contribution that completely eliminates the quadratic UV divergence and other associated power-law types of divergences when applied to late-time and wave function/adiabatic renormalization schemes, as will be covered in the latter portion of this paper. The next subsections will show how specific combinations of counter-terms corresponding to the six operators mentioned in the third-order perturbed action that are addressed may be used to define the cumulative counter-terms in both schemes. With these connections, we will manage to fully eradicate the quadratic UV and other associated power-law types of divergences from the rectified primordial scalar power spectrum formula after just one loop.

\item[\ding{43}] On the other hand, the coarse grain and smoothing of the logarithmic IR divergence are associated with the counter-term contribution of the third-order perturbed action.  The one counter term $\delta_{{\bf Z}^{\bf IR}}$ can be used to clearly show that it is the direct outcome of the computation made utilizing this specific component. The higher even loop diagrams that arise in the perturbative expansion are consistent with the logarithmic IR divergence being smoothed by transferring it to a higher order by adding this factor to the 1PI one-loop corrected two-point amplitude calculation. Specifically, we will discuss the power spectrum renormalization approach in greater depth in the second half of this work.

\end{itemize}

\section{Renormalized scalar power spectrum formulas}
\label{appC}

\subsection{One-loop renormalized contribution} \label{appC1} 

\textcolor{black}{In this section, we outline the proposed simplifications to interpret the cumulative one-loop contribution. We begin by reminding the expression of the desired one-loop renormalized power spectrum introduced back in equation (\ref{renormpowerspectrum}):}
\bea 
\overline{\Delta_{\zeta,\textbf{EFT}}^{2}(k)}&=&\bigg[\Delta_{\zeta,\textbf{Tree}}^{2}(k_{*})\bigg]_{\textbf{SRI}}\times\left(1-\overline{{\bf W}}_{\bf C,*}-\overline{{\bf W}}_{\bf B,*}-\overline{{\bf W}}_{\bf SRI,*}-\overline{{\bf W}}_{\bf USR,*}-\overline{{\bf W}}_{\bf SRII,*}\right)\nonumber\\
&&\quad\quad\quad\quad\quad\quad\quad\quad\quad\quad\quad\quad\quad\quad\quad\quad\times\left(1+\overline{{\bf W}}_{\bf C}+\overline{{\bf W}}_{\bf B}+\overline{{\bf W}}_{\bf SRI}+\overline{{\bf W}}_{\bf USR}+\overline{{\bf W}}_{\bf SRII}\right)\nonumber\\
&=&\bigg[\Delta_{\zeta,\textbf{Tree}}^{2}(k_{*})\bigg]_{\textbf{SRI}}\times \bigg(1+\overline{{\bf X}}_{\bf Loop}\bigg),\eea
\textcolor{black}{where the one-loop contribution is given by:}
\bea \label{totrenormalized}
\overline{{\bf X}}_{\bf Loop}&=&\bigg(\overline{{\bf W}}_{\bf C}-\overline{{\bf W}}_{\bf B}-\overline{{\bf W}}_{\bf SRI}-\overline{{\bf W}}_{\bf USR}-\overline{{\bf W}}_{\bf SRII}-\overline{{\bf W}}_{\bf C,*}-\overline{{\bf W}}_{\bf B,*}-\overline{{\bf W}}_{\bf SRI,*}-\overline{{\bf W}}_{\bf USR,*}-\overline{{\bf W}}_{\bf SRII,*}\nonumber\\
&&-\overline{{\bf W}}_{\bf C}\overline{{\bf W}}_{\bf C,*}-\overline{{\bf W}}_{\bf B}\overline{{\bf W}}_{\bf B,*}-\overline{{\bf W}}_{\bf SRI}\overline{{\bf W}}_{\bf SRI,*}-\overline{{\bf W}}_{\bf USR}\overline{{\bf W}}_{\bf USR,*}-\overline{{\bf W}}_{\bf SRII}\overline{{\bf W}}_{\bf SRII,*}\nonumber\\
&&-\overline{{\bf W}}_{\bf B}\overline{{\bf W}}_{\bf C,*}-\overline{{\bf W}}_{\bf SRI}\overline{{\bf W}}_{\bf C,*}-\overline{{\bf W}}_{\bf USR}\overline{{\bf W}}_{\bf C,*}-\overline{{\bf W}}_{\bf SRII}\overline{{\bf W}}_{\bf C,*}
\nonumber\\
&&-\overline{{\bf W}}_{\bf C}\overline{{\bf W}}_{\bf B,*}-\overline{{\bf W}}_{\bf SRI}\overline{{\bf W}}_{\bf B,*}-\overline{{\bf W}}_{\bf USR}\overline{{\bf W}}_{\bf B,*}-\overline{{\bf W}}_{\bf SRII}\overline{{\bf W}}_{\bf B,*}
\nonumber\\
&&-\overline{{\bf W}}_{\bf C}\overline{{\bf W}}_{\bf SRI,*}-\overline{{\bf W}}_{\bf B}\overline{{\bf W}}_{\bf SRI,*}-\overline{{\bf W}}_{\bf USR}\overline{{\bf W}}_{\bf SRI,*}-\overline{{\bf W}}_{\bf SRII}\overline{{\bf W}}_{\bf SRI,*}\nonumber\\
&&-\overline{{\bf W}}_{\bf C}\overline{{\bf W}}_{\bf USR,*}-\overline{{\bf W}}_{\bf B}\overline{{\bf W}}_{\bf USR,*}-\overline{{\bf W}}_{\bf SRI}\overline{{\bf W}}_{\bf USR,*}-\overline{{\bf W}}_{\bf SRII}\overline{{\bf W}}_{\bf USR,*}
\nonumber\\
&&-\overline{{\bf W}}_{\bf C}\overline{{\bf W}}_{\bf SRII,*}-\overline{{\bf W}}_{\bf B}\overline{{\bf W}}_{\bf SRII,*}-\overline{{\bf W}}_{\bf SRI}\overline{{\bf W}}_{\bf SRII,*}-\overline{{\bf W}}_{\bf USR}\overline{{\bf W}}_{\bf SRII,*}\bigg).\eea
\textcolor{black}{We now introduce the following expressions for the various terms inside the above equation: }
\bea \overline{{\bf W}}_{\bf C}&=&\frac{\bigg[  \Delta_{\zeta,\textbf{Tree}}^{2}(k)\bigg]_{\textbf{SRI}}}{\bigg[  \Delta_{\zeta,\textbf{Tree}}^{2}(k_{*})\bigg]_{\textbf{SRI}}}\times\overline{{\bf W}}_{\bf C,*},\\
\overline{{\bf W}}_{\bf B}&=&\frac{\bigg[  \Delta_{\zeta,\textbf{Tree}}^{2}(k)\bigg]_{\textbf{SRI}}}{\bigg[  \Delta_{\zeta,\textbf{Tree}}^{2}(k_{*})\bigg]_{\textbf{SRI}}}\times\overline{{\bf W}}_{\bf B,*},\\
\overline{{\bf W}}_{\bf SRI}&=&\frac{\bigg[  \Delta_{\zeta,\textbf{Tree}}^{2}(k)\bigg]_{\textbf{SRI}}}{\bigg[  \Delta_{\zeta,\textbf{Tree}}^{2}(k_{*})\bigg]_{\textbf{SRI}}}\times\overline{{\bf W}}_{\bf SRI,*},\\
\overline{{\bf W}}_{\bf USR}&=&\frac{\bigg[  \Delta_{\zeta,\textbf{Tree}}^{2}(k)\bigg]_{\textbf{SRI}}}{\bigg[  \Delta_{\zeta,\textbf{Tree}}^{2}(k_{*})\bigg]_{\textbf{SRI}}}\times\overline{{\bf W}}_{\bf USR,*},\\
 \overline{{\bf W}}_{\bf SRII}&=&\frac{\bigg[  \Delta_{\zeta,\textbf{Tree}}^{2}(k)\bigg]_{\textbf{SRI}}}{\bigg[  \Delta_{\zeta,\textbf{Tree}}^{2}(k_{*})\bigg]_{\textbf{SRI}}}\times\overline{{\bf W}}_{\bf SRII,*}.\eea
Using this new conversion nomenclature the loop contribution can be further recast in the following simplified form:
\bea \overline{{\bf X}}_{\bf Loop}&=&\sum^{8}_{i=1}\overline{{\bf Y}}_i=\bigg(\overline{{\bf Y}}_1+\overline{{\bf Y}}_2+\overline{{\bf Y}}_3+\overline{{\bf Y}}_4+\overline{{\bf Y}}_5+\overline{{\bf Y}}_6+\overline{{\bf Y}}_7+\overline{{\bf Y}}_8\bigg),\eea
where the quantities $\overline{{\bf Y}}_i\;\forall\;i=1,2,\cdots,8$ are defined by the following expressions:
\bea \overline{{\bf Y}}_1:&=&\left\{\frac{\bigg[  \Delta_{\zeta,\textbf{Tree}}^{2}(k)\bigg]_{\textbf{SRI}}}{\bigg[  \Delta_{\zeta,\textbf{Tree}}^{2}(k_{*})\bigg]_{\textbf{SRI}}}-1\right\}\times \overline{{\bf W}}_{\bf C,*}, \eea\bea
\overline{{\bf Y}}_2:&=&\left\{\frac{\bigg[  \Delta_{\zeta,\textbf{Tree}}^{2}(k)\bigg]_{\textbf{SRI}}}{\bigg[  \Delta_{\zeta,\textbf{Tree}}^{2}(k_{*})\bigg]_{\textbf{SRI}}}-1\right\}\times \overline{{\bf W}}_{\bf B,*}, \eea\bea
\overline{{\bf Y}}_3:&=&\left\{\frac{\bigg[  \Delta_{\zeta,\textbf{Tree}}^{2}(k)\bigg]_{\textbf{SRI}}}{\bigg[  \Delta_{\zeta,\textbf{Tree}}^{2}(k_{*})\bigg]_{\textbf{SRI}}}-1\right\}\times \overline{{\bf W}}_{\bf SRI,*}, \eea\bea
\overline{{\bf Y}}_4:&=&\left\{\frac{\bigg[  \Delta_{\zeta,\textbf{Tree}}^{2}(k)\bigg]_{\textbf{SRI}}}{\bigg[  \Delta_{\zeta,\textbf{Tree}}^{2}(k_{*})\bigg]_{\textbf{SRI}}}-1\right\}\times \overline{{\bf W}}_{\bf USR,*}, \eea\bea
\overline{{\bf Y}}_5:&=&\left\{\frac{\bigg[  \Delta_{\zeta,\textbf{Tree}}^{2}(k)\bigg]_{\textbf{SRI}}}{\bigg[  \Delta_{\zeta,\textbf{Tree}}^{2}(k_{*})\bigg]_{\textbf{SRI}}}-1\right\}\times \overline{{\bf W}}_{\bf USR,*}, \eea\bea
\overline{{\bf Y}}_6:&=&\left\{\frac{\bigg[  \Delta_{\zeta,\textbf{Tree}}^{2}(k)\bigg]_{\textbf{SRI}}}{\bigg[  \Delta_{\zeta,\textbf{Tree}}^{2}(k_{*})\bigg]_{\textbf{SRI}}}-1\right\}\times \overline{{\bf W}}_{\bf SRII,*}, \eea\bea  
\overline{{\bf Y}}_7:&=&-\left\{\frac{\bigg[  \Delta_{\zeta,\textbf{Tree}}^{2}(k)\bigg]_{\textbf{SRI}}}{\bigg[  \Delta_{\zeta,\textbf{Tree}}^{2}(k_{*})\bigg]_{\textbf{SRI}}}\right\}\times\bigg(\overline{{\bf W}}^2_{\bf C,*}+\overline{{\bf W}}^2_{\bf B,*}+\overline{{\bf W}}^2_{\bf SRI,*}+\overline{{\bf W}}^2_{\bf USR,*}+\overline{{\bf W}}^2_{\bf SRII,*}\bigg), \eea\bea
\overline{{\bf Y}}_8:&=&-\left\{\frac{\bigg[  \Delta_{\zeta,\textbf{Tree}}^{2}(k)\bigg]_{\textbf{SRI}}}{\bigg[  \Delta_{\zeta,\textbf{Tree}}^{2}(k_{*})\bigg]_{\textbf{SRI}}}\right\}\nonumber\\
&&\quad\quad\quad\quad\quad\quad\quad\quad\quad\times\bigg(\overline{{\bf W}}_{\bf B,*}\overline{{\bf W}}_{\bf C,*}+\overline{{\bf W}}_{\bf SRI,*}\overline{{\bf W}}_{\bf C,*}+\overline{{\bf W}}_{\bf USR,*}\overline{{\bf W}}_{\bf C,*}+\overline{{\bf W}}_{\bf SRII,*}\overline{{\bf W}}_{\bf C,*}
\nonumber\\
&&\quad\quad\quad\quad\quad\quad\quad\quad\quad+\overline{{\bf W}}_{\bf C,*}\overline{{\bf W}}_{\bf B,*}+\overline{{\bf W}}_{\bf SRI,*}\overline{{\bf W}}_{\bf B,*}+\overline{{\bf W}}_{\bf USR,*}\overline{{\bf W}}_{\bf B,*}+\overline{{\bf W}}_{\bf SRII,*}\overline{{\bf W}}_{\bf B,*}
\nonumber\\
&&\quad\quad\quad\quad\quad\quad\quad\quad\quad+\overline{{\bf W}}_{\bf C,*}\overline{{\bf W}}_{\bf SRI,*}+\overline{{\bf W}}_{\bf B,*}\overline{{\bf W}}_{\bf SRI,*}+\overline{{\bf W}}_{\bf USR,*}\overline{{\bf W}}_{\bf SRI,*}+\overline{{\bf W}}_{\bf SRII,*}\overline{{\bf W}}_{\bf SRI,*}\nonumber\\
&&\quad\quad\quad\quad\quad\quad\quad\quad\quad+\overline{{\bf W}}_{\bf C,*}\overline{{\bf W}}_{\bf USR,*}+\overline{{\bf W}}_{\bf B,*}\overline{{\bf W}}_{\bf USR,*}+\overline{{\bf W}}_{\bf SRI,*}\overline{{\bf W}}_{\bf USR,*}+\overline{{\bf W}}_{\bf SRII,*}\overline{{\bf W}}_{\bf USR,*}
\nonumber\\
&&\quad\quad\quad\quad\quad\quad\quad\quad\quad+\overline{{\bf W}}_{\bf C,*}\overline{{\bf W}}_{\bf SRII,*}+\overline{{\bf W}}_{\bf B,*}\overline{{\bf W}}_{\bf SRII,*}+\overline{{\bf W}}_{\bf SRI,*}\overline{{\bf W}}_{\bf SRII,*}+\overline{{\bf W}}_{\bf USR,*}\overline{{\bf W}}_{\bf SRII,*}\bigg)\nonumber \eea\bea
&=&-2\times\left\{\frac{\bigg[  \Delta_{\zeta,\textbf{Tree}}^{2}(k)\bigg]_{\textbf{SRI}}}{\bigg[  \Delta_{\zeta,\textbf{Tree}}^{2}(k_{*})\bigg]_{\textbf{SRI}}}\right\}\nonumber\\
&&\quad\quad\quad\quad\quad\quad\quad\quad\quad\times\bigg(\overline{{\bf W}}_{\bf B,*}\overline{{\bf W}}_{\bf C,*}+\overline{{\bf W}}_{\bf SRI,*}\overline{{\bf W}}_{\bf C,*}+\overline{{\bf W}}_{\bf USR,*}\overline{{\bf W}}_{\bf C,*}+\overline{{\bf W}}_{\bf SRII,*}\overline{{\bf W}}_{\bf C,*}
\nonumber\\
&&\quad\quad\quad\quad\quad\quad\quad\quad\quad+\overline{{\bf W}}_{\bf SRI,*}\overline{{\bf W}}_{\bf B,*}+2\overline{{\bf W}}_{\bf USR,*}\overline{{\bf W}}_{\bf B,*}+\overline{{\bf W}}_{\bf SRII,*}\overline{{\bf W}}_{\bf B,*}
\nonumber\\
&&\quad\quad\quad\quad\quad\quad\quad\quad\quad+\overline{{\bf W}}_{\bf USR,*}\overline{{\bf W}}_{\bf SRI,*}+\overline{{\bf W}}_{\bf SRII,*}\overline{{\bf W}}_{\bf SRI,*}+\overline{{\bf W}}_{\bf SRII,*}\overline{{\bf W}}_{\bf USR,*}\bigg).\eea
Here it is important to note that at the SRI phase the power spectrum at any scale $k$ can be written as:
\bea \bigg[\Delta^{2}_{\zeta,{\bf Tree}}(k)\bigg]_{\bf SRI}
&=&\bigg[\Delta^{2}_{\zeta,{\bf Tree}}(k_*)\bigg]_{\bf SRI}\times \bigg(1+\bigg(\frac{k}{k_s}\bigg)^2\bigg)\quad\quad{\rm where}\quad\quad \bigg[\Delta^{2}_{\zeta,{\bf Tree}}(k_*)\bigg]_{\bf SRI}=\left(\frac{2^{2\nu-3}H^{2}}{8\pi^{2}M^{2}_{ pl}\epsilon c_s}\left|\frac{\Gamma(\nu)}{\Gamma\left(\frac{3}{2}\right)}\right|^2\right)_*.\quad\quad\eea
In the super-horizon scale it is expected to have:
\bea \bigg[\Delta^{2}_{\zeta,{\bf Tree}}(k)\bigg]_{\bf SRI}
&\approx&\bigg[\Delta^{2}_{\zeta,{\bf Tree}}(k_*)\bigg]_{\bf SRI}=\left(\frac{2^{2\nu-3}H^{2}}{8\pi^{2}M^{2}_{ pl}\epsilon c_s}\left|\frac{\Gamma(\nu)}{\Gamma\left(\frac{3}{2}\right)}\right|^2\right)_*.\quad\quad\eea
As an immediate consequence of this, we have the following facts which appear at the one-loop level:
\bea \overline{{\bf Y}}_1\approx 0,~\overline{{\bf Y}}_2\approx 0,~\overline{{\bf  Y}}_3\approx 0,~\overline{{\bf Y}}_4\approx0,~\overline{{\bf Y}}_5\approx0,~\overline{{\bf Y}}_6\approx0.\eea
This implies that the one-loop contribution cancels and the loop correction at the minimal level is quantified by the two-loop amplitudes $\overline{{\bf Y}}_7$ and $\overline{{\bf Y}}_8$ which further fix the following structure of the loop contribution factor:
\bea \overline{{\bf X}}_{\bf Loop}&=&\bigg(\overline{{\bf Y}}_7+\overline{{\bf Y}}_8\bigg)\nonumber\\
&=&-\left\{\frac{\bigg[  \Delta_{\zeta,\textbf{Tree}}^{2}(k)\bigg]_{\textbf{SRI}}}{\bigg[  \Delta_{\zeta,\textbf{Tree}}^{2}(k_{*})\bigg]_{\textbf{SRI}}}\right\}\times\bigg[\bigg(\overline{{\bf W}}^2_{\bf C,*}+\overline{{\bf W}}^2_{\bf B,*}+\overline{{\bf W}}^2_{\bf SRI,*}+\overline{{\bf W}}^2_{\bf USR,*}+\overline{{\bf W}}^2_{\bf SRII,*}\bigg)\nonumber\\
&&\quad\quad\quad\quad\quad+2
\bigg(\overline{{\bf W}}_{\bf B,*}\overline{{\bf W}}_{\bf C,*}+\overline{{\bf W}}_{\bf SRI,*}\overline{{\bf W}}_{\bf C,*}+\overline{{\bf W}}_{\bf USR,*}\overline{{\bf W}}_{\bf C,*}+\overline{{\bf W}}_{\bf SRII,*}\overline{{\bf W}}_{\bf C,*}
\nonumber\\
&&\quad\quad\quad\quad\quad\quad\quad\quad\quad+\overline{{\bf W}}_{\bf SRI,*}\overline{{\bf W}}_{\bf B,*}+2\overline{{\bf W}}_{\bf USR,*}\overline{{\bf W}}_{\bf B,*}+\overline{{\bf W}}_{\bf SRII,*}\overline{{\bf W}}_{\bf B,*}
\nonumber\\
&&\quad\quad\quad\quad\quad\quad\quad\quad\quad+\overline{{\bf W}}_{\bf USR,*}\overline{{\bf W}}_{\bf SRI,*}+\overline{{\bf W}}_{\bf SRII,*}\overline{{\bf W}}_{\bf SRI,*}+\overline{{\bf W}}_{\bf SRII,*}\overline{{\bf W}}_{\bf USR,*}\bigg)\bigg],\eea

\textcolor{black}{We can now use the above simplified loop contribution to calculate the renormalized power spectrum back from equation (\ref{renormpowerspectrum}). }

\subsection{Quantities to describe the cosmological flow equations} \label{appC2} 

\textcolor{black}{In this section we introduce the explicit definitions of the various factors appearing inside the various versions of the cosmological flow equations that later helps to compute and interpret them. The symbols $\alpha_{\bf C}$, $\alpha_{\bf B}$, $\alpha_{\bf SRI}$, $\alpha_{\bf USR}$, $\alpha_{\bf SRII}$, and ${\cal A}_{\bf C}$, ${\cal A}_{\bf B}$, ${\cal A}_{\bf SRI}$, ${\cal A}_{\bf USR}$ and, ${\cal A}_{\bf SRII}$ inside the equations (\ref{renormflowtilt}, \ref{renormflowalpha}, \ref{renormflowbeta}) after renormalization and, also in equations (\ref{resumflowtilt}, \ref{resumflowalpha}, \ref{resumflowbeta}) after resummation, are defined by the following expressions:}
\bea \alpha_{\bf C}&=& \frac{4}{3}
\Bigg(1+\frac{2}{15\pi^2}\frac{1}{c^2_{s}k^2_c}\bigg(-\left(1-\frac{1}{c^2_{s}}\right)\epsilon_c+6\frac{\bar{M}^3_1}{ HM^2_{ pl}}-\frac{4}{3}\frac{M^4_3}{H^2M^2_{ pl}}\bigg)\Bigg)\times \left(\frac{\epsilon_*}{\epsilon_c}\right),\\
\alpha_{\bf B}&=&\frac{4}{3}
\Bigg(1+\frac{2}{15\pi^2}\frac{1}{c^2_{s}k^2_b}\bigg(-\left(1-\frac{1}{c^2_{s}}\right)\epsilon_b+6\frac{\bar{M}^3_1}{ HM^2_{ pl}}-\frac{4}{3}\frac{M^4_3}{H^2M^2_{ pl}}\bigg)\Bigg)\times \left(\frac{\epsilon_*}{\epsilon_b}\right),\\
\alpha_{\bf SRI}&=&\frac{4}{3}
\Bigg(1+\frac{2}{15\pi^2}\frac{1}{c^2_{s}k^2_*}\bigg(-\left(1-\frac{1}{c^2_{s}}\right)\epsilon_*+6\frac{\bar{M}^3_1}{ HM^2_{ pl}}-\frac{4}{3}\frac{M^4_3}{H^2M^2_{ pl}}\bigg)\Bigg),\\
\alpha_{\bf USR}&=&\frac{1}{4}\bigg[\bigg(\frac{\Delta\eta(\tau_{e})}{\tilde{c}^{4}_{s}}\bigg)^{2}\bigg(\frac{k_{e}}{k_{s}}\bigg)^6  - \left(\frac{\Delta\eta(\tau_{s})}{\tilde{c}^{4}_{s}}\right)^{2}\bigg],\\ 
\alpha_{\bf SRII}&=&\Bigg(1+\frac{2}{15\pi^2}\frac{1}{c^2_{s}k^2_*}\bigg(-\left(1-\frac{1}{c^2_{s}}\right)\epsilon_*+6\frac{\bar{M}^3_1}{ HM^2_{ pl}}-\frac{4}{3}\frac{M^4_3}{H^2M^2_{ pl}}\bigg)\Bigg)=\frac{3}{4}\alpha_{\bf SRI},\eea
and
\bea
{\cal A}_{\bf C}&=&\bigg[\frac{1}{\delta_{\bf C}}\bigg\{\left(\frac{k_{b}}{k_*}\right)^{\delta_{\bf C}}-\left(\frac{k_{c}}{k_*}\right)^{\delta_{\bf C}}\bigg\}+\frac{1}{\left(\delta_{\bf C}+2\right)}\bigg\{\left(\frac{k_{b}}{k_*}\right)^{\delta_{\bf C}+2}-\left(\frac{k_{c}}{k_*}\right)^{\delta_{\bf C}+2}\bigg\}\bigg]\nonumber\\
&=&\Bigg[\frac{1}{\delta_{\bf C}}\bigg\{\exp(-\delta_{\bf C}\Delta N_{\bf B})-\exp(-\delta_{\bf C}(\Delta N_{\bf B}+\Delta N_{\bf B}))\bigg\}\nonumber\\
  &&\quad\quad\quad\quad\quad\quad\quad\quad\quad+\frac{1}{\delta_{\bf C}+2}\bigg\{\exp(-(\delta_{\bf C}+2)\Delta N_{\bf B})-\exp(-(\delta_{\bf C}+2)(\Delta N_{\bf B}+\Delta N_{\bf B}))\bigg\}\Bigg],\\
{\cal A}_{\bf B}&=&\frac{1}{(\delta_{\bf B} +2)}\times\bigg[\, _2F_1\left(\frac{\delta_{\bf B}+2}{2},\frac{1}{\epsilon_b-1}-1;\frac{\delta_{\bf B}+4}{2};-1\right)\nonumber\\
 &&\quad\quad\quad\quad\quad\quad\quad\quad\quad\quad\quad\quad\quad\quad\nonumber\\
 &&\quad\quad\quad\quad\quad\quad\quad-\left(\frac{k_{b}}{k_{*}}\right)^{\delta_{\bf B}+2}\, _2F_1\left(\frac{\delta_{\bf B}+2}{2},\frac{1}{\epsilon_b-1}-1;\frac{\delta_{\bf B}+4}{2};-\left(\frac{k_{b}}{k_{*}}\right)^{2}\right)\bigg]\nonumber\\
 &=&\frac{1}{(\delta_{\bf B} +2)}\bigg[\, _2F_1\left(\frac{\delta_{\bf B}+2}{2},\frac{1}{\epsilon_b-1}-1;\frac{\delta_{\bf B}+4}{2};-1\right)\nonumber\\
 &&\quad\quad\quad\quad\quad\quad-\exp(-(\delta_{\bf B}+2)\Delta N_{\bf B})\, _2F_1\left(\frac{\delta_{\bf B}+2}{2},\frac{1}{\epsilon_b-1}-1;\frac{\delta_{\bf B}+4}{2};-\exp(-2\Delta N_{\bf B})\right)\bigg],\\
 {\cal A}_{\bf SRI}&=&\ln\left(\frac{k_s}{k_*}\right)=N_s-N_*=\Delta N_{\bf SRI},\eea\bea
 {\cal A}_{\bf USR}&=&-\ln\left(\frac{k_e}{k_s}\right)=-(N_e-N_s)=-\Delta N_{\bf USR},\\
 {\cal A}_{\bf SRII}&=&-\ln\left(\frac{k_{\rm end}}{k_e}\right)=-(N_{\rm end}-N_e)=-\Delta N_{\bf SRII}.\eea

\textcolor{black}{Also, we mention again that the two quantities $\delta_{\bf C}$ and $\delta_{\bf B}$ inside the above expressions reads as:}
\bea
\delta_{\bf C}&\equiv&\left(3-2\nu+\frac{2\epsilon_c}{\epsilon_c-1}\right),\\
\delta_{\bf B}&\equiv&\left(3-2\nu+\frac{2}{\epsilon_b-1}\right).
\eea

\newpage
\bibliography{Refs}
\bibliographystyle{utphys}

\end{document}